\documentclass[12pt,a4paper,oldfontcommands,hidelinks]{memoir}  
\usepackage{fixltx2e}
\chapterstyle{southall} 

\usepackage{amsmath,amsfonts,amssymb,amsthm,latexsym,mathtools} 
\usepackage{listings}			
\usepackage[mathscr]{euscript} 	
\usepackage{bbm} 			
\usepackage{relsize}			
\usepackage{arydshln}   		
\usepackage{subfig}  			
\usepackage{cancel} 			
\usepackage{extarrows}		
\usepackage[square, comma, numbers, sort&compress]{natbib}
\newcommand{\TODOcite}[1]{[\textcolor{Red}{TODO}]} 	

\ifpdf
	\usepackage[pdftex]{graphicx}	
	\usepackage[dvipsnames,usenames,pdftex]{color}
	\usepackage[pdftex,bookmarks]{hyperref}	
	\hypersetup{
 			pdfstartview=FitH, 		
			bookmarksopen=false, bookmarksnumbered=true,
	 		linkcolor=BrickRed,		
	 		citecolor=OliveGreen, 	
	 		filecolor=MidnightBlue,	
	 		urlcolor=MidnightBlue,	
			pdfauthor = {Simon J. Tyler},
			pdftitle = {Studies of low-energy effective actions %
						in supersymmetric field theories},
			pdfsubject = {Physics},
			pdfkeywords = {supersymmetry, effective action},
			pdfcreator = {LaTeX with hyperref package},
			pdfproducer = {pdflatex}
	}
\else
	\usepackage[dvips]{graphicx}
	\usepackage[dvipsnames,usenames,dvips]{color}
	\usepackage[dvips,bookmarks,pagebackref]{hyperref}
\fi
\maxtocdepth{subsection}		
\setsecnumdepth{subsection}		
\graphicspath{{figures/}} 		
\usepackage{figlatex}

\newcommand{\WZ}{{\mathrm{WZ}}} 	
\newcommand{\AV}{{\mathrm{AV}}} 	
\newcommand{\KS}{{\mathrm{KS}}} 	
\newcommand{\SW}{{\mathrm{SW}}} 	
\newcommand{\BI}{{\mathrm{BI}}} 	
\newcommand{\BG}{{\mathrm{BG}}} 	
\newcommand{\vb}{{\bar{v}}}
\newcommand{\ub}{{\bar{u}}}
\newcommand{\wb}{{\bar{w}}}
\newcommand{\lt}{{\tilde{\lambda}}}

\newcommand{\vt}{{\tilde v}}
\newcommand{\vbt}{{\bar{\tilde v}}}
\newcommand{\SQED}{{\mathrm{SQED}}}				
\newcommand{\dpm}{\bem \d_+ & 0 \\ 0 & \d_- \eem}	
\newcommand{\At}{{\tilde A}}
\newcommand{\Bt}{{\tilde B}}
\newcommand{\Ct}{{\tilde C}}
\newcommand{\JbJm}{\bem\Jb\\\J\eem}
\newcommand{\BoxV}{{\Box_{\rm v}}}
\newcommand{\Mg}{{M^2}}				
\newcommand{\MgT}{{{M^2}^\rmT}}		
\newcommand{\Mv}{{M^2_{v}}}			
\newcommand{\MgS}{{M^2_{v}}}
\newcommand{\Mf}{{\cM^2}}			
\newcommand{\MfT}{{{\cM^2}^\rmT}}	
\newcommand{\MfW}{{\bm\cM^2}}		
\newcommand{\MfWT}{{{\bm\cM^2}^\rmT}}	
\newcommand{\FP}{{\rm FP}}

\newcommand{\dssig}{{\bm \sigma}}
\newcommand{\MSb}{{$\overline{\text{MS}}$} } 	
\newcommand{\mub}{\bar{\mu}}		
\newcommand{\muh}{\hat{\mu}} 		
\newcommand{\zb}{\bar{\zeta}}
\newcommand{\rt}{\tilde{\rho}}
\newcommand{\qbar}{{\bar q}}
\newcommand{\qm}{{q^{-1}}}
\newcommand{\qbarm}{{\bar q^{-1}}}
\newcommand{\hb}{{\bar h}}
\newcommand{\cMg}{\cM_{(g,1)}}
\newcommand{\cMh}{\cM_{(h,q)}}
\newcommand{\Tg}{T^{(g,1)}}
\newcommand{\Th}{T^{(h,q)}}
\newcommand{\Ghl}{{\stackrel{\hookleftarrow}{G}}}
\newcommand{\Ghr}{{\stackrel{\hookrightarrow}{G}}}
\newcommand{\Gb}{{\scG}}
\newcommand{\Gbhl}{{\stackrel{\hookleftarrow}{\scG}}{}}
\newcommand{\Gbhr}{{\stackrel{\hookrightarrow}{\scG}}{}}
\def\BetaMp{\big|(q-\qm)(\f_1-\frac\rmi{\sqrt3}\f_2)\big|^2}
\def\BetaMm{\big|(q-\qm)(\f_1+\frac\rmi{\sqrt3}\f_2)\big|^2}
\def\BetaMa{|(q-\qm)\f_1+(q+\qm)\f_2|^2}
\def\BetaMb{|(q+2\qm)\f_1+q\f_2|^2}
\def\BetaMc{|(q+2\qm)\f_1-q\f_2|^2}
\def\BetaMat{|(q-\qm)\f_1-(q+\qm)\f_2|^2}
\def\BetaMbt{|(2q+\qm)\f_1+\qm\f_2|^2}
\def\BetaMct{|(2q+\qm)\f_1-\qm\f_2|^2}
\def\BetaMaZ{f_q|2\f_2|^2} 
\def\BetaMbZ{f_q|3\f_1+\f_2|^2} 
\def\BetaMcZ{f_q|3\f_1-\f_2|^2}
\def\BetaMaz{|2\f_2|^2} 
\def\BetaMbz{|3\f_1+\f_2|^2} 
 
\newcommand {\cD}{{\cal D}}

\newcommand {\cF}{{\cal F}}
\newcommand {\cG}{{\cal G}}
\newcommand {\cH}{{\cal H}}
\newcommand {\cI}{{\cal I}}

\newcommand {\cK}{{\cal K}}
\newcommand {\cL}{{\cal L}}
\newcommand {\cM}{{\cal M}}
\newcommand {\cN}{{\cal N}}

\newcommand {\cP}{{\cal P}}

\newcommand {\cR}{{\cal R}}

\newcommand {\cW}{{\cal W}}

\newcommand{\bH}{{\bf H}}

\newcommand{\bfs}{\mathbf{s}}
\newcommand{\bfx}{\mathbf{x}}
\newcommand{\dsA}{{\mathbb A}}
\newcommand{\dsa}{{\mathbf A}} 
\newcommand{\dsC}{{\mathbb C}}
\newcommand{\dsE}{{\mathbb E}}
\newcommand{\dsF}{{\mathbb F}}
\newcommand{\dsG}{{\mathbb G}}
\newcommand{\dsH}{{\mathbb H}}
\newcommand{\dsI}{{\mathbb I}}
\newcommand{\dsJ}{{\mathbb J}}

\newcommand{\dsL}{{\mathbb L}}

\newcommand{\dsR}{{\mathbb R}}

\newcommand{\dsZ}{{\mathbb Z}}
\def\ds1{\ensuremath{\mathbbm{1}}}

\def\a{\alpha}
\def\b{\beta}
\def\c{\chi}
\def\d{\delta}
\def\e{\epsilon}
\def\f{\phi}
\def\g{\gamma}
\def\G{\Gamma}
\def\h{\eta}

\renewcommand{\i}{\iota}

\renewcommand{\j}{\psi}
\def\k{\kappa}
\def\l{\lambda}
\def\m{\mu}
\def\n{\nu}
\def\p{\pi}
\def\q{\theta}
\def\r{\rho}
\def\s{\sigma}
\def\t{\tau}

\def\w{\omega}
\def\x{\xi}
\def\z{\zeta}
\def\D{\Delta}
\def\F{\Phi}
\def\J{\Psi}
\def\L{\Lambda}
\def\O{\Omega}

\def\S{\Sigma} 		

\def\X{\Xi}
\def\eps{\varepsilon}
\def\vf{\varphi}

\newcommand{\abar}{{\bar a}}
\newcommand{\bbar}{{\bar b}}

\newcommand{\Bb}{{\bar B}}

\newcommand{\Db}{{\bar D}}

\newcommand{\Nb}{{\bar N}}
\newcommand{\Wb}{{\bar W}}
\newcommand{\cDb}{{\bar\cD}}
\newcommand{\cNb}{{\bar\cN}}
\newcommand{\cWb}{{\bar\cW}}
\newcommand{\Fb}{{\bar\Phi}}
\newcommand{\fb}{{\bar\f}}
\newcommand{\Jb}{{\bar\Psi}}
\newcommand{\jb}{{\bar\psi}}
\newcommand{\vfb}{{\bar\varphi}}
\newcommand{\qb}{{\bar\q}}

\newcommand{\ct}{\tilde{c}}

\newcommand{\st}{\tilde{\sigma}}
\newcommand{\Gt}{\tilde{\Gamma}}

\newcommand{\scF}{\mathscr{F}}
\newcommand{\scG}{\mathscr{G}}

\newcommand{\scR}{\mathscr{R}}

\newcommand{\scJ}{\mathscr{J}}

\newcommand{\rmd}{{\rm d}}
\newcommand{\rme}{{\rm e}}
\newcommand{\rmi}{{\rm i}}
\newcommand{\rmr}{{\rm r}}

\newcommand{\rmT}{{\rm T}} 		

\newcommand{\da}{{\dot{\alpha}}}
\newcommand{\db}{{\dot{\beta}}}
\newcommand{\dg}{{\dot{\gamma}}}
\newcommand{\ab}{{\alpha\beta}}
\newcommand{\ada}{{\alpha\dot\alpha}}
\newcommand{\adb}{{\alpha\dot\beta}}
\newcommand{\daa}{{\dot\alpha\alpha}}

\newcommand{\dadb}{{\dot\alpha\dot\beta}}
\newcommand{\ba}{{\beta\alpha}}
\newcommand{\bda}{{\beta\dot\alpha}}
\newcommand{\bdb}{{\beta\dot\beta}}
\newcommand{\dba}{{\dot\beta\alpha}}
\newcommand{\dbb}{{\dot\beta\beta}}
\newcommand{\dbda}{{\dot\beta\dot\alpha}}

\newcommand{\1}{{\underline{1}}}

\newcommand{\au}{{\underline{a}}}

\newcommand{\iu}{{\underline{i}}}
\newcommand{\iui}{{\underline{i}i}}

\newcommand{\ju}{{\underline{j}}}

\newcommand{\ku}{{\underline{k}}}

\newcommand{\rmI}{{\rm I}}
\newcommand{\rmII}{{\rm II}}

\def\half{\ensuremath{\frac{1}{2}}}		
\def\third{\ensuremath{\frac{1}{3}}}	
\def\quart{\ensuremath{\frac{1}{4}}}

\newcommand{\non}{\nonumber}
\newcommand{\bem}{\begin{pmatrix}}
\newcommand{\eem}{\end{pmatrix}}

\ifpdf	
	 
\else
	
\fi
\newcommand{\bm}[1]{\boldsymbol{#1}}
\def\un{\underline}
\def\Bar#1{\overline{#1}} 
\def\deq{\triangleq}  
\newcommand{\vect}[1]{\ensuremath{\!\vec{\,#1}}} 
\newcommand{\oLRa}[1]{\overset{\text{\tiny$\leftrightarrow$}}{#1}}

\newcommand{\pd}{\partial}
\newcommand{\ddt}[1]{\ensuremath{\frac{\rmd #1}{\rmd t}}}
\newcommand{\acom}[2]{\ensuremath{\left\{ #1 , #2 \right\}}}
\newcommand{\com}[2]{\ensuremath{\left[ #1 , #2 \right]}}
\newcommand{\ket}[1]{\ensuremath{\left| #1 \right>}}

\newcommand{\expt}[1]{\ensuremath{\left< #1 \right>}}

\newcommand{\vac}{0}  
\newcommand{\VEV}[1]{\ensuremath{\left<\vac| #1 |\vac\right>}}
\renewcommand{\Re}{\ensuremath{\text{Re}}}

\def\Tr{{\rm Tr}} 					
\def\tr{{\rm tr}}					
\def\trF{{\rm tr_F}}				
\def\trR{{\rm tr_R}}				
\def\trad{{\rm tr_{\ad}}} 			
\def\ad{{\rm ad}} 					
\newcommand{\AD}[1]{{\rm L}_{#1}} 	
\def\Det{{\rm Det}}
\def\sDet{{\rm sDet}}

\def\diag{{\rm diag}}

\def\sgn{{\rm sgn}}

\def\rnk{{\rm rnk}}
\def\const{{\rm const}}
\def\unren{{\rm unren}} 
\def\wrt{\mbox{with respect to\ }}
\def\ie{{\rm i.e., }}
\def\cc{{\rm c.c.\ }}

\newcommand{\ns}{{\ensuremath{\text{(no sum)}}}}
\newcommand{\Li}{\ensuremath{\mathrm{Li}}}  	
\newcommand{\su}{\ensuremath{{\mathfrak su}}} 	

\newcommand{\so}{\ensuremath{{\mathfrak so}}}
\newcommand{\ord}[1]{\ensuremath{{\rm O}\!\!\,\left(#1\right)}}
	
\newcommand{\lfrac}[2]{\ensuremath \raisebox{-.18em}{$\shortmid$}
 \hspace{-.235em}\genfrac{}{}{}{}{#1\,}{\,#2}\hspace{-.235em}
 \raisebox{.24em}{$\shortmid$}}
\newcommand{\rfrac}[2]{\ensuremath \raisebox{.24em}{$\shortmid$}
 \hspace{-.235em}\genfrac{}{}{}{}{\,#1}{#2\,}\hspace{-.235em}
 \raisebox{-.18em}{$\shortmid$}}
\newcommand{\lrFrac}[3]{\ensuremath \raisebox{-.18em}{$\shortmid$}
 \hspace{-.235em}\genfrac{}{}{}{}{\,#1\,\smash{\raisebox{-.47ex}{$|$}}\,#2\,}
 {\,#3\,}\hspace{-.235em}\raisebox{-.18em}{$\shortmid$}}
\newcommand{\lrfrac}[3]{\ensuremath \raisebox{-.18em}{$\shortmid$}
 \hspace{-.235em}\genfrac{}{}{}{}{\,#1\,
 \smash{\raisebox{-.33ex}{$\shortmid$}}\,#2\,}
 {\,#3\,}\hspace{-.235em}\raisebox{-.18em}{$\shortmid$}}
 
\def\intx{\int\!\!{\rmd}^4x\,}
\def\intdx{\int\!\!{\rmd}^dx\,}
\def\intk{\int\!\!\frac{\rmd^dk}{(2\p)^d}\,}
\newcommand{\intmtm}[1]{\int\!\!\frac{\rmd^d#1}{(2\p)^d}\,}
\def\intrho{\int\!\!{\rmd}^4\rho\,}
\def\intz{\int\!\!{\rmd}^8z\,}

\def\intc{\int\!\!{\rmd}^6z\,}
\def\intac{\int\!\!{\rmd}^6\bar z\,}

\usepackage{navigator}				
\embeddedfile{SelfDualSQEDIntegrals}{SelfDualSQEDIntegrals.cdf}
\embeddedfile{WessZuminoEffPotential}{WessZuminoEffPotential.cdf}

\title{Studies of low-energy effective actions in supersymmetric field theories}
\author{Simon James Tyler}
\date{January 2013}

\begin{document}
\frontmatter 
\phantomsection
\addcontentsline{toc}{chapter}{Title Page}
\maketitle
\thispagestyle{empty} 
\setcounter{page}{-1}
{\centering 
\vfill
\includegraphics[clip=true, trim=0 70 0 0, scale=0.6]%
				{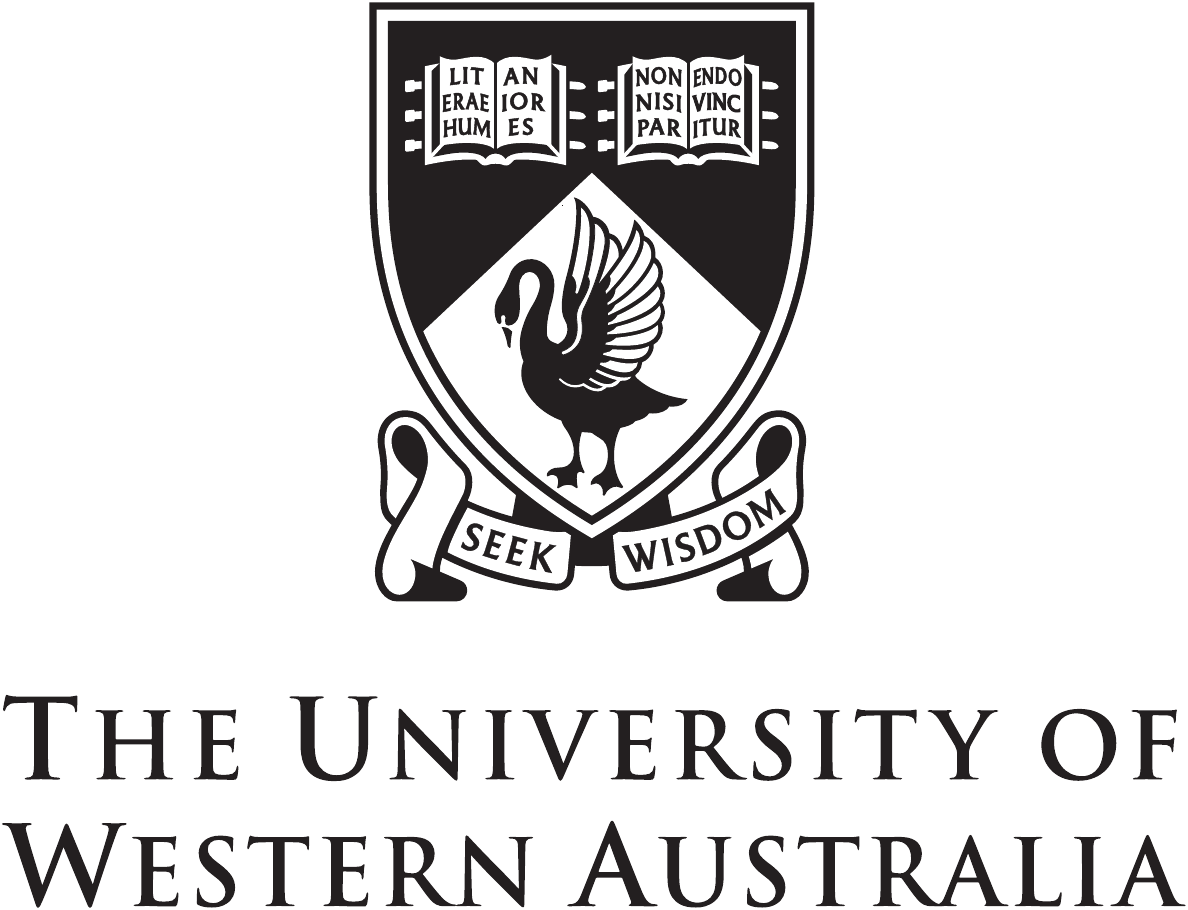}
\vfill
This thesis is presented for the degree of\\
Doctor of Philosophy\\
of The University of Western Australia\\
School of Physics.
\vfill} \newpage
\thispagestyle{empty} 
{\let\thefootnote\relax\footnotetext{Supervisor: 
	Winthrop Professor Sergei M.\ Kuzenko}}
{\let\thefootnote\relax\footnotetext{Co-supervisor: 
	Winthrop Professor Ian N.\ McArthur}}

\cleardoublepage
\phantomsection
\begin{abstract}
\addcontentsline{toc}{chapter}{Abstract}

This thesis examines low-energy effective actions of 
supersymmetric quantum field theories.
These effective actions contain information about the low-energy field content
and dynamics of quantum field theories
and are essential for understanding
their phenomenological and theoretical properties.

In chapters \ref{Ch:Neq1Quant} to \ref{Ch:BetaDef}, 
the covariant background field method is used to 
investigate quantum corrections to sectors of the low-energy effective actions
for a variety of supersymmetric field theories at one- and two-loops.
We start by looking at the background field quantisation of a general 
$\cN=1$ super-Yang-Mills theory and rederiving the well known one-loop
finiteness conditions.
This is followed by a reexamination of the effective potential of 
the simplest supersymmetric quantum field theory, 
the Wess-Zumino model.
Next, the two-loop Euler-Heisenberg effective action is constructed for 
$\cN=1$ supersymmetric quantum electrodynamics.
This is a natural object to study in the progression of 
such two-loop Euler-Heisenberg calculations 
and is only the second such result using superfields.
The theory is renormalised and the self-dual limit of the renormalised 
effective action is given explicitly in terms of digamma functions. 
The final quantum effective action studied is the two-loop K\"ahler potential 
for $\b$-deformed $\cN=4$ super-Yang-Mills theory.
This sector of the effective action is purely a product of the deformation
and its finiteness is demonstrated in a general background before 
specialising to give explicit results for two special cases.

Chapter \ref{ch:Goldstino} studies spontaneously broken supersymmetry and, 
in particular, the pure Goldstino action.
This is a universal sector of the low-energy effective action 
of any theory with spontaneously broken supersymmetry.
A very general approach to constructing explicit field redefinitions is
used to relate all known models of the Goldstino and to study both their
nonlinear supersymmetries and their previously unnoticed trivial symmetries.
This approach is also used to construct the most general pure Goldstino action
and to examine its nonlinear supersymmetry transformations.
Finally, a new embedding of the Goldstino into a complex linear superfield is
presented. Its interactions to matter and gravity are examined and
compared to existing Goldstino superfield constructions.
\end{abstract}

\cleardoublepage
This thesis is based in part on the following five published papers:
\vspace{5mm}

\begin{itemize}
\item[\cite{Kuzenko2007a}]
S.~M. Kuzenko and S.~J. Tyler, ``{Supersymmetric Euler-Heisenberg \linebreak 
  effective action: Two-loop results},''
  \href{http://dx.doi.org/10.1088/1126-6708/2007/05/081}{{\em JHEP} {\bfseries
  2007} 05, (2007) 081}, \linebreak
  \href{http://arxiv.org/abs/hep-th/0703269v2}{{\ttfamily
  [arXiv:hep-th/0703269]}}.
\item[\cite{Tyler2008}]
S.~J. Tyler, ``{Two loop Kahler potential in beta-deformed N=4 SYM theory},''
  \href{http://dx.doi.org/10.1088/1126-6708/2008/07/061}{{\em JHEP} {\bfseries
  2008} 07, (2008) 24}, \href{http://arxiv.org/abs/0805.3574}{{\ttfamily
  [arXiv:0805.3574]}}.
\item[\cite{KuzenkoTyler2010}]
S.~M. Kuzenko and S.~J. Tyler, ``{Relating the Komargodski-Seiberg and
  Akulov-Volkov actions: Exact nonlinear field redefinition},''
  \href{http://dx.doi.org/10.1016/j.physletb.2011.03.020}{{\em Phys. Lett. B}
  {\bfseries 689} 4, (2010) 319--322},
  \href{http://arxiv.org/abs/1009.3298}{{\ttfamily [arXiv:1009.3298]}}.
\item[\cite{KuzenkoTyler2011}]
S.~M. Kuzenko and S.~J. Tyler, ``{On the Goldstino actions and their
  symmetries},'' \href{http://dx.doi.org/10.1007/JHEP05(2011)055}{{\em JHEP}
  {\bfseries 2011} 5, (2011) 32},
  \href{http://arxiv.org/abs/1102.3043}{{\ttfamily [arXiv:1102.3043]}}.
\item[\cite{KuzenkoTyler2011a}]
S.~M. Kuzenko and S.~J. Tyler, ``{Complex linear superfield as a model for
  Goldstino},'' \href{http://dx.doi.org/10.1007/JHEP04(2011)057}{{\em JHEP}
  {\bfseries 2011} 4, (2011) 8},
  \href{http://arxiv.org/abs/1102.3042}{{\ttfamily [arXiv:1102.3042]}}.
\end{itemize}

\vspace{5mm}
Permission has been granted to include this work:

\vspace{1cm}
Simon J. Tyler
\vspace{0.5cm}


\vspace{0.8cm}
Sergei M. Kuzenko
\vspace{0.5cm}


\cleardoublepage\tableofcontents

\chapter*{Acknowledgements}
\addcontentsline{toc}{chapter}{Acknowledgements}
First and foremost, I wish to thank my supervisor W/Prof.\ Sergei Kuzenko. 
I thank him for his guidance and deep knowledge which have helped me learn 
supersymmetry and quantum field theory. 
As well as for his suggestions of research problems, his collaboration 
and his patience when my calculations didn't proceed 
in the manner originally envisioned.

I'm indebted to my co-supervisor W/Prof.\ Ian McArthur, 
for always finding time to listen to me 
and providing sage advice on how to proceed.
It was Ian who introduced me to the principle of conservation of difficulty, 
which is almost unavoidable in theoretical physics research.

Computer aided calculation is sometimes 
a way to bypass the principle of conservation of difficulty
or at least provide a useful sanity check.
It is in this area that I'm thankful for the friendship and guidance
of A/Prof.~Paul Abbott and his expertise in Mathematica.
I also wish to thank Paul for the great teaching opportunities in his 
computational physics courses. 

Thanks to Dr.\ Gabriele Tartaglino-Mazzucchelli, for being such a great friend.
Gab, you've always listened and made me feel better about myself.
Thanks for the great music, pizza and games of pool!

To the supersymmetric PhDs who came before me, 
Dr.\ Darren Grasso and Dr.\ Shane McCarthy. 
Thank you for your advice and friendship.

Fellow supersymmetric students, Tim Gargett, Joseph Novak and Ryan Mickler.
Thanks for the good discussions and support. 
Ryan, one day we'll finish that research project!

To my nonsupersymmetric UWA physics friends,
John Adams, Karen Livesey, Peter Metaxas, Kim Kennewell, Rhet Magaraggia, 
Alison Fowler, Stuart Napier, Nils Ross, Matt Ambrose, Zoe Budrikis, 
James Anstie, Milan Maric, Mike Eilon and many others.
Thanks for making me feel welcome at UWA. 
Especially to Karen and Pete for being perfect office mates; 
to Pete for the great squash games;
and to John for letting me crash at your place 
for that last couple of months in Perth.


I am very grateful to have a loving and supportive family.
I have never doubted your support in anything that I do.
Mom, now that I've finished, 
I will finally give you a page of my scribbles to put on your wall.

Finally, a big thank you to my soon-to-be wife Fiona. 
For being interested in what I'm working on, 
even though I never describe it the same way twice.
For putting up with me being locked in the study for so many evenings.
For being such a great person to know and be around.

\mainmatter 
\chapter{Introduction}\label{Ch:Intro}

Throughout the 20th century, symmetry groups and their representations 
have played an ever increasing r\^ole in physics.%
\footnote{For a comprehensive review of the history, see \cite{Bonolis2005}.}
This started with the use of the Poincar\'e group in special relativity,%
\footnote{In the past, this group was often referred to as the 
inhomogeneous Lorentz group.} 
then continued with group representations underlying much of the structure 
of the spectra calculated with quantum mechanics.  
The unification of relativity and quantum mechanics gave birth to 
quantum field theory (QFT) 
and the use of QFT in particle physics culminated
in the 1970s with the establishment of the standard model of particle physics. 

The standard model describes all directly observed matter and 
interactions apart from gravity and at its heart
is the gauge group $SU(3) \times SU(2) \times U(1)$
which fixes the structure of the fundamental forces. 
The standard model 
has been so successful at describing observable particle physics that, 
40 years later, no significant deviations from it have been recorded.
At CERN in Geneva, the Large Hadron Collider (LHC), 
the world's largest and most powerful particle accelerator, 
is being slowly brought up to full energy.
Its primary goal is the observation of the last unobserved fundamental particle
needed in the standard model, the Higgs particle. 
Recently, the two main experiments on the LHC announced the observation of 
a Higgs-like particle \cite{ATLASHiggs2012,CMSHiggs2012}, 
although futher observations are needed to understand its properties in detail.

There are reasons to expect that the standard model can not be 
a complete theory of particle physics and the LHC experiments
also hope to observe other signals or particles that reveal 
the so called ``beyond the standard model physics''.%
\footnote{For some quick reviews in increasing order of required knowledge, 
see \cite{Womersley2005, Ellis2007Nature, Lykken2010}.}
%
%
Most ``beyond the standard model'' models build upon a framework that somehow
increases the amount of fundamental symmetry in the theory.
They either increase the existing types of symmetry, such as 
flavour symmetry, 
gauge symmetry 
or the spacetime symmetry and dimension. 
Supersymmetry is a popular framework for quantum field theories that 
adds a completely new type of symmetry that
relates fermionic (matter-like) and bosonic (force-like) particles.
Supersymmetric quantum field theory is the central focus of this thesis.

\section{Supersymmetry}
Supersymmetry in four dimensions was independently discovered 
three times in the early 1970s
\cite{Gol'fand1971},
\cite{Volkov1972,Volkov1973},
\cite{Wess1974a,Wess1974b} %
and each time with different motivations.
Gol'fand and Likhtman 
wanted to examine the field theory consequences of their novel extension 
of the Poincar\'e algebra \cite{Likhtman2001}. 
Their stated aim in \cite{Gol'fand1971} was to try to understand why only 
a subset of all possible interactions occur in nature.
Volkov and Alkulov \cite{Volkov1972,Volkov1973}
tried to explain the then apparent masslessness of neutrinos 
as a consequence of them being Goldstone particles for a spontaneously broken 
fermionic symmetry -- supersymmetry.
Finally, Wess and Zumino \cite{Wess1974a,Wess1974b} 
knew of world-sheet (two-dimensional) supersymmetry 
\cite{Raymond1971,NeveuSchwarz1971,neveu1971reformulation}
that had recently been discovered in string theory
and looked for a realisation of it in four dimensional field theory. 
It was at this point that supersymmetry became a very active field research.
For more details of the history of supersymmetry see, e.g.,
\cite{SupersymmetryDecade,ManyFacesOfSusy,KaneShifman2000,%
	  Volkov1994,Likhtman2001,Zumino2006}.


Currently, there is no firm evidence for supersymmetry as 
a fundamental symmetry of nature.
However, supersymmetry has many aspects that are attractive to 
both phenomenological and theoretical quantum field theorists.%
\footnote{Supersymmetry also occurs in many quantum mechanical systems. 
It gives a basis and language for the ``factorisation method'' 
used by Schr\"odinger \cite{Schrodinger1940} to solve 
the quantum harmonic oscillator and Coulomb problems  
and later generalised by Infeld and Hull \cite{Infeld1951} to 
classify the solvable 1D potentials.
Supersymmetry also arises as an approximate symmetry 
in some condensed matter and nuclear physics systems.
See \cite{Cooper1995} 
for more examples and references.
} 
On the phenomenological side, supersymmetry is the most popular ingredient in
``beyond the standard model'' candidates, providing possible solutions
to the hierarchy problem, dark matter, gauge unification, etc\dots 
For more details, see, e.g., 
\cite{KaneSUSYWhatWhyWhen, Murayama2000, BaerTata2006}.
On the theory side, there are numerous reasons to study supersymmetric theories.
\begin{itemize}
\item
Supersymmetry is one of the few ways to successfully
unify internal symmetries with Poincar\'e symmetry, 
something that is highly constrained due to the Coleman-Mandula theorem
\cite{Coleman1967,Haag2005, WeinbergQFT3}. 
\item
The gauge field needed to make the supersymmetry transformations local
is the gravitino and its superpartner the graviton, so, local supersymmetry
implies a supersymmetric version of gravity, known as supergravity. 
This was realised early in the history of supersymmetry 
\cite{Freedman1976,Volkov1994}.
\item
Due to the matching of bosonic and fermionic degrees of freedom 
and the constrained couplings,
supersymmetric quantum field theories have better ultraviolet (UV) behaviour 
than their non-supersymmetric counterparts. 
In fact, there are various non-renormalisation theorems that prove that
some sectors of supersymmetric theories 
are completely free from quantum corrections, see, e.g.,
\cite{Grisaru1979,Seiberg1993}.
\item
As the amount of supersymmetry increases, 
the theory becomes better behaved until you reach
maximally supersymmetric ($\cN=4$) super-Yang-Mills theory, 
which was the first known UV finite, 
four dimensional quantum field theory 
\cite{Sohnius1981,howe1984miraculous,brink1983ultra,Buchbinder1997}. 
In maximally supersymmetric ($\cN=8$) supergravity, 
there is some evidence that it is UV finite 
and it is currently an active area of research, 
e.g., \cite{Kallosh2008,Kallosh2010,BCDJH2009}.
\item
The high amount of symmetry also allows the calculation of many exact results, 
such as the Seiberg-Witten potential for $\cN=2$ gauge theories
\cite{SeibergWitten1994,SeibergWitten1994a},
the Dijkgraaf-Vafa effective glueball superpotential
\cite{Veneziano1982,DijkgraafVafa2002,DijkgraafVafa2002a,DijkgraafVafa2002b},
the Novikov-Shifman-Vainshtein-Zakharov exact beta function
\cite{NSVZ1983,NSVZ1986,Shifman1999Exact},
and so on.
The NSVZ exact beta function was key to the original construction 
\cite{Leigh1995} of the marginal deformations 
of $\cN=1$ superconformal theories, 
including the $\b$-deformed theory studied in chapter \ref{Ch:BetaDef}.
\item
Supersymmetric gauge theories provide a concrete example of the strong-weak
duality that generalises the classical electromagnetic duality 
\cite{MontonenOlive1977,Seiberg1995}. 
These types of dualities and their string theory analogues
have become a powerful tool in studying strongly coupled systems.
\item
Just as the target space of a (nonsupersymmetric) nonlinear sigma model (NLSM) 
is a Riemannian manifold, 
the target space of a $\cN=1$ NLSM is a K\"ahler manifolds
and the target space of a $\cN=2$ NLSM is hyper-K\"ahler.
The target spaces of $\cN=1$ and $\cN=2$ NLSMs coupled to supergravity are
Hodge-K\"ahler and quaternionic-K\"ahler manifolds respectively.
This link between supersymmetric field theory and geometry has been a powerful
tool in both understanding the structure of the field theories and in new 
geometric constructions.
See, e.g., the reviews \cite{Bagger1984a,Kuzenko2010a}.
\item
Finally, supersymmetry is intimately connected with string theory,
which is a very large area of research that is tied to the hope
of finding a successful theory of quantum gravity.
Not only was world-sheet supersymmetry discovered at the same time as 
four-dimensional supersymmetry 
\cite{Raymond1971,NeveuSchwarz1971,neveu1971reformulation}, 
but the AdS/CFT correspondence (an example of a gravity/gauge duality) 
links superstring theory in special backgrounds to 
superconformal Yang-Mills theories 
(see \cite{Maldacena2003} and references therein).
It was the existence of such a AdS/CFT correspondence that fuelled the interest
in the $\b$-deformed theory studied in chapter \ref{Ch:BetaDef}.
\end{itemize}

\section{Effective actions}
In quantum field theories, all information about the dynamics of the
quantised fields is encoded in the effective action.
The effective action is often calculated perturbatively using Feynman diagrams,
which can be organised by their number of loops. 
Each loop gives a factor of $\hbar$, 
so the loop expansion can be thought of as an asymptotic series in quantum corrections.
In calculating quantum corrections to scattering amplitudes, 
the effective action is also expanded in the number of external fields,
which, diagrammatically, is just number of external lines on the Feynman graphs.
However, to understand the low-energy dynamics and vacuum structure 
of phenomenologically interesting theories 
and to test many of the exact results and conjectures mentioned above, 
the low-energy effective action is the relevant object.
The low-energy effective action
sums over diagrams containing all number of external lines,
but assumes that they carry a small amount of momentum 
so that the derivative expansion 
(an expansion in the number of derivatives acting on the external fields) 
becomes viable.
As the number of loops increases, this calculation becomes 
prohibitively difficult using diagrammatic methods
and the powerful, covariant background field method becomes the 
calculational tool of choice.

The general low-energy effective action for a $\cN=1$ super-Yang-Mills theory
contains the following structures 
(where the notation follows that of \cite{BK} and will be described later)
\begin{align*}
	&\intz \Big( K(\Fb, \F)  + F(\F, \Fb, D^2\F, \Db^2\Fb) \Big)
	+ \Big(\intc W(\F) + \cc\!\Big)\\
	&+ \Big(\intc f_{IJ}(\F) \cW^{I\a}\cW^J_\a   + \cc\!\Big)
	+ \intz \cW^2\cWb^2  \L\Big({\cD^2\cW^2},{\cDb^2\cWb^2}\Big)\ .
\end{align*}
These are, in order, the K\"ahler potential, the auxiliary potential,
the superpotential, the effective gauge kinetic term
and the Euler-Heisenberg-type effective action.
Chapters \ref{Ch:Neq1Quant} to \ref{Ch:BetaDef} are mainly concerned
with examining the above sectors of the effective action.

Since supersymmetry is not manifest at everyday energies, 
or even current particle accelerator energies, 
if it is a fundamental symmetry of nature then it must be spontaneously broken.
Following the general theory of spontaneously broken global symmetries
\cite{Coleman1969,Callan1969,Ivanov1977,Ivanov1978}
there must be a spin-half field associated with the broken supersymmetry 
generators. This field is now known as the Goldstino.
The pure Goldstino action is 
{a universal sector of the low-energy effective action}
of theories with broken supersymmetry
and, up to a field redefinition, it is unique. 
Central to the calculations in section \ref{sect:Goldstino} 
is the general form of the pure Goldstino action,
\begin{equation} \tag{\ref{eqn:GoldStruct}}
  S_{\text{Goldstino}} \sim \intx \sum_{n=0}^4  \k^{2n-2}\l^n\bar\l^n\pd^n\,,
\end{equation}
where $\k$ is a coupling constant of dimension $\text{(length)}^2$.
There are many different superfield embeddings and geometric constructions
that form natural choices for describing the Goldstino.
The primary aim of chapter \ref{ch:Goldstino} is to explicitly
find the relationship between them.


\section{Supersymmetry algebra} 
In this thesis, a basic knowledge of supersymmetry and superspace is assumed,
such as that found in the introductory textbooks 
\cite{BK,WessBagger1992,West-Book,GGRS1983}.
For later reference we need the $\cN=1$ supersymmetry algebra%
\footnote{Throughout this thesis we will follow the conventions and notation 
of the book by I.\ Buchbinder and S.\ Kuzenko, \cite{BK}.} 
written using two-component Weyl spinors 
(van der Waerden's notation) 
\begin{align}\label{defn:SusyAlgebra}
\begin{gathered}
    \acom{Q_\a}{Q_\b} = \acom{\bar{Q}_\da}{\bar{Q}_\db} = 0\,,\quad 
	\acom{Q_\a}{\bar{Q}_\da}  = 2 P_\ada = 2\s^a_\ada P_a \,, \\
	\com{Q_\a}{P_b} = \com{\bar{Q}_\da}{P_b} = 0\ .
\end{gathered}
\end{align}
This is combined with the Lorentz algebra, 
under which the supercharges ${Q_\a,\bar{Q}_\da}$ transform in the spinor 
representation and the momentum $P_a$ transforms in the vector representation.
We will also need the algebra of superspace covariant derivatives
\begin{align}\label{defn:SusyDAlgebra}
\begin{gathered}
    \acom{D_\a}{D_\b} = \acom{\bar{D}_\da}{\bar{D}_\db} = 0\,,\quad 
	\acom{D_\a}{\bar{D}_\da}  = -2 P_\ada = -2\s^a_\ada P_a \,, \\
	\com{D_\a}{P_b} = \acom{D_\a}{Q_\b} 
		= \acom{D_\a}{\bar{Q}_\db} = 0 \text{ and c.c.}
\end{gathered}
\end{align}

\section{Structure}
The structure of this thesis is as follows:

In chapter \ref{Ch:Neq1Quant} we perform the background field quantisation
of a general $\cN=1$ super-Yang-Mills theory. 
This helps set up the notation for the rest of the thesis
and the quantisation underlies the following three chapters.
For the general theory, 
we derive the propagators and low-energy one-loop effective action
in the 't Hooft gauge and examine the conditions for one-loop finiteness.

In chapter \ref{Ch:WZ} we review the calculations of 
\cite{Buchbinder1993,Buchbinder1994a,Buchbinder1996}, 
which pertain to the simplest four dimensional supersymmetric theory, 
the Wess-Zumino (WZ) model \cite{Wess1974a}.
In this unpublished work, we 
include the one-\  and two-loop K\"ahler potential calculations,
and calculate the full auxiliary potential, 
which was previously only known to the leading order.

In chapter \ref{Ch:SQED} we present the 
two-loop Euler-Heisenberg effective action for $\cN=1$ 
supersymmetric quantum electrodynamics, as published in \cite{Kuzenko2007a}.
This calculation was the second supersymmetric two-loop Euler-Heisenberg
effective action known and was a good test of both 
the background field method as presented in \cite{Kuzenko2003a} and 
the functional structures that occur in self-dual backgrounds \cite{Dunne2004}. 
We also present the (previously unpublished) results for the
one-loop K\"ahler potential in a general, two parameter $R_\x$-gauge.


Chapter \ref{Ch:BetaDef} contains the calculation of the 
2-loop K\"ahler potential for $\b$-deformed $\cN=4$ SYM. 
This calculation was published in \cite{Tyler2008} 
and is a continuation of the work in \cite{Kuzenko2005b,Kuzenko2007}.
As the undeformed $\cN=4$ SYM theory has 
no corrections to the K\"ahler potential,
the primary purpose of this chapter is to examine the properties of the 
low-energy effective-action that are interesting consequences of the 
$\b$-deformation.

Chapter \ref{ch:Goldstino} concentrates on nonlinearly realised (broken) 
supersymmetry and the Goldstino action. This is a universal sector
of low-energy effective actions in theories with broken supersymmetry.
In this chapter we are only concerned with the general form of the Goldstino 
actions and not concerned about deriving them from some more complete model.
The chapter starts with a quick review of supersymmetry breaking
and the supersymmetric sigma-model. 
Then section \ref{sect:Goldstino}, which is based on the papers
\cite{KuzenkoTyler2010,KuzenkoTyler2011},
shows the equivalence of all of the 
Goldstino actions found in the literature by constructing 
explicit field redefinitions that map them to the Akulov-Volkov action.
Section \ref{sect:mCL}, based on \cite{KuzenkoTyler2011a},
constructs and investigates a new embedding
of the Goldstino in a constrained modified complex linear superfield.
It is shown how it relates to all of the previously known 
constrained superfield realisations of the Goldstino.
This section expands upon the discussion in \cite{KuzenkoTyler2011a}
by explicitly providing the previously unpublished
component reduction of the new Goldstino coupled to an arbitrary matter sector.

There are five 
appendices. 
Appendix \ref{A:Propagators} 
investigates the heat kernels used in the first
few chapters. It contains a derivation for the heat kernels that occur
in $\cN=1$ super-Yang-Mills theories in a covariantly constant background
and a derivation of the heat kernel for the Wess-Zumino model in 
the effective potential limit.
Appendix \ref{A:WZCompEffPot} reproduces the component calculations
for the one-loop effective potential of the Wess-Zumino model.
Appendix \ref{A:IntegerRelations} details the mathematics and some computer code
used in finding the closed form for the two-loop, self-dual integrals 
of section \ref{sect:SelfDualSQED}.
Appendix \ref{A:2LoopVac} examines the two-loop integrals needed for the 
K\"ahler potential calculations of this thesis. 
It provides a new perspective on the differential equations that the 
two-loop ``fish'' Feynman diagram satisfies and contains the compact 
closed form first presented in \cite{Tyler2008}.
Finally, appendix \ref{A:Golden}
contains results necessary for the investigation
of the Goldstino actions of chapter \ref{ch:Goldstino}.
It includes a general basis for all Goldstino actions 
and the composition rules for the general field redefinition used in section 
\ref{sect:Goldstino}.


\chapter[Quantisation of \texorpdfstring{$\cN=1$}{N=1} SYM]%
		{Quantisation of a general \texorpdfstring{$\cN=1$}{N=1} 
			super-Yang-Mills theory}
\label{Ch:Neq1Quant}

In this chapter we examine the background field quantisation 
of a general $\cN=1$ super-Yang-Mills  (SYM) theory.
We discuss the choice of background fields, gauge fixing conditions 
and how they effect what we can reasonably calculate.
%
%
The subsequent three chapters examine loop corrections to
particular supersymmetric theories and will draw heavily on
the scaffolding provided in this chapter.
It is for this reason that we keep our discussion 
as general as possible for as long as possible.
This chapter does not contain any original results, but hopefully 
presents the quantisation and one-loop low-energy effective action 
of $\cN=1$ SYM in an original and useful manner.

The plan for this chapter is as follows.
In the first section we write down the most general renormalisable 
$\cN=1$ super-Yang-Mills theory that can be constructed from 
chiral superfields and gauge superfields.
We then split the fields 
into background and quantum parts 
and examine gauge transformations of these superfields.
In the third section we gauge fix the gauge symmetry of the quantum fields 
and write down the quadratic and interaction terms of the resultant action.
From the quadratic terms we calculate the Hessian of the action which is 
inverted in the fourth section to give the propagators for the theory. 
Explicit, heat kernel, calculations of the propagators in 
a covariantly constant background are given in 
appendix \ref{A:Propagators}.
In the last section, we calculate the one-loop K\"ahler potential 
and the $F^4$ corrections to the
general theory and reproduce the well known one-loop finiteness conditions, 
c.f., \cite{West-Book}.


\section{Classical theory}
Within the $\cN=1$ superspace approach, introduce the 
gauge covariant derivatives 
$\cD_A=(\cD_a,\cD_\a,\cDb^\da)$
for the simple (or $U(1)$) compact gauge group%
\footnote{The extension to reductive Lie groups 
(the direct product of simple and $U(1)$ components) is straightforward.
The prime difference is that the normalised trace \eqref{normalised trace}
becomes the sum of normalised traces for each component of the group
\[
	\tr(\frac1{g^2}T_\m T_\n) 	
	= \sum_{\scR_i\in\scR}\frac1{g_i^2} T(\scR_i)^{-1} 
	\tr_{\scR_i}(T^{\scR_i}_\m T^{\scR_i}_\n) \ .
\]
where the coupling constant can differ for each component of the group.
Our group theory conventions and results 
are given as needed later in the thesis.
}
$G$
\begin{equation} \label{defn:Neq1CovD}
 \cD_A=D_A+\rmi\G_A\,,\quad \G_A=\G_A^{\m}T_\m\,,\quad (T_\m)^\dag=T_\m \,,
\end{equation}
where the $T_\mu$ are the generators of some representation $\scR$ of 
the gauge group $G$.
In general they satisfy the algebra
\begin{align}\label{Neq1algBig} 
	[\cD_A,\cD_B\}=T_{AB}{}^{C}\cD_C+\rmi\cF_{AB} \,,
\end{align}
where $\cF$ is the field strength
and the torsion comes from the algebra of flat covariant derivatives
\eqref{defn:SusyDAlgebra}
\begin{align} 
	[D_A,D_B\}=T_{AB}^{~~C}D_C \,,\quad 
	T_{\a\db}^{~~c}=T_{\db\a}^{~~c}=-2\rmi\s_{\a\db}^c \,,
\end{align}
and all other index combinations for the torsion are zero,
see \cite{BK} for more details.

In order to reduce the number of independent components 
in the connection $\G_A$ to that of a vector multiplet,
covariant constraints are applied to the field strengths.
This was first done in \cite{Wess1977,Wess1977Salamanca} 
and generalised to $\cN$-extended theories in \cite{Grimm1978,Sohnius1978a}.
For $\cN=1$ SYM, the constraints are 
the representation preserving constraints, $\cF_\ab=\cF_\dadb=0$, 
and the conventional constraint, $\cF_\adb=0$. 
These constraints break \eqref{Neq1algBig} into the algebra
\begin{subequations}\label{Neq1alg}
\begin{align}
 \{\cD_\a,\cD_\b\}&=\{\cDb_\da,\cDb_\db\}=0\,,\;& 
 \{\cD_\a,\cDb_\da\}&=-2\rmi\cD_\ada=-2\rmi\s^a_\ada\cD_a\,,\label{Neq1alg1}\\
 [\cD_\a,\cD_\bdb]&=2\rmi\eps_\ab\cWb_\db\,,&
 [\cDb_\da,\cD_\bdb]&=2\rmi\eps_\dadb\cW_\b\,,\\
 [\cD_\ada,\cD_\bdb]
	&=-\eps_\ab\cDb_{(\da}\cWb_{\db)}-\eps_\dadb\cD_{(\a}\cW_{\b)}
	\hspace{-30cm}& 
	&=\rmi\cF_{\ada,\bdb} \ .
\end{align}\end{subequations}
The constraints are equivalent to the first line of the above algebra 
and are consistent with the Bianchi identities
\begin{align} \label{Neq1BI}
	\cDb_\da\cW_\a=\cD_\a\cWb_\da=0\,,\quad \cD^\a\cW_\a=\cDb_\da\cWb^\da \ .
\end{align}
So we see that for $\cN=1$ SYM, all components of the super field strength
can be derived from the spinor field strength
\begin{align} \label{cWfromcD}
	\cW^\a	=-\frac1{4\rmi} [\cDb_\da,\cD^\daa]
			=-\frac18[\cDb_\da,\{\cDb^\da,\cD^\a\}]\,,
\end{align}
and its complex conjugate. 

Now we can construct the classical, pure $\cN=1$ SYM action
\begin{equation}\label{pureSYMclass}
\begin{aligned}
	S[W]
	&=\frac1{2g^2}\intc\tr\cW^\a\cW_\a+\frac1{2g^2}\intac\tr\cWb_\da\cWb^\da \,,
\end{aligned}
\end{equation}
where the trace is the normalised trace over any faithful representation $\scR$
\begin{align}\label{normalised trace} 
	\tr(T_\m T_\n) 	= T(\scR)^{-1} \tr_{\scR}(T^{(\scR)}_\m T^{(\scR)}_\n)
	= T(\scR)^{-1} g^{(\scR)}_{\m\n} \ .
\end{align}
The Dynkin index $T(\scR)$ is defined by 
choosing the basis $\{T^{(\scR)}_\m\}$ to be orthogonal so that 
\( g^{(\scR)}_{\m\n} = T(\scR) \d_{\m\n} \).
We keep the general metric 
and the distinction between covariant and contravariant 
indices since, in the low-energy background field calculations below, 
the natural basis is the Cartan-Weyl basis which is not orthogonal.

The pure SYM action decomposes into four terms
\begin{align}  \non
\!	\intc\tr\cW^2
	&=\frac12\intx\tr\Big(\frac12(\cD\cW)^2
	+4\rmi\cW^\a\cD_\ada\cWb^\da
	-\cF_{ab}\cF^{ab}-\rmi \cF_{ab}\tilde\cF^{ab} \Big)\Big|_{\q=0} ,
\end{align}
which correspond respectively to the
(non-dynamical) term for the auxiliary field,
the kinetic term for the gaugino, 
the ($\cN=0$) Yang-Mills kinetic term
and the Pontryagin index, which is a topological surface term.
This decomposition combined with the Bianchi identity \eqref{Neq1BI}
and integration by parts gives a quick demonstration that
\begin{align*} 
	\intc\tr\cW^2 = \intac\tr\cWb^2 + \text{surface terms}\ .
\end{align*}

We introduce a multiplet of covariantly chiral (antichiral) matter superfields 
$\F$ ($\F^\dag$), that obey the constraints 
\begin{align}
\cDb_\da\F=\cD_\a\F^\dag=0\,,
\end{align}
and transform in some representation of the gauge group 
($\F^\dag$ transforms in the conjugate representation)
\begin{gather}\label{eqn:GaugeXformPhi_Quant}
\F'=\rme^{\rmi \cK}\F\,, \quad (\F^\dag)'=\F^\dag\rme^{-\rmi \cK} \,,\\
\cD_A'=\rme^{\rmi \cK}\cD_A\rme^{-\rmi \cK}\,,\quad \cK=\cK^\m T_\m=\cK^\dag \ .
\end{gather}

The most general renormalisable SYM action is then
\begin{gather}\label{classNeq1Act}
	S_{\rm class}
	=\frac{1}{g^2}\intc\tr\cW^\a\cW_\a + \intz\F^\dag\F  
	+ \Big( \intc P(\F)+\cc\Big) ,
\intertext{where the (chiral) superpotential $P(\F)$ is}
	\label{classNeq1SuperPot}
	\quad P(\F)=\frac12 m_{ij}\F^i\F^j+\frac16\l_{ijk}\F^i\F^j\F^k \ .
\end{gather}
The mass $m$ and coupling constants $\l$ 
are invariant tensors of the gauge group.

\section{Background-field splitting}
We split the fields in the background-vector, quantum-chiral representation
\cite{Grisaru1979}
\begin{equation}
\begin{aligned}
	\F&\to{\bf\F}=\F+\vf \,,&
	\F^\dag&\to{\bf\F}^\dag\rme^{gv}=(\F^\dag+\vf^\dag)\rme^{gv} \,,\\
	\cDb_\da&\to\cDb_\da \,,&
	\cD_\a&\to\rme^{-gv}\cD_\a\rme^{gv} \ ,
\end{aligned}
\end{equation}
where the lowercase letters correspond to the quantum fields.
We note that the above splitting implies that the field strength splits as
\begin{equation} 
\begin{aligned}
	\cW^\a\to{\bf W}^\a
	&=\cW^\a-\frac{g}8\cDb^2\frac{1-\exp(-\AD{gv})}{\AD{gv}}\cD^\a v\,,  
\end{aligned}
\end{equation}
where
\begin{equation}
\AD{x}y=[x,y]  \quad\text{and}\quad
\frac{1-\exp(-\AD{x})}{\AD{x}}=\sum_{n=0}^\infty\frac{1}{(n+1)!}(-\AD{x})^n\,.
\end{equation}

The background-quantum split leads to an extra gauge invariance
\begin{subequations}\label{KLgaugeXform1}
\begin{align}
\mbox{$\cK$}: &\left\{\quad 
\begin{gathered}
	{\bf\F}		\to \rme^{\rmi \cK}{\bf\F}\,, \quad
	{\bf\F}^\dag\to {\bf\F}^\dag\rme^{-\rmi \cK}\,,\quad
	\cD_A		\to \rme^{\rmi \cK}\cD_A\rme^{-\rmi \cK} \\
	v			\to \rme^{\rmi \cK}v\rme^{-\rmi \cK}\,, \quad
	{\bf W}_\a	\to \rme^{\rmi \cK}{\bf W}_\a\rme^{-\rmi \cK}\,, \quad
	\cK^\dag=\cK  
\end{gathered}\right. \\
\mbox{$\L$}:&\left\{\quad 
\begin{gathered}
	{\bf\F}		\to \rme^{\rmi\L}{\bf\F}\,,\quad
	{\bf\F}^\dag\to {\bf\F}^\dag\rme^{-\rmi\L^\dag}\,,\quad
	\cD_A		\to \cD_A \\
	\rme^{gv}	\to \rme^{\rmi\L^\dag}\rme^{gv}\rme^{-\rmi\L}\,,\quad
	{\bf W}_\a 	\to \rme^{\rmi\L}{\bf W}_\a\rme^{-\rmi\L}\,,\quad
	\cDb_\da\L=0 \ .
\end{gathered}\right.
\end{align}
\end{subequations} 
The relation and geometric origins of the two gauge parameters $\cK$ and $\L$
is explained, e.g., in section 3.6 of \cite{BK}.
It is straightforward to check that all of 
the terms in the background-quantum split action,
\begin{equation}\label{splitNeq1Act}
	S_{\rm split}
	=\frac1{g^2}\intc\tr{\bf W}^\a{\bf W}_\a 
	+ \intz{\bf\F}^\dag\rme^{gv}{\bf\F} 
	+ \Big\{\intc P({\bf\F}) +\cc\Big\}\,,
\end{equation}
are invariant under both sets transformations. 
We note that the gauge invariance of the tensors $m$ and $\l$ 
means that the $\rme^{gv}$ term 
drops out of the antichiral integral (in the `c.c.' term).

In the gauge transformations \eqref{KLgaugeXform1}, 
there is still freedom on how to split the transformations 
between $\F$ and $\vf$ in the $\bf\F$ terms.  
This is fixed by requiring that the $\L$ 
transformations only effect the quantum fields,
\begin{subequations}\label{KLgaugeXform2}
\begin{align}
\mbox{Background}: \quad {\F}&\to\rme^{\rmi \cK}{\F}\,,& 
	\cD_A&\to\rme^{\rmi \cK}\cD_A\rme^{-\rmi \cK}\,,
	\label{KLgaugeXform2a} \\ \label{KLgaugeXform2b}
\mbox{Quantum}:\quad \vf&\to\rme^{\rmi\L}(\F+\vf)-\F\,,& 
	\rme^{gv}&\to\rme^{\rmi\L^\dag}\rme^{gv}\rme^{-\rmi\L}\,,
\end{align}
\end{subequations}
where we ignored the quantum fields in the background gauge transformations,
since they are integrated out of the effective action.  
It is the gauge invariance 
under the quantum gauge transformations \eqref{KLgaugeXform2b} 
that needs to be gauge fixed in order to quantise the theory.

\section{Constructing the perturbation theory}\label{sect:Neq1Pert}
To construct the perturbation theory, 
we need to split the action \eqref{splitNeq1Act} into terms that are 
independent, linear, quadratic and higher order in the quantum fields
\begin{equation}\label{splitNeq1Act2}
	S_{\rm split} 
	= S^{(0)}+S^{(1)}+S^{(2)}+S^{\rm (int)}\,, \quad
 	S^{\rm (int)}=\sum_{n\geq3}S^{(n)} \ .
\end{equation}
Since all of the splittings are of the form $\J\to\J+O({\rm quantum})$, 
the terms of zero$^{\rm th}$ order in the quantum fields 
are exactly the classical action \eqref{classNeq1Act}.
The linear terms can not contribute to the one-particle irreducible (1PI) 
diagrams that sum to the effective action, so we will ignore them.
The terms quadratic in quantum fields that come from the matter action are
\begin{equation}\label{phi:quad:action}
\begin{aligned}
	S_{\rm mat}^{(2)}
	&=	\intz\left(\vf^\dag\vf+g(\vf^\dag v\F
	+	\F^\dag v\vf)+\frac{g^2}2\F^\dag v^2\F\right) \\
	&+	\left\{\frac12\intc(m_{ij}+\l_{ijk}\F^k)\vf^i\vf^j+\cc  \right\}\,,
\end{aligned}
\end{equation}
and the interaction terms are
\begin{align}\label{Int34Phi}
	S_{\rm mat}^{(\rm int)}
	&=	\intz\Bigg(\vf^\dag(\rme^{gv}-1)\vf
	+	\big(\vf^\dag (\rme^{gv}-1-gv)\F+\cc\big) \\ \non
	&+	\F^\dag (\rme^{gv}-1-gv-\frac12g^2v^2)\F\Bigg)
	+	\left\{\frac16\intc\l_{ijk}\vf^i\vf^j\vf^k+ \cc  \right\}\ .
\end{align}
The second term in $S_{\rm mat}^{(2)}$ mixes the matter and gauge 
quantum superfields.
As discussed below, this mixing can be removed by choosing 
a supersymmetric $R_\x$ gauge or by a change of variables in the path integral.
The term in $S_{\rm mat}^{(2)}$ that's quadratic in $v$ can be written as
a mass term for the vector multiplet: 
\begin{align}\begin{aligned}
	\frac{g^2}2\intz\F^\dag v^2\F 
	&=\frac1{2}\intz v^i_{~j}(\Mg)^j_{~k}v^k_{~i}
	 =\frac1{2}\intz v^\m \Mg_{(\m\n)} v^\n\,,
\end{aligned}\end{align}
where
\begin{align}\label{M2def}
	(\Mg)^i_{~j}=g^2\F^i\Fb_j \,,\qquad
	\Mg_{\m\n}=g^2 \F^\dag T_{\m}T_{\n}\F \ .
\end{align}
To fit this mass term into the gauge field action below, we want 
\( \tr(v\Mg v)=v^\m \Mg_{\m\n} v^\n, \) 
where we define $\Mg$ as an operator acting on an element of the Lie algebra by
\( \Mg v = T_\m \Mg^\m_{~\n} v^\n \).
This means that we must have
\( \Mg_{\m\n} = T(\scR)^{-1}g^{(\scR)}_{\m\l}(\Mg)^\l_{~\n}\ .\)

Now we examine the terms that come from the pure SYM action.  
First we expand the total field strength as
\begin{align}
	{\bf W}_\a
	&=	\cW_\a-\frac g8\cDb^2\cD_\a v+\frac{g^2}{16}\cDb^2[v,\cD_\a v]
	-	\frac{g^3}8\cDb^2 R_\a(v) \,, 
\end{align}
where the remainder $R_\a(v)$ contains terms that are 
cubic or higher in the quantum field.
So, the pure SYM action expands as
\begin{gather}
\frac1{g^2}\tr\intc{\bf W}^2=
	\frac1{g^2}\tr\intc{\cW}^2
	+ S^{(1)}_{\rm SYM} 
	+ S^{(2)}_{\rm SYM} + S^{(\rm int)}_{\rm SYM}\,,
\end{gather} 
with the kinetic term
\begin{align}\non
S^{(2)}_{\rm SYM}
  	&=	\frac12\tr\intz v\left(-\BoxV+\frac1{16}\{\cD^2,\cDb^2\}\right)v 
  	+	\text{surface terms} \,,
\end{align}
where 
\begin{align} \label{defn:BoxV}
	\BoxV \deq \cD^a\cD_a-\cW^\a\cD_\a+\cWb_\da\cDb^\da \ .
\end{align}
The interaction terms are
$S^{(\rm int)}_{\rm SYM}=S^{(\rm 3)}_{\rm SYM}+S^{(\rm 4)}_{\rm SYM}+\dots$,
with
\begin{align}\label{Int34SYM}
	S^{(3)}_{\rm SYM}
	&=	\frac{g}2\tr\!\intz \com{v}{\cD^\a v} \left(\frac13\com{\cW_\a}{v}
	+	\frac1{8}\cDb^2\cD_\a v \right) \,, \\ \non
	S^{(4)}_{\rm SYM}
	&=	-\frac{g^2}8\tr\!\intz \com{v}{\cD^\a v}
		\left(\frac1{3}\com{\com{\cW_\a}{v}}{v}
	+	\frac1{6}\com{\cDb^2\cD_\a v}{v}
	+	\frac{1}{8}\cDb^2\com{v}{\cD_\a v} \right) .
\end{align}
We don't explicitly write the higher order interactions 
since they only become important at three and higher loops.

The Faddeev-Popov-DeWitt quantisation procedure \cite{Faddeev1967,Dewitt1967II}
fixes the quantum gauge symmetry \eqref{KLgaugeXform2b} 
by effectively introducing three new terms in to the action,
\begin{gather} 
	S_{\rm FP} 
  	=	g\,\tr\left(\intc\ct\,\d_\L\c+\cc\right)\!\Big|_{\rmi\L\to c}~,\\
	S_{\rm GF} = -\frac1\a\tr\intz\c^\dag\c ~,\quad
	S_{\rm NK} = -\frac1\a\tr\intz b^\dag b~, \label{QuadActNK}
\end{gather}
where $\c=\c(v,\vf,\cD,\F)$ is a covariantly chiral gauge fixing condition, 
$c$ and $\ct$ are covariantly chiral Faddeev-Popov ghosts 
(covariantly chiral anticommuting scalars in the adjoint), 
$b$ is the Nielson-Kallosh ghost
\cite{Nielsen1978,Nielsen1978a,Kallosh1978,Nielsen1979}
(a covariantly chiral anticommuting scalar in the adjoint)
and $\a$ is a gauge fixing parameter. 
We choose a supersymmetric $R_\x$ 
\cite{Ovrut1982,Marcus1983,Binetruy1983,Banin2002,Banin2003a} 
type of gauge fixing condition
\begin{align}\label{RxiGauge}
	\c^\m = -\frac14\cDb^2\big(v^\m
	+	\frac{g}{\x}(\frac1{\Box_-}\vf^\dag)
		T(\scR)g_{(\scR)}^{\m\n}T^{(R)}_\n \F\big)\,,\quad
	\c = \c^\m T^{(\scR)}_\m\,,	
\end{align}
where $\scR$ is the arbitrary representation used in the trace in 
\eqref{QuadActNK}, 
note that for an orthogonal basis $T(\scR)g_{(\scR)}^{\m\n}=\d^{\m\n}$.
The (anti)chiral d'Alembertian operator introduced above is
\begin{align} \label{defn:Box+-}
	\Box_+ &= \cD^a\cD_a-\frac12(\cD^\a\cW_\a)-\cW^\a\cD_\a\,,&
	 \Box_+\f &= \frac1{16}\cDb^2\cD^2\f \,, \\
	\Box_- &= \cD^a\cD_a+\frac12(\cDb_\da\cWb^\da)+\cWb_\da\cDb^\da\,,&
	\Box_-\fb &= \frac1{16}\cD^2\cDb^2\fb \,,
\end{align}
with ($\fb$) $\f$ an (anti)chiral superfield.
The gauge fixing term then becomes
\begin{align}\label{eqn:Neq1Rxi:S_GF} \begin{aligned}
	S_{\rm GF}
	&=	\frac{-1}{\a}\intz\Big(\frac1{32} \tr( v\{\cD^2,\cDb^2\}v )
	+	\frac{g}\x(\F^\dag v \vf+\vf^\dag v\F) \\
	&+	\frac{g^2}{\x^2}T(\scR)g_{(\scR)}^{\m\n} 
		\vf^\dag T_\m\F\F^\dag	T_\n\frac1{\Box_+}\vf + \ord{\cD \F} \Big) \ .
\end{aligned}\end{align}

The $\vf$-$v$ mixing terms in \eqref{phi:quad:action} are only cancelled
by the gauge fixing term \eqref{eqn:Neq1Rxi:S_GF} if $\a\x=1$ \emph{and} 
we ignore all derivatives of the background chiral field.  
The condition that $\cD_A \F = 0$ can be quite restrictive and will
be discussed further at the end of this subsection.

If we want to include the derivatives of $\F$, 
for example to calculate the auxiliary field's potential 
(see chapter \ref{Ch:WZ}),
then we can not cancel the mixing terms by a simple choice of gauge.
In this case, the most practical choice is 
the supersymmetric covariant gauge where $\x\to\infty$, 
which includes both the Feynman ($\a=1$) and Landau ($\a\to0$) gauges
and we have to account for the mixing by making the
change of variables $\vf\to\vf+F(v,\F)$ in the path integral.
This effectively dresses the $v$-$v$ propagator and makes it non-local 
(in the 't~Hooft gauge, 
it is the chiral propagator that is dressed and non-local,
see \eqref{chiralPropNasty}). 
It also introduces some more interaction terms.
On the other hand, the ghost sector is simpler 
and we don't get a non-chiral mass term for $\vf$.

Such a procedure is used in section \ref{sect:SQED-1loopMatter} 
to calculate the chiral field's two-point function and K\"ahler potential 
for $\cN=1$ SQED in the Feynman gauge.
In the calculation of the two-point function, the non-local $v$-$v$ propagator
is expanded in powers of the external chiral fields to give diagrams
that only consist of simpler, local propagators. 
The results of section \ref{sect:SQED-1loopMatter}
can be generalised to any $\cN=1$ SYM theory.
Generalising the Feynman gauge calculation of the K\"ahler potential to
higher loops and more general $\cN=1$ SYM theories is, however, problematic.
The non-local propagators have to be kept 
and it is doubtful whether useful closed form results for the Feynman diagrams 
with those propagators can be obtained.


The bulk of this chapter is dedicated to calculating the low energy sectors
of the effective action, so we will ignore any derivatives of the background
chiral field and will use the supersymmetric 't~Hooft gauge where $\a=\x=1$.

Define the mass matrix
\begin{align} 
	\label{defn:ChMass}
	\cP_{ij}&=m_{ij}+\l_{ijk}\F^k \,,\\
\intertext{its conjugage $\bar\cP^{ij}$, and}
	\label{defn:MfMass}
	\Mf &= [(\Mf)^i_{~j}]
		= g^2 T(\scR)g_{(\scR)}^{\m\n} T_\m\F\F^\dag T_\n \ .
\end{align}
The physical (non-ghost), gauged fixed quadratic terms are now written as
\begin{align}\label{QuadActPhys}\begin{aligned}
	S^{(2)}_{\rm phys}
	&= \intz\left(\frac12\tr \, v\big(-\BoxV+\Mg \big)v
	+ \vf^\dag\frac{\Box_+-\Mf}{\Box_+}\vf\right) \\
	&+ \left\{\frac12\intc\vf^i \cP_{ij}\vf^j+\cc  \right\} \ .
\end{aligned}\end{align}

To find the Fadeev-Popov ghost action we need to know the variation of the
gauge fixing functions under the quantum gauge transformations 
\eqref{KLgaugeXform2b}.  
The variation of the quantum chiral fields is 
\begin{equation} 
	\d_\L\vf
	= \rmi\L(\F+\vf)~,\qquad \d_\L\vf^\dag
	=-\rmi(\F^\dag+\vf^\dag)\L^\dag \,,
\end{equation}
and the variation of the quantum gauge fields is
\begin{equation} 
	\d_\L(gv)=-\frac\rmi2\AD{gv}\left(
	 \L+\L^\dag-\coth(\frac12\AD{gv})(\L^\dag-\L)\right) \ .
\end{equation}
So, the Faddeev-Popov ghost action becomes
\begin{align*} 
	S_{\rm FP}
	&=	\intz\tr\Bigg((\ct-\ct^\dag)\AD{\frac12gv}(c^\dag-c)-
	 	(\ct+\ct^\dag)(\AD{\frac12gv}\coth\AD{\frac12gv})(c+c^\dag)\\
	&\quad	+g^2\Bigg(\ct\,T_\m\d^{\m\n}\Big(
		\frac1{\Box_-}((\F^\dag+\vf^\dag)c^\dag)T^{(R)}_\n\F\Big)
	+ \cc\Bigg)\,,
\end{align*}
which has the quadratic part
\begin{equation} \label{QuadActFP}
S^{(2)}_{\rm FP}=\intz\tr\left(c^\dag\big(\ds1-\Mg\Box_+^{-1}\big)\ct
 -\ct^\dag\big(\ds1-\Mg\Box_+^{-1}\big)c \right)~,
\end{equation}
where $\Mg$ is defined in \eqref{M2def}. 
The interaction terms are 
$S^{(\rm int)}_{\rm FP}=S^{(3)}_{\rm FP}+S^{(4)}_{\rm FP}+\dots$
where the cubic and quartic terms simplify to 
\begin{align}\label{Int34FP}
	S^{(3)}_{\rm FP}&=g\intz\left(\frac12\tr\,(\ct-\ct^\dag)[v,c^\dag-c]
	 -g\big(\frac1{\Box_-}\vf^\dag c^\dag\big)\ct\F
	 +g\F^\dag\ct^\dag\frac1{\Box_+}c\vf\right) \,,	\non\\
	S^{(4)}_{\rm FP}&=-\frac{g^2}{12}\intz\tr(\ct+\ct^\dag)[v,[v,c+c^\dag]] \ .
\end{align}

We now have explicit expressions for all of the quadratic terms in the action
(\ref{QuadActNK}, \ref{QuadActPhys}, \ref{QuadActFP})
and for all of the interaction terms up to quartic order
(\ref{Int34Phi}, \ref{Int34SYM}, \ref{Int34FP}).
The former can be combined into a single symmetric Hessian $\dsH$,
\begin{align*}
	S^{(2)}	
	&=	\frac12\J^\rmT\cdot\dsH\cdot\J \\
	&=	\smash[b]{-\frac12 v\cdot H_v\cdot v
 	+	\frac12\big(\vf^\rmT,\vf^\dag\big)\cdot H_\vf
 			\cdot {\bem\vf\\\bar\vf\eem}}\\
 	\\&+\smash[t]{\frac12\big(b^\rmT,b^\dag\big)\cdot H_{\rm NK}
 			\cdot {\bem b\\\bar b\eem}
 	+	\frac12\big(c^\rmT,\ct^\rmT,c^\dag,\ct^\dag\big)\cdot H_{\rm FP}
 			\cdot {\bem c\\\ct\\\bar c\\\bar\ct\eem}} \,,
\end{align*}
where the inner product ``$\cdot$'' is 
a summation over group and flavour indices
as well as the integration over the appropriate superspace.
In the first line we introduced 
\(
	\J^\rmT=\big(v,(\vf^\rmT,\vf^\dag),
	(c^\rmT,\ct^\rmT,c^\dag,\ct^\dag),(b^\rmT,b^\dag)\big)
\) 
and the block diagonal Hessian
$\dsH=\diag(H_v,H_\vf,H_{\rm FP},H_{\rm NK})=\dsH^\rmT$.
The individual blocks of the Hessian are
\begin{subequations}\label{Hessians}
\begin{align}
	\label{HessianVect}
	H_v&=(\BoxV-\MgS)\d^8~,\qquad 
	H_{\rm NK}=\bem 0 &\frac{\cDb^2}{-4} \\-\frac{\cD^2}{-4} & 0 \eem \dpm \,,\\
	\label{HessianFP}
	H_{\rm FP}&=\bem 0 & \dssig\frac{\Box_+-\MgT}{\Box_+}
			\frac{\cDb^2}{-4} \\
		\dssig\frac{\Box_--\Mg}{\Box_-}\frac{\cD^2}{-4} & 0 \eem\dpm \,,\\
	\label{HessianMat}
	H_\vf&=\bem \cP & \frac{\Box_+-\MfT}{\Box_+}\frac{\cDb^2}{-4} \\
		\frac{\Box_--\Mf}{\Box_-}\frac{\cD^2}{-4} & \bar\cP \eem \dpm~,
\end{align}\end{subequations}
where we've introduced the notations $\dssig=\rmi\s_2=\smash{\bem0&1\\-1&0\eem}$
and $\MgS=\frac12(\Mg+\MgT)$.

The background field method, combined with the loop expansion, 
gives the one-particle irreducible (1PI) effective action as 
\[
	\G[\cD_A,\F]=\sum_{n=0}^\infty\G^{(n)}[\cD_A,\F]
 	= S_{\text{class}}[\cD_A,\F]+\tilde\G[\cD_A,\F] \ .
\]
The loop corrections $\tilde\G$ can be calculated from the path integral
\begin{align} \label{genEffectiveAction}
	\exp(\rmi\Gt)
	&= 	\cN\int\cD\J\exp\big(\rmi S^{(2)}[\J]
	+ 	\rmi S^{(\rm int)}[\J]\big)\big|_{\rm 1PI} \\
&=\exp(\rmi\G^{(1)})\exp\big(\rmi S^{(\rm int)}[\frac{\d}{\rmi\d\dsJ}]\big)
  \exp\big(\frac\rmi2\dsJ^\rmT\cdot\dsG\cdot\dsJ\big)
  \big|_{\genfrac{}{}{0pt}{}{\rm 1PI}{\dsJ\to0}} \ ,
\end{align}
where $\dsJ$ is a current that couples to $\J$ and 
$\dsG=-\dsH^{-1}$ contains all of the propagators of the theory.
The one-loop effective action $\G^{(1)}$ 
is defined by the functional superdeterminant%
\footnote{We use the notation that $\Det$ ($\Tr$)
correspond to functional determinants (traces) combined
with possible determinants (traces) over the gauge and/or colour indices.
We reserve $\det$ and $\tr$ for non-functional determinants
and traces only.
}
\begin{align}\begin{aligned}
	\exp(\rmi\G^{(1)})&=\cN\,\sDet^{-\frac12}(\dsH) \\
	&=\cN\,\Det^{-\frac12}(H_v)\Det^{-\frac12}(H_\vf)
	 \Det^{\frac12}(H_{\rm FP})\Det^{\frac12}(H_{\rm NK}) \,,
\end{aligned}\end{align}
where the background independent normalisation factor, $\cN$,
is chosen such that $\tilde\G[0,0]=0$. This means that 
$\cN\sim\sDet^{-\frac12}(\dsH_0)$, 
where $\dsH_0=\dsH|_{\text{background}\to0}$,
is absorbed into the functional determinant to make its argument dimensionless.

All of the above functional determinants and Green's functions 
are straightforward to calculate, except for those coming from 
the matter sector, which has the Hessian $H_\vf$.  
The difficulty stems from the fact that, in general, 
the mass matrices $\cP$ and $\Mf$ do not commute.  
For simple theories such as SQED this is not too much 
of a problem, since, as in section \ref{sect:SQED-1loopMatter},
the two-by-two matrices can be explicitly dealt with.
For more general theories, this difficulty is significant.
As noticed, e.g., in \cite{Grisaru1996},
the problem disappears if we choose the only non-zero components $\F$ to 
be massless after gauge symmetry breaking.  
These fields then satisfy their classical equations of motion 
up to derivative terms.
This idea can also be run backwards, e.g., \cite{Nibbelink2005}.
The principle observation is that if the background fields $\F$ satisfy 
their classical equations of motion 
$-\frac{1}{4}\cDb^2\Fb_i = \cP_{,i}$
then gauge invariance of the classical action 
implies that the mass matrices commute up to derivative terms,
$\cP\Mf=\Mf\bar\cP=0+O(\cD\F)$ .
Care needs to be taken with this argument since if the vacuum condition
$\cP_{,i}(\F)=0$ has only discrete or trivial solutions, then we are not left
with any dynamical chiral background fields.%
\footnote{This error was made in the first version of \cite{Nibbelink2005}, 
where they calculated the two-loop K\"ahler potential for a general, 
non-renormalisable $\cN=1$ SYM theory assuming $\cD_\a\F = \cP_{,i}(\F) = 0$. 
Their general result was then applied to the case of massive SQED with an 
on-shell background where these conditions imply that either the mass
or the background fields vanish, 
making the calculated K\"ahler potential ambiguous at best. 
In the revised version of \cite{Nibbelink2005} the example was changed to 
massless SQED, which does not suffer such problems.
I would like to thank Prof.~Kuzenko for bringing this issue to my attention.}

The condition $\cP\Mf=\Mf\bar\cP=0+O(\cD\F)$, although restrictive, 
is natural for some theories: such as $\cN=1$ SYM with massless quarks 
and no chiral vertices, so that $\cP=0$; Wess-Zumino theories (not gauged), 
where $\Mf=0$; and most importantly on the Coulomb branch of $\cN=2$ theories
where we have a true example of $\cP\Mf=0$.  

Finally, we note that ignoring all covariant derivatives of the 
background chiral fields, 
which is equivalent to setting $[\cD_A,\F]=0$ during the calculation,
is not as innocuous as it may seem.
It implies the non-derivative condition $[\cW_A, \F]=0$, which says that 
for a background gauge field strength that breaks the gauge group,
the background chiral fields must form a trivial representation
for any group generators in the broken directions.
Once again, for $\cN=2$ SYM theories on the Coulomb branch, this condition
is naturally satisfied since both $\F$ and $\cW_A$ come from a single
$\cN=2$ background field strength that lies in the Cartan subalgebra.

\section{Propagators and Feynman rules}\label{sect:PropsFeynRules}
To find the propagators we need to invert the Hessians \eqref{Hessians}.
In this section we only do this formally 
and try to get them in the form of some operator or matrix acting on
$(\Box-m^2)^{-1}$ where $\Box$ is a d'Alembertian-type operator.  
Then, in appendix \ref{A:Propagators}, 
we use heat kernel and propertime methods to obtain closed form results in 
the covariantly constant (Euler-Heisenberg) approximation.
Remember that we are working in the supersymmetric 't~Hooft gauge 
and ignoring all derivatives of the chiral background fields, 
but are not yet making any other assumptions.

In the 't~Hooft gauge, the vector Hessian \eqref{HessianVect} 
is particularly simple and easy to (formally) invert
\begin{gather}
	G_v(z,z')=-\frac1{\BoxV-\MgS}\d^8(z,z') \ .
\end{gather}
The other Hessian in \eqref{HessianVect} is that of the Nielson-Kallosh
ghost. This is also easily inverted by noting that
$H_{\rm NK} \cdot H_{\rm NK}=-\diag(\Box_+\d_+,\Box_-\d_-)$, so
\begin{align} 
	G_{\rm NK}
	= \bem0&\frac{\cDb^2}{-4}\frac1{\Box_-}\\
			\frac{\cD^2}{+4}\frac1{\Box_+}&0\eem\dpm ,
\end{align}
but as the Nielson-Kallosh ghosts only couple to the background gauge fields,
this propagator is not needed.

To find the other propagators we introduce 
the Hessian for a massless covariantly chiral superfield
\begin{align}\label{defn:cH}
 	\cH &= \bem 0 & \frac{\cDb^2}{-4} \\ \frac{\cD^2}{-4} & 0 \eem\dpm.
\end{align}
Noting that $\cH\cdot\cH=\diag(\Box_+\d_+,\Box_-\d_-)$, 
we see that its inverse is
\begin{align}\label{defn:cH-inverse}
	\cH^{-1} 
		&= \bem0&\frac{\cDb^2}{-4}\frac1{\Box_-}\\
				\frac{\cD^2}{-4}\frac1{\Box_+}&0\eem\dpm .
\end{align}
This is used to simplify the inversion of other Hessians through the trick
$H^{-1}=(H\cdot\cH\cdot\cH^{-1})^{-1}=\cH\cdot(H\cdot\cH)^{-1}$.
In particular, for the Fadeev-Popov ghosts, the Hessian \eqref{HessianFP}
is inverted to give
\begin{align}
	G_\FP&=\bem0&\dssig\frac{\cDb^2}{-4}\frac{1}{\Box_--\Mg} \\
		\dssig\frac{\cD^2}{-4}\frac{1}{\Box_+-\MgT} & 0 \eem\dpm .
\end{align}

The inversion of the chiral matter fields' Hessian \eqref{HessianMat}
is a little harder. Using the above trick, we see that we need the inverse of
\begin{align} 
	\tilde H_\vf= H_\vf\cdot\cH=
 		\bem\Box_+-\MfT &  \cP\frac{\cDb^2}{-4}\\
 			\bar \cP\frac{\cD^2}{-4} & \Box_--\Mf \eem \dpm~.
\end{align}
We proceed by using the block matrix formula
\begin{align} \label{eqn:BlockMatrixInverse}
 \bem A&B\\C&D\eem^{-1}=\bem (A-BD^{-1}C)^{-1}
	&-(A-BD^{-1}C)^{-1}BD^{-1}\\-(D-CA^{-1}B)^{-1}CA^{-1}
	&(D-CA^{-1}B)^{-1}\eem,
\end{align}
that holds whenever the matrices $A$ and $D$ have inverses.
We also use a compact notation for 
the left, right and central fractions of noncommutative terms,
defined by sliding the numerators off the fraction in the direction 
that's allowed by the lips of the vinculum, i.e.,  
\begin{align}
	\lfrac{A}{B}=AB^{-1}\,,\quad \rfrac{A}{B}=B^{-1}A 
	\quad \text{and} \quad \lrFrac{A}{B}{C}=AC^{-1}B \ .
\end{align}
So the propagator, 
$G_\vf=-\cH\cdot\tilde H_\vf^{-1}$, is
\begin{align}\label{chiralPropNasty}
	G_\vf= 
		\bem -H_+^{-1} & 0 \\ 0 & -H_-^{-1} \eem
		\bem -\lfrac{\bar\cP\Box_+}{\Box_+-\MfT}
		& 	\frac{\cDb^2}{-4}
		\\	\frac{\cD^2}{-4}
		& 	 -\lfrac{\cP\Box_-}{\Box_--\Mf} \eem\!\dpm ,
\end{align}
where we define
\begin{equation}
\begin{aligned} 
	H_+ &= \Box_+-\Mf-\lrFrac{\bar\cP}{\cP}{\Box_+-\MfT}\Box_+ \,,\\ 
	H_- &= \Box_--\MfT-\lrFrac{\cP}{\bar\cP}{\Box_--\Mf}\Box_- \ .
\end{aligned}
\end{equation}
In the above equations, the term 
$\frac1{16}\lrfrac{\cP\cDb^2}{\bar\cP\cD^2}{\Box_+-\Mf}$ (and its transpose)
can be viewed as the result of treating the mass terms $\cP$ and $\bar\cP$
as interactions. Then the propagator \eqref{chiralPropNasty} is simply
the dressed propagator obtained from the infinite sum of these 
two-point mass interactions with the undressed propagators being of form
$(\Box_\pm-\cM^2)^{-1}$.

Now we make the assumption $\cP\Mf=\Mf\bar\cP=0$ that was discussed at the end 
of section \ref{sect:Neq1Pert}. 
This greatly simplifies the chiral matter propagators
\begin{align}
G_\vf&=\bem\lfrac{\bar\cP}{\Box_+-\cP\bar\cP}
 &\frac{-1}{\Box_+-(\Mf+\bar\cP \cP)}\frac{\cDb^2}{-4}\\
 \frac{-1}{\Box_--(\Mf+\bar\cP \cP)^\rmT}\frac{\cD^2}{-4}
 &\lfrac{\cP}{\Box_--\bar\cP \cP}\eem\dpm 	\non\\
&=\bem\lfrac{\bar\cP}{\Box_+-\MfWT}
 &\frac{\cDb^2}{-4}\frac{-1}{\Box_--\MfW}\\
  \frac{\cD^2}{-4}\frac{-1}{\Box_+-\MfWT}
 &\lfrac{\cP}{\Box_--\MfW}\eem\dpm ,
\label{chiralPropSimp}
\end{align}
where we have introduced the combined mass-squared matrices
\begin{align}\label{defn:combinedmass}
	\MfW = \Mf + \bar\cP \cP\,,\quad \MfWT = \MfT + \cP \bar\cP \ .
\end{align}

The Feynman rules for constructing the Feynman diagrams
can be read off, as usual, from equation \eqref{genEffectiveAction}.
In the following chapters we will provide the Feynman rules for 
the specific models as we need them.

\section{One-loop effective action}\label{sect:Neq1-1Loop}
The simplification of the functional determinants in the one-loop effective
action is very similar to the simplification used to find 
the propagators in the previous section. 
So we will not give too many details.

As shown above, the one-loop effective action can be written as
\begin{align*} 
\G^{(1)}&=\G_v^{(1)}+\G_\vf^{(1)}+\G_{\rm FP}^{(1)}+\G_{\rm NK}^{(1)}\\
 		&=\frac\rmi2\Tr\log H_v+\frac\rmi2\Tr\log H_\vf
 -\frac\rmi2\Tr\log H_{\rm FP}-\frac\rmi2\Tr\log H_{\rm NK} \  ,
\end{align*}
where the traces are a combination of the functional 
trace over the appropriate superspace 
and a normal trace over the colour and flavour indices.

The only trace over full superspace is that for the gauge multiplet
\begin{align} 
	\G_v^{(1)} = \frac\rmi2\Tr\log(\BoxV-\MgS) \ . 
\end{align}
All other traces are over the chiral subspaces and we use the general notation
\begin{align} \label{eqn:1loopChiralDet}
	\G^{(1)}_{m^2,\cR} &= \frac\rmi4\log\Det 
		\bem \Box_+-m^2{}^\rmT & 0 \\ 0 & \Box_--m^2 \eem  \non\\
	&= \frac\rmi4 \tr_\cR\left( \Tr_+ \log(\Box_+-m^2) 
	 + \Tr_- \log(\Box_--m^2) \right) \ ,
\end{align}
for some mass-squared matrix $m^2$ 
in the representation $\cR$ of the gauge group.
With this notation and using the trick of inserting $\ds1=\cH\cdot\cH^{-1}$,
we have
\begin{align*} 
	\G^{(1)}_{\rm NK} 	&= -\G^{(1)}_{0,\ad} \,, &
	\G^{(1)}_{\rm FP} 	&= 2\G^{(1)}_{0,\ad} - 4\G^{(1)}_{\Mg,\ad} \,, &
	\G^{(1)}_{\vf} 		&= \G^{(1)}_{\Mf,R} - \G^{(1)}_{0,R} 
						+  \G^{(1)}_{\tilde\vf} \,,
\end{align*}
where 	
\begin{align} 
	&\G^{(1)}_{\tilde\vf} 		\non
	= \frac\rmi4\Tr\log\bem 
	\Box_+-\MfT-\frac1{16}\lrfrac{\cP\cDb^2}{\bar\cP\cD^2}{\Box_--\Mf} & 0 \\
	0&\Box_-\!-\Mf-\frac1{16}\lrfrac{\bar\cP\cD^2}{\cP\cDb^2}{\Box_+-\MfT}\eem\\
\intertext{which reduces, in the case of $\cP\cM^2=0$, to}
	&\G^{(1)}_{\tilde\vf}\Big|_{\cP\cM^2=0}= 
	\frac\rmi4\Tr\log	\bem \Box_+-\MfWT &  0\\ 0 & \Box_--\MfW \eem
	= \G^{(1)}_{\MfW,R} \ .
\end{align}

In total, this yields
\begin{align} \label{eqn:1LoopNeq1Total}
	\G^{(1)}
	&=\G_v^{(1)}+\G^{(1)}_{\Mf,R} + \G^{(1)}_{\tilde\vf} - 4\G^{(1)}_{\Mg,\ad}
	+ \G^{(1)}_{0,\ad} - \G^{(1)}_{0,R} \ .
\end{align}
In the subsections below we will calculate using 
regularisation by dimensional reduction (DRed)
\cite{Siegel1979,Siegel1980,Jack1994}. 
The only changes that need to be made to the preceding discussion
are that the space-time integrals 
(including the space-time parts of the superspace integrals) 
now only range over the first $d=4-2\eps$ dimensions. 
Details of this are discussed in appendix \ref{A:Propagators}.

\subsection{K\"ahler potential}\label{ssect:Neq1Kahler}
In the K\"ahler approximation we turn off the background gauge fields
so that all derivatives become the flat superspace derivatives.
This means that $\G_v^{(1)}\to0$ as there are no remaining superspace
derivatives to annihilate the Grassmann delta function, and 
the $\G^{(1)}_{0,\cR}$ become background independent and so can be dropped.
With the assumption that $\cP\cM^2=0$, we see that the remaining terms are
\begin{align*} 
	\G^{(1)}_{\text{K\"ahler}} = \intz K^{(1)}
	&=\frac\rmi2\Tr_+\Big(
	\log(\Box-\Mf P_+) + \log(\Box-\MfW P_+)\\ &- 4\log(\Box-\Mg P_+)
	\Big) . 
\end{align*}
The functional trace $\rmi\Tr_+\log(\Box-m^2P_+)=\intz\scJ(m^2)$
can be evaluated by moving to momentum space,
\begin{align} \label{defn:scJ_Integral}
	\scJ(m^2) 
	&= -\rmi\m^{2\eps}\intk\frac1{k^2}\log(k^2+m^2-\rmi\e) 
	= \frac{m^2}{(4\p)^2}\frac{\G(\eps-1)}{\eps-1}
		\frac{(4\p\m^2)^\eps}{m^{2\eps}} \non\\
	&= \frac{m^2}{(4\p)^2}\left(\frac1\eps+2-\log\frac{m^2}{\mub^2}
		+\ord\eps\right) \,,
\end{align}
where $\mu$ is the DRed minimal subtraction (MS) renormalisation point and 
$\mub^2=4\p\rme^{-\g}\m^2$ is the modified 
minimal subtraction (\MSb) renormalisation point. 
The standard $\rmi\e$, $\e\to0_+$, term ensures that 
the correct causal (Feynman) boundary condition has been used in the propagator.
So the one loop K\"ahler potential is
\begin{align} \label{Neq1 1loop Kahler}
	K^{(1)} &= \frac12\Big(
	\trR\scJ(\Mf) + \trR\scJ(\MfW) - 4\trad\scJ(\Mg)	\Big)  \non \\
	&= \frac12\trR\scJ(\bar\cP \cP) - \trad\scJ(\Mv) \ ,
\end{align}
where to get the the last line we used
$\MfW{}^{n} = \Mf{}^{n} + (\bar\cP \cP)^n$
and $\trad\Mg{}^n=\trad\MgS{}^n=\tr_R\Mf{}^n$.
This result matches that of \cite{Nibbelink2005}.

Note that in particular cases, e.g., where the superpotential mass matrix $\cP$
is of full rank, it is possible to calculate the K\"ahler potential without 
assuming the $\cP\cM^2=0$ condition. 
At one-loop, this reduces to factoring polynomials in $k^2$ with 
coefficients constructed from the mass matrices so that the integrand
becomes a simple sum of $\log(k^2+m_i^2)$ terms. 
Massive $\cN=1$ SQED is such a case and is examined in in 
section \ref{sect:SQED-1loopMatter}. 
In the general case, see e.g., equation (4.17) of \cite{Grisaru1996}, 
the mass matrices are not explicitly known and the final results 
can not be expressed as explicitly as those in 
section \ref{sect:SQED-1loopMatter}.

Looking at the divergent terms in the above 
and writing out the masses explicitly, we have
\begin{equation}\label{1LKahlerDiv}
	K_{\mbox{\small div}}^{(1)}
	=\frac{1/\eps}{(4\p)^2}	\left(
	 \half\bar\l^{ijl}\l_{jik} - g^2 C_2^{(R)}\d_k^l\right)\F^k\Fb_l \,,
\end{equation} 
where the quadratic Casimir for any irreducible representation
is defined by 
\(
 {C}_2^{(R)} 
 	= T^{(R)}_\m {T(\ad)g_{(\ad)}^{\m\n}} T^{(R)}_\n 
 \, .
\)

\subsection{\texorpdfstring{$F^2$}{F2} 
	and \texorpdfstring{$F^4$}{F4} corrections}
\label{ssect:Neq1F2F4}

We now look at the low energy effective action for the gauge field.
This can be written in terms of the heat kernels of appendix \ref{A:Propagators}
using the generic result
\begin{align}\label{eqn:1loopHeatKernel}
	\Tr\log(\hat\square-m^2)
	=-\Tr \int_0^\infty\frac{\rmd s}{s}K(s)\rme^{-\rmi s (m^2-\rmi\e)}\,,
\end{align}
where $K(s)=K(z,z'|s)\big|_{z'\to z}$ is the coincidence limit of 
the heat kernel of some d'Alembertian $\hat\square$.
We only want to calculate terms up to and including 4th order in the 
field strength and without derivatives 
(i.e., those proportional to $\intc \cW^2$ and $\intz \cW^2 \bar\cW^2$).
At higher loops this means we can use the limits given in \eqref{HK_DWeq0},
but at one-loop we need to be careful with the chiral trace
and include the first two of terms in the expansion 
of the chiral heat kernel. 
The appropriate limits of the heat kernels 
(\ref{KeqExpBox}, \ref{defn:ChiralHKs}) are
\begin{align} \label{eqn:HK_DWeq0Plus}
\begin{aligned}
	K(s)	&=\frac{\rmi\m^{2\eps}s^4}{(4\p\rmi s)^{d/2}}\cW^2\cWb^2 \,, \\
	K_+(s)	&=\frac{-\rmi\m^{2\eps}s^2}{(4\p\rmi s)^{d/2}}
			\left(1-\frac{s^2}{12}\frac{\cDb^2}{-4}\cWb^2+\cdots\right)\cW^2 \,,
\end{aligned}
\end{align}
and similarly for $K_-(s)$.

In the expression for the one-loop effective action \eqref{eqn:1LoopNeq1Total}, 
the contribution from the vector multiplet, $\G_v^{(1)}$, 
is the only one that uses the heat kernel $K(s)$. 
It evaluates to
\begin{equation}
\begin{aligned} \label{eqn:1loopFullDetF2}
	\G_v^{(1)}&=\frac{\m^{2\eps}}{2(4\p\rmi)^{d/2}}\intz
	\trad\int_0^\infty\rmd s s^{1+\eps} \rme^{-\rmi \Mv s}\cW^2\cWb^2 \\
	&=\frac{\G(2+\eps)}{32\p^2}\intz
	\trad\frac{\cW^2\cWb^2}{(\Mv)^2}\left(\frac{4\p\m^2}\Mv\right)^{2\eps} ,
\end{aligned}
\end{equation}
which is UV finite as $\eps\to0$. 
We note that if there are any zero eigenvalues of the above mass matrices
then is necessary to introduce an IR regulator.
Two possible options are a mass term 
or a cut-off on the upper limit of the propertime integral. 
If we use dimensional regularisation (instead of dimensional reduction)
then any scale free momentum integral 
is automatically set to zero, 
in particular \cite{collins1986renormalization,Smirnov2006}
\begin{align*} 
	  \frac1{\G(n)}\int_0^\infty \rmd s\, s^{n-3+\eps} 
	= -\rmi\int\frac{\rmd^dk}{\p^{d/2}} \frac1{k^{2n}} 
	= 0 \ .
\end{align*}
This can be seen as a cancellation between the 
UV and IR sectors of the integral, 
which can be problematic in certain situations
and is briefly discussed in section \ref{ssect:Neq1MatterRenorm}.

With the assumption that $\cP\cM^2=0$, 
all the other terms in \eqref{eqn:1LoopNeq1Total} 
are of the form \eqref{eqn:1loopChiralDet},
which, using the chiral heat kernel of \eqref{eqn:HK_DWeq0Plus},
evaluates to
\begin{align} \label{eqn:1loopChiralDetF2}
	\G_{m^2,\scR}^{(1)} 
	&=-\frac{\G(\eps)}{4(4\p)^2}\left(
		\intc\tr_\scR\cW^2\Big(\frac{4\p\m^2}{m^2{}^\rmT}\Big)^{2\eps}
		+ \cc \right) \\\non
	&+\frac{\G(2+\eps)}{48(4\p)^2}\intz\tr_\scR\cW^2\cWb^2
		\left(\frac1{m^4{}^\rmT}\Big(\frac{4\p\m^2}{m^2{}^\rmT}\Big)^{2\eps}
				+\frac1{m^4}\Big(\frac{4\p\m^2}{m^2}\Big)^{2\eps}\right) ,
\end{align} 
where, for the sake of conciseness, we've given only the final, 
integrated form and assumed that any mass matrices with zero eigenvalues
have been appropriately modified by some IR regularisation.
%
%
The UV divergences in the above do not depend on the masses, 
\begin{align} 
	\G^{(1)}_{m^2, \scR}\big|_{\rm div} 
	=\frac{-1/\eps}{32\p^2}\intc\trR\cW^2 +\ord{\eps}^0 \,,
\end{align}
and so, writing the final trace in a normalised form, we find the 
total one-loop divergence to be
\begin{equation} 
	\G^{(1)}\big|_{\rm div} 
	= \G^{(1)}_{(R)}\big|_{\rm div}-3\G^{(1)}_{(\ad)}\big|_{\rm div}
	= \frac{1/\eps}{32\p^2}\left(3 T(\ad)-T(R)\right)\intc\tr\cW^2 \ .
\end{equation}

The finite one-loop corrections to the $F^2$ and $F^4$ terms can be easily
found be expanding \eqref{eqn:1loopFullDetF2} and \eqref{eqn:1loopChiralDetF2}
in $\eps$ and substituting the results into the general 
one-loop effective action equation \eqref{eqn:1LoopNeq1Total}.

\subsection{One-loop finiteness}
In the above two subsections we have recovered the well known conditions 
\cite{Jones1984,Parkes1984,West-Book} 
for the one-loop UV finiteness of the effective action 
for a general $\cN=1$ SYM theory 
with the matter fields in the representation $R$:
\begin{equation}\label{1loopfinite}
	(\bar\l\l)^i{}_j=2g^2 C_2^{(R)}\d^i_j 
	\quad \mbox{and} \quad
	3 T(\ad) = T(R) \ .
\end{equation}
Although the equations above were derived assuming $\cP\cM^2=0$,
they are actually true for a more general background. 
This can be checked, e.g., by calculating the quadratic terms in 
\eqref{eqn:1LoopNeq1Total} instead of the full K\"ahler potential.

$\cN=2$ SYM can be written as a $\cN=1$ SYM theory 
with a specific matter sector. 
This is obtained from the above by choosing the matter representation
\begin{align*}
	R \to \ad\otimes R\otimes R_c
	\quad\text{with}\quad
	\F^i\to (g^{-1}\F^\m, Q^i,\tilde{Q}^{\tilde i}) \,,
\end{align*}
with the specific superpotential \eqref{classNeq1SuperPot}
\[ \cP(\F)\to\cP(\F,Q,\tilde{Q})=\tilde{Q}(m+\F^\m T_\m) Q \ . \] 
On the Coulomb branch only $\F$ is given 
a nonvanishing background, so the mass condition $\cP\cM^2=0$ 
is automatically satisfied.  
Writing $\F^{(\scR)}=T^{(\scR)}_\m\F^\m$, 
the one-loop K\"ahler potential becomes
\begin{align}
	K_{\cN=2}^{(1)}=\tr_R\scJ(\bar\F^{(R)}\F^{(R)}) 
		-\trad\scJ(\bar\F^{(\ad)}\F^{(\ad)}) \ .
\end{align}
Thus $\cN=2$ SYM has a finite one-loop K\"ahler
potential if $\trR\bar\F\F=\trad\bar\F\F$, i.e., if $T(R)=T(\ad)$. 
The same finiteness condition is found on the gauge side.  
This is, of course, automatically satisfied for $\cN=4$ where $R=\ad$.

A full classification of all $\cN=1$ one-loop finite theories 
for simple groups has been given in 
\cite{Hamidi1984,Jiang1987,Jiang1989}
and for $\cN=2$ theories in \cite{Koh1984} which was extended to pseudoreal
representations and semisimple groups in 
\cite{Derendinger1984,Jiang1984,Jiang1985}.


\chapter[Wess-Zumino model]{Loop corrections to the Wess-Zumino model}
\label{Ch:WZ}
The Wess-Zumino model \cite{Wess1974a} was the second %
known four-dimensional field theory with a linear realisation of supersymmetry.
It is also the simplest supersymmetric field theory, in the sense that
it has the simplest particle content and interactions.%
\footnote{Other criteria for simplicity do exist, e.g., 
\cite{Arkani-Hamed2008}.}
In this way, it is analogous to the $\phi_4^4$ or $\phi_6^3$ 
non-supersymmetric field theories that are commonly used in QFT textbooks as
testbeds for quantising relativistic fields. 
The Wess-Zumino model is a testbed 
for examining quantisation of supersymmetric field theories.

We include an analysis of its one- and two-loop effective actions 
in this thesis for three main reasons:
1) it provides a good comparison to the more complicated K\"ahler potential 
corrections that we wish to calculate;
2) there is a disagreement in the literature on the coefficient
of the leading term in the one-loop auxiliary potential 
which warrants further investigation; 
3) the complete one-loop auxiliary potential has never before 
been derived using superfields and is the primary new result of this chapter.
This result is made possible by the calculations in appendix \ref{sect:WZProp},
where, following the methods of \cite{Buchbinder1993,Buchbinder1994a},
we present a clean derivation of the heat kernel for the Wess-Zumino model 
with a background superfield that is constant in spacetime.
We then compare the one-loop auxiliary potential 
with the existing component results of the one-loop effective potential, 
which are rederived in appendix \ref{A:WZCompEffPot}.

\section{The model}\label{sect:WZ-Model}
The Wess-Zumino model is obtained from \eqref{classNeq1Act} by turning off the
gauge fields and choosing only a single chiral field $\F$
\begin{align} \label{Action:WZ}
	S_{\WZ}[\F,\Fb] &= \intz \Fb\F 
		+ \intc\cP(\F) + \intac\bar\cP(\Fb) \,,\\
	\cP(\F) &= \frac m2 \F^2 + \frac \l6 \F^3 \ .
\end{align}
As it is not a gauge theory, the background field method yields
an identical perturbation theory to the standard construction of the effective
action (see, e.g., section 4.3 of \cite{BK}).
The functional integral equation for the effective action is
\begin{align} \label{WZEffectiveAction}
	\rme^{\frac\rmi\hbar\tilde\G[\F,\Fb]}
	= \cN \int\cD\vf\cD\bar\vf \rme^{\frac\rmi\hbar S^{(\J)}[\vf,\bar\vf]
	+\rmi\hbar^\half S_{\rm int}[\vf,\bar\vf]
	-\rmi\hbar^{-\half}\left(\vf\cdot\frac{\d\tilde\G}{\d\F}
		+\bar\vf\cdot\frac{\d\tilde\G}{\d\Fb}\right) } \,, 
\end{align}
where $\tilde\G$ was defined in \eqref{genEffectiveAction}
and we've introduced
\begin{align} 
\label{WZDefnOfPsi}
    \J	&=m+\l\F=\cP^{''}(\F) \,, \\
\label{WZFreeAction}
	S^{(\J)}[\vf,\bar\vf]&=\intz \bar\vf\vf 
		+ \half\left(\intc\J\vf^2+\cc \right) \,, \\
\label{WZIntAction}
	S_{\rm int}[\vf,\bar\vf]&=\frac\l6\intc\vf^3 +\cc
\end{align}
Equation \eqref{WZIntAction} shows that the only interactions are the
chiral cubic vertex and its complex conjugate.

To find the propagators for the model, we note that
the Hessian for the free action \eqref{WZFreeAction} is
\begin{align}\label{eqn:WZ_Hessian}
	H^{(\J)} &= \bem \J&-\quart\Db^2\\-\quart D^2&\Jb\eem\dpm .
\end{align}
Just as in section \ref{sect:PropsFeynRules}, 
we can invert the Hessian by writing \newline
\(G^{(\J)}=-H^{(0)}\cdot(H^{(\J)}\cdot H^{(0)})^{-1}\)
and using the block matrix inverse \eqref{eqn:BlockMatrixInverse} to get
\begin{align*}
	G^{(\J)} = 
	\bem
	\frac1{16}\Db^2\frac{1}{\Box-\frac1{16}\Jb D^2\frac1\Box\J\Db^2} 
		\Jb\frac1\Box D^2 \!\!\! && \!\!\!
	\frac14\Db^2\frac{1}{\Box-\frac1{16}\Jb D^2\frac1\Box\J\Db^2} \\
	\frac14 D^2\frac{1}{\Box-\frac1{16}\J \Db^2\frac1\Box\Jb D^2} \!\!\!&&\!\!\!
	\frac1{16} D^2\frac{1}{\Box-\frac1{16}\J \Db^2\frac1\Box\Jb D^2}
		\J \frac1\Box \Db^2
	\eem \dpm \,,
\end{align*}
where all derivatives act on all terms to the right and the inverses
should be understood as a power (geometric) series.
Now, writing 
\[\dpm = -\frac14\bem \Db^2 && 0\\0 && D^2 \eem\d^8\,,\]
the Green's function becomes
\begin{align*}
	G^{(\J)} &= \frac1{16}\sum_{n=0}^\infty\bem
	\Db^2(\frac1\Box\frac{\Jb D^2}{-4}\frac1\Box\frac{\J\Db^2}{-4})^n
		\frac{-1}{\Box}\frac{\Jb D^2}{-4}\Db^2 \!\!\!\!\!\!\!\!\!&&
	\Db^2(\frac1\Box\frac{\Jb D^2}{-4}\frac1\Box\frac{\J\Db^2}{-4})^nD^2
	\!\!\!\\\!\!\!\!\!\!
	D^2(\frac1\Box\frac{\J\Db^2}{-4}\frac1\Box\frac{\Jb D^2}{-4})^n \Db^2&&
	\!\!\!\!\!\!
	D^2(\frac1\Box\frac{\J\Db^2}{-4}\frac1\Box\frac{\Jb D^2}{-4})^n
		\frac{-1}{\Box}\frac{\J \Db^2}{-4} D^2
	\eem \frac{-1}\Box\d^8 \\
	&= \frac1{16}\sum_{n=0}^\infty\bem
	\Db^2(-\frac1\Box\frac{\Jb D^2}{-4}-\frac1\Box\frac{\J\Db^2}{-4})^n \Db^2 
	\!\!\! && \!\!\!
	\Db^2(-\frac1\Box\frac{\Jb D^2}{-4}-\frac1\Box\frac{\J\Db^2}{-4})^n D^2	\\ 
	D^2(-\frac1\Box\frac{\Jb D^2}{-4}-\frac1\Box\frac{\J\Db^2}{-4})^n \Db^2 
	\!\!\! && \!\!\!
	D^2(-\frac1\Box\frac{\Jb D^2}{-4}-\frac1\Box\frac{\J\Db^2}{-4})^n D^2
	\eem \frac{-1}\Box\d^8 \ .
\end{align*}
Resumming the series, we recover the result of \cite{Buchbinder1994a} 
and see that the Green's function can be written in the form
\begin{align} \label{WZ-Green's Function}
		G^{(\J)}(z,z')
	&= \bem  G^{(\J)}_{++} & G^{(\J)}_{+-} \\ 
			 G^{(\J)}_{-+} & G^{(\J)}_{--} \eem
	 = \frac1{16}\bem \Db^2\Db'^2 & \Db^2 D'^2 \\ D^2\Db'^2 & D^2D'^2\eem
	G_V^{(\J)}(z,z')\,,
\end{align}
where the auxiliary Green's function $G_V^{(\J)}$ is defined by
\begin{align} \label{WZ-AuxGreen’s Function}
	\D G_V^{(\J)} = -\d^8\,, \quad\text{with}\quad
	\D = \Box-\quart\J\Db^2-\quart\Jb D^2 \ .
\end{align}
The heat kernel representation of $G_V^{(\J)}$
is studied in section \ref{sect:WZProp}.

As seen in chapter \ref{Ch:Neq1Quant}, 
the one-loop effective action can be written as the functional determinant
\begin{align} 
	\G^{(1)} = \frac\rmi2\log\Det (H^{(\J)}/H^{(0)}) 
	=\frac\rmi2\Tr\log (H^{(\J)}/H^{(0)}) \ .
\end{align}
The argument of the $\log$ can be written as
\begin{align}
	(H^{(0)})^{-1}\cdot H^{(\J)}
	=\left(1+\frac1\Box
		\bem 0&-\quart\Db^2\Jb\\-\quart D^2\J&0 \eem\right)\dpm\,,
\end{align}
where, for the rest of this chapter, we use the convention that all derivatives 
act on all terms to the right unless bracketed or otherwise indicated.
Since only the diagonal terms survive the trace, we obtain
\begin{align} \label{WZ-1Loop-EffPot1}
	\G^{(1)}
	&=\frac\rmi4\Tr_+\log\left(1
		-\frac{\Db^2}{4\Box}\Jb\frac{D^2}{4\Box}\J\right) + \cc
\end{align}
By using the fact that the chiral trace is equivalent to the chiral projection
of the full trace $\Tr_+F_{++}=\Tr(F_{++}P_+)$ and the cyclicity of the 
functional trace (i.e., integration by parts) 
we obtain two useful forms for the one-loop effective action
\begin{subequations}\label{WZ-1Loop-FunctionalForms}
\begin{align}
 \G^{(1)}
\label{WZ-1Loop-DirectEffPot}
 &=\frac\rmi4\Tr\sum_{n=1}^\infty\frac{-1}{n}\left(
 (P_+\Jb\frac1\Box\J)^nP_+ +\cc \right) \\ 
\label{WZ-1Loop-HKEffPot}
 &=\frac\rmi2\Tr\sum_{n=1}^\infty\frac{-1}{n}\left(\frac{1}{\Box}
 \Jb\frac{D^2}{4}+\frac{1}{\Box}\J\frac{\Db^2}4\right)^n
 =\frac\rmi2\Tr\log(\frac\D\Box) \ .
\end{align}
\end{subequations}
The first form lends itself to a direct expansion of one-loop effective
potential and leads to a calculation similar to the graphical expansion 
undertaken in \cite{Pickering1996}. 
The second form, 
which can also be derived starting from \eqref{WZ-Green's Function},
is used for the heat kernel based calculations of \cite{Buchbinder1994a}.

The full quantum effective action takes the generic form 
\cite{Buchbinder1994a, BK}
\begin{align}
\begin{aligned}
	\G[\F,\Fb]&=\intz\dsL(\F,D_A\F,D_AD_B\F,\dots,\Fb,D_A\Fb,D_AD_B\Fb,\dots)\\
	&+\left(\intc\dsL_c(\F)+\cc\right)\,,
\end{aligned}
\end{align}
where $\dsL=\Fb\F+\ord\hbar$ is the effective super Lagrangian 
and $\dsL_c=\cP+\ord\hbar$ is the effective superpotential. 
For fields constant in spacetime the effective super Lagrangian decomposes into
\begin{align} 
\label{defn:WZ-KahlerPot}
	\dsL\big|_{\pd_a\F=\pd_a\Fb=0}&=K(\F,\Fb)
	+F(\F,D_\a\F,D^2\F,\Fb,\Db_\da\Fb,\Db^2\Fb)\,,
\intertext{where }
	K&=\Fb\F+\sum_{n=1}^{\infty}\hbar^n K^{(n)}\,, \\ 
\intertext{is the K\"ahler potential and}
\label{defn:WZ-AuxPot}
	F&=\sum_{n=1}^{\infty}\hbar^n F^{(n)}\,,
	\quad F\big|_{D_\a\F=\Db_\da\bar\F=0}=0\,,
\end{align}
is called the auxiliary field's effective potential. 
This is an appropriate name since, when reduced to components
in the $\pd_a\F=\pd_a\Fb=0$ background,
the auxiliary potential is of at least third order \cite{Buchbinder1994a,BK}
in the auxiliary field $f = -\frac14 D^2\F\big|_{\q=0}$. 

It is shown in section 3 of \cite{Buchbinder1994a} 
that a general term in the one-loop effective action is invariant under 
$\J\to\rme^{-2\rmi\a}\J$, $\Jb\to\rme^{2\rmi\a}\Jb$,
so each term contains an equal number of $\J$ and $\Jb$.
Their argument also shows that the one-loop effective action is
constructed from an even number of $\J\Db^2+\Jb D^2$ operators, so
all terms in the one-loop effective potential with 
$2, 6, 10,\dots$ Grassmann derivatives must vanish. 
This means that the leading order term in the one-loop auxiliary potential 
is the four-derivative term, as is also clear from 
\eqref{WZ-1Loop-DirectEffPot}.

Finally, we note that the above result 
and the following integration by parts identities
\begin{align} \label{WZ-Pot-IBP}
\begin{aligned}
	(D^2\J)^n \J^k &= -k (D^\a\J)(D_\a\J) (D^2\J)^{n-1} \J^{k-1}
		+ \text{surface terms} \,, \\
	(\Db^2\Jb)^n \Jb^k &= -k(\Db_\da\Jb)(\Db^\da\Jb)(\Db^2\Jb)^{n-1}\Jb^{k-1}
		+ \text{surface terms} \,,
\end{aligned}
\end{align}
imply that up to surface terms, 
the one-loop auxiliary potential can always be reduced to the form
\begin{align*}
	F(\F,D_\a\F,D^2\F,\Fb,\Db_\da\Fb,\Db^2\Fb)
	\approx
	(D\J)^2(\Db\Jb)^2\tilde{F}(\F,D^2\F,\Fb,\Db^2\Fb) \, .
\end{align*}

In the rest of this chapter we calculate the K\"ahler potential up to two-loops 
and the 
auxiliary potential at one-loop.
The one-loop K\"ahler potential 
and leading contribution to the auxiliary potential
were first calculated using superfields 
and functional techniques in \cite{Buchbinder1993,Buchbinder1994a} 
followed by the supergraph results of \cite{Pickering1996}.
The two-loop K\"ahler potential was first calculated using superfields in 
\cite{Buchbinder1996}. 

When all fields are massive, the (chiral) superpotential does not receive any
quantum corrections, this was one of the earliest 
supersymmetric nonrenormalisation theorems 
\cite{Wess1974b,Iliopoulos1974,Ferrara1974,Delbourgo1975,
West1976,Weinberg1976,Grisaru1979}. 
When there are massless fields present, 
finite corrections to the superpotential can exist \cite{GGRS1983}.
In the massless Wess-Zumino model, 
the first correction to the superpotential is at two-loops. 
It was originally calculated in components by 
Jack, Jones and West in \cite{West1991} 
and then using superfield methods by 
Buchbinder, Kuzenko and Petrov in \cite{Buchbinder1994,BK}.
We will not repeat the superpotential calculation in this thesis.

\section{K\"ahler potential}\label{sect:WZ-Kahler}
In the K\"ahler approximation ($D_A\F=D_A\Fb=0$) the Green's function becomes
quite simple and may be compared with \eqref{chiralPropSimp}
\begin{align} \label{WZ-KahlerProp}
	G^{(\J)}(z,z') = \frac{-1}{\Box-\Jb\J}
		\bem -\Jb & -\frac14\Db^2 \\ -\frac14D^2 & -\J \eem \dpm \ .
\end{align}
It is straight forward to check that, in the K\"ahler approximation,
\eqref{WZ-KahlerProp} really is the inverse of \eqref{eqn:WZ_Hessian}.
This form of the Green's function is the easiest to use for loop calculations,
but in the one-loop case we have more options.

\subsection{One-loop}
To calculate the one-loop K\"ahler potential we have many possible pathways.
We may use the appropriate limit ($\cP\to\J$, $\Mv\to0$) 
of \eqref{Neq1 1loop Kahler} or directly evaluate any of the three traces
(\ref{WZ-1Loop-EffPot1}, \ref{WZ-1Loop-DirectEffPot}, \ref{WZ-1Loop-HKEffPot})
by going to momentum space. 
For example, starting with \eqref{WZ-1Loop-DirectEffPot}, we commute all
of the projection operators to the far right of the trace to find
\begin{align*} 
	\G^{(1)} &= \frac{\rmi}{4}\Tr\sum_{n=1}^\infty\frac{-1}{n}
		\Big(\frac{\Jb\J}{\Box}\Big)^n\big(P_+ + P_-\big)
	= \frac{\rmi}{2}\Tr_+\log\Big(1-\frac{\Jb\J}{\Box}\Big)
		\frac1\Box \ . 
\end{align*}
The trace is now in the form of those seen in section \ref{ssect:Neq1Kahler},
so it can be evaluated in momentum space \eqref{defn:scJ_Integral}, 
to give the one-loop K\"ahler potential
\begin{align} \label{WZ-1loop-Kahler}
	K^{(1)} = \frac12\scJ(\Jb\J)
	\approx \frac{\Jb\J}{32\p^2}
		\left(\frac1\eps+2-\log\frac{\Jb\J}{\mub^2}+\ord\eps\right) \ .
\end{align}

Alternatively, we can start with \eqref{WZ-1Loop-HKEffPot} and write
the one-loop effective action as
\begin{align} \label{WZ-1Loop-HKEffAct}
	\G^{(1)} 
	= -\frac\rmi2\Tr\log(G_V^{(\J)})
	= -\frac\rmi2\int_0^\infty\frac{\rmd s}{s}\Tr U_V^{(\J)}(s) \ ,
\end{align}
where the Green's function 
$G_V^{(\J)}$ and its heat kernel expansion
are studied in section \ref{sect:WZProp}.
In the K\"ahler approximation (see subsection \ref{ssect:WZProp:Kahler})
we have 
\begin{align} 
\begin{aligned}
	U_V^{(\J)}(z,z|s) &= (\cos su - 1)(P_++P_-)\d^4(\q-\q')
		U(x,x'|s)\big|_{z'\to z} \\
		&= \frac2\Box(\cos su - 1) U(x,x'|s)\big|_{x'\to x}
	\,,
\end{aligned}
\end{align}
where $u^2=\Jb\J\Box$ and 
$U(x,x'|s)$ is the bosonic heat kernel \eqref{FreeKGHeatKern}.
This can be evaluated by either using the momentum space representation
of the $\d$-function and the first line of \eqref{FreeKGHeatKern}
or by following \cite{Buchbinder1994a} and using the defining equation
for the bosonic heat kernel
\begin{align} \label{WZ-BosonicHK-DefRel}
	\Box U(x,x'|s) = -\rmi\frac{\pd}{\pd s} U(x,x'|s)\,,
\end{align}
to perform the propertime integral without going via momentum space.
Both methods yield the result \eqref{WZ-1loop-Kahler}. 
See section 2 of \cite{WZNotebook} for more details.

\subsection{Two-loop}
\begin{figure}[t!]
  \centering
  \subfloat{\label{fig:WZ2Loop1}
	(\ref{fig:WZ2Loop}a) \includegraphics[scale=0.65]{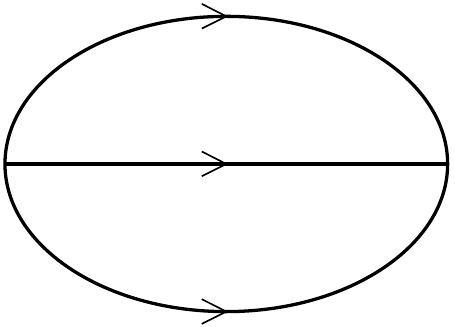}}
  \subfloat{\label{fig:WZ2Loop2}
	(\ref{fig:WZ2Loop}b) \includegraphics[scale=0.65]{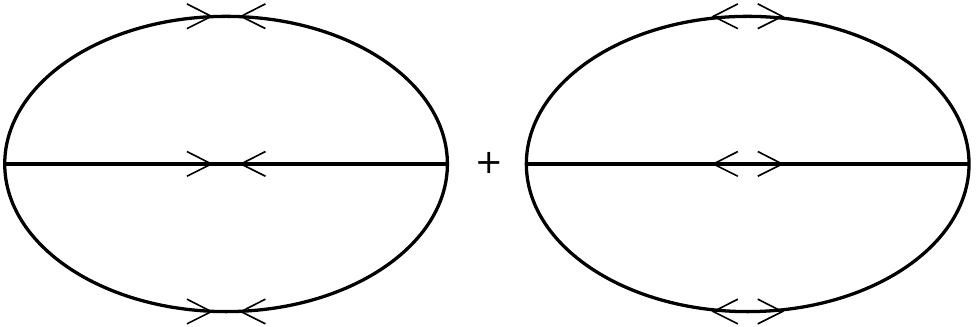}}
  \caption{Two-loop graphs for the Wess-Zumino model}
  \label{fig:WZ2Loop}
\end{figure}

In the Wess-Zumino model the only interactions are 
the (anti)chiral cubic vertices.
This means that there are no figure-eight diagrams, 
as they require a quartic vertex. 
There are two types of fish diagrams, see figure \ref{fig:WZ2Loop}.
All of the loops in figure \ref{fig:WZ2Loop2} are purely chiral (antichiral)
and thus vanish identically (in the K\"ahler approximation) due to 
the lack of Grassmann derivatives to remove the Grassmann delta-functions.
This means that the only non-zero two-loop diagram is figure \ref{fig:WZ2Loop1},
which has the analytic expression
\begin{align} \label{defn:WZ2Loop-1}
	\G^{(2)} 
	&= -\frac{\bar\l\l}{6} \intac\!\!\intc\!' \big(G_{-+}(z,z')\big)^3 \\
	&= -\frac{\bar\l\l}{6} \intz\!\!\intx\!' 
		\Big( \frac{-1}{\Box-\Jb\J}\frac{D^2\Db^2}{16}\d^8(z,z')\Big)^2
		\frac{-1}{\Box-\Jb\J}\frac{\Db^2D^2}{16}\d^8(z,z') \non 	\,,
\end{align}
where we used integration by parts to get to the second line.
Moving to momentum space, the two-loop K\"ahler potential becomes
\begin{align} \label{defn:WZ2Loop-2}
	\!\! K^{(2)} 
	&= -\frac{\bar\l\l}{6} \mu^{4\eps} \intk\intmtm{l}
		\frac1{k^2+\Jb\J}\frac1{l^2+\Jb\J}\frac1{(k+l)^2+\Jb\J} .
\end{align}
This can be written in terms of the two-loop integral
studied in appendix \ref{A:2LoopVac}. 
Using the explicit form \eqref{eqn:2LoopExpansion} we have
\begin{align} \label{WZ2LoopKahler}
	K^{(2)} 
	&= \frac{|\l\J|^2}{2(4\p)^4} 
		\left(
		-\frac1{2\eps^2}+\frac1\eps\log\Big(\frac{\J\Jb}{\muh^2}\Big)
		-\log^2\Big(\frac{\J\Jb}{\muh^2}\Big)
		- k 
		\right)\,,
\end{align}
where 
$k=\frac{\z(2)}2+\frac54+\sqrt3 N(\p/3) \approx 0.9005$
and $\muh^2$ is defined in \eqref{defn:muh}.

\subsection{Renormalisation}
In minimal subtraction, we use counterterms to remove the divergences only.
In the Wess-Zumino model, the non-renormalisation theorem means 
that the only counterterm that occurs is for the kinetic term 
and affects a wave-function renormalisation.
\begin{align} 
	S_{\text{ct}}
		= \left(Z^{(1)} + Z^{(2)} + \dots\right) \intz \Fb\F \,,
\end{align}
At one-loop, the counterterm is read from \eqref{WZ-1loop-Kahler} to be
\begin{align} 
	Z^{(1)} = \frac{-\bar\l\l}{2(4\p)^2}\left(\frac1\eps+z_1\right) \,,
\end{align}
where we've included the finite correction $z_1$.
Since $z_1=0$ in both the minimal subtraction (MS) and on-shell (OS) 
renormalisation schemes (described below), 
we will set it to zero for the rest of this analysis.
So, the renormalised one-loop result is
\begin{align} \label{WZ-1loop-Kahler-ren}
	K^{(1)}_{\text{ren}} = K^{(1)}+K^{(1)}_{\text{ct},0} 
	= \frac{\Jb\J}{32\p^2}\left(2-\log\frac{\Jb\J}{\mub^2}\right) \ .
\end{align}
The one-loop counterterm contributes to the 2-loop K\"ahler potential 
through the one-loop diagram with a two-point counterterm insertion
\begin{align*} 
	\G^{(2)}_{\text{ct},1} 
	= \raisebox{-0.4\height}{\includegraphics[scale=0.6]{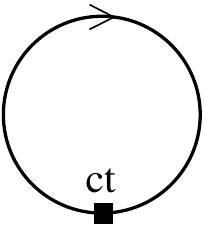}}
	= \frac\rmi2\frac{\bar\l\l}{\eps(4\p)^2}\intz G_{+-}(z,z)
	= -\frac12\frac{\bar\l\l}{\eps(4\p)^2}\intz J(\Jb\J)\ .
\end{align*}
The integral $J(m^2)$ is described in appendix \ref{A:2LoopVac} 
and its $\eps$-expansion given in \eqref{defn:J_Integral} yields
\begin{align} 
	K^{(2)}_{\text{ct},1} 
	= &\frac{|\l\J|^2}{2(4\p)^4}\Bigg(\frac{1}{\eps^2}
		+\frac1\eps\left(1-\log\frac{\Jb\J}{\mub^2}\right) \\ \non
		&+\frac12\left(2+\z(2)+\log^2\frac{\Jb\J}{\mub^2}
			-2\log\frac{\Jb\J}{\mub^2}\right)
		+\ord\eps\Bigg) .
\end{align}
This is then combined with \eqref{WZ2LoopKahler}
to give the renormalised 2-loop K\"ahler potential
\begin{align} \label{WZ-2LoopKahler-MS}
	K^{(2)}_{\text{ren}}
	&= K^{(2)} + K^{(2)}_{\text{ct},1} + K^{(2)}_{\text{ct},0} \\ \non
	&= \frac{-|\l\J|^2}{4(4\p)^4}\Bigg(
		\log^2\Big(\frac{\Jb\J}{\mub^2}\Big)
		- 4\log\Big(\frac{\Jb\J}{\mub^2}\Big)
		+ {5}-{\z(2)}+2\sqrt3N(\frac\p3)+z_2
		\Bigg) ,
\end{align}
where we've chosen the two-loop counterterm with a finite part 
$z_2$ that vanishes in MS,
\begin{align} 
	Z^{(2)} = -\frac{|\l|^4}{2(4\p)^4}
	\Big(\frac{1}{2\eps^2}-\frac{1}{2\eps}-z_2\Big)\ .
\end{align}
Up to the constant $z_2$, equation \eqref{WZ-2LoopKahler-MS} 
matches the minimal subtraction result given in \cite{Nibbelink2005}.

We can now enforce the physical/on-shell renormalisation conditions
\begin{align} \label{WZ-onshell-renorm-cond}
	\frac{\pd^2 K}{\pd\F\pd\Fb}\Big|_{\F=\F_0} = 1 \ .
\end{align}
This requires that 
\begin{align*} 
	\mub^2=|m+\l\F_0|^2=|\J_0|^2
	\quad\text{and}\quad
	z_2=\frac12+\tfrac{\z(2)}{2}-\sqrt3N(\frac\p3) \ .
\end{align*}
Then the physical K\"ahler potential up to two-loops is
\begin{align} 
\begin{aligned}
	K_{\text{phys}} &= \Jb\J\Bigg(\frac1{\bar\l\l}
	+\frac{1}{2(4\p)^2}\Big(2-\log\frac{\Jb\J}{\Jb_0\J_0}\Big) \\\non
	&-\frac{\bar\l\l}{4(4\p)^4}\Big(6-4\log\frac{\Jb\J}{\Jb_0\J_0}
		+\log^2\frac{\Jb\J}{\Jb_0\J_0}\Big)
	\Bigg)\ ,
\end{aligned}
\end{align}
where for massive theories we may choose $\F_0=0$ which implies $\J_0=m$
and for massless theories we note that $\J=\l\F$ and $\J_0=\l\F_0$.

Finally, we compare with the existing two-loop results for this calculation.
The first was the 1984 component calculation in \cite{Miller1984,Fogleman1984},
followed by a superfield calculation in 1989 \cite{SantosDos1989}%
\footnote{The ICTP preprint (ICTP/85/205) of this paper was released in 1985.}. 
In the latter, they calculated with the spurion background field 
$\F = \f + f\q^2$ which breaks the explicit supersymmetry of their result. 
The results of these papers
are expressed in terms of three different squared masses
$|m+\l \f|^2+s|\l f|$ for $s=-1,0,+1$, and the terms that are quadratic
in the auxiliary fields integrate up to the K\"ahler potential
(see the end of section \ref{ssect:WZ-FullAuxPot} for further discussion).
For now, we note that the counterterms given in the papers match 
those above for particular choices of the finite constants $z_1$ and $z_2$.

As for pure superfield calculations, 
we have already seen that the MS result \eqref{WZ-2LoopKahler-MS} 
matches the result given in \cite{Nibbelink2005}. 
The earlier superfield calculation, \cite{Buchbinder1996}, matches ours 
up to their result for the unrenormalised two-loop K\"ahler potential,
which seems to have an overall sign error. %
This error propagated into their renormalisation analysis, 
so that their renormalised results and counterterms do not match those above.

\section{Auxiliary potential}
In this section we study the auxiliary potential \eqref{defn:WZ-AuxPot},
primarily focusing on the calculation of the 
leading, four-derivative term for which conflicting results were given in
\cite{Buchbinder1994a,Buchbinder1993,Pickering1996,Pletnev1999}.
The general arguments given in section \ref{sect:WZ-Model} 
and dimensional analysis 
show that 
we can always reduce the 4-derivative term to the form
\begin{align} \label{WZ-4Deriv-Zeta}
	F\Big|_{\text{4-deriv.}} 
	= \z\, \frac{(D^\a\J D_\a\J)(\Db_\da\Jb\Db^\da\Jb)}{(4\p)^2(\Jb\J)^2}\,,
\end{align}
up to surface terms and for some numerical constant $\z$.

In \cite{Pickering1996} 
diagrammatical methods were used to find 
$\z_{\text{PW}}=\frac1{384}=\frac1{3\times2^7}$. 
The same result was given in \cite{Pletnev1999}  
on the basis a covariant expansion using a symbol operator technique.
In subsection \ref{ssect:WZ-4D-DirectExpansion}, we reproduce this 
result by direct expansion of \eqref{WZ-1Loop-DirectEffPot}.

On the other hand, in \cite{Buchbinder1994a,Buchbinder1993} 
$\z$ was left as an unevaluated integral.
Unfortunately, due to small typographical errors 
that differ between the two versions,
the integrals given for $\z$ are actually IR divergent.
In subsection \ref{ssect:WZ-4D-HK}, 
we use the results of appendix \ref{sect:WZProp} to show that
the heat kernel calculation of \cite{Buchbinder1993,Buchbinder1994a} 
yields $\z_{BKY} = -\frac1{64}$.

The reason that these two methods yield different results is 
because the final, unregularised integral is only conditionally convergent. %
As there are no renormalisation or symmetry conditions 
to impose upon the 4-derivative term,
there is no way to fix the ambiguity in the result.
However, if we use dimensional regularisation 
we consistently find the result to be $\z=\z_{PW}=\frac1{384}$, 
which also matches the component results.
Other regularisation schemes, such as analytic, zeta function and propertime 
cut-off regularisations do not fix the conditional convergence.

Finally, in subsection \ref{ssect:WZ-FullAuxPot}, 
we use dimensional regularisation and the heat kernel representation
to calculate the surprisingly simple, full one-loop auxiliary potential.

Throughout the rest of this section, we use the following notation for 
the various combinations of derivatives of the background fields
\begin{align} \label{def:WZPropShortHands_ChWZ}
	&a=(D^\a\J)(D_\a\J)\,,\quad  \abar=(\Db_\da\Jb)(\Db^\da\Jb)\,,\quad
	b=(D^2\J)\,,\quad  \bbar=(\Db^2\Jb)\,, \non\\
	&\qquad\qquad u^2=\Jb\J\Box\,,\quad
	\scF^2=\bbar b/64\,,\quad \scG^2=u^2+\scF^2 \, .
\end{align}
For more details, see appendix \ref{sect:WZProp}.

\subsection{Four-derivative term via direct expansion}
\label{ssect:WZ-4D-DirectExpansion}
In this subsection we expand the expression for the one-loop effective action
\eqref{WZ-1Loop-DirectEffPot} and only keep the 4-derivative terms.
From \eqref{WZ-1Loop-DirectEffPot}, we see that we need to examine the term
\begin{align} 
	T_n \deq (P_+\Jb\frac1\Box\J)^nP_+\,,
\end{align}
and its complex conjugate,
remembering that all derivatives, unless otherwise indicated, 
act on all terms to the right.
Since we're in the effective potential approximation 
we can commute all of the $\Box^{-1}$ terms to the left.  
As we only want to let a total of 4 derivatives hit the backgrounds fields, 
most of the chiral projectors will go straight through to the right.
However, there must be a first (from the right) 
chiral projector that hits a field, so we sum over all possibilities
\begin{align} T_n
	&= \Box^{-n}\sum_{j=0}^{n-1}(P_+\Jb\J)^{n-j-1}
	 \frac{\Db^2D^2}{16\Box}(\Jb\J)^{j+1}P_+\\ \non
	&= \Box^{-n}\sum_{j=0}^{n-1}(P_+\Jb\J)^{n-j-1}
	\Bigg( \com{\Db^2D^2}{(\Jb\J)^{j+1}}\frac{1}{16\Box} \\ \non 
	&\quad+ \com{\Db^2D^\a}{(\Jb\J)^{j+1}}\frac{D_\a}{8\Box}
	+\com{\Db^2}{(\Jb\J)^{j+1}}\frac{D^2}{16\Box} \\ \non
	&\quad+\com{\Db^\da D^\a}{(\Jb\J)^{j+1}}\frac{\Db_\da D_\a}{4\Box}
	+\com{\Db_\da}{(\Jb\J)^{j+1}}\frac{\Db^\da D^2}{8\Box} \Bigg)P_+  \\
	&\deq	T^{(1)}_n + T^{(2)}_n + T^{(3)}_n + T^{(4)}_n + T^{(5)}_n \ .
\end{align}
We evaluate each term, $T_n^{(1,\dots,5)}$, separately.  
Note that for $n=1$, only the first term exists, 
but it is a total derivative and can thus be ignored.  
We will see that all of the terms have $n$ dependent coefficients 
that are automatically zero when $n$ is too small for that term to be generated.

\subsubsection{Evaluation of \texorpdfstring{$T^{(1)}_n$}{T^1_n}}
Since all four derivatives come from a single $P_+$, the rest of the 
projection operators commute through to the right,
\begin{align*} 
	T_n^{(1)} &= \frac{1}{\Box^n}\sum_{j=0}^{n-1}\frac{(\J\Jb)^{n-j-1}}{16\Box}
			\com{\Db^2 D^2}{(\J\Jb)^{j+1}}P_+ \\
	&= \frac{(\J\Jb)^{n-2}}{16\Box^{n+1}}\sum_{j=0}^{n-1} (j+1)^2
			(\Jb\bbar + j\abar)(\J b + j a) P_+ \ .
\end{align*}
Performing the simple sums of polynomials,
we find
\begin{align*} 
	T_n^{(1)} &= \frac{(\J\Jb)^{n-2}}{16\Box^{n+1}} n (n+1) \Big(
		  \abar a\frac{(n-1)(3n^2-2)}{15} \\
		&\qquad + (\abar\J b+a\Jb\bbar)\frac{(n-1)(3n+2)}{12}  
			+ \Jb\bbar\J b\frac{2n+1}{6}	 \Big)P_+ \\
		&= \abar a \frac{(\J\Jb)^{n-2}}{16\Box^{n+1}} \frac{n(n^4-1)}{30} P_+ 
			+ \text{surface terms} \ .
\end{align*}

\subsubsection{Evaluation of \texorpdfstring{$T^{(2)}_n$}{T^2_n}}
The first projection operator provides three derivatives to give
\begin{align*} 
	T_n^{(2)} &= \frac{1}{8\Box^{n+1}}\sum_{j=0}^{n-1}(j+1)^2
		(P_+\Jb\J)^{n-j-1}\Jb^{j-1}(\Jb\bbar+j\abar)\J^j(D^\a\J)D_\a P_+ \ .
\end{align*}
Since $P_+D_\a P_+=0$, 
the final $D_\b$ to hit a field must come from the next projector on the right. 
This yields
\begin{align*} 
	T_n^{(2)} &= -\frac{(\Jb\J)^{n-2}}{8\Box^{n+1}}\sum_{j=0}^{n-1}(j+1)^2
		(\Jb\bbar+j\abar)(\J b + (j+1) a) P_+ \\
		&= \frac{-(\Jb\J)^{n-2}}{8\Box^{n+1}}
		\frac{n(n+1)}{60} \Big(\bar{a}a (n-1)\big(12 n^2+15 n+2\big)
			+5\bar{a}\J b (n-1)(3n+2) \\
		&\qquad + 15 a\Jb\bar{b} n(n+1) + 10\Jb\bar{b}\J b (2n+1) \Big)P_+\\
		&= -\abar a \frac{(\Jb\J)^{n-2}}{8\Box^{n+1}}
		\frac{(n-2)(n-1)n(n+1)(2n-1)}{60}P_+ + \text{surface terms} \ .
\end{align*}

\subsubsection{Evaluation of \texorpdfstring{$T^{(3)}_n$}{T^3_n}}
Although only two derivatives come from the first $P_+$, because $P_+D^2P_+=0$
the rest of the derivatives must come from the next projection operator,
so the evaluation of $T_n^{(3)}$ is very similar to $T_n^{(2)}$. 
The result is
\begin{align*} 
	T_n^{(3)} &= \frac{(\Jb\J)^{n-2}}{16\Box^{n+1}}
		\frac{n(n+1)(n+2)}{60} 
		\Big(3\bar{a}a (n-1)(4n+2) + 15\bar{a}\J b (n-1) \\
		&\qquad + 5 a\Jb\bar{b} (3n+1) + 20\Jb\bar{b}\J b \Big)P_+\\
		&= \abar a \frac{(\Jb\J)^{n-2}}{16\Box^{n+1}}
		\frac{(n-2)(n-1)n(n+1)(n+2)}{30}P_+ + \text{surface terms} \ .
\end{align*}

\subsubsection{Evaluation of \texorpdfstring{$T^{(4)}_n$}{T^4_n}}
One $D_\a$ and one $\Db_\da$ from the first projection operator hit fields
leaving
\begin{align*} 
	T_n^{(4)} &= \frac{1}{4\Box^{n+1}}\sum_{j=0}^{n-1}(j+1)^2
		(P_+\Jb\J)^{n-j-1}(\Jb\J)^{j}(\Db^\da\Jb)(D^a\J)\Db_\da D_\a P_+ \ .
\end{align*}
Since $\Db_\da D_\a P_+ = -2\rmi\pd_\ada P_+$, the next derivative can come
from any of the remaining projection operators. 
We sum over all possibilities and, 
after some work (see \cite{WZNotebook} for details), get the result 
\begin{align*} 
	T_n^{(4)} &= -a\frac{(\Jb\J)^{n-2}}{8\Box^{n+1}}
	\sum_{j=0}^{n-1}\sum_{k=0}^{n-j-2}
	(j+1)^2 (\Jb\bbar+(j+k+1)\abar)P_+ \\
	&= -a\frac{(\Jb\J)^{n-2}}{8\Box^{n+1}}\frac{(n-1)n(n+1)}{12}
	\big(n \Jb\bbar + \frac\abar{10}(8n^2-5n-2)\big)P_+ \\
	&= \abar a \frac{(\Jb\J)^{n-2}}{8\Box^{n+1}} 
	\frac{(n-2)(n-1)n(n+1)(2n-1)}{120}P_+ + \text{surface terms} \ .
\end{align*}

\subsubsection{Evaluation of \texorpdfstring{$T^{(5)}_n$}{T^5_n}}
The evaluation of $T_n^{(5)}$ is similar to that of 
$T_n^{(4)}$, so we leave the details to \cite{WZNotebook}.
The final result is
\begin{align*} 
	T_n^{(5)}
	&= -\frac{(\Jb\J)^{n-2}}{8\Box^{n+1}}\frac{n(n+1)(n+2)}{60}
	\big(\abar a (16 n^2 -13 n - 3) \\
	&\qquad	+ 5 \abar \J b (3n+1)
	+ 20 \Jb\bbar a (n-1) + 20 \Jb\bbar\J b \big)P_+ \\
	&= -\abar a \frac{(\Jb\J)^{n-2}}{8\Box^{n+1}} 
	\frac{(n-2)(n-1)n(n+1)(n+2)}{60}P_+ + \text{surface terms} \ .
\end{align*}

\subsubsection{Total}
Combining all of the above, we find that 
\begin{align} \label{WZ-Tn-noIBP}
	T_n &= -\frac{(\Jb\J)^{n-2}}{16\Box^{n+1}}\frac{n(n+1))}{12}\Big(
	\abar a (8n^3+5n^2-11n-2) \\\non &\qquad
	+ 2\abar\J b (3n^2+5n+4) + 2a\Jb\bbar (5n^2+3n-8) 
	+ 2\Jb\bbar\J b (4n+5) \Big) P_+ \,,
\end{align}
which becomes remarkably simple after integration by parts
\begin{align} \label{WZ-Tn-IBP}
	T_n &= \abar a \frac{(\Jb\J)^{n-2}}{16\Box^{n+1}}
		\frac{n^2(n^2-1)}{12}P_+ 
		+ \text{surface terms}\ .
\end{align}

We can now calculate the 4-derivative correction to the auxiliary potential
\begin{align*} 
	\G^{(1)}_{\text{4-deriv}} 
	&= \frac\rmi4 \Tr\sum_{n=1}^\infty\frac{-1}{n} \Big(T_n P_+ + \cc\Big) \,,
\end{align*}
Using \eqref{WZ-Tn-IBP} and 
moving to momentum space to diagonalise the trace, we have
\begin{align} \label{WZ-4D-DirectExp}
	\G^{(1)}_{\text{4-deriv}} 
	&= \frac1{(4\p)^2}\intz\frac{\abar a}{32}
		\int_0^\infty\rmd k\,\frac{k^{3}}{(k^2+\Jb\J)^4} \ .
\end{align}
Performing the momentum integral yields a result of the form
\eqref{WZ-4Deriv-Zeta} with $\z=\frac1{384}$, 
in agreement with calculations of \cite{Pickering1996} and \cite{Pletnev1999}.

If, instead, we used \eqref{WZ-Tn-noIBP}, then, provided we 
integrated by parts before performing the momentum integral, 
we obtain the same result. 
However, if we leave the integration by parts until last, then 
each of the four terms in the momentum integral are IR divergent.
In which case, the momentum integrals can be performed if, e.g.,
we regularise with dimensional regularisation.
The result is
\begin{align} \label{WZ-4D-Direct-EpsExp}
	\G^{(1)}_{\text{4-deriv}} 
	&= \frac{(4\p\m^2)^\eps}{\G(2-\eps)(4\p)^2}\intz\frac{1/96}{(\Jb\J)^2}
	\Bigg( \frac{\abar a}{2}\Big(\frac1\eps-\log(\Jb\J)-\frac{13}{2}\Big)
	\!\!\\ \non
	&+(\abar\J b+\cc)\Big(\frac1\eps-\log(\Jb\J)+1\Big)
	-\frac{\Jb\bbar\J b}{2}\Big(\frac5\eps-5\log(\Jb\J)-9\Big) 
	\Bigg).
\end{align}
Integrating by parts, the $\frac1\eps$ and $\log$ terms cancel
and we once again recover the result 
\eqref{WZ-4Deriv-Zeta} with $\z=\frac1{384}$.

\subsection{Four-derivative term via the heat kernel}\label{ssect:WZ-4D-HK}
In this subsection we re-derive the result of 
\cite{Buchbinder1994a,Buchbinder1993} and obtain $\z_{BKY}$ 
in terms of an integral constructed from 
\begin{align} \label{defn:BKY-dsJ}
	\dsJ(s) \deq \frac2s \int_0^\infty \sin(p)\rme^{-p^2/s} \rmd p
	= \sqrt{\frac\p s}\rme^{-s/4}\mathrm{erfi}(\frac{\sqrt{s}}{2})\,,
\end{align}
where $\mathrm{erfi}(z)=-\rmi\,\mathrm{erf}(\rmi z)$ 
is the so-called ``imaginary error function''.
We then show that the integrals involved are only conditionally convergent
and repeat the calculation using dimensional regularisation to obtain
an unambiguous result.

The one-loop effective action is written in terms of the heat kernel as
\begin{align} 
	\G^{(1)} = \frac\rmi2\Tr\log(\frac\D\Box)
	= -\frac\rmi2\int_0^\infty\frac{\rmd s}{s}\Tr U_V^{(\J)}(s) \,,
\end{align}	
where $U_V^{(\J)}(s)$ is studied in appendix \ref{sect:WZProp}. 
In the effective potential limit, where $\pd_a\F=\pd_a\Fb=0$, 
it reduces to
\begin{align}
	\G^{(1)} = -\frac\rmi2\intz\int_0^\infty\frac{\rmd s}{s}
		\Big(A(s)+\tilde{A}(s)\Big)U(x,x'|s)\Big|_{x'\to x} \ .
\end{align}
Using the results and notation of section \ref{ssect:WZProp:4DExp}, 
we can read off the four-derivative terms in the integrand,
\begin{align} \label{WZ-AAt-4D}
	\hspace{-1.0em} 
	A(s)+\At(s)\mathop{\Big|}_{\text{\smash{\clap{4-deriv}}}}
	&= \frac{s\abar a}{512u^3}\Big(\big(7-\frac{10}{3}s^2u^2\big)\sin(su)
		+ su\big(s^2u^2-7\big)\cos(su)\Big) \non\\
	&+ \frac{s(\Jb\bbar a+\cc)}{64u^3}\Big(su\cos(su)
		- \big(1-\frac{s^2u^2}{3}\big)\sin(su)\Big) \\ \non
	&+ \frac{s \Jb\bbar\J b}{64u^3}\Big(\sin(su)-su\cos(su)\Big)   \ .
\end{align}

A general term in \eqref{WZ-AAt-4D} is of the form $\dsA=su^{-3}\dsa(su)$ 
and its contribution to the effective potential is
\begin{align} \label{WZ-G1_dsA0}
	\G^{(1)}\big|_\dsA &= -\frac\rmi2\intz\int_0^\infty\frac{\rmd s}{s}
		\frac{s^4}{(su)^3}\dsa(su) U(x,x'|s)\Big|_{x'\to x} .
\end{align}
By using the $d$-dimensional momentum space representation of $U(x,x'|s)$ 
\begin{align*} 
	U(x,x'|s)=\m^{2\eps}\int\frac{\rmd^dk}{(2\pi)^d}
		\rme^{-\rmi k^2 s+ \rmi k(x-x')}\,,
\end{align*}
integrating out the angular parts of the momentum integral,
Wick rotating and rescaling the propertime integral, we obtain
\begin{align} \label{WZ-G1_dsA}
	\!\G^{(1)}\big|_\dsA
	&= \frac{\m^{2\eps}}{\G(2-\eps)(4\p)^{d/2}}
	\int\!\!\frac{\rmd^8z}{(\Jb\J)^{2+\eps}}
	\int_0^\infty\frac{\rmd s}{s^{1-2\eps}}
		\int_0^\infty\frac{\rmd p}{p^{2\eps}}
		\dsa(p)\rme^{-\frac{p^2}{s}}\,, \!
\end{align}
where we've defined $p = s|k|\sqrt{\Jb\J}$. 

Removing the dimensional regularisation, 
it is now straight forward to use the definition \eqref{defn:BKY-dsJ}
to perform the momentum integral in \eqref{WZ-G1_dsA}
to write the four derivative contribution as 
\begin{align*} 
	\G^{(1)}_{\text{4-deriv}} &= \frac1{64(4\p)^2}\int\frac{\rmd^8z}{(\Jb\J)^2}
	\int_0^\infty\frac{\rmd s}{s}\Bigg( 
	\frac{s \Jb\bbar\J b}{4}\Big((s+2)\dsJ(s)-2\Big) \\
	&- \frac{s(\Jb\bbar a+\cc)}{24}\Big(\big(s^2+4s+12\big)\dsJ(s)-2(s+6)\Big)
	 \non\\
	&+ \frac{s\abar a}{384}\Big(\big(3s^3+2s^2+44s+168\big)\dsJ(s)
		-2\big(3s^2+8s+84\big)\Big)\Bigg)  .
\end{align*}
Each of the three terms in the above propertime integral are IR divergent, 
but the divergences cancel when combined using integration by parts 
\eqref{WZ-Pot-IBP}. This gives a result of the form \eqref{WZ-4Deriv-Zeta}
with $\z=\z_{BKS}$ defined by the integral
\begin{align} \label{WZ-zetaBKS}
	\z_{BKS} = 
	\frac1{1024}\int_0^\infty\!\!{\rmd s}\Bigg(1-\dsJ(s)
	+\frac{s}{2}(\dsJ(s)+4)-\frac{s^2}{4}(5\dsJ(s)+1)+\frac{s^3}{8}\dsJ(s) 
	\Bigg)\,,
\end{align}
which can be compared with equation (5.15) of 
\cite{Buchbinder1994a,Buchbinder1993}.
The integral can be evaluated and yields the value $\z_{BKS} = -\frac1{64}$,
which clearly does not match the result of the previous section.

Alternatively we can integrate by parts first and use the expression
\begin{align} \label{WZ-AAtIBP-4D}
	A(s)+\At(s)\mathop{\Big|}_{\text{\clap{4-deriv}}}
	&= \frac{s\abar a}{1536 u^3}
	\Big( 3(1+2s^2u^2)\sin(su)-(3-s^2u^2)su\cos(su) \Big),
\end{align}
which holds up to surface terms. 
Proceeding to evaluate \eqref{WZ-G1_dsA} without regularisation, 
as in the last paragraph, 
we find the four-derivative correction \eqref{WZ-4Deriv-Zeta} with
\begin{align*} 
	\z = 
	\frac1{1024}\int_0^\infty\!\!{\rmd s}\Bigg(\dsJ(s) - 1
	+\frac{s}{2}(3\dsJ(s)+8/3)-\frac{s^2}{4}(3\dsJ(s)+1/3)
	+\frac{s^3}{8}\frac13\dsJ(s) 
	\Bigg) . 	
\end{align*}
This result is different from \eqref{WZ-zetaBKS} 
and evaluates to the numerical value of $\frac1{192}$ which agrees with neither
$\z_{BKS}$ or $\z_{PW}$.

The problem lies in the fact that the unregularised ($\eps\to0$)
integrals are only conditionally convergent 
and not invariant under the rescaling required to obtain \eqref{WZ-G1_dsA}.  
If we don't perform the rescaling then it makes sense to try 
to exchange the order of the  propertime and momentum integrals, 
which is useful since it leads to simpler intermediate expressions that
are free from the error functions above.
However, when the order of integration is exchanged and the integrals are not
regularised, the result changes. 
This is a clear sign of conditional convergence.

If we keep the dimensional regularisation in \eqref{WZ-G1_dsA} then we
consistently get the correction \eqref{WZ-4Deriv-Zeta} 
with $\z=\z_{PW}=\frac1{384}$. 
We demonstrate this with two possible order of operations.
First, we start with \eqref{WZ-AAtIBP-4D} and perform the propertime
integral to get
\begin{align*} 
	\G^{(1)}_{\text{4-deriv}} 
	&= \frac{\m^{2\eps}}{\G(2-\eps)(4\p)^{d/2}}
	\intz\frac{\abar a}{32}
	\int_0^\infty{\!\!\rmd k}\,\frac{k^{3-2\eps}}{(k^2+\Jb\J)^4} \ .
\end{align*}
This momentum integral is clearly equivalent to \eqref{WZ-4D-DirectExp}
and converges for $-2<\eps<2$, 
so it does not need dimensional regularisation.
We recover the result \eqref{WZ-4Deriv-Zeta} with $\z=\frac1{384}$. 
However, if we start with \eqref{WZ-AAt-4D} and leave the integration by parts
until the very end, then we definitely need the dimensional regularisation.
Once again, for simplicity, performing the propertime integral first, we find
\begin{align*} 
	\G^{(1)}_{\text{4-deriv}} 
	= \frac{(4\p\m^2)^\eps}{\G(2-\eps)(4\p)^2}
	\intz \frac{\Jb\J}{8} &\int_0^\infty \!\! \frac{\rmd k}{k^{1+2\eps}}
	\Big(\frac{\abar a}{12}\frac{5\Jb\J-4k^2}{(k^2+\Jb\J)^4} \\
	&- \frac{a\Jb\bbar+\cc}{3(k^2+\Jb\J)^3}
	+ \frac{b\bbar}{4(k^2+\Jb\J)^2}\Big)\ .
\end{align*}
The momentum integrals are IR divergent 
(i.e., in dimensional regularisation, they converge for $-2<\eps<0$)
and we get the $\eps$-expansion
\begin{align*} 
	\G^{(1)}_{\text{4-deriv}} 
	&= \frac{(4\p\m^2)^\eps}{\G(2-\eps)(4\p)^2}\intz \frac{1/96}{(\Jb\J)^2} 
	\Big(\frac{\abar a}{2}\Big(-\frac5{\eps} + 5\log(\Jb\J)-\frac{21}{2}\Big)\\
	&+ (a\Jb\bbar+\cc)\Big(\frac2\eps-2\log(\Jb\J)+3\Big)
	-\frac{3 \Jb \bbar \J b}{2}\Big(\frac1\eps-\log(\Jb\J)+1\Big) \Big) .
\end{align*}
Although the coefficients in the above are different to those in 
\eqref{WZ-4D-Direct-EpsExp}, integrating by parts still yields 
\eqref{WZ-4Deriv-Zeta} with $\z=\frac1{384}$.  

We note that in \cite{Buchbinder1994a,Buchbinder1993}, 
the action of $A(s)+\At(s)$ on $U(x,x'|s)$ 
was not evaluated by going to momentum space, 
but rather by series expansion and using \eqref{WZ-BosonicHK-DefRel}.
This leads to essentially identical results and problems to those 
discussed above. See \cite{WZNotebook} for more details of this 
and for the calculations using other regularisation schemes.

\subsection{The full, one-loop auxiliary potential}\label{ssect:WZ-FullAuxPot}
In the previous subsections, we've seen that the most robust and compact 
way to calculate the leading correction to the auxiliary potential
is to use the dimensionally regularised heat kernel, 
integrate by parts first, then perform the propertime integral 
and finally the momentum space integral.
We'll now follow this procedure 
to calculate the full one-loop auxiliary potential.

The first step is to use integration by parts to get $A(s)+\At(s)$ into
a usable form.  Starting with the results \eqref{WZ-HKC-Solutions}
we find, after some work,
\begin{align*} 
	\J C(s) + \Jb \Ct(s)
	&\approx -2\rmi\Jb\J\frac{\sin(su)}{u} 
		-\rmi\frac{\abar a}{\bbar b}\Bigg(
		\Big(\frac{s^2u^2-1}{2u}-\frac{u}{\scF^2}\Big)\sin(su) \\
		&-\frac{3s}{2}\cos(su) 
		+\frac{\scG}{\scF}\Big(
		\frac{\cos(s\scF)\sin(s\scG)}{\scF}+\frac{\sin(s\scF)\cos(s\scG)}{\scG}
		\Big)\Bigg),
\end{align*}
which can then be integrated using \eqref{Aeqn} to get
\begin{align} \label{eqn:AplusAtIBP}
	A(s)+\At(s)
	&\approx 2\Jb\J\frac{\cos(su)-1}{u} \\\non
	&+\frac{\abar a}{\bbar b}\Bigg(
	\frac{s^2}{2}\Big(\cos(su)+\frac{\sin(su)}{su}\Big)
	+\frac{\cos(s\scF)\cos(s\scG)-\cos(su)}{\scF^2}\Bigg).
\end{align}
The first term is derivative free and corresponds 
to the K\"ahler approximation discussed in sections 
\ref{sect:WZ-Kahler} and \ref{ssect:WZProp:Kahler}.
The second term contains all of the terms that generate the auxiliary potential,
starting with four derivative term \eqref{WZ-AAtIBP-4D}. 

Equation \eqref{eqn:AplusAtIBP} is an amazingly simple expression, 
considering the complexity of the results found in appendix \ref{sect:WZProp},
and is quite easily integrated to give the low-energy effective action.
The general structure is
\begin{align*} 
	\G^{(1)}
	&= \frac{\m^{2\eps}(4\p)^{-d/2}}{\G(2-\eps)}
	\intz\!\!\int_0^\infty\!\!\rmd k\, k^{3-2\eps}\!
	\int_0^\infty\!\frac{\rmd s}{s}\big(A(-\rmi s,u)+\At(-\rmi s,u)\big)
	\rme^{-k^2s}\ . 
\end{align*}
Performing the propertime integral yields
\begin{align*} 
	\G^{(1)}
	&= \frac{\m^{2\eps}(4\p)^{-d/2}}{\G(2-\eps)}
	\intz\!\!\int_0^\infty\!\!\rmd k\, k^{3-2\eps}
	\Bigg[2\Jb\J\frac{\log(1+\Jb\J/k^2)}{2k^2\Jb\J} \\
	&+ \frac{\abar a}{\bbar b} \Bigg( 
	\frac{-1}{(k^2+\Jb\J)^2} 
	+\frac{2\log(\frac{\Jb\J}{k^2}+1)
	- \log\!\big(\big(\frac{\Jb\J}{k^2}+1\big)^2-\frac{4\scF^2}{k^2}\big)}{4\scF^2}
	\Bigg)\Bigg] \, .
\end{align*}
Factorising the final logarithm term,
the momentum integral can then be evaluated to get
\begin{align*} 
	\G^{(1)}
	&= \frac{\m^{2\eps}\G(\eps)}{(4\p)^{d/2}\G(2-\eps)}
	\intz\!\!
	\Bigg[\frac{\G(1-\eps)}{2(1-\eps)}(\Jb\J)^{1-\eps}
	- \frac{\abar a}{2\bbar b} \Bigg( 
	\frac{\G(2-\eps)}{(\Jb\J)^{\eps}} \\
	&+\frac{\G(1-\eps)}{(2-\eps)}\frac{2(\Jb\J)^{2-\eps}
		-(\Jb\J+2\scF)^{2-\eps}-(\Jb\J-2\scF)^{2-\eps}}{4\scF^2}
	\Bigg)\Bigg] \,.
\end{align*}
Expanding around $d=4$ and simplifying we get our result
\begin{align}
	\G^{(1)} = \intz \big( K^{(1)} + F^{(1)} \big)\,,
\end{align}
where the K\"ahler potential $K^{(1)}$ was given in \eqref{WZ-1loop-Kahler}
and the auxiliary potential is
\begin{align} \label{WZ-fullAuxPot1}
\begin{aligned}
	(4\p)^{2}F^{(1)}
	= \frac{1}{4}\frac{\abar a}{\bbar b}\Bigg(&3 
	- \Big(1+\frac{16\Jb^2\J^2}{\bbar b}\Big)
	\log\Big(1-\frac{\bbar b}{16\Jb^2\J^2}\Big) \\
	&-\frac{16\Jb\J}{\sqrt{\bbar b}}\coth^{-1}
	\Big(\frac{4\Jb\J}{\sqrt{\bbar b}}\Big)\Bigg).
\end{aligned}
\end{align}
This has the series expansion
\begin{align} 
	(4\p)^{2}F^{(1)}
	&= \frac{\abar a}{4}\sum_{n=1}^{\infty}
	\frac{1}{n(n+1)(2n+1)}\frac{(\bbar b)^{n-1}}{(4\Jb\J)^{2n}} \\\non
	&= \frac{\abar a}{\Jb^2\J^2}\Big(
	\frac1{384}+\frac1{30720}\frac{\bbar b}{(\Jb\J)^{2}}
	+\frac1{1376256}\frac{(\bbar b)^2}{(\Jb\J)^{4}}+\dots\Big)\,,
\end{align}
where the natural expansion parameter is the dimensionless
\begin{equation}
	p^2=\frac{\bbar b}{(4\Jb\J)^2}\, .
\end{equation}

Using integration by parts to remove $\abar a$ from the auxiliary potential 
essentially requires that we integrate $F^{(1)}$ with respect to $p$ twice.
This yields an expression with dilogarithms
\begin{align} \label{WZ-fullAuxPot2}
\begin{aligned}
	F^{(1)}	&= \frac{\Jb\J}{36(4\p)^{2}}\Bigg(
	8+3p\Li_2(p)-3p\Li_2(-p) \\
	&-\frac{1}{2p^2}
	\Big((p+1)\big(11p^{2}+7p+2\big)\log(p+1) + \big(p\to-p\big)\Big)
	\Bigg)\ .
\end{aligned}
\end{align}
This is reminiscent of \cite{DeWit1996,Banin2003c,Banin2003a} where, 
for a $\cN=2$ SYM theory written in terms of $\cN=1$ superfields,
the one-loop K\"ahler potential was twice integrated to recover the
$\cN=2$ non-holomorphic potential. Their results were also expressed using
dilogarithms.


\subsection{Component projections and comparisons}\label{ssect:WZCompProj}
Choosing the background superfield $\F=\f+\q^2f$ and projecting to components
we find the above result for $K^{(1)} + F^{(1)}$ is in complete agreement with
the old component results 
\cite{Fujikawa1975,O'Raifeartaigh1976a,Huq1977,Amati1982,Grisaru1983a,
	Miller1983a,Miller1983,Fogleman1983}, 
where the contributions coming from the K\"ahler and auxiliary potentials 
are mixed in the single momentum integral
\begin{align}\label{WZCompEffPot-MtmInt}
	V^{(1)} = \frac{\rmi}{2}\int\!\!\frac{\rmd^dk}{(2\p)^d}\log
		\Big(1-\frac{|\l f|^2}{(k^2+|m+\l\f|^2)^2}\Big)\,,
\end{align}
a result that is reproduced in appendix \ref{A:WZCompEffPot}.

In fact, the above component projection may be reversed
and the entire superfield result for K\"ahler and auxiliary potentials 
can be recovered from a component calculation in the background $\F=\f+\q^2f$.
The projection of the K\"ahler potential is%
\footnote{The K\"ahler potential will actually always be a function of
$\J=m+\l\F$ and $\Jb=\bar m+\bar\l\Fb$ and never $\F$ or $\Fb$ alone.}
\begin{align} \label{WZ-KahlerProj-p^2}
	\int\!\!\rmd^4\q K(\Fb,\F)\Big|_{
		\genfrac{}{}{0pt}{}{\F=\f+\q^2f}{\pd\f=\pd f=0}}
	= \bar{f}f \pd_\f\pd_\fb K(\fb,\f)\,,
\end{align}
so the K\"ahler potential can be recovered (up to a K\"ahler transformation)
by integrating the term quadratic in the auxiliary field.
Assuming%
\footnote{
The two-loop results of \cite{Fogleman1983,Fogleman1984} 
contains a term that is cubic in the auxiliary fields so can not come from
a the projection of a function of $p^2$ like \eqref{WZ-AuxProj-p^2}. 
However, the most general structure that can occur
\begin{align*} 
	\intz\tilde{F}(\J,\Jb,b,\bbar)\Big|_{
		\genfrac{}{}{0pt}{}{\F=\f+\q^2f}{\pd\f=\pd f=0}}
	= \intx |\l f|^2 \pd_{m,\bar m}\tilde{F}(m+\l\f,\bar m+\bar\l\fb,
											-4\l f,-4\l\bar f)\,,	
\end{align*}
still allows for reconstruction of the full auxiliary potential.
}
that the projection of the auxiliary potential can be written as
\begin{align} \label{WZ-AuxProj-p^2}
	\int\!\!\rmd^4\q F(\Fb,\F,a,\abar,b,\bbar)%
		\Big|_{\genfrac{}{}{0pt}{}{\F=\f+\q^2f}{\pd\f=\pd f=0}}
	= |\l f|^2 g(p_|^2)\,,
\end{align}
where $p_|^2=\frac{|\l f|^2}{|m+\l\f|^4}$, 
and $g(p_|^2)=\sum_{n=1}^{\infty}g_n\, p_|^{2n}$,
then we can recover the auxiliary potential from the terms in the component 
potential that are of quartic and higher order in the auxiliary field.
A quick calculation yields
\[
	F(\Fb,\F,\abar,a,\bbar,b)=\frac{\abar a}{(4\Jb\J)^2}\frac{g(p^2)}{p^2} \ .
\]

This recovery of the superfield results from the component expressions 
is easily performed for the one-loop results 
and provides a good check on our results.
In principle it is also simple at any loop order.
However, in practice it is not so straightforward. 
The two main obstacles at higher loops are:
1) for the K\"ahler potential, the renormalisation scheme is important
and finite counterterms may need to be chosen in order to match the results;
2) for the auxiliary potential, it is not always simple
to expand the component expression in the auxiliary fields to obtain the
separation 
$\bar{f}f \pd_\f\pd_\fb K(\fb,\f) + |\l f|^2 g(p_|^2)$.
The existing two-loop calculations give evidence of these possible difficulties.

There are three calculations of the two-loop component effective potential 
for the Wess-Zumino model in the literature, however, 
none of them are completely satisfactory.
The first calculation \cite{Miller1984} 
only gives the result implicitly as the sum of 
eleven unevaluated two-loop integrals.
The other calculations \cite{Fogleman1983,Fogleman1984,SantosDos1989} 
actually evaluate the loop integrals and give the combined result 
in a modified minimal subtraction scheme.
Both final results contain a non-elementary function named $J(z)$ 
coming from the two-loop integral that should be
equivalent (up to $\log$ and polynomial terms) to \eqref{defn:2Loop_dsI} 
for the case of only two different squared masses with ratio $z$. 
The finite part of the two-loop integral was not well understood at that time,
so the function $J(z)$ was not explicitly defined in \cite{Fogleman1984}
and the series expansion given in \cite{SantosDos1989}
only matches my version for $z=1$.
Both calculations have terms that are either independent or linearly dependent
on the auxiliary fields. This is incompatible with the projection from 
superfields and so puts their results in question.
The unrenormalised two-loop K\"ahler potential derived from appendix C of 
\cite{Fogleman1984}, agrees with mine up to a finite term proportional to the 
classical action.
The unrenormalised two-loop K\"ahler potential derived from \cite{SantosDos1989}
does not contain a $\log^2(\Jb\J)$ term, so can not be correct.
Finally, the coefficient of the leading (4-derivative) term 
in the two-loop auxiliary potential derived from the two calculations 
do not match each other.

%

%

\chapter{Supersymmetric quantum electrodynamics}\label{Ch:SQED}

In this chapter we examine the low energy effective action of 
$\cN=1$ supersymmetric quantum electrodynamics (SQED)
obtained by integrating out the matter fields up to two loops
in a covariantly constant gauge background.
%
%
We also examine the matter sector of the one-loop effective action
which is needed for the two-loop renormalisation.

The Euler-Heisenberg (EH) effective Lagrangian \cite{Heisenberg1936} 
was the second extension of Maxwell's equations that arose in the mid 1930s.
The first being the Born-Infeld action \cite{Born1934} which is a non-linear
action designed to address the problem of the self-energy of the electron
whilst maintaining the symmetries of the electromagnetic field. 
The Born-Infeld action has turned out to play a big role in 
the low-energy effective actions of string theories 
(see \cite{Tseytlin1999} and references within)
and it also occurs naturally in the class of self-dual $U(1)$ gauge theories.
In subsection \ref{ssect:BI}
we shall see how it can be related to broken supersymmetry.
The EH action, on the other hand, 
is the seminal work on low-energy quantum effective actions. 
Euler and Heisenberg obtained their effective Lagrangian
by solving the Dirac equation in a constant electromagnetic field 
then working backwards to the Lagrangian.  
Their work was clarified, and also extended to the case of 
scalar matter fields, by Weisskopf in \cite{Weisskopf1936}.
A good historical discussion of these early calculations is available in 
the article ``The Heisenberg-Euler Effective Action: 75 years on''
\cite{Dunne2012}.

In the elegant paper \cite{Schwinger1951}, Schwinger moved the calculations
of \cite{Heisenberg1936,Weisskopf1936}
to the language of quantum field theory and rederived the results 
using functional techniques 
and propertime representations of the Green's functions.
Schwinger's paper, which starts with 
``\emph{the elementary remark that the extraction of gauge invariant results 
from a formally gauge invariant theory is ensured if one employs methods
of solution that involve only gauge covariant quantities}'', 
formed the basis of the background field method 
used throughout this thesis, 
see references and discussion in \cite{Kuzenko2003a}.

The one-loop results for spinor and scalar QED 
\cite{Heisenberg1936,Weisskopf1936,Schwinger1951} 
were extended to two-loops in 1975 by Ritus \cite{Ritus1975,Ritus1977}.
Further two-loop analysis has been performed by many groups 
using a variety techniques, see, e.g., 
\cite{Dittrich1985,Reuter1996,Fliegner1997,Kors1998}.
The calculation of the one-loop EH effective action for supersymmetric SQED 
(and SYM) has been performed many times 
\cite{Shizuya1984, Ohrndorf1986a,McArthur1997,
		Pletnev1999,Buchbinder1999,Kuzenko2003}
but, prior to the paper that this chapter is based on, \cite{Kuzenko2007a}, 
had only been extended to two-loops for the $\cN=2$ case in \cite{Kuzenko2003}.
Below, we reproduce and elaborate the $\cN=1$ calculation of \cite{Kuzenko2007a}
and so, unlike the other chapters in this thesis, we regularise the 
loop integrals with a propertime cut-off.

\section{Classical action and quantisation}\label{sect:SQEDQuant}
The classical action for $\cN=2$ SQED is
\begin{align} \label{defn:SQED2-Classical-Action}
\begin{aligned}
	S_{\SQED}^{\cN=2} 
	&= \frac1{e^2}\intz\Fb\F+\frac1{e^2}\intc W^\a W_\a \\
		&+ \intz\big(\bar Q_+ Q_+ + \bar Q_- Q_-\big)
		+ \Big(\intc\big(Q_+ \F Q_-\big) + \cc\Big)\,,
\end{aligned}
\end{align}
where $W_\a$ and $\F$ form the $\cN=2$ gauge multiplet
and the covariantly chiral $Q_\pm$ form the hypermultiplet.
The action for $\cN=1$ SQED is obtained from \eqref{defn:SQED2-Classical-Action}
by making the field $\F$ non-dynamical and fixing its value to the mass $m$
\begin{align} \label{defn:SQED1-Classical-Action}
\begin{aligned}
	S_{\SQED}^{\cN=1} 
	&= \frac1{e^2}\intc W^\a W_\a \\
		&+ \intz\big(\bar Q_+ Q_+ + \bar Q_- Q_-\big)
		+ \Big(m\intc\big(Q_+Q_-\big) + \cc\Big) \ .
\end{aligned}
\end{align}
The matter fields $Q_\pm$ are in representations of the $U(1)$ gauge group
with $\pm1$ charge respectively. 
The gauge fields $W_\a$ and $\F$ have charge zero.
This needs to be taken into account when using 
the covariant $\cD$-algebra \eqref{Neq1alg}.

As discussed in chapter \ref{Ch:Neq1Quant}, when quantising 
using the background field method and 't Hooft gauge, 
in order to get simple, local chiral propagators, 
we need the condition $\cP\Mf=\Mf\bar\cP=0$ to hold.
In the $\cN=1$ SQED case
this means we must have either vanishing
mass, $m=0$, or vanishing background matter fields, $Q_\pm=0$. 
For the EH calculations in the following sections, the latter condition holds.
In the final section of this chapter 
(section \ref{sect:SQED-1loopMatter})
we calculate the one-loop matter sector of the effective action,
so unless we restrict our attention to the massless case, the propagators
are necessarily not minimal.
Nevertheless, at one-loop, 
we can extract the two-point function and the K\"ahler potential,
both of which can be evaluated in an arbitrary $R_\xi$ gauge.
At higher loops, local propagators only occur when expanding in the 
field strength to calculate $n$-point functions.

We follow the background field quantisation procedure given 
in chapter \ref{Ch:Neq1Quant} 
and note that since there are no background matter fields, 
the 't Hooft gauge reduces to the standard Fermi-Feynman gauge.
Also, since we have an abelian gauge group, the ghost dynamics completely 
decouple from the background and can be ignored.
The result is the following action that is to be used for loop calculations
\begin{align} \label{eqn:Neq1SQED-quant-action}
\begin{aligned}
	S^{(2)}+S^{(\rm int)}
	&= -\frac12\intz v \square v
		+ \intz\Big(\bar q_+ \rme^{ev} q_+ + \bar q_- \rme^{-ev} q_-\Big) \\
		&+ \Big( m \intc q_+q_- + \cc \Big) \ .
\end{aligned}
\end{align}
It yields the propagators
\begin{subequations}\label{eqn:Neq1SQED-props}
\begin{alignat}{3} 
	\rmi\expt{v(z)v(z')} &= -G_0(z,z') &
	&= \raisebox{-0.4\height}{\includegraphics[scale=0.5]{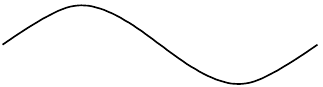}}\\
	\rmi\expt{q_+(z)q_-(z')} &= -m G_+(z,z') &
	&= \raisebox{0.1\height}{\includegraphics[scale=0.5]{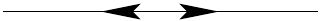}}\\
	\rmi\expt{\bar q_-(z) \bar q_+(z')} &= -m G_-(z,z') &
	&= \raisebox{0.1\height}{\includegraphics[scale=0.5]{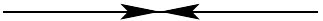}}\\
	\rmi\expt{q_+(z)\bar q_+(z')} &= G_{+-}(z,z') &
	&= \raisebox{0.1\height}{\includegraphics[scale=0.5]{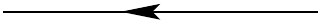}}\\
	\rmi\expt{\bar q_-(z) q_-(z')} &= G_{-+}(z,z') &
	&= \raisebox{0.1\height}{\includegraphics[scale=0.5]{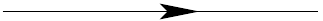}} \ ,
\end{alignat}
\end{subequations}
where the gauge field propagator, $G_0(z,z')$, 
is $\d^4(\q-\q')$ times the bosonic propagator discussed in
section \ref{sect:BosonicProp} in the massless limit.
The interactions are easily read from \eqref{eqn:Neq1SQED-quant-action} 
by expanding in the quantum fields,
\linebreak
$S^{(int)} = \sum_{n=1}^\infty \sum_{\pm} S^{(int)}_{n,\pm}$
where
\begin{align} 
	S^{(int)}_{n,\pm} 
	= \frac{(\pm e)^n}{n!}\intz \qbar_\pm q_\pm v^n
	= \raisebox{-0.5\height}{\scalebox{0.6}{
  		\begin{picture}(0,0)%
\includegraphics{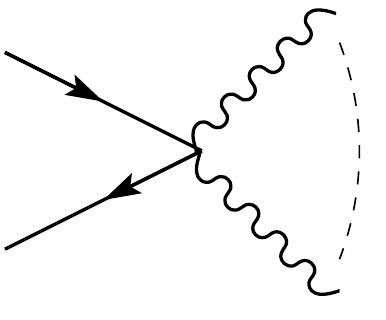}%
\end{picture}%
\setlength{\unitlength}{4144sp}%
\begingroup\makeatletter\ifx\SetFigFont\undefined%
\gdef\SetFigFont#1#2#3#4#5{%
  \reset@font\fontsize{#1}{#2pt}%
  \fontfamily{#3}\fontseries{#4}\fontshape{#5}%
  \selectfont}%
\fi\endgroup%
\begin{picture}(1747,1452)(3129,-3757)
\put(3376,-2536){\makebox(0,0)[lb]{\smash{{\SetFigFont{14}{16.8}{\rmdefault}{\mddefault}{\updefault}{\color[rgb]{0,0,0}$\pm$}%
}}}}
\put(3376,-3661){\makebox(0,0)[lb]{\smash{{\SetFigFont{14}{16.8}{\rmdefault}{\mddefault}{\updefault}{\color[rgb]{0,0,0}$\pm$}%
}}}}
\put(4861,-3031){\makebox(0,0)[lb]{\smash{{\SetFigFont{14}{16.8}{\rmdefault}{\mddefault}{\updefault}{\color[rgb]{0,0,0}$n$}%
}}}}
\end{picture}%
}} \quad .
\end{align}
Only the $n=1,2$ interactions can occur in two-loop 1PI diagrams 
and the quartic ($n=2$) interaction does not contribute 
in the Fermi-Feynman gauge.
The arrows in the above Feynman rules show the ``flow of chirality'' 
in the diagrams.

\section{One-loop Euler-Heisenberg effective action}
As mentioned above, the one-loop Euler-Heisenberg effective action for 
SQED has been calculated in many other places 
\cite{Shizuya1984, Ohrndorf1986a, McArthur1997,
		Pletnev1999, Buchbinder1999, Kuzenko2003}. 
We repeat the calculation here only for the sake of completeness 
and in order to establish some notation. 

As seen in section \ref{sect:Neq1-1Loop}, 
the unrenormalised one-loop effective action can be written as
\begin{align} \label{eqn:SQED-1LoopEA}
	\G_\text{unren}^{(1)} = \rmi\log\Det_+(\square_+-m^2) 
	= -\rmi\int_{s_0}^\infty\frac{\rmd s}{s}
		\Tr_+ K_+(s) \rme^{-\rmi(m^2-\rmi\e)s}\ ,
\end{align}
where $\e\to0$ is the standard $\rmi\e$ condition and
$K_+(s)$ is the 
four dimensional
chiral heat kernel \eqref{chiral_heat_kernel} 
evaluated in the coincidence limit, i.e.,
$K_+(s)=K_+(z,z|s)$.
To evaluate this coincidence limit, the first result we need is
\begin{align} 
	U(s) \z^2 U(-s)\Big|_{\z\to0} =
	\z(s)^2\Big|_{\z\to0} = 2 W^2 \,\frac{\cos(sB)-1}{B^2} \ .
\end{align}
Since the above expression contains a factor of $W^2$,
it prevents any further terms arising from the action of $U(s)$ 
from contributing in the coincidence limit $K_+(s)$.
Then we simply need the result for the Lorentz determinant 
\begin{align} \tag{\ref{eqn:LorentzDet-CovConstBG}}
	\det\!\Big(\frac{s\cF}{\sinh s\cF}\Big)^{\half}
	= \frac{s \l_+}{\sin(s \l_+)}\frac{s \l_-}{\sin(s \l_-)}
	= -\frac{s^2}2\frac{B^2-\Bb^2}{\cos(sB)-\cos(s\Bb)}\,,
\end{align}
to get the limit for the heat kernel
\begin{align} 
	K_+(s) &= -\frac{\rmi s^2}{(4\p\rmi s)^{2}} W^2    \non 
		\frac{B^2-\Bb^2}{\cos(sB)-\cos(s\Bb)}	\frac{\cos(sB)-1}{B^2} \\
		&= \frac{\rmi}{(4\p)^{2}} W^2
		\Big(1+\Bb^2\frac{B^{-2}(1-\cos(sB))-\cc}
		                 {\cos(sB)-\cos(s\Bb)}\Big) \, .
\end{align}
The second line has been arranged 
to isolate the divergent part from the 
unrenormalised one-loop effective action
\begin{align} \label{eqn:SQED-1LoopEA-unren}
	\G_{\text{unren}}^{(1)} 
	&= \frac{1}{(4\p)^2}\intc W^2 
		\int_{s_0}^\infty\frac{\rmd s}{s} \rme^{-\rmi(m^2-\rmi\e)s} \\ 
	&-\frac1{(4\p)^2}\intz W^2\Wb^2\int_0^\infty\frac{\rmd s}{s}
		\frac{B^{-2}(1-\cos(sB))-\cc}
		                 {\cos(sB)-\cos(s\Bb)}
		 \rme^{-\rmi(m^2-\rmi\e)s}\ .	\non
\end{align}
The integral in the second line is finite, 
so we have removed the regularisation 
and have also used the fact that for an on-shell background, 
$B^2=\frac14 D^2W^2$.
The propertime integral in the first line of \eqref{eqn:SQED-1LoopEA-unren} 
is divergent as $s_0\to0$ and is absorbed into the renormalisation of $e^2$.
We leave further discussion of the renormalisation to 
section \ref{sect:SQED_Renorm}.

\section{Two-loop Euler-Heisenberg effective action}\label{sect:SQED_2Loop}
\begin{figure}[t!]
  \centering
  \hfill
  \subfloat{\label{fig:SQED2Loop1}
	(\ref{fig:SQED2Loop}I) \includegraphics[scale=1]{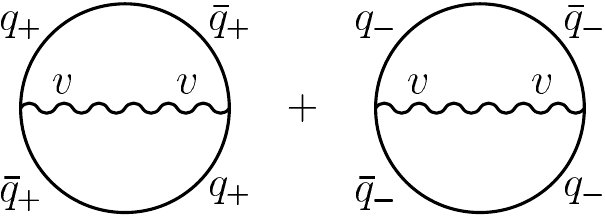}}
  \hfill
  \subfloat{\label{fig:SQED2Loop2}
	(\ref{fig:SQED2Loop}II) \includegraphics[scale=1]{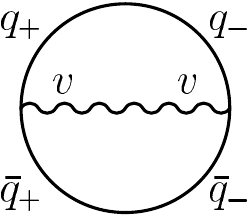}}
  \hfill
  \caption{Two-loop graphs for $\cN=1$ SQED}
  \label{fig:SQED2Loop}
\end{figure}

As shown in figure \ref{fig:SQED2Loop}, there are two non-zero
two-loop 1PI supergraphs%
\footnote{There is a third 1PI supergraph, 
the so-called `figure eight' graph, whose contribution is easily seen
to be zero in the Fermi-Feynman gauge.} 
that contribute to the two-loop effective action.
The first graph contributes
\begin{align} \label{eqn:SQED-2loopI}
	\G^{(2)}_{\rmI} = e^2 \intz\!\!\intz'\,
	 G_0(z,z')G_{-+}(z,z')\,G_{-+}(z',z)\,,
\end{align}
whilst the contribution from the second diagram is
\begin{align} \label{eqn:SQED-2loopII}
	\G^{(2)}_{\rmII} = - e^2 m^2 \intz\!\!\intz'\,
	  G_0(z,z')G_+(z,z')\,G_-(z',z) \ .
\end{align}

In $\cN=2$ SQED there is a third diagram that comes from the chiral vertex
$\intc\big(Q_+ \F Q_-\big)$ and its complex conjugate. This extra diagram
gives the identical contribution as $\G^{(2)}_{\rmI}$, 
except for the replacement $G_{-+}(z',z) \to G_{+-}(z',z)$, which 
as shown in \cite{Kuzenko2003}, can be combined with $\G^{(2)}_{\rmI}$
to give a dramatic simplification in the two-loop calculation.
For more details see \cite{Kuzenko2003,Kuzenko2007a}.

We can now use the delta function in the gauge propagator to 
integrate out the $\q$ and $\q'$ Grassmann variables in $\G^{(2)}_{\rmI}$. 
Shifting the remaining spatial integration variables 
via the rule $\{x,x'\}\to\{x,\rho\}$, 
where $\r$ is defined in \eqref{defn:SusyIntervals}, we have
\begin{align} \label{eqn:SQED-2loopIa}
\begin{aligned}
	\!\G^{(2)}_{\rmI} = \rmi e^2 \intz\!\!\int\!\!\rmd^4\r\!\! 
	&\int_{0+}^\infty\!\!\rmd s\,\rmd t\,\rmd u\, K_{\text{bos}}(\r|u) \\
	&\times K_{-+}(z,z'|s)\,K_{-+}(z',z|t)
	 \rme^{-\rmi(m^2-\rmi\e)(s+t)}\Big|_{\z\to0}\,, \!\!\!
\end{aligned}
\end{align}
where the propertime integrals are taken to be cutoff at small propertimes.
In \eqref{eqn:SQED-2loopIa} we use the bosonic heat kernel in four dimensions
\begin{align} \tag{\ref{FreeKGHeatKern}}
	K_{\text{bos}}(\r|u) &= \frac{\rmi}{(4\p\rmi u)^{2}} 
		\rme^{\rmi\frac{\r^2}{4u}} \ .
\end{align}
So, we see that to calculate \eqref{eqn:SQED-2loopIa}, we first need 
to calculate the antichiral-chiral heat kernel \eqref{K-+} 
in the Grassmann coincidence limit $(\zeta_\a,\ \bar\z_\da)\to 0$.

The easiest way to calculate the coincidence limit is to break the 
exponential in \eqref{K-+} into two parts, the first being 
\begin{align*} 
	U(s)\exp\Big(\frac\rmi4\rt\cF\coth s\cF\rt\Big)U(-s)\Big|_{\z\to0}
\end{align*}
which is evaluated using the result
\begin{align*} 
	\rt_\ada(s)\Big|_{\z\to0} = \r_\ada
	+ 4\int_0^s\Big(\frac{\rme^{\rmi \t N}-\ds1}{N}\Big)_{\!\!\a}^{\;\b} W_\b
		\Wb_\db \Big(\rme^{\rmi \t \Nb}\Big)^{\!\!\db}_{\;\da} \ \rmd t \ .
\end{align*}
The second term in the exponential, \eqref{R(z,z')}, 
is most simply evaluated when combined with the $I(z,z')$ term, because
\begin{align} 
	&U(s)\Big(\rme^{R(z,z')}I(z,z')\Big) \\ \non 
	&= \rme^{R(z,z')+\rmi\int_0^sU(t)\Big(
	2\z^2\Wb^2-\z^2\zb\Nb\Wb + \frac\rmi2\big(\z\r\Nb\Wb+\z N\r\Wb\big)
	\Big)U(-t)\rmd t}I(z,z')\,,
\end{align}
the coincidence limit of which is straightforward, but tedious, to calculate.
The combined result is
\begin{align} 
	K_{-+}(z,z'|s)\Big|_{\z=0} &= \frac{\rmi}{(4\pi\rmi s)^{2}}
	\det\!\left(\frac{sF}{\sinh sF}\right)^{\!\half}
	\exp \Big(	\frac\rmi{4}\rho\,F\!\coth(sF)\,\rho \non \\ 
	&-\rmi W^\b \rho_a f^{a}_{~\b\db}(s)\Wb^\db-\rmi W^2\Wb^2 f(s)
	\Big)	I(z,z')\Big|_{\z=0} \,,
\end{align}
where
\begin{align}\label{defn:SQED f(s)}
	f^a{}_{\bdb}(s) &= \half(\ds1-\coth(s F))^a{}_b
	\Big(\frac{\rme^{-\rmi sN}-\ds1}{N}\s^b\rme^{-\rmi s\Nb}
	+\s^b\frac{\rme^{-\rmi s\Nb}-\ds1}{\Nb}\Big)_{\b\db}\,, \non \\
	f(s)&=4\frac{B\sin(sB)\sin^2(s\Bb/2)-\Bb\sin(s\Bb)\sin^2(sB/2)}
	{B^2\Bb^2(\cos(sB)-\cos(s\Bb)} \ .
\end{align}

The coincidence limit of $K_{-+}(z',z|t)$ is simply obtained 
from the above result via the obvious replacements $z\leftrightarrow z'$ 
and $s\to t$.  
Then, by pushing the parallel displacement operator 
through to the left using \eqref{eqn:CovTaylor-OnShell}, 
we can combine the two heat kernels to get
\begin{align}
	\G^{(2)}_{\rmI} 
	&= \frac{e^2}{(4\pi)^6}\intz\!\int\rmd^4\r\! 
	\int\limits_{0+}^\infty \frac{\rmd s\rmd t\rmd u}{(stu)^{2}} 
	P_+P_- \\\non
	&\times\rme^{\frac\rmi4\rho A\rho
	-\rmi W^\b \rho_a \left(f^{a}_{~\bdb}(s)-f^{a}_{~\bdb}(t)\right)\Wb^\db
	-\rmi W^2\Wb^2 \left(f(s)+f(t)\right)}\rme^{-\rmi (m^2-{\rmi}\e)(s+t)}~,
\end{align}
where the parallel displacement operators have annihilated each other, 
in accordance with \eqref{eqn:IAnnihilate}.
In the above we've introduced the symbols
\begin{align} \label{SQED-Ppm-defn}
 	P_\pm &= \frac{s\l_\pm}{\sinh(s\l_\pm)}\frac{t\l_\pm}{\sinh(t\l_\pm)}\,, \\
 	\label{SQED-A-defn}
	A &= F\coth(sF)+F\coth(tF)+\frac{1}{u} \ ,
\end{align}
where the $P_\pm$ come from the determinant \eqref{eqn:LorentzDet-CovConstBG}.

All $\rho$ dependence is now explicit in the exponential, 
so we can complete the square and perform the Gaussian integral which yields
\begin{align}\label{SQED-rho-int}
	\frac{-1}{(4\pi\rmi)^{2}} 
	\int\!\!\rmd^4\r\, 
	\rme^{\frac\rmi4\rho A\rho- 
	\rmi W^\b \rho_a \left(f^{a}{}_{\bdb}(s)-f^{a}{}_{\bdb}(t)\right)\Wb^\db}
	=\frac\rmi{\sqrt{\det A}}\rme^{-\rmi W^2\Wb^2\scF(s,t,u)} \,, 
\end{align}
where 
\begin{align} \label{eqn:scFdef}
	\scF(s,t,u)=\quart\left(f^{a}{}_{\g\dg}(s)-f^{a}{}_{\g\dg}(t)\right) 
	{(A^{-1})_a}^{b} \left(f_b{}^{\dg\g}(s)-f_b{}^{\dg\a}(t)\right) \ .
\end{align}
Recalling the eigenvalues of $F$, \eqref{eqn:covConstF-eigenvalues}, 
we find 
\begin{align} 
	\frac1{\sqrt{\det A}} &= \frac1{(a_++u^{-1})(a_-+u^{-1})} \,, \\ 
	\label{SQED-apm-defn}
	a_\pm &= \l_\pm\coth(s\l_\pm)+\l_\pm\coth(t\l_\pm)~. 
\end{align}
Equation \eqref{eqn:scFdef} can be evaluated with the help of 
\eqref{defn:cN}, \eqref{eqn:covConstOnShell-TraceN} and the identity
\begin{align}
	\left(\coth(s\l_\pm)+1\right)\left(\coth(t\l_\pm)-1\right)=
	-\frac{\rme^{\rmi\frac{B\pm\Bb}2(s-t)}}
	{\sin(s\frac{B\pm\Bb}2)\sin(t\frac{B\pm\Bb}2)}~.
\end{align}
After some work it yields
\begin{align}
	\scF(s,t,u)=\frac{\scF_+}{a_++u^{-1}}+\frac{\scF_-}{a_-+u^{-1}} ~,
\end{align}
with $\scF_+ \xrightarrow{\Bb \to -\Bb} \scF_-$ and, taking 
advantage of the integrand's $s\leftrightarrow t$ symmetry,
\begin{align}\label{defn:scF+}\!\!\!
	&\scF_+ = 2\frac{B^2\sin^2(\frac{s\Bb}2)+(B\leftrightarrow\Bb)
	+2B\Bb\cos(s\frac{B+\Bb}2)\sin(\frac{sB}2)\sin(\frac{s\Bb}2)}%
		{B^2\Bb^2\sin^2(s\frac{B+\Bb}2)}\\\non
	&-\frac{B^2\cos(B\frac{s-t}2)\sin(\frac{s\Bb}2)\sin(\frac{t\Bb}2)+
	\mbox{\scriptsize $(B\leftrightarrow\Bb)$}+
		2B\Bb\cos(\frac{sB+t\Bb}2)\sin(\frac{tB}2)\sin(\frac{s\Bb}2)}
		{\half B^2\Bb^2\sin(s\frac{B+\Bb}2)\sin(t\frac{B+\Bb}2)} .
\end{align}

Since $W_\a W_\b W_\g=0$ we get a simple, terminating expansion for the 
remaining exponential in $\G^{(2)}_{\rmI}$,
\begin{align*} 
	\rme^{-\rmi W^2\Wb^2(f(s)+f(t)+\scF(s,t,u))}
	=1-\rmi \, W^2\Wb^2\Big(f(s)+f(t)+\scF(s,t,u)\Big) \ .	
\end{align*}
Here the first term does not contribute to the Euler-Heisenberg sector of the 
effective action (it actually leads to  higher derivative quantum corrections),
so the penultimate form for the first supergraph is
\begin{align}\label{eqn:SQED-2loopIb}
	\G^{(2)}_{\rmI} &= \frac{e^2}{(4\pi)^4} 
	\intz W^2\Wb^2\!\!\int\limits_{0+}^{\infty}\! 
	\frac{\rmd s\rmd t\rmd u}{(stu)^{2}}\frac{P_+}{a_++u^{-1}} 
	\frac{P_-}{a_-+u^{-1}}\times \\\non
	&\times\left(f(s)+f(t)+\frac{\scF_+}{a_++u^{-1}}
	+\frac{\scF_-}{a_-+u^{-1}}\right)
	\rme^{-\rmi (m^2 -\rmi\e) (s+t)}\ .
\end{align}

The second supergraph is identical to one calculated in \cite{Kuzenko2003}, 
we repeat the calculation here for the sake of completeness.
Since the propagators in \eqref{eqn:SQED-2loopII} 
have fewer derivatives than those in \eqref{eqn:SQED-2loopI}, 
the calculation of this supergraph is a lot simpler. 

Performing the same steps as those leading up to \eqref{eqn:SQED-2loopIa},
we find
\begin{align} \label{eqn:SQED-2loopIIa}
\begin{aligned}
	\!\G^{(2)}_{\rmII} = \rmi m^2 e^2 \intz\!\!
	\int\!\!\rmd^4\r\!\! 
	&\int_{0+}^\infty\!\!\rmd s\,\rmd t\,\rmd u\, K_{\text{bos}}(\r|u) \\
	&\times K_{+}(z,z'|s)\,K_{-}(z',z|t)
	 \rme^{-\rmi(m^2-\rmi\e)(s+t)}\Big|_{\z\to0}\,, \!\!\!
\end{aligned}
\end{align}
The Grassmann coincidence limit of the chiral heat kernel 
is easily calculated to be
\begin{align} 
	K_+(z,z'|s)\Big| 
	=  \frac{-4\rmi W^2}{(4\p\rmi s)^{2}} 
		\det\!\left(\!\frac{sF}{\sinh sF}\!\right)^{\!\!\half}
		\frac{\sin^2(sB/2)}{B^2}\rme^{\frac\rmi{4}\rho\,F\!\coth(sF)\,\rho}
		I(z,z')\Big| 
		\,,
\end{align}
with the antichiral heat kernel 
obtained through simple replacements that yield
\begin{align} 
	K_-(z',z|t)\Big| 
	=  \frac{-4\rmi \Wb'^2}{(4\p\rmi t)^{2}} 
		\det\!\left(\!\frac{tF}{\sinh tF}\!\right)^{\!\!\half}
		\frac{\sin^2(t\Bb/2)}{\Bb^2}\rme^{\frac\rmi{4}\rho\,F\!\coth(tF)\,\rho}
		I(z',z)\Big| 
		\ .
\end{align}
The parallel displacement propagator in $K_-$ can be pushed through
to the left with its only affect being $\Wb'\to\Wb$. 
It can then annihilate with the corresponding term in $K_+$, 
so that the effective action contribution becomes
\begin{align} \label{eqn:SQED-2loopIIaa}
	\!\G^{(2)}_{\rmII} = m^2\frac{e^2}{(4\p)^6} 
	&\intz\!W^2\Wb^2
	\int\!\!\rmd^4\r\,\rme^{\frac{\rmi}{4}\r A \r} 
	\int\limits_{0+}^\infty\frac{\rmd s\rmd t\rmd u}{(stu)^{2}} P_+ P_-\\\non
	&\times\Big(\frac{\sin(sB/2)\sin(t\Bb/2)}{(sB/2)(t\Bb/2)}\Big)^2
	 \rme^{-\rmi(m^2-\rmi\e)(s+t)}\Big|_{\z\to0} \ .
\end{align}
The Gaussian $\r$-integral can be performed as in \eqref{SQED-rho-int}, 
and we get the penultimate form for the second supergraph
\begin{align} \label{eqn:SQED-2loopIIb}
	\!\G^{(2)}_{\rmII} = \frac{e^2}{(4\p)^4} 
	\intz\!W^2\Wb^2
	&\int\limits_{0+}^\infty\!\frac{\rmd s\rmd t\rmd u}{(stu)^{2}} 
	\frac{P_+P_-\,T(s,t)\rme^{-\rmi(m^2-\rmi\e)(s+t)}}%
		{(a_++u^{-1})\,(a_-+u^{-1})}	\ ,
\end{align}
where
\begin{align} \label{SQED-T(s,t)-defn}
	T(s,t) = \frac{-8\rmi m^2}{B^2\Bb^2}
	\Big(\sin^2\big(\frac{sB}{2}\big)\sin^2\big(\frac{t\Bb}{2}\big) 
		+ s\leftrightarrow t\Big)\ .
\end{align}

Finally, the two propertime $u$-integrals in 
\eqref{eqn:SQED-2loopIb} and \eqref{eqn:SQED-2loopIIb}
can be evaluated in closed form 
and are identical to those considered by Ritus \cite{Ritus1975}.
Their direct evaluation gives 
\begin{align}
	&\int\limits_{0}^{\infty}\frac{\rmd u}{u^2}\frac1{(a_++u^{-1})(a_-+u^{-1})}
		=\frac1{a_+-a_-}\log\left(\frac{a_+}{a_-}\right) \,, \\
	&\int\limits_{0}^{\infty}\frac{\rmd u}{u^2}\frac1{(a_++u^{-1})(a_-+u^{-1})}
		\left(\frac{\scF_+}{a_++u^{-1}}+\frac{\scF_-}{a_-+u^{-1}}\right)
		\non \\ &\qquad \qquad 
		=\frac1{a_+-a_-}\left(\frac{\scF_-}{a_-}-\frac{\scF_+}{a_+}\right)+
	 \frac{\scF_+-\scF_-}{(a_+-a_-)^2}\log\left(\frac{a_+}{a_-}\right) \ .
\end{align}

We can now write down the the complete, 
unrenormalised two-loop effective action
\begin{align} \label{eqn:SQED-2loop-unren}
	\G^{(2)}_{\text{unren}} &= \frac{e^2}{(4\p)^4} 
	\intz W^2\Wb^2 \!
	\int_{0+}^\infty\frac{\rmd s\rmd t}{(st)^{2}}\frac{P_+P_-}{a_+-a_-} 
	F(s,t) 	\rme^{-\rmi(m^2-\rmi\e)(s+t)} \,, \\
\intertext{\smash{where}}
\label{SQED-defn-F(s,t)}
	F(s,t) &= \frac{\scF_-}{a_-}-\frac{\scF_+}{a_+} 
	+\Big(f(s)+f(t)+T(s,t)+\frac{\scF_+-\scF_-}{a_+-a_-}\Big)
	\log\big(\frac{a_+}{a_-}\big)  \ .
\end{align}

\section{Renormalisation}\label{sect:SQED_Renorm}
As previously mentioned, we have regularised the divergences  
by using a propertime cut-off.  
These cut-off dependent divergences are then removed in the standard way,
by adding counterterms to the original action.
Since the use of the background field method gives us the
freedom to rescale the quantum fields \cite{Abbott1982,Abbott1981}, 
and gauge invariance implies that
the background gauge field $W^\a$ is not renormalised,%
\footnote{
Normally it is the combination $e V$ that is renormalisation invariant, 
but we have absorbed the charge into the field strength.}
the counterterm action takes the simple form
\begin{align} \label{eqn:SQED-CTAct}
	S_\text{CT} = \frac{1}{e^2} (Z_e-1)\intc \,W^2 
	+ (Z_Q-1) \left( m \intc\, q_+ q_- + \cc \right) .
\end{align}
We note that the first term above is proportional to the classical 
action, $\G^{(0)}=\frac{1}{e^2}\intc W^2$.
The counterterm coefficients are related to the multiplicative 
renormalisation of charge and mass via
\begin{align}  
	e^2 = Z_e e_0^2\,, \qquad
	m^2 = Z_m m_0^2 = Z_Q^{-2}m_0^2\,, 
\end{align}
where we have used the fact that the $\cN=1$ nonrenormalisation theorem 
\cite{Wess1974b,Iliopoulos1974,Ferrara1974,Delbourgo1975,
West1976,Weinberg1976,Grisaru1979}
implies that $Z_m^\half \,Z_Q=1$.
The renormalisation constants are expanded with respect to the 
fine structure constant, $\a=e^2/8\p$,
\begin{align}
	Z_e = 1 + \a Z_e^{(1)} + \a^2 Z_e^{(2)} + \dots\,,\qquad
	Z_Q = 1 + \a Z_Q^{(1)} 
			+ \dots\,. 
\end{align}
It is worth noting that in (S)QED, 
since $\a$ is the only coupling constant, 
an expansion in $\a$ is equivalent to the loop expansion. 

Each term in the loop expansion of the effective action is 
constructed from both the standard diagrams computed 
in the sections above and from diagrams with counterterm insertions.
There is a freedom in how much of the finite part of $\G_\text{unren}^{(n)}$ is 
to be removed by the counterterm contribution $\G_\text{ct}^{(n)}$.
This corresponds to the freedom of choosing the finite part of the charge 
and matter renormalisation and
can be fixed by either choosing a consistent subtraction scheme, 
for example a (modified) minimal subtraction, or
by enforcing some renormalisation conditions.

We choose to work with physical parameters and thus calculate the counterterms 
using physical renormalisation conditions.
Following \cite{Vainshtein1985,Shifman1986}, 
we define the physical charge squared as 
the reciprocal of the coefficient in front of the $W^2$ term.
This clearly leads to the correct charge in the gauge-matter coupling.  
The physical mass is harder to define from 
within the Euler-Heisenberg sector of the effective action.  
The standard way to proceed is to use a separate
calculation of, for example, the K\"ahler potential and 
use the physical renormalisation conditions in that sector to 
find the correct mass renormalisation.  
We could also calculate the one-loop two-point function 
as done in \cite{Kuzenko2007a} 
and repeated below in section \ref{sect:SQED-1loopMatter}.
However, as shown by Lebedev and Ritus \cite{Ritus1978,Lebedev1984} 
and discussed in \cite{Dunne2004}, 
the physical mass can be extracted from the Euler-Heisenberg sector alone. 

First we examine the one-loop renormalisation.  
Adding the one-loop counterterm contribution to \eqref{eqn:SQED-1LoopEA-unren}
yields
\begin{align}
	\G^{(1)} &= \G^{(1)}_\text{ct} + \G_\text{unren}^{(1)} 
		=  \frac{\a}{e^2}\left(Z_e^{(1)} 
			+ \frac{1}{2\p} E_1(\rmi m^2 s_0) \right) \intc\, W^2 \\ \non
	 	&-\frac1{(4\p)^2}\intz W^2\Wb^2\int_0^\infty\frac{\rmd s}{s}
		\frac{B^{-2}(1-\cos(sB))-\cc}{\cos(sB)-\cos(s\Bb)}
		 \rme^{-\rmi(m^2-\rmi\e)s}\,,
\end{align}
where the exponential integral, $E_1$, 
is defined by \cite{DLMF, WolframFun}\footnote%
{The permalinks for definition in the two sites are
\url{http://dlmf.nist.gov/8.19.E3} and
\url{http://functions.wolfram.com/06.34.02.0001.01} respectively.}
\begin{align} \label{defn:ExpIntegral}
	E_n(z) = \int\limits_1^\infty\!\rmd t \frac{\rme^{-z t}}{t^n}\,,\quad
	n= 0, 1, 2, \dots\,, \quad \Re(z) > 0\,, 
\end{align}
with $E_1(z) = -\log(z\rme^\g)+\ord{z}$ where $\g$ is the Euler-Mascheroni 
constant. It is clear that the renormalisation condition implies
\begin{align}\label{SQED-Ze1}
	Z_e^{(1)} = - \frac{1}{2\p} E_1(\rmi m^2 s_0)
	\approx \frac{1}{2\p}\big(\log(m^2)+\log(\rmi s_0\rme^\g) + \ord{s_0}\big)\,,
\end{align}
so that the renormalised one-loop quantum correction is
\begin{align}\label{eqn:SQED-full1loop} 
\G^{(1)} = \frac1{(4\p)^2}\intz W^2\Wb^2\int_0^\infty\frac{\rmd s}{s}
		\frac{B^{-2}(\cos(sB)-1)-\cc}{\cos(sB)-\cos(s\Bb)}
		 \rme^{-\rmi(m^2-\rmi\e)s} \ .
\end{align}

Now we examine the two-loop renormalisation.
The two-loop counterterm contributions, read from \eqref{eqn:SQED-CTAct}, are
\begin{align} 
	\G^{(2)}_\text{ct} = \a^2 Z_e^{(2)}\, \G^{(0)}
		+\rmi\, \a m^2 Z_Q^{(1)} \Big(\Tr_+G_+ + \Tr_-G_-\Big) \ . 
\end{align}
This can be reduced to a more useful form by noting 
$\Tr_+G_+ = \Tr_-G_-$, see \cite{BK}, 
and that
\begin{align}
	 \frac\pd{\pd m^2}\G^{(1)}_\text{unren}
	 = -\rmi\frac\pd{\pd m^2}\int\limits_{0}^\infty\!\frac{\rmd s}s\Tr_+K_+ 
		\rme^{-\rmi (m^2 -{\rm i}\e)s}
	 = \rmi \, \Tr_+G_+ \ .
\end{align}
Then, using $\G^{(1)}_\text{unren } =\G^{(1)}-\G^{(1)}_\text{ct}$ 
combined with the fact
\begin{align*}
	m^2\frac{\pd}{\pd m^2}E_1(\rmi m^2s_0) = -\rme^{-\rmi m^2 s_0} \,,
\end{align*}
we have
\begin{align}
	\G^{(2)}_\text{ct} 
	= \a^2 \left(Z_e^{(2)}-\frac1\p Z_Q^{(1)}\rme^{-\rmi m^2 s_0}\right)\G^{(0)}
		+ 2\a Z_Q^{(1)} m^2 \frac\pd{\pd m^2}\G^{(1)}\ .
\end{align}

A close examination of the propertime integrand in the unrenormalised two-loop 
effective action (\ref{eqn:SQED-2loop-unren}, \ref{SQED-defn-F(s,t)})
shows that the only divergences that occur are in the 
$f(s)$ and $f(t)$ terms when $t$ or $s$ go to zero respectively.
We can separate off the divergent contribution 
by adding and subtracting the limit
\begin{align}\label{SQED:Ftilde}	
\begin{aligned}
	\tilde F(s)
	&\deq \lim_{t\to 0} t \,
		\frac{P_+P_-}{a_+-a_-}\frac{F(s,t)}{(st)^{2}} \\
	&= -2 \frac{B^2-\Bb^2}{B^2\Bb^2}\frac{B\sin(sB)\sin^2(s\Bb/2) -
		(B\leftrightarrow\Bb)}{(\cos(sB)-\cos(s\Bb))^2}\,,
\end{aligned}
\end{align}
and similarly for $\tilde{F}(t)$, to give
\begin{align}\label{SQED:unrenorm.2} 
	\G^{(2)}_\unren &= \frac{e^2}{(4\pi)^4}\intz W^2\Wb^2 \Bigg(
		2 E_1(\rmi m^2s_0)\,\int\limits_0^{\infty}\!\rmd s\, 
			\tilde{F}(s)\,\rme^{-\rmi (m^2-{\rm i}\e)s} \\\non
	&+	
		\int\limits_0^{\infty}\!\!
		\rmd s\rmd t \left(\frac{P_+P_-}{a_+-a_-}\frac{F(s,t)}{(st)^{2}}
			-\frac{\tilde{F}(s)}{t} - \frac{\tilde{F}(t)}{s} \right)
		\rme^{-\rmi (m^2-{\rm i}\e)(s+t)} \Bigg).
\end{align}
Then, motivated by the form of $\G^{(2)}_\text{ct}$ and by previous 
renormalisations of two-loop Euler-Heisenberg effective actions
we note that
\begin{align} \label{SQED:mass-correction}
\frac{1}{(4\pi)^2}\intz \, W^2\Wb^2\int\limits_0^{\infty}\!\!
	\rmd s\, \tilde{F}(s)\,\rme^{-\rmi (m^2-\rmi\e) s}
	= m^2\frac{\pd}{\pd m^2}\G^{(1)}\ .
\end{align}

We can now combine $\G^{(2)}_\text{unren}$ with $\G^{(2)}_\text{ct}$
and see that the renormalisation condition on the gauge kinetic term
fixes $Z_e^{(2)}$ in terms of $Z_Q^{(1)}$.
However, demanding the two-loop effective action to be finite
leaves freedom in choosing the finite part of $Z_Q$:
\begin{align}\label{SQED-ZQ1}
	Z_e^{(2)} &=  \frac{1}{\p}  Z_Q^{(1)}\rme^{-\rmi m^2 s_0}\,,&
	Z_Q^{(1)} &= -\frac{1}{2\p} E_1(\rmi m^2 s_0) + Z_{Q,\text{finite}}^{(1)}\ .
\end{align}
Thus we see that in SQED, the one-loop mass renormalisation 
is the sole cause of the two-loop charge renormalisation.  
As discussed earlier, the finite correction $Z_{Q,\text{finite}}^{(1)}$
is not easily fixed from the Euler-Heisenberg sector alone.
In section \ref{sect:SQED-1loopMatter} we will show that 
$Z_{Q,\text{finite}}^{(1)}$ is actually zero when using the 
physical renormalisation conditions.

Since we are using an `on-shell' renormalisation 
\cite{Dittrich1985,Coquereaux1980}, the appropriate renormalisation equation 
is the Callan-Symanzik equation \cite{Callan1970,Symanzik1970}.  
The renormalisation group functions are defined by
\begin{align}
	\b_\text{CS} = m\frac{\rmd\a}{\rmd m} 
		= \a\frac{\rmd\log Z_e}{\rmd\log m}\,,\qquad
	\g_m = \frac{\rmd\ln Z_Q}{\rmd\ln m}
		= \frac{\rmd\ln Z_m^{-\half}}{\rmd\ln m} \ .
\end{align}
In QED it can be shown \cite{Dittrich1985,Coquereaux1980} 
that the $\b$-function for dimensional regularisation 
with minimal subtraction coincides with the above 
$\b$-function to $\ord{\a^3}$ and the proofs also hold for SQED.
Then, using the basic results
\begin{align}
	\frac{\pd E_n(z)}{\pd z} = - E_{n-1}(z)\,, 
	\quad\text{and}\quad
	E_0(z) = \frac1z\rme^{-z}\,,
\end{align}
it is simple to calculate
\begin{align} 
	\b_\text{CS} = \frac{\a^2}\p\left(1+\g_m\right) 
		+ \ord{\a^3} \,,\qquad 
	\g_m = \frac\a\p + \ord{\a^2}\,. 
\end{align}
These results coincide with the known $\b$ and $\g$ functions, e.g., 
\cite{Vainshtein1985,Shifman1986}.
Given that only the one-loop effective action contributes directly to the $F^2$ 
term \cite{Vainshtein1985,Shifman1986,NSVZ1983} 
it must be that all higher contributions to the 
charge renormalisation are due to the mass renormalisation.  
Therefore, following the arguments of \cite{Vainshtein1985,Shifman1986}, we 
expect that the above $\b$-function is an exact result.

We can now write the renormalised low energy effective action to two loops,
\begin{align} \label{SQED-fullEA}
	\G&[W,\bar{W}] 
	= \G^{(0)}+\G^{(1)}+\G^{(2)}
	= \frac1{e^2}\intc \,W^2 \\ 
	&+ \frac1{(4\p)^2}\intz W^2\Wb^2\int_0^\infty\!\!\rmd s
		\frac{B^{-2}(\cos(sB)-1)-\cc}{s(\cos(sB)-\cos(s\Bb))}
		(1+\d{m}^2\pd_{m^2})
		 \rme^{-\rmi m^2 s}  \non \\ \non 
	&+ \frac{e^2}{(4\pi)^4}\intz  W^2\Wb^2\!\!
		\int_0^{\infty}\!\!
		\rmd s\rmd t\left(F(s,t)-\frac{\tilde{F}(s)}{t}
			-\frac{\tilde{F}(t)}{s} \right) \!
		\rme^{-\rmi m^2(s+t)} .
\end{align}
where $\d{m^2}=2 m^2 \a Z_{Q,\text{finite}}^{(1)} + \ord{\a}^2$
and $B$ is now understood as
\begin{align}
B^2 = \frac{1}{4} D^2 W^2\,, 
\end{align}
with the vector multiplet not subject to any constraints.
This is the final form of our Euler-Heisenberg-type calculation, 
and using it allows one to compute, by standard means, 
quantities of interest, such as the vacuum non-persistence amplitude 
\cite{Schwinger1951, Ritus1975, Ritus1977}.

\section{The limit of a self-dual background}\label{sect:SelfDualSQED}
In this section we examine the self-dual limit of the Euler-Heisenberg 
effective action calculated above.
Ten years ago \cite{Dunne2001,Dunne2002,Dunne2002a,Dunne2004}
it was noted that when the background field is self-dual 
the propertime integrals in the Euler-Heisenberg effective actions 
for scalar and spinor QED can be integrated in closed form. 
This worked at both one and two-loops, with the results  
written completely in terms of the function
\begin{align}\label{defn:xi}
	\x(x) 	&= \half\int\limits_0^\infty\!\rmd s
				\left(\frac1{s^2} - \frac1{\sinh^2s}\right)\rme^{-2xs} 
			= x\int_0^\infty\!\!\rmd{s}\,
				\left(\coth(s) - \frac1s\right)\rme^{-2xs}	\non \\
			&= -x\left(\psi(x)-\log(x)+\frac1{2x}\right)\,, 
\end{align}
and its derivatives.
In \eqref{defn:xi}, 
$\j(x) = \G'(x) / \G(x)$ is the digamma function \cite{DLMF, WolframFun}
and $x \sim m^2/F$ is the ratio of the mass squared to the single 
invariant of the self-dual field strength.
The function $\x(x)$ has the following weak and strong field expansions
\begin{align}
\label{eqn:XiLargeFieldExpansion}
	\x(x) &= \sum_{n=1}^\infty \frac{B_{2n}}{2n}x^{1-2n}\,,& &x\gg1\,,\\
\label{eqn:XiSmallFieldExpansion}
	\x(x) &= \frac12 + x(\g+\log(x))
		-\sum_{n=2}^\infty\z(n)(-x)^n\,,& &x\ll1\,,
\end{align}
where $B_{n}$ are the Bernoulli numbers, $\g$ is the Euler-Mascheroni constant,
and $\z$ is the Riemann zeta function.

As discussed in \cite{Kuzenko2004a} and references therein, 
the effective action for a supersymmetric theory becomes
trivial in the case of a self-dual background.  
Yet we can still impose a relaxed form of self-duality which allows us to 
retain a holomorphic-like sector of the effective action.  
If we write the full supersymmetric Euler-Heisenberg effective action as 
\begin{align}\label{SQED-EH-EA-structure}
	\G = \frac{1}{e^2}  \intc \,W^2
		+ \intz \, W^2{\bar W}^2 \, \Omega\big(B^2,  \Bb^2\big) \,.
\end{align}
and impose the relaxed self-duality conditions
\begin{align}\label{defn:selfdualconds} 
	W_\a \neq 0\,, \quad D_\a W_\b = 0\,,\quad 
	\Db_\da\Wb_\db = \Db_{(\da}\Wb_{\db)} \neq 0 \,,
\end{align}
then we can track the following sector
\begin{align}
 \intz W^2\Wb^2 \, \Omega\big(0,\Bb^2\big)  \ .
\end{align}
It should be noted that although the conditions \eqref{defn:selfdualconds}
are inconsistent with the structure of 
a single, real vector multiplet,\footnote%
{Equivalently, the relaxed self-dual condition 
it is inconsistent with the Bianchi identity \eqref{Neq1BI}.}
their use is perfectly justified as long 
as we realise we are only calculating the above sector.
At the end of the calculation we can remove the self-duality condition 
and have a well defined sector of the effective action.
This was the approach taken in \cite{Kuzenko2004c} to calculate this sector
for $\cN=4$ $SU(N)$ SYM in a $U(1)$ background.
Since we already have the full two-loop Euler-Heisenberg effective action, 
we can simply take its limit as
$B \to 0$ ($\Bb\neq0$) to obtain the above sector. 
Further discussion and the form of the heat kernels in the self-dual limit
can be found in appendix \ref{ssect:Self-dual}.

Taking the self-dual limit of 
the renormalised one-loop effective action \eqref{eqn:SQED-full1loop}
and Wick rotating the propertime integral we get
\begin{align} \non 
	\G_\text{SD}^{(1)}
	= \frac{x^2}{(4\p)^2}\intz \frac{W^2\Wb^2}{m^4}\int_0^\infty\!\!\rmd{s}\,
		s\left(\frac1{s^2}-\frac1{\sinh^2(s)}\right)\rme^{-2sx}\,,
\end{align}
where we have written the field strength in terms of $x=m^2/\Bb$, 
a natural dimensionless variable.  
This is clearly seen to be proportional to the first derivative of 
$\x(x)$ \eqref{defn:xi}, so we find
\begin{align}
	\G_\text{SD}^{(1)}
	= -\frac{1}{(4\p)^2} \intz \frac{W^2\Wb^2}{m^4}\,x^2\x'(x) \ .
\end{align}

Dimensional arguments tell us that the self-dual sector of the two-loop
effective action must take the same form.
We will split the two-loop effective action into parts, writing
\begin{align}
	\G_\text{SD}^{(2)}
	= \frac{e^2}{(4\p)^4} \intz \frac{W^2\Wb^2}{m^4} 
	  \Big(I_f + I_\scF + I_\rmII + I_{\d m} \Big) \,,
\end{align}
where $I_f$ and $I_\scF$ are the terms from $\G^{(2)}_\rmI$ that are associated
with $f(s)$ and $\scF(s)$ respectively
and $I_\rmII$ is the contribution from $\G^{(2)}_\rmII$.
The final term is an optional finite mass correction that it proportional to
the \eqref{SQED:mass-correction} and comes from choosing
$Z^{(1)}_{Q,\text{finite}}\neq0$ in \eqref{SQED-ZQ1}.

The integral $I_{\d m}$, is most simply found by evaluating 
$\d m^2 \pd_{m^2} \G_\text{SD}^{(1)}$ to find
\begin{align}\label{eqn:SQED-intdm}
	I_{\d m} = -4\p Z_{Q,\text{finite}}^{(1)} x^3 \x''(x)\ .
\end{align}

The integral $I_\rmII$ has already been calculated in \cite{Kuzenko2004a},
following their lead, we take the limit of $B\to0$, Wick rotate 
and write $\Bb=m^2/x$ to get
\begin{align*}
	I_\rmII &= 8x^3\int_0^\infty\!\!\rmd{s}\rmd{t}\,
		s^2\big(\coth(s)-\coth(s+t)\big)\rme^{-2x(s+t)} \ . 
\end{align*}
In the first term, the $s$ and $t$ integrals are not entangled and
the $t$ integral is trivial. The second term can be simplified with 
the change of variables
\begin{align}\label{defn:2DPropTimeCOV}
	\int_0^\infty\!\!\rmd{s}\rmd{t}\, f(s,t)
	= \frac12\int_{-1}^1\rmd\a\int_0^\infty\rmd\t \, \t\,
		f\Big(\frac{(1+\a)\t}2,\frac{(1-\a)\t}2\Big)\ .
\end{align}
Performing the $t$-integral in the first term 
and the $\a$-integral in the second, we find
\begin{align*}
	I_\rmII &= 4x^2\int_0^\infty\!\!\rmd{s}\,s^2\coth(s)\rme^{-2xs}
		- \frac{8}{3}x^3\int_0^\infty\!\!\rmd\t\,\t^2\coth(\t)\rme^{-2x\t}\ .
\end{align*}
The two integrals can be combined by integrating the first one by parts
and the result is easily integrated into its final form 
\begin{align}\label{eqn:SQED-intII}
	I_\rmII &= \frac43 x^2 \int_0^\infty\!\!\rmd{s}\,s^3 
						\mathrm{csch}^2(s)\rme^{-2xs}
			= \frac13\big(1+x^2\x'''(x)\big)\ .
\end{align}

The $f$ dependent terms in the integrand $F(s,t)$ 
generate all of the divergences in the unrenormalised 
two-loop effective action \eqref{SQED:unrenorm.2}. 
So, we define $I_f$ to be the contribution from $f$ minus its divergent part,
i.e.,
\begin{align*}
	I_f &= 2m^4\int_0^\infty\!\!\!\rmd{s}\rmd{t}
		\left(\frac{P_+P_-}{a_+-a_-}\frac{f(s)}{s^2t^2}
			\log\left(\frac{a_+}{a_-}\right) - \frac1{t}\tilde{F}(s)\right)
		\rme^{-\rmi m^2(s+t)}\Bigg|_{B\to0} \\\non
		&= 4x^2\int_0^\infty\!\!\!\rmd{s}\rmd{t} 
			\frac{s^2}{\sinh^2s}\left(\coth s-\frac1s\right)
			\left(\coth(s+t)+\frac1t-\coth t\right)\rme^{-2x(s+t)} \ .
\end{align*}
Apart from the $\coth(s+t)$ term, the integral factorises as
\begin{align*}
	4x^2\int_0^\infty\!\!\!\rmd{s}\rmd{t} 
			\frac{s^2}{\sinh^2s}\left(\coth s-\frac1s\right)
			\left(\frac1t-\coth t\right)\rme^{-2x(s+t)}
	= -2x^2\x(x)\x''(x) .
\end{align*}
By repeated integration by parts, the $\coth(s+t)$ term can be 
reduced to a combination of surface terms in one of the integration variables, 
so the result is the linear combination
\begin{align*}
	4x^2\int_0^\infty\!\!\!\rmd{s}\rmd{t} 
			\frac{s^2}{\sinh^2s}
			\frac{\coth(s)-s^{-1}}{\tanh(s+t)}\rme^{-2x(s+t)}
	= -x^3\x''(x) -2x^2\x'(x)+2x\x(x)\ .
\end{align*}
Putting it together, we get the result
\begin{align}\label{eqn:SQED-intf}
	I_f &= 2x \x(x) - 2 x^2(\x'(x)+\x(x)\x''(x)) - x^3 \x''(x) \ .
\end{align}

The final contribution, $I_\scF$, 
comes from the $\scF_\pm$ dependent terms in $F(s,t)$,
\begin{align}\label{eqn:SQED-intscF-0}\!\!
	I_\scF &= m^4\int_0^\infty\!\!\!\rmd{s}\rmd{t}\,
		{s^2t^2}\frac{P_+P_-}{a_+-a_-}
		\left(\frac{\scF_-}{a_-}-\frac{\scF_+}{a_+}
		+\frac{\scF_+-\scF_-}{a_+-a_-}\log\Big(\frac{a_+}{a_-}\Big)\!\right)
		\rme^{-\rmi m^2(s+t)}\Bigg|_{B\to0} \non \\
	&= 2x^2\int_0^\infty\!\!\!\rmd{s}\rmd{t}\,
		{\sinh^2(s+t)}\!\left(\frac{s^2}{\sinh^2s}
		-\frac{st\cosh(s-t)}{\sinh s \sinh t}\right)\rme^{-2x(s+t)} . \!\!
\end{align}
By using the identity $\cosh(s+t)-\cosh(s-t)=2\sinh(s)\sinh(t)$, 
we can separate out a term proportional to $I_\rmII$,
\begin{align*}
	I_\scF	&= \frac12 I_\rmII 
		+2x^2\int_0^\infty\!\!\!\rmd{s}\rmd{t}\,
		{\sinh^2(s+t)}\left(\frac{s^2}{\sinh^2s}
			-\frac{st\cosh(s+t)}{\sinh s \sinh t}\right)\rme^{-2x(s+t)} .
\end{align*}
We have been unable to find a way to directly integrate the remaining integral.
However, by assuming that it can be written as a quadratic combination
of $\x(x)$ and its derivatives, we can use high precision numerical integration
combined with integer relation algorithms to guess its form.
A more detailed description of this process and some code to perform
the procedure is presented in appendix \ref{A:IntegerRelations}.
The result is
\begin{align}\label{eqn:SQED-intscF}
	I_\scF	&= \frac12 I_\rmII 
		+ \frac14 + x^2 -x^3\x''(x) - (\x(x)-x\x'(x)+x)^2\ .
\end{align}
It was derived using small integer values for $x$ and is thus
accurate in the strong field limit.  
It can also be checked in the weak field limit by comparing 
the expansion of the result using \eqref{eqn:XiSmallFieldExpansion}
against the expansion 
of the original integral 
\eqref{eqn:SQED-intscF-0} using the formula 
\begin{align}
	\int_0^\infty\!\!\!\rmd{s}\rmd{t}\, \frac{s^n t^m}{(s+t)^l}\rme^{-2x(s+t)}
	= \frac{n!m!(n+m-l+1)!}{(n+m+1)!(2x)^{n+m-l+1}}\ .
\end{align}
This check is easy using a computer algebra system and has been performed
to the $100^{\text{th}}$ order in the field strength. 
This means that we can be quite confident in equation \eqref{eqn:SQED-intscF}.

Putting all the above results together we have
\begin{align}
	\G_\text{SD}^{(2)}  
	&= \frac{e^2}{(4\p)^4}\intz\frac{W^2\Wb^2}{m^4}
	   \Big(\frac12\left(1+x^2\x'''(x)\right) 	\\\non
		& + \frac14 - 2(1+2\p Z_{Q,\text{finite}}^{(1)}) x^3 \x''(x)
		- 2 x^2\x(x)\x''(x) - (\x(x)-x\x'(x))^2\Big) ,
\end{align}
where the first line is exactly the $\cN=2$ result \cite{Kuzenko2004a}.



	
\section{One-loop matter sector in the Feynman gauge}
	\label{sect:SQED-1loopMatter}

In the previous sections we have examined the Euler-Heisenberg effective
action for $\cN=1$ SQED, i.e., the low-energy effective action 
in a gauge field background. In this section we examine the pure matter
section of the effective action. This is interesting in its own right,
but is primarily used here to calculate the physical renormalisation constant 
for the matter fields.  

We start by looking at the quantisation of the theory
in the Fermi-Feynman gauge with background matter fields 
and writing the general form for the one-loop effective action.
This discussion is easily generalised to an arbitrary $\cN=1$ gauge theory.
We then use this result to find the one-loop corrections to both 
the two-point function and the K\"ahler potential.
From these we can find the desired one-loop mass renormalisation.
Finally, we discuss the K\"ahler potential in a more general $R_\xi$ gauge.

\subsection{Quantisation in Fermi-Feynman gauge}
Starting from the classical action \eqref{defn:SQED1-Classical-Action}
we perform the background-quantum splitting
\begin{align}
	V\to e\,v~,\qquad Q_\pm \to Q_\pm + q_\pm \ .
\end{align}
Then, introducing the matrix notation
\begin{align} 
	\hat{v}=e\s_3v\,, \qquad 
	\hat{m}=\s_1m\,, \qquad 
	q^{\rm T}=(q_+,q_-)\,,
\end{align}
and the gauge invariant quantities
\begin{align}\label{SQED:MvandKappa} 
	M_v^2 = e^2 Q^\dag Q\,, \quad
	\k = e^2 Q^{\rmT} \s_1 Q\,, \quad
	\bar\k = e^2 Q^\dag \s_1 \bar{Q}~,
\end{align}
the resulting quadratic quantum action, in the Fermi-Feynman gauge,
takes the form
\begin{align}\label{SQED:QuadMatter}\begin{aligned}
	S_\text{quad} &= -\half\intz\, v\left(\square-M_v^2\right)v \\
	&+ \intz\left(q^\dag q+q^\dag\hat{v}Q+Q^\dag\hat{v}q\right)
	&+\half\left(\intc\,q^{\rm T}\hat{m}q+\cc\right) \ .
\end{aligned}\end{align}

As briefly discussed in section \ref{sect:Neq1Pert},
the mixing terms in $S_\text{quad}$ can be 
eliminated from the path integral by a change of variables in the path integral.
To see this, we define
\[
	\mathbf{q}^\rmT=(q,\bar{q}) 
	\,\text{,}\quad
	\scJ^\rmT=\big(\frac{\Db^2}{-4}(Q^\dag\hat{v}), 
		\frac{D^2}{-4}(Q^\rmT\hat{v})\big)
	\quad\text{and}\quad
	H_\phi = \bem \hat{m} & \frac{\Db^2}{-4}\\ \frac{D^2}{-4} & \hat{m} \eem \,,
\]
then rewrite the second line of \eqref{SQED:QuadMatter} in the functional form
\begin{align*}
	\frac12 \mathbf{q}^\rmT\cdot H_{q} \cdot\mathbf{q}
	+\frac12 \mathbf{q}^\rmT\cdot\scJ
	+\frac12 \scJ^\rmT\cdot\mathbf{q}\ .
\end{align*}
The chiral fields' Hessian, $H_{q}$, is simple to invert
\begin{align}\label{SQED:matrixchprop}
	{\bf G} &= -H_{q}^{-1} 
	= \begin{pmatrix} G_{++} & G_{+-} \\ G_{-+} & G_{--}\end{pmatrix} 
	= \begin{pmatrix} \hat{m}\frac{\Db^2}{4} & \frac{\Db^2D^2}{16}\ds1 
 		\\ \frac{D^2\Db^2}{16}\ds1 & \hat{m}\frac{D^2}{4}
 	  \end{pmatrix}G \,,\\
\label{basechprop}
 G(z,z') &= -\frac{1}{\square-m^2}\d^8(z,z') 
		 = G_{\rm bos}(x,x') \,\d^4(\q-\q') \,,
\end{align}
where $G_{\rm bos}$ is the bosonic Green's function 
discussed in section \ref{sect:BosonicProp}.
Inspired by \cite{ostrovsky1988covariant}, 
we make the change of variables
\begin{align}
	\mathbf{q} &\to \mathbf{q} - \mathbf{G}\cdot\scJ \\ \non
	\implies
	q(z) &\to q(z) - \int\rmd^8z'\left(
		G_{++}(z,z')\hat{v}(z')\bar{Q}(z')+G_{+-}(z,z')\hat{v}(z')Q(z')\right)
		\,,
\end{align}
to remove the mixing terms. This has the cost of introducing the new term
\begin{align}
	\frac12\scJ^\rmT\cdot\mathbf{G}\cdot\scJ 
	&= \frac{e^2}{2} \intz\!\rmd^8z' v(z) \D(z,z') v(z')\,,
\end{align}
where
\begin{align}\non
	\D(z,z') = e^2\Big(
		 &Q^\dag(z) G_{+-}(z,z')Q(z') + Q^\rmT(z) G_{-+}(z,z')\bar{Q}(z') \\
		-&Q^\dag(z) G_{++}(z,z')\bar{Q}(z') - Q^\rmT(z) G_{--}(z,z')Q(z')
		\Big)\ .
\end{align}
At one-loop, this is the only cost of the change of variables,
since its Jacobian is obviously equal to unity 
and the new interaction terms are only important at higher loops.
So the the final form for the quadratic part of the classical action is
\begin{align}
	S_\text{quad} &= -\frac12 v \cdot (\square - M_v^2 - \D) \cdot v + 
	\frac12 \mathbf{q}^\rmT\cdot H_{q} \cdot\mathbf{q}\ .
\end{align}
The gauge field's propagator $(\square - M_v^2 - \D)^{-1}$ can be 
expanded in powers of $\D$  which is equivalent to expanding in the external
fields $Q$ and $\bar{Q}$. This expansion is best thought of as a 
dressing of the standard propagator $(\square - M_v^2)^{-1}$
and is exactly the right way to view the calculation of the two-point 
function below. 
Since the components of $\bf G=-H_{q}^{-1}$ in \eqref{SQED:matrixchprop} are 
background independent, the one-loop effective action is 
calculated purely from the gauge field's Hessian,
\begin{align} 
	\G_\unren^{(1)}=\frac{\rmi}2\Tr\ln\left(\square-M_v^2-\D\right) \ . 
\end{align}
Note that this is an exact one-loop result for the matter sector, 
as we are yet to make any approximations.

\subsection{Two-point function}
To calculate the two-point function we keep only terms quadratic in the 
external chiral fields.
This can be achieved by expanding the logarithm to first order,
\begin{align}
	\G_\unren^{(1)} \approx \frac{\rmi}2\Tr\left(\ln(\square)
		-\frac1\square M_v^2-\frac1\square\D\right) \ . 
\end{align}
Due to a lack of spinor derivatives to annihilate the Grassmann delta function, 
the first two terms above evaluate to zero.  
Similarly the last two terms in $\D$ also do not contribute.
This leaves
\begin{align} \non
	\!\G_\unren^{(1)}\approx {\rmi e^2}{}\intz\rmd^8z'
	G_{0}(z,z')\!
	\left(Q^\dag(z')G_{\rm bos}(z',z)Q(z)
	 \right) , 
\end{align}
where $G_0$ is the massless limit of \eqref{basechprop}.
Integrating out the Grassmann delta function in $G_0$, 
making the change of variables $x'\to\rho=x-x'$ 
and writing the Green's functions in their propertime representation 
\eqref{FreeKGHeatKern} yields
\begin{align} 
	\!\G_\unren^{(1)}\approx \frac{\rmi e^2}{(4\p)^4}\intz\rmd^4\rho\,
	Q^\dag(z') Q(z) 
	\int_0^\infty\frac{\rmd{s}\rmd{t}}{s^2t^2}
	\rme^{\rmi\frac{s+t}{4st}\rho^2-\rmi s m^2} \Big|_{\q'=\q}\  .
\end{align}
We note that the above expression involves a single Grassmann integral, 
although it is non-local in space-time, 
in accordance with the $\cN=1$ non-renormalisation theorem.  

We now check that is is equivalent to the standard
momentum space representation for the two-point function, see, e.g.,
\cite{GGRS1983,West-Book}.
First, expand the fields in their Fourier components
\begin{align*}
	Q(z)=\int\frac{\rmd^4k}{(2\p)^4}Q(k,\q)\rme^{-\rmi k x} \,,	\quad
	Q^\dag(z')\big|_{\q'=\q}
		=\int\frac{\rmd^4p}{(2\p)^4}Q^\dag(p,\q)\rme^{-\rmi p (x-\rho)}\ .
\end{align*}
This allows us to write the two point function as
\begin{align} \non
	\!\G_\unren^{(1)}\approx \frac{\rmi e^2}{(4\p)^4}\!
	\int\!\!\frac{\rmd^4k\,\rmd^4p\,\rmd^4\q}{(2\p)^8}
	Q^\dag(p,\q) Q(k,\q)  \!\int\!\!\rmd^4x\rmd^4\rho\!\!
	\int\limits_0^\infty\!\frac{\rmd{s}\rmd{t}}{s^2t^2}
	\rme^{\rmi\frac{s+t}{4st}\rho^2-\rmi(k+p)x+\rmi p \rho -\rmi s m^2}.
\end{align}
The $x$-integral yields a momentum $\d$-function and the $\r$-integral
can be computed by completing the square and using \eqref{SQED-rho-int}.
The result is
\begin{align}
	\G_\unren^{(1)}
	&\approx - e^2 \int \! \frac{\rmd^4p\,\rmd^4\q }{(2\p)^4}
		Q^\dag(p,\q) Q(-p,\q) A(p^2)\,,
\end{align}
where the integral $A(p^2)$ is the standard one-loop propagator diagram
with one massive and one massless internal edge
\begin{align*}
	A(p^2) = \rmi\int\!\!\frac{\rmd^4k}{(2\p)^4}\frac{1}{(k^2+m^2)(k+p)^2}
	= \frac{1}{(4\p)^2}\int_0^\infty\!\!\!\frac{\rmd{s}\rmd{t}}{(s+t)^2} 
		\rme^{-\frac{\rmi st}{s+t}p^2 - \rmi s m^2} .
\end{align*}
The evaluation of this integral is straightforward.
Introducing a UV cut-off, $s_0$, in the $s$-integral, we find
\begin{align}\label{SQED:A(p2)}\begin{aligned}
	A(p^2) &= \frac{E_2(\rmi s_0 m^2)-E_2(\rmi s_0 (m^2+p^2))}{\rmi s_0 p^2} \\
	&\approx - \log(\rmi s_0 \rme^\g m^2) + 1 
		- \frac{m^2+p^2}{p^2}\log\left(\frac{m^2+p^2}{m^2}\right) + O(s_0)\,,
\end{aligned}\end{align}
where the exponential integrals are defined in \eqref{defn:ExpIntegral}.
To renormalise at zero external momentum, we need the result
\begin{align}\label{SQED:A(0)}
	A(0) &= E_1(\rmi s_0 m^2) \approx -\log(\rmi s_0 m^2 \rme^{\g}) + O(s_0)\ .
\end{align}

\subsection{K\"ahler potential}
To compute the K\"ahler potential, 
it suffices to choose $Q$ and $Q^\dagger$ to be constant, 
then $\D$ reduces to
\begin{align}\non
	\D(z,z')
	=-\frac{1}{\square-m^2}\left(\frac1{16}M_v^2\{D^2,\Db^2\}-m\k\frac{D^2}4
	-m\bar\k\frac{\Db^2}4\right)\d^8(z,z')\,, 
\end{align}
where $\k$ and $\bar\k$ are defined in \eqref{SQED:MvandKappa} above. 
The effective action is then
\begin{align} 
	\G_\unren^{(1)}
	= \frac\rmi2\Tr\log\Bigg(1+\frac{1}{16}\,
	\frac{M_v^2 \{D^2,\Db^2\}-4m \k D^2-4 m\bar\k \Db^2}
		{(\square-m^2)(\square-M_v^2)} \Bigg) \ .
\end{align}
The logarithm can be factorised using
\begin{gather*}
\begin{aligned}	
	1+&X\Db^2D^2+YD^2\Db^2+ZD^2+\bar{Z}\Db^2 \\
	  &= (1+ND^2)(1+U\Db^2D^2+VD^2\Db^2)(1+\bar{N}\Db^2)\,, 
\end{aligned}\\
	N = (1+16\square Y)^{-1}Z\,,\quad 
	V = Y\,,\quad
	U = X-\bar{Z}(1+16\square Y)^{-1}Z\,, 
\end{gather*}
for constant, matrix coefficients.
Evaluating the trace by going to momentum space 
gives the K\"ahler potential as
\begin{align}\label{SQED:FF-Kahler} 
	\!\!K^{(1)}_\unren
	= - \frac\rmi2\int\!\!\frac{\rmd^4k}{(2\p)^4} \frac1{k^2}
		\log\!\left[\!\frac{(k^2(k^2+m^2)+m^2M_v^2)^2
		+k^2m^2\k\bar\k}{k^4(k^2+M_v^2)^2}\right] \!. \!\!\!\!\ 
\end{align}
This can be compared with the calculation given in \cite{Grisaru1996}.
Although we can factorise the above quartic in $k^2$ 
and thus expand the logarithm and perform the momentum integration, 
it is not very enlightening.

However, it is interesting to examine the result in the general, 
two parameter $R_\xi$ gauge defined in \eqref{RxiGauge}. 
The calculation is very similar to the above,
except that now the chiral and ghost fields couple 
to the background matter fields.
The final (unrenormalised) result is
\begin{align}\label{SQED:Gen-Kahler}\non 
	K^{(1)}_{R_\xi}
	&= -\frac\rmi2\int\!\!\frac{\rmd^{4}k}{(2\p)^{4}}\frac{1}{k^2}
	\log\!\Bigg[ k^2 \Bigg(\!k^2\left((k^2(k^2+m^2)
			+\a m^2 M_v^2)^2+\a^2 k^2 m^2\k\bar{\k}\right) \\ \non
	& +\x^{-1}(2k^2+\x^{-1}M_v^2)
		\Big(\x^{-1}(2k^2+\x^{-1}M_v^2)(k^2 M_v^4+m^2\k\bar{\k}) 
			-2\a k^2m^2\k\bar{\k} \\ 
	& +2k^2M_v^2(k^2(k^2+m^2)+\a m^2 M_v^2) \Big)
	\Bigg) (k^2 + M_v^2)^{-2}(k^2 + \x^{-1} M_v^2)^{-4}\Bigg]\,.
\end{align}
It is straight forward to check that this reduces to \eqref{SQED:FF-Kahler} 
in the Fermi-Feynman gauge (where $\x\to\infty$ and $\a\to1$).
Other interesting gauges are
the Landau gauge 
\begin{align}\label{SQED:Landau-Kahler} 
	K^{(1)}_{R_\xi}\Big|_{\genfrac{}{}{0pt}{}{\x\to\infty}{\a\to0}}
	&= -\rmi\int\!\!\frac{\rmd^{4}k}{(2\p)^{4}}\frac{1}{k^2}
	\log\!\Bigg[\frac{k^2+m^2}{k^2+M_v^2}\Bigg]\,,
\end{align}
and the 't Hooft gauge 
\begin{align}\label{SQED:tHooft-Kahler}
\!\!
	K^{(1)}_{R_\x}\Big|_{\genfrac{}{}{0pt}{}{\x\to1}{\a\to1}}
	&= -\frac\rmi2\int\!\!\frac{\rmd^{4}k}{(2\p)^{4}}\frac{1}{k^2}
	\log\!\Bigg[\frac{k^2\big(k^2(k^2+m^2+M_v^2)^2 + m^2 \k\bar\k \big)}
				   {(k^2 + M_v^2)^{4}}\Bigg]\,.\!\!
\end{align}
It is also worth noting that if the chiral fields are massless, 
then they are also on-shell up to terms that don't contribute to the 
K\"ahler approximation, so we expect the one-loop effective action
to be gauge independent -- and this is exactly what we see
\begin{align}\label{SQED:Massless-Kahler} 
	K^{(1)}_{R_\xi} \Big|_{m^2\to0}
	&= -\rmi\int\!\!\frac{\rmd^{4}k}{(2\p)^{4}}\frac{1}{k^2}
	\log\!\Bigg[\frac{k^2}{k^2+M_v^2}\Bigg]\,.
\end{align}
This result is familiar from older calculations of the K\"ahler potential
\cite{Buchbinder1994a, Pickering1996} and 
can also be obtained from the appropriate limit of \eqref{Neq1 1loop Kahler}.
Finally, 
it is one of explicit examples given at the end of \cite{Nibbelink2005}.

\subsection{Renormalisation}\label{ssect:Neq1MatterRenorm}
All of the above polynomials in $k^2$ can be factorised, 
the logarithms expanded and, if dimensionally regularised,
the integrals can be evaluated as a sum of the integrals $\scJ(x)$ 
defined in \eqref{defn:scJ_Integral}.
However, to compare with the one-loop mass renormalisation found
in section \ref{sect:SQED_Renorm}, 
we need to use regularisation by a propertime cut-off. 
For this, all we need is the quadratic part of the K\"ahler potential. 
Expanding the logarithm in $K^{(1)}_{R_\xi}$ up to first order in $M_v^2$,
we find
\begin{align}\label{SQED:QuadMatterTerm}
	K^{(1)}_{R_\xi}
	&= -\rmi M_v^2 \int\!\!\frac{\rmd^{4}k}{(2\p)^{4}}\frac{1}{k^2}
	\Bigg(\frac{\a-1-2/\x}{k^2} - \frac{\a-2/\x}{k^2+m^2}\Bigg)
	+ \ord{M_v^4,\k\bar\k}\ .
\end{align}
The first term above has infrared divergences, this is not surprising
since we are essentially looking at the zero momentum, 
weak field strength limit. 
In general, neither the calculation of the two-point function 
(with non-vanishing external momentum), 
nor of the general K\"ahler potential
have any (field dependent) IR divergent terms.
If we were to naively use dimensional regularisation, 
then the first term in \eqref{SQED:QuadMatterTerm} would be set to zero 
and the second term would provide a gauge dependent divergence, 
leading to a gauge dependent anomalous dimension, which is not a good thing.
However, if we realise that we are only interested in UV divergences,
then we see that the UV limit of the integrand is 
$-k^{-4}$, which is gauge independent.
The IR divergence vanishes for the class of gauges where $\a - 2/\x = 1$,
then the integrand becomes $-k^{-2}(k^2+m^2)^{-2}$.
This class of gauges includes the Fermi-Feynman gauge,
which has been long known \cite{Grisaru1979} 
to have comparatively good infrared behaviour.

Finally, 
the propertime regularised expression for \eqref{SQED:QuadMatterTerm} is
\begin{align*}
	K^{(1)}_{R_\xi}
	&= \frac{M_v^2}{(4\p)^2} \int_{\rmi s_0}^{\infty}\!\frac{\rmd s}{s}
	\Bigg(\!\Big(\a-1-\frac2\x\Big)\rme^{-m^2_\text{IR}s} 
	- \Big(\a-\frac2\x\Big)\rme^{-m^2s}\Bigg) + \ord{M_v^4,\k\bar\k}\,,
\end{align*}
where we've also introduced $m^2_\text{IR}$ as an IR regulator. 
The integrals can be performed in terms of 
exponential integrals \eqref{defn:ExpIntegral}
\begin{align*}
	K^{(1)}_{R_\xi}
	&= \frac{M_v^2}{(4\p)^2}
	\Bigg(\!\Big(\a-1-\frac2\x\Big)E_1(\rmi s_0 m_\text{IR}^2)
	- \Big(\a-\frac2\x\Big)E_1(\rmi s_0 m^2)\Bigg) + \ord{M_v^4,\k\bar\k}\ .
\end{align*}
In the Fermi-Feynman gauge ($\a=1$ and $\x\to\infty$) this reduces to
\begin{align*}
	K^{(1)}_{\text{unren}}
	&= \frac{-M_v^2}{(4\p)^2} E_1(\rmi s_0 m^2) + \ord{M_v^4,\k\bar\k}
	\approx \frac{M_v^2}{(4\p)^2} \log(\rmi s_0 m^2 \rme^{\g}) 
	+ \ord{M_v^4,\k\bar\k} \,,
\end{align*}
where the last expression is also the general form of the UV divergence.

Enforcing the physical renormalisation condition
\begin{align}
	\frac{\pd^2 K}{\pd Q\pd Q^\dag}\Big|_{Q=0} = \ds1 \,, 
\end{align}
yields the matter renormalisation constant
\begin{align} 
	Z_Q	= 1 - \frac{\a}{2\p}E_1(\rmi s_0 m^2)+\ord{\a^2} 
		= 1 - \a Z_Q^{(1)} +\ord{\a^2} \ ,
\end{align}
which matches that of equation \eqref{SQED-ZQ1} provided the previously unfixed
finite term $Z_{Q,\text{finite}}^{(1)}$ is actually zero.

\chapter[Beta-deformed N=4 SYM]{\texorpdfstring{$\b$-deformed $\cN=4$}%
		{Beta-deformed N=4} super-Yang-Mills}
\label{Ch:BetaDef}


The marginal deformations \cite{Leigh1995}%
\footnote{%
The results of Leigh and Strassler have been further clarified and 
extended using different techniques in \cite{Kol2002, Green2010}.
}
of $\cN=4$ supersymmetric Yang-Mills theory (SYM) 
are a class of $\cN=1$ superconformal
field theories which enjoyed a lot of attention
in the first decade of the 21st century.  
In particular, the $\b$-deformed theory has been the subject of 
intense investigations, since its supergravity dual was found
in \cite{Lunin2005}.  
Many aspects of the $\b$-deformed theory have
been studied at both the perturbative and nonperturbative level.
In this chapter we concentrate only on perturbative aspects.

An important observation of \cite{Leigh1995} is that the 
renormalisation group beta function vanishes 
(the deformation becomes exactly marginal) 
subject to a single, loop corrected, constraint on the deformed couplings.  
The nature of this constraint has been examined in both the perturbative
and nonperturbative windows
using a range of methods and in a variety of limits, e.g.,  
\cite{Freedman2005,Penati2005,Rossi2005,Rossi2006,Mauri2005,
Khoze2006,Georgiou2006,Ananth2006,Chu2007,Oz2008}
and is still a topic of ongoing discussion
\cite{Elmetti2007a,Elmetti2007b,Elmetti2007c,Bork2008,Kazakov2007}.
Despite this wealth of knowledge about the requirements for conformal invariance
in $\b$-deformed theories, the exact functional nature of the 
quantum corrections has received less attention 
\cite{Kuzenko2005b,Kuzenko2007,Dorey2004}.
The purpose of this chapter is to continue in the vein of
\cite{Kuzenko2005b,Kuzenko2007} and investigate the structure of
the two-loop K\"ahler potential in the $\b$-deformed theory.

The K\"ahler potential is a supersymmetric generalisation 
of the effective potential \cite{Coleman1973} and thus it can be used to 
examine the renormalisation effects and vacuum structure of a quantised theory.
Superfield calculations of the one-loop K\"ahler potential in $\cN=1$ 
superspace are presented 
in \cite{Buchbinder1993,Pickering1996,Grisaru1996} and in section 
\ref{ssect:Neq1Kahler}.
Both the one- and two-loop corrections to the Wess-Zumino model 
are discussed in section \ref{sect:WZ-Kahler}.
A computation of the two-loop K\"ahler potential of a
general, non-renormalisable $\cN=1$ theory was presented in 
\cite{Nibbelink2005}.
Although this calculation includes the case of $\b$-deformed $\cN=4$ SYM, 
at the time of writing \cite{Tyler2008}, the work of \cite{Nibbelink2005}
had some technical problems
(see the conclusion of \cite{Tyler2008} for a discussion).

In $\cN=1$ theories the K\"ahler potential is a particularly interesting
sector of the low energy effective action in that it is not constrained 
by holomorphy in the way that the superpotential and gauge potential are. 
This is not the case for finite $\cN=2$ theories where the low-energy
non-renormalisation theorems \cite{Dine1997, Buchbinder1997, Lowe2008} 
imply that holomorphic $\cF(\cW)$ term receives no loop corrections
and the $\cH(\cW,\cWb)$ is one-loop exact.%
\footnote{There is some evidence \cite{Kuzenko2003,Kuzenko2004b,Kuzenko2004c} 
that 
the non-renormalisation results are not as strong as originally thought.}
As there is no correction to the one-loop K\"ahler potential in $\cN=4$ SYM,
the K\"ahler potential of $\b$-deformed $\cN=4$ 
SYM is a product purely of the deformation.
It is for this reason that we find the K\"ahler potential a particularly
interesting object to examine in the $\b$-deformed SYM theory.
%

\section{Classical action}\label{sect:Beta:ClassAct}
The classical action for $\b$-deformed $\cN=4$ SYM is
\begin{align} \label{defn:BetaDef-Act}
\begin{aligned}
	S &= \intz\sum_{i=1}^4\trF\F_i^\dag \F^i + \frac1{g^2}\intc\trF\cW^2 \\
	&+ \Big(\!h\intc\trF\,\F_1\!\com{\F_2}{\F_3}_q + \cc\!\!\Big) ,
\end{aligned}
\end{align}
where $q=\rme^{\rmi\p\b}\in\dsC$ and the $q$-deformed commutator
is defined by
\begin{align} \label{defn:q-deformed-commutator}
	\com{\F_2}{\F_3}_q = q\F_2\F_3-q^{-1}\F_3\F_2 \ .
\end{align}
The superfields $\cW_\a$ and $\F_i$ are covariantly chiral 
and are in the adjoint representation of ${SU}(N)$, 
which we write in \eqref{defn:BetaDef-Act} using matrices in the 
fundamental representation.\footnote{%
We note that since there are terms that are trilinear in the group generators,
we can not naively use an arbitrary representation 
and the normalised trace used for the bilinear $\cN=1$ gauge kinetic terms
of chapter \ref{Ch:Neq1Quant}.
This is because the ratio of the Dynkin and anomaly indices, $T(R)/A(R)$,
depends on the representation $R$. That said, any difference in 
normalisation can be absorbed into $q_+=(q+q^{-1})/2$.}
In the limit of vanishing deformation, $h \to g$ and $q \to 1$, 
the action \eqref{defn:BetaDef-Act} becomes that of $\cN=4$ SYM.

Although there is a cyclical symmetry%
\footnote{This symmetry is a remnant of the the $SU(3)$ subgroup 
of the $SU(4)$ R-symmetry that exists in the undeformed theory.} 
in the $\F_i$ of \eqref{defn:BetaDef-Act},
it helps to think of the action as a $\cN=2$ gauge multiplet constructed
from $(\F_1,\cW_\a)$ and a \emph{deformed hypermultiplet} $(\F_2,\F_3)$ 
\cite{Kuzenko2007}.  
Then, if we quantise using a covariantly constant 
$\cN=2$ SYM background in the Cartan subalgebra 
and a vanishing background deformed hypermultiplet, 
we are automatically on the Coulomb branch of the theory.
Such a background will, in general, break the gauge symmetry to the 
maximal torus, $SU(N)\to U(1)^{N-1}$.

This choice of background also means that all aspects of the deformation 
are captured in the hypermultiplet propagators and chiral cubic vertices.
This is particularly important as it greatly simplifies the calculation of terms 
in higher loop contributions to the effective action \cite{Kuzenko2007}.
As mentioned above, in the undeformed $\cN=4$ theory, the K\"ahler
potential and $\cF^4$ terms do not receive any quantum corrections.
This means that if we wish to calculate such terms in the $\b$-deformed theory,
we need only calculate the diagrams that are effected by the deformation
and subtract from them their undeformed counterparts.
Calculating the two-loop $\cF^4$ correction,  
only required four diagrams \cite{Kuzenko2007}.
For the two-loop K\"ahler potential only two nonvanishing diagrams remain 
\cite{Tyler2008}.

As we are only interested in calculating the K\"ahler potential
we make the particularly simple quantum-background split
\begin{equation}\label{Beta:qtm.bgnd.split}
	\F_1 \to \F+\vf_1\,,\quad 
	\F_{2,3} \to \vf_{2,3} \,,\quad
	\cD_\a \to \rme^{-gv} D_\a \rme^{gv}\,,\quad 
	\cDb_\da \to \Db_\da \,.
\end{equation}
We then quantise in the `t Hooft gauge, as in chapter \ref{Ch:Neq1Quant},
to obtain the quadratic terms in the action
\begin{subequations}\label{Beta:quad.class.act}\begin{align}
	S_{\rm YM}^{(2)}
		&=\intz\trF\left(-\half v(\square - |\cMg|^2)v 
	  	+ \vf_1^\dag\frac{\square-|\cMg|^2}{\square}\vf_1\right)\\
	S_{\rm hyp}^{(2)}
		&=\intz\trF\left(\vf_2^\dag\vf_2+\vf_3^\dag\vf_3\right)
	  	+\Big(\intc\trF\vf_3\cMh\vf_2+
	  	\cc \Big)\\
	S_{\rm gh}^{(2)}
		&=\intz\trF\left(c^\dag\frac{\square-|\cMg|^2}{\square}\ct
	  	-\ct^\dag\frac{\square-|\cMg|^2}{\square}c\right) ,
\end{align}\end{subequations}
where all of the masses defined in section \ref{sect:Neq1Pert} 
have been written using the mass operators introduced in \cite{Kuzenko2005b}, 
which are elegantly defined by their action on a 
Lie algebra valued superfield $\S$ in the fundamental representation:
\begin{equation}\label{Beta:mass.op.defn}\begin{split}
	\cMh\S&=h({q}\F\S-\qm\S\F)-h\frac{{q}-\qm}{N}\trF(\F\S)\ds1\\
	\cMh^\dag\S&=\hb(\qbar\F^\dag\S-\qbarm\S\F^\dag)
		-\hb\frac{\qbar-\qbarm}{N}\trF(\F^\dag\S)\ds1\,.
\end{split}\end{equation}
The relevant interactions for the two-loop diagrams of interest are the
cubic couplings
\begin{subequations}\label{Beta:cubic-couplings}\begin{align} 
	S_{\rmI}^{(3)}&=
		h\intc\trF\left(q\vf_1\vf_2\vf_3-\qm\vf_1\vf_3\vf_2\right)+\cc \non \\
 		&= - \intc\,\Th_{\m\n\k}\,\vf_1^\m\vf_2^\n\vf_3^\k - \cc  \\
	S_{\rmII}^{(3)}&=
		g\intz\trF\left(\vf_i^\dag[v,\vf_i]\right)
		= - \intz\,\Tg_{\m\n\k}\,\vfb^\m_iv^\n\vf^\k_i\,,
\end{align}\end{subequations}
where, following \cite{Kuzenko2007}, 
we introduce the $q$-deformed adjoint generators%
\begin{align}\begin{aligned}
	h\com{T_\m}{T_\n}_q &= T_\k(\Th_\m)^\k{}_\n  
						+ h\frac{q-q^{-1}}N g_{\m\n}^{(F)} \ds1  \\ 
	\implies
	\Th_{\m\n\k} &= g^{(F)}_{\n\l}(\Th_\m)^\l{}_\k 
				  =  h\trF(q T_\m T_\n T_\k - \qm T_\n T_\m T_\k )\,,
\end{aligned}\end{align}
which enjoy the algebraic properties
\begin{equation}
	\Th_{\m\n\k} = \Th_{\n\k\m} = -T_{\n\m\k}^{(h,1/q)}\,, \quad 
	(\Th_{\m\n\k})^* = T_{\n\m\k}^{(\hb,\qbar)} \ .
\end{equation}
Note that the deformed generators can also be used to give the mass operator
the compact representation $\cMh = \F^\k T_\k^{(h,q)}$.

The propagators for the action \eqref{Beta:quad.class.act} that are used in
the two-loop calculation below are 
\begin{subequations}\begin{align}
	\rmi \big<v(z)v^\rmT(z')\big> 
		&=-G_{(g,1)}(z,z')  \\
	\rmi \big<\vf_1(z)\vf_1^\dag(z')\big>
		&=\frac{\Db^2D^2}{16}G_{(g,1)}(z,z') \\
	\rmi \big<\vf_2(z)\vf_2^\dag(z')\big>
		&=\frac{\Db^2D^2}{16}\Ghl_{(h,q)}(z,z')  \\
	\rmi \big<\vfb_3(z)\vf_3^\rmT(z')\big>
		&=\frac{D^2\Db^2}{16}\Ghr_{(h,q)}(z,z') \,,
\end{align}\end{subequations}
where all of the fields are treated as adjoint column-vectors, 
in contrast to the Lie-algebraic notation used in defining the action.
In this chapter, unlike chapter \ref{Ch:Neq1Quant}, 
we find it convenient to treat the three $\vf_i$ separately 
and not as a column vector.
Thus we don't use a symmetric, blocked mass matrix
and we have to use the ``left'' and ``right'' propagators defined by
\begin{subequations}\begin{align}
	\left(\square-\cMh^\dag\cMh\right)\Ghr_{(h,q)}(z,z')&=-\d^8(z,z')\\
	\left(\square-\cMh\cMh^\dag\right)\Ghl_{(h,q)}(z,z')&=-\d^8(z,z')\,,
\end{align}\end{subequations}
with the usual, causal boundary conditions.  
As we only have flat derivatives, the above propagators are 
most simply expressed by moving to momentum space.
In the limit of vanishing deformation the mass matrices commute 
so that the left and right Green's functions coincide: 
$\Ghr_{(g,1)}=\Ghr_{(g,1)}=G_{(g,1)}$.

Throughout this chapter we will use regularisation by dimensional reduction
\cite{Siegel1979} and since we only go to two loops 
and do not have a vector (gauge) field background,
we do not worry about any possible inconsistencies
\cite{Siegel1980,Avdeev1981}.
This is merely a convenience, as none of the results in this paper
rely on the choice of regularisation scheme 
and can all be argued at the level of the integrands.

\subsection{Cartan-Weyl basis and the mass operator}\label{ssect:Beta:CWbMO}
The properties of the mass matrices defined in \eqref{Beta:mass.op.defn} 
play a central role in our computations. 
Since the the background $\F$ is in the Cartan subalgebra,
a natural choice of basis for our gauge group is
the Cartan-Weyl basis, see e.g., \cite{Barut1986}.
In this subsection we introduce some notation and a few results that will
be used in the loop calculations.

Any element in $\su(N)$ can be expanded in the Cartan-Weyl basis,
\begin{equation} 
	\j = \j^\m T_\m = \j^{ij}E_{ij}+\j^IH_I  \quad  i \neq j\,,
\end{equation}
where $T_\m$ is the basis for the fundamental representation used above and 
we choose our Cartan-Weyl basis as the set
\begin{equation}
	\left\{ E_{ij}, H_I \quad \Big| \quad 
	i\neq j=1,\dots, N\,, \quad I=1,\dots, N-1 \right\} ,
\end{equation}
where $E_{ij}$ are the elementary matrices
\begin{align} 
	(E_{ij})_{kl}=\d_{ik}\d_{jl}\,,
\end{align}
and the generators of the Cartan subalgebra can be chosen to be
\begin{align} 
	H_I=\frac1{\sqrt{I(I+1)}}\sum_{i=1}^{I+1}\big(1-i\d_{i(I+1)}\big)E_{ii} \ . 
\end{align}
The Cartan-Weyl basis satisfies%
\footnote{Due to our choice of normalisation 
the Cartan metric is just the Kronecker delta, thus
we can raise and lower the group indices with impunity.
} 
\begin{equation}
	\trF E_{ij}E_{kl}=\d_{il}\d_{jk} \,,
	\quad \trF H_I H_J = \d_{IJ} 
	\quad\text{and}
	\quad \trF E_{ij}H_K=0
\end{equation}

Since the background is chosen to lie in the Cartan subalgebra, 
\[ 
	\F = \f^I H_I \deq \F^i E_{ii} \,,
\]
the mass matrix is block diagonal when written in the Cartan-Weyl basis
\begin{equation}
	\cMh = \bem \cMh^{ijkl} & 0\\0&\cMh^{IJ} \eem
		 = \bem m^{ki}\d_{il}\d_{jk} & 0\\0&\cMh^{IJ} \eem \quad \ns,
\end{equation}
where the masses $m^{ij}$ are defined by
\begin{equation} \label{Beta:weyl.mass}
	m^{ij} = h(q\F^i-\qm\F^j) \ .
\end{equation}
The mass matrix in the Cartan subalgebra is symmetric, but in general not 
diagonal,  we find
\begin{equation}
	\cMh^{IJ} = h(q-\qm)\f^K \trF\big(H^IH^JH^K\big) = \cMh^{JI} \ .
\end{equation}
In the limit of vanishing deformation the above expression is obviously zero,
and we will denote that limit of the masses in \eqref{Beta:weyl.mass} by
\begin{equation}
m^{ij}\underset{h=g}{\xrightarrow{q=1}}m_0^{ij}=g(\F^i-\F^j)\ .
\end{equation}
It is now straightforward to calculate the mass squared matrix, 
it is also block diagonal and has the non-zero components
\begin{align}\label{Beta:mass2}
	\left(\cMh^\dag\cMh\right)^{ijkl}&=\left(\cMh\cMh^\dag\right)^{ijkl}
		=|m^{ki}|^2\d_{il}\d_{jk}  \\ \non
	\left(\cMh^\dag\cMh\right)^{IJ}&=\left(\cMh\cMh^\dag\right)^{JI}
		=|h(q-\qm)|^2\fb^L\f^M\h^{IJLM}\,, 
\end{align}
where
\begin{equation}\label{Beta:eta.defn} 
	\h^{IJLM} = \trF\big(H^IH^JH^LH^M\big)-\frac1N\d^{IL}\d^{JM} \ .
\end{equation}

To proceed in the one and two-loop calculations below, we will need to 
assume that the eigenvalues and eigenvectors of the mass squared matrix are
known, that is, we know a unitary matrix $U$ such that
\begin{equation}\label{Beta:mass.eigensystem}
\left(U^\dag\cMh^\dag\cMh U\right)^{IJ}=|m_I|^2\d^{IJ} \quad \ns \ .
\end{equation}
We also need the trace of the mass squared operator.  This requires the
trace of $\h^{IJLM}$, which can be found using the
completeness relation for the Cartan subalgebra.  The final expression
is simplified by using the tracelessness of $\F$ to get
\begin{align}\label{Beta:mass.trace}\non
	\tr|\cMh|^2&=|h|^2\left(\sum_{i\neq j}|m^{ij}|^2+\sum_I|m_I|^2\right)\\
	&=N|h|^2\left(|q|^2+|\qm|^2-\frac2{N^2}|q-\qm|^2\right)\tr\F^\dag\F \ .
\end{align}

\section{One-loop K\"ahler potential}\label{sect:Beta:1loop}
From the quadratic terms defined in \eqref{Beta:quad.class.act} we can 
read off, see e.g., section \ref{sect:Neq1-1Loop}, 
the one-loop effective action as
\begin{equation}\label{Beta:oneloop.ea.1} \G^{(1)}=
	\rmi\Tr\ln(\square - |\cMh|^2 P_+)-\rmi\Tr\ln(\square - |\cMg|^2 P_+)\,,
\end{equation}
where $\Tr$ is both a matrix trace 
and a functional trace over full superspace.
The evaluation of the functional trace is described in 
subsection \ref{ssect:Neq1Kahler}, 
factoring out the integral over full superspace we get the 
one-loop K\"ahler potential 
(which can also be derived from \eqref{Neq1 1loop Kahler})
\begin{align} 
	K^{(1)} &= \trad\scJ(|\cMh|^2) -\trad\scJ(|\cMg|^2) \,,
\end{align}
where the dimensionally regularised (DRed) integral $\scJ(m^2)$ 
and its $\eps$-expansion are
(see subsection \ref{ssect:Neq1Kahler})
\begin{align} 
	\scJ(m^2) 
	&= -\rmi\m^{2\eps}\intk\frac1{k^2}\log(k^2+m^2-\rmi\e)\,,\quad\e\to0 \non\\
	&= \frac{m^2}{(4\p)^2}\left(\k_M-\log\frac{m^2}{M^2}
		+\ord\eps\right) \,,\\
	\label{Beta:KappaM}
	\k_{M} &= \frac1\eps+2-\log\frac{M^2}{\mub^2} \,,
	\quad \mub^2=4\p\rme^{-\g}\m^2\,,
\end{align}
with $M^2>0$ an arbitrary mass scale.
For other regularisation schemes, we find similar expressions for $\k_M$.
The matrix trace can be converted to a sum of eigenvalues using the results of 
subsection \ref{ssect:Beta:CWbMO}, 
to get the one-loop K\"ahler potential in the form
\begin{align}\label{Beta:1loop.eqn1}
	K^{(1)} &= \sum_I\scJ(|m_I|^2)
  	+\sum_{i\neq j}\left(\scJ(|m^{ij}|^2)-\scJ(|m_0^{ij}|^2)\right) \ .
\end{align}

As a check on the above result, we note that it is zero in the limit of 
vanishing deformation. 
The $\k_M$-dependent terms are proportional to the trace of the difference 
of the deformed and undeformed mass matrix
\begin{equation} 
	\sum_I|m_I|^2+\sum_{i\neq j}\left(|m^{ij}|^2-|m_0^{ij}|^2\right)=
	\trad\left(|\cMh|^2-|\cMg|^2\right) .
\end{equation}
So, using the trace formula \eqref{Beta:mass.trace},
it is easily seen that the above
term is zero if the well known one-loop finiteness condition holds 
\cite{Freedman2005,Penati2005,Rossi2005},
\begin{equation}\label{Beta:finite.cond}
	2g^2 = |h|^2 
		\left(|q|^2+|\qm|^2-\frac2{N^2}|q-\qm|^2\right)
		\deq 2|h|^2f_q \ .
\end{equation}

If we enforce the finiteness condition and choose $M^2$ to be any 
nonvanishing field dependent mass term 
then we get the explicitly superconformal result
\begin{align}\begin{aligned}
	(4\p)^2K_{\text{finite}}^{(1)}
	&= \sum_{i\neq j}\left(
		|m_0^{ij}|^2\log\frac{|m_0^{ij}|^2}{M^2}
	  -	|m^{ij}|^2\log\frac{|m^{ij}|^2}{M^2}\right) \\
	&-\sum_I|m_I|^2\log\frac{|m_I|^2}{M^2} \ .
\end{aligned}\end{align}
We emphasise that this result is independent of the choice of $M^2$.

\section{Two-loop K\"ahler potential}\label{sect:Beta:2loop}
\begin{figure}[t]
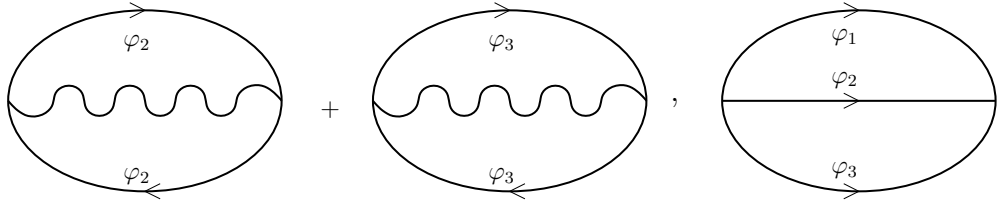

  \centering
  \hfill\raisebox{-0.5\height}{\scalebox{0.8}{
  		\input{figures/BetaDefGamma1.pspdftex}}} \ ,
  \hfill\raisebox{-0.5\height}{\scalebox{0.8}{
  		\input{figures/BetaDefGamma2.pspdftex}}}\hspace{20mm}
  \caption{The two-loop diagrams contributing to the K\"ahler potential,
	$\G_1$ and $\G_2$ respectively.
	The arrows show the flow of chirality around the loop, while the
	fields label the propagators.  The wavey line corresponds to 
	the $\cN=1$ gauge superfield.}
  \label{Beta:2loop.fig}
\end{figure}

In the $\b$-deformed theory there are only four two-loop diagrams that differ 
from the undeformed theory \cite{Kuzenko2007}, 
but some simple $D$-algebra shows that only two give non-zero contributions 
to the K\"ahler potential, these are given in figure \ref{Beta:2loop.fig}.
Both are of the sunset type and have the generic group theoretic structure
\begin{equation}
	\G=\k\intz\rmd^8z'G^{\m\n}\trad\left(
	T_\m^{(h,\qm)}\hat G_{(h,q)}T_\n^{(\hb,\qbarm)}\check{G}'_{(h,q)}\right)\,,
\end{equation}
where $G$ is an undeformed Green's function and, in general,
$\hat G$ and $\check G$ denote spinor derivatives of deformed Green's functions.
This decomposes in the Cartan-Weyl basis into three terms,
\begin{align}\label{Beta:decomp.CW} 
	\G =
	&{}\k |h|^2 \intz\rmd^8z'\Big(
	G^{ijji}\big(q\qbar\hat{G}^{j kkj}_{(h,q)}\check{G}'^{ikki}_{(h,q)}
	 +(q\qbar)^{-1}\hat{G}^{k iik}_{(h,q)}\check{G}'^{kjjk}_{(h,q)}\big)\non\\
	&+(q(H_K)_{jj}-q^{-1}(H_K)_{ii})(\qbar(H_L)_{jj}-\qbar^{-1}(H_L)_{ii})\times
	\non\\\non&\qquad\qquad\qquad\times
	\big(G^{KL}\hat{G}^{ijji}_{(h,q)}\check{G}'^{ijji}_{(h,q)}
	+ G^{jiij}\hat{G}^{KL}_{(h,q)}\check{G}'^{jiij}_{(h,q)}
	+ G^{ijji}\hat{G}^{jiij}_{(h,q)}\check{G}'^{LK}_{(h,q)}\big)\Big.\\
   &+ |q-q^{-1}|^2 G^{IJ}\hat{G}^{MN}_{(h,q)}\check{G}'^{LK}_{(h,q)}
      \trF(H_IH_KH_M)\trF(H_JH_LH_N) \Big) \\\non
   =&{} \G_A+\G_B+\G_C \ .
\end{align}
We should note that if the vertices are undeformed, \ie 
$T^{(h,q)}_\m \to T^{(g,1)}_\m = g T^{(\text{ad})}_\m$, 
then the final term, $\G_C$, is zero.

For an arbitrary background in the Cartan subalgebra
$\G_A$ is easy to evaluate as all of its Green's functions are diagonal.  
To evaluate the other terms, which involve sums over the Cartan subalgebra,
we will use the unitary matrices defined in 
\eqref{Beta:mass.eigensystem} to diagonalise the Green's functions, 
\begin{equation}
	(H_I)_{jj} \Ghl_{(h,q)}^{IJ} (H_J)_{ii}
	=(H_I)_{ii} \Ghr_{(h,q)}^{IJ} (H_J)_{jj}
	=(\bar\bH_K)_{ii}G_{(h,q)}^{(K)}(\bH_K)_{jj} \ .
\end{equation}
The modified generators are defined by
\begin{equation}
	\bH_I=U_I{}^J H_J \,,\quad 
	\bar\bH_I=H_J (U^\dag)^J_{~I} \ .
\end{equation}
In the following subsections, these modified generators will be 
combined into coefficients for the scalar loop integrals.  
Alternatively, as was done, e.g., in \cite{Nibbelink2005}, 
we could reabsorb the diagonalising unitary matrices
back into the loop integrals to get a matrix valued expression.
Although this does make some expressions look a bit neater and keep
all of the field dependence in the now matrix valued loop integrals, 
to evaluate the these expressions we would still have to diagonalise the 
mass matrices. 

\subsection{Evaluation of \texorpdfstring{$\G_\rmI$}{Gamma 1}}
The first diagram we evaluate has the analytic expression
\begin{align}\label{Beta:gammaI} 
\begin{aligned} 
	\G_{\rmI} &= -\frac1{2^8}\intz\rmd^8z'G^{\m\n}_{(g,1)}(z,z')\times \\
		&\trad\Big( T^{(h,\qm)}_\m\Db^2D^2\Ghl_{(h,q)}(z,z')
    T^{(\hb,\qbarm)}_\n D'^2\Db'^2\Ghr_{(h,q)}(z',z)\Big) \ .
\end{aligned}
\end{align}
For a nonzero result to occur when integrating over $\rmd^4\q'$, 
all Grassmann derivatives have to hit the 
Grassmann delta functions contained in the deformed propagators.  
Then, writing $\scG$ for the remaining bosonic parts 
of the propagators, shifting the $x'$ integration variable to $\r=x-x'$
and using \eqref{Beta:decomp.CW} we obtain
\begin{align}
	K_{\rmI} =& -|h|^2\intrho\Big\{
		\Gb^{ijji}_{(g,1)}\big(
			q\qbar{\Gb}^{jkkj}_{(h,q)}{\Gb}'^{ikki}_{(h,q)}
			+(q\qbar)^{-1}{\Gb}^{kiik}_{(h,q)}{\Gb}'^{kjjk}_{(h,q)}\big)\non\\
	&+(q(H_K)_{jj}-q^{-1}(H_K)_{ii})(\qbar(H_L)_{jj}-\qbar^{-1}(H_L)_{ii})
	\times\Big.\non\\\non&\qquad\qquad\times
	\big(\Gb^{KL}_{(g,1)}\Gb^{ijji}_{(h,q)}\Gb'^{ijji}_{(h,q)}
	+\Gb^{jiij}_{(g,1)}\Gbhl^{KL}_{(h,q)}\Gb'^{jiij}_{(h,q)}
	+\Gb^{ijji}_{(g,1)}\Gb^{jiij}_{(h,q)}\Gbhl'^{KL}_{(h,q)}\big)\\%
	&+ |q-q^{-1}|^2 \Gb^{IJ}_{(g,1)}\Gbhl^{MN}_{(h,q)}\Gbhr'^{LK}_{(h,q)}
      \trF(H_IH_KH_M)\trF(H_JH_LH_N) \Big\} \non\\
	=&{} K_{\rmI A}+K_{\rmI B}+K_{\rmI C} \ .
\end{align}
The above has been slightly simplified 
by using the symmetries of the propagators.

Now, as all of the propagators in $K_{\rmI A}$ are already diagonal, 
we can move straight to momentum space and perform the $\r$ integral to get
\begin{align*} 
\begin{aligned} K_{\rmI A} &= -|h|^2\sum_{i\neq j\neq k}
	\int\frac{\rmd^dk\rmd^dp}{(2\p)^{2d}}
 	\frac1{k^2+|m_0^{ij}|^2} \times \\
 	&\times\left(\frac{|q|^2}{p^2+|m^{kj}|^2}\frac1{(k+p)^2+|m^{ki}|^2} 
  		+  \frac{|\qm|^2}{p^2+|m^{ik}|^2}\frac1{(k+p)^2+|m^{jk}|^2}\right) .
\end{aligned}
\end{align*}
Then, using the results and notation of appendix \ref{A:2LoopVac} we have
\begin{equation} K_{\rmI A}=
	 |h|^2\sum_{i\neq j\neq k}\left(
	 |q|^2I(|m_0^{ij}|^2,|m^{ki}|^2,|m^{kj}|^2)
	+\frac1{|q|^2}I(|m_0^{ij}|^2,|m^{ik}|^2,|m^{jk}|^2)\right) .
\end{equation}

To evaluate $K_{\rmI B}$ we diagonalise the propagators, as described above.
The result is
\begin{align}
K_{\rmI B}&=-|h|^2\sum_{i\neq j,K}\intrho
	(q(\bar{\bH}_K)_{jj}-\frac1q(\bar{\bH}_K)_{ii})
	(\qbar({\bH}_K)_{jj}-\frac1\qbar({\bH}_K)_{ii})\times
	\non\\&\times
	\big(\Gb^{(K)}_{(g,1)}\Gb^{ijji}_{(h,q)}\Gb'^{ijji}_{(h,q)}
	+\Gb^{jiij}_{(g,1)}\Gb^{(K)}_{(h,q)}\Gb'^{jiij}_{(h,q)}
	+\Gb^{ijji}_{(g,1)}\Gb^{jiij}_{(h,q)}\Gb'^{(K)}_{(h,q)}\big) \\	\non
	&= |h|^2 \sum_{i\neq j, K}\varpi_{\qbar,Kij}\left(
	 I(0,|m^{ji}|^2,|m^{ji}|^2) 
	 + 2 I(|m_0^{ij}|^2,|m^{ij}|^2,|m_K|^2)
	  \right) ,
\end{align}
where $\varpi_{q,Kij}$ is defined by
\begin{equation}
	\varpi_{q,Kij} = \left(q(\bH_K)_{ii}-\qm(\bH_K)_{jj}\right)
		\left(\qbar(\bar\bH_K)_{ii}-\qbarm(\bar\bH_K)_{jj}\right) \ .
\end{equation}

Similarly we evaluate $K_{\rmI C}$ to find
\begin{align} 
	K_{\rmI C} = |h(q-\qm)|^2\sum_{I,J}\bm\h_{IJ} I(0,|m_I|^2,|m_J|^2) \,,
\end{align}
where $\bm\h$ is closely related to $\h$, 
defined in \eqref{Beta:eta.defn},
\begin{equation}
	\bm\h_{IJ} = \trF(\bar{\bH}_I\bar{\bH}_J{\bH}_I{\bH}_J )
		-\frac1N (U^TU)_{IJ}(U^\dag U^*)_{JI}\,,\quad\ns\ .
\end{equation}

Note that in general the coefficients $\varpi_{q,Kij}$ and $\bm\h_{IJ}$ are
functions of ratios of the background dependent masses.
\subsection{Evaluation of \texorpdfstring{$\G_\rmII$}{Gamma 2}}
The second diagram,
\begin{align}\label{Beta:gammaII}
	\G_\rmII 
  	&=\frac1{2^9}\intz\rmd^8z'G^{\m\n}_{(g,1)}(z,z')\trad\Big(
  	\Big.\non\\&\Big.\qquad\phantom{+}
    T^{(g,1)}_\m \Db^2D^2\Ghl_{(h,q)}(z,z')
    T^{(g,1)}_\n D'^2\Db'^2\Ghl_{(h,q)}(z',z)
  \Big.\\\non&\Big.\qquad+
	T^{(g,1)}_\m \Db^2D^2\Ghr_{(h,q)}(z,z')
	T^{(g,1)}_\nu D'^2\Db'^2\Ghr_{(h,q)}(z',z)\Big) ,
\end{align}
is simpler to evaluate due to the lack of deformed vertices.  
Following the same procedure as above we find 
$K_{\rmII} = K_{\rmII A}+K_{\rmII B}$, with
\begin{equation}
	K_{\rmII A}=-g^2 \sum_{i\neq j\neq k}\left(
		I(|m_0^{ji}|^2,|m^{kj}|^2,|m^{ki}|^2)+
		I(|m_0^{ji}|^2,|m^{ik}|^2,|m^{jk}|^2)\right) ,
\end{equation}
and
\begin{equation}
K_{\rmII B}=-g^2 \sum_{i\neq j, K}\varpi_{1,Kij}\left(
	     I(0,|m^{ji}|^2,|m^{ji}|^2) 
	 + 2 I(|m_0^{ij}|^2,|m^{ij}|^2),|m_K|^2\right) .
\end{equation}
\subsection{Finiteness and conformal invariance}
Combining the two diagrams we see that the two-loop K\"ahler potential, 
like its one-loop counterpart,  is written as the difference 
of terms that cancel in the limit of vanishing deformation:
\begin{align}\label{Beta:K2.all} 
\begin{aligned}
	K^{(2)}&=\sum_{i\neq j\neq k}
   \big((|hq|^2-g^2)\, I(|m_0^{ij}|^2,|m^{ki}|^2,|m^{kj}|^2) \\
	 &\qquad+(|h\qm|^2-g^2)I(|m_0^{ij}|^2,|m^{ik}|^2,|m^{jk}|^2)\big) \\
	 &+\sum_{i\neq j, K}\left(|h|^2\varpi_{\qbar,Kij}-g^2\varpi_{1,Kij}\right)
	 \\&\qquad\times
	 \left(I(0,|m^{ji}|^2,|m^{ji}|^2) 
	 + 2 I(|m_0^{ij}|^2,|m^{ij}|^2,|m_K|^2)\right)\\
	 &+|h(q-\qm)|^2\sum_{I,J}\bm\h_{IJ} I(0,|m_I|^2,|m_J|^2) \ .
\end{aligned}
\end{align}
As described in appendix \ref{A:2LoopVac}, 
the two-loop integral, $I(x,y,z)$, can be decomposed as
\begin{equation} \tag{\ref{eqn:2LoopExpansion}}
	I(x,y,z)=\i(x)+\i(y)+\i(z)+\dsI(x,y,z)\,,
\end{equation}
where the $\i$ terms include all of the divergences and renormalisation point
dependence, and $\dsI$, defined in \eqref{defn:2Loop_dsI}, is a 
function of mass ratios only.
Since the masses are disentangled in the $\i$ terms, 
the sums can be simplified by using the following identities:%
\footnote{%
	Note that using \eqref{Beta:varpisum} 
	it becomes possible to perform the sum over $K$
	in the first term of the middle line of \eqref{Beta:K2.all}.
}
\begin{subequations}\label{Beta:sums}
\begin{align}
\label{Beta:varpisum} 
	\sum_K\varpi_{q,Kij} &= \frac{1}{N}\sum_{i\neq j}\varpi_{q,Kij}
		= |q|^2+\frac1{|q|^2}-\frac1N|q-\qm|^2 \deq 2g_q \\
\label{Beta:etasumJ}	
	\sum_J\bm\h_{IJ} &= \sum_J\h_{IIJJ}=\frac{N-2}N \ .
\end{align}
\end{subequations}
The result is that all $\i$ dependence can be collected into
\begin{equation}\begin{split} K_\i^{(2)}=&
N\left(|h|^2\left(|q|^2+|\qm|^2-\frac2{N^2}|q-\qm|^2\right)-2g^2\right)\\&
\times\left(\sum_{i\neq j}\left(\i(|m_0^{ij}|^2)+2\i(|m^{ij}|^2)\right)
+2\sum_{I}\i(|m_I|^2)\right)\ .
\end{split}\end{equation}

From the above expression and the trace formulae given in subsection 
\ref{ssect:Beta:CWbMO} we may read off the quadratic terms in the 
K\"ahler potential:
\begin{align*} 
	K^{(2)}_{\text{quad}} 
		\propto 4N^2(|h|^2f_q-g^2)(2|h|^2f_q+g^2)\tr\F^\dag\F \,,	
\end{align*}
where the constant of proportionality is a number that is 
subtraction scheme dependent and
$f_q$ is the function 
that occurs in the one-loop finiteness condition \eqref{Beta:finite.cond}.
The above prefactor is, for good reason, reminiscent of the general
expression for the two-loop anomalous dimension given in, for example, 
\cite{West1984,Parkes1985,Hamidi1984,Jack1996}.

So, as expected, the two-loop K\"ahler potential is finite and independent of
the renormalisation point if the one-loop finiteness condition, 
\eqref{Beta:finite.cond}, is satisfied.  
It is interesting to note that the `meaning'
of \eqref{Beta:finite.cond} is different at one and two-loops.  
At one-loop it implies that the trace of the mass matrix 
is invariant under the deformation, 
while at two loops it implies that 
the coefficients of the scalar diagrams sum to zero.

If we enforce the finiteness condition, \eqref{Beta:finite.cond}, 
then we get the explicitly superconformal two-loop K\"ahler potential 
by making the replacements
$g^2 \to |h|^2f_q$ and $I \to \dsI$ in \eqref{Beta:K2.all}.

\section{Special backgrounds and explicit masses}\label{sect:Beta:simpBG}
In the above analysis the background superfield pointed in an arbitrary 
direction in the Cartan subalgebra of $\su(N)$.
In order to make our previous analysis concrete 
we now choose the specific background
\begin{equation}\label{Beta:special-background}
	\F = \sqrt{N(N-1)}\f_1H_{N-1}+\sqrt{(N-1)(N-2)}\f_2H_{N-2} \ .
\end{equation}
The characteristic feature of this background is that it leaves the 
subgroup $U(1)^2 \times SU(N-2)$ of $SU(N)$ unbroken.  
The two $U(1)$s are associated with the generators $H_{N-1}$
and $H_{N-2}$.  In the limit $\f_2\to0$, we obtain the background previously
used for the calculation of the two-loop K\"ahler potential in 
\cite{Kuzenko2007}. 

There are twelve different, nonzero masses that occur with this background.
There are nine deformed masses:
\begin{subequations}
\begin{align}
	m_1^2&=|m^{ij}|^2=|m_I|^2=|h(q-\qm)|^2|\f_1+\f_2|^2\,,  \\
	m_2^2&=|m^{i(N-1)}|^2=|h(q-\qm)\f_1+h(q+(N-2)\qm)\f_2|^2\,,  \\
	 m_{\tilde 2}^2&=|m^{(N-1)j}|^2=m_2^2\big|_{q\to\qm}\,,  \\
	m_3^2&=|m^{iN}|^2=|hq\f_2+h(q+(N-1)\qm)\f_1|^2\,,  \\
	 m_{\tilde 3}^2&=|m^{Nj}|^2=m_3^2\big|_{q\to\qm}\,,  \\
	m_4^2&=|m^{(N-1)N}|^2=|h(N-2)q\f_2-h(q+(N-1)\qm)\f_1|^2\,,  \\
	 m_{\tilde 4}^2&=|m^{N(N-1)}|^2=m_4^2\big|_{q\to\qm}\,,  \\
	m_\pm^2&=\half|h(q-\qm)|^2\left(a+c\pm\sqrt{(a-c)^2+4|b|^2}\right)\,,
\end{align}
where the indices $i,j$ and $I$ range from $1$ to $(N-2)$ and $(N-3)$ 
respectively, and their three undeformed counterparts:
\begin{align}
	m_{02}^2&=|m_0^{i(N-1)}|^2=|m_0^{(N-1)j}|^2=g^2(N-1)^2|\f_2|^2\,, \\
	m_{03}^2&=|m_0^{iN}|^2=|m_0^{Nj}|^2=g^2|N\f_1+\f_2|^2\,, \\
	m_{04}^2&=|m_0^{(N-1)N}|^2=|m_0^{N(N-1)}|^2=g^2|N\f_1-(N-2)\f_2|^2\ .
\end{align}\end{subequations}
The quantities $a$, $b$ and $c$ come from the Cartan subalgebra block of the 
mass matrix, which is diagonal except for the bottom $2\times2$ block: 
\begin{align}\label{Beta:Cartan.mass.bg}
 (\cMh^\dag\cMh)_{IJ} = |h(q-\frac1q)|^2
	\left(\begin{array}{ccc;{1pt/3pt}cc}
		|\f_1+\f_2|^2 & & & &\\ 
		& \ddots & & & \\
		& & |\f_1+\f_2|^2 & & \\\hdashline[1pt/3pt]
		& & & a & b \\
		& & & b^* & c	 
		\end{array}\right)\,, 
\end{align}
where the explicit values for $a$, $b$ and $c$ are
\begin{subequations}
\begin{align}\label{Beta:Cartan.mass.abc}
	a&=\big((N-3)^2+1-2/N\big)\fb_2\f_2+\fb_1\f_1-(N-3)(\fb_2\f_1+\fb_1\f_2) \\
	b&=\sqrt{1-2/N}\big((3-N)\fb_2\f_2+\fb_1\f_2-(N-2)\fb_2\f_1\big) \\
	c&=(N-2)^2\fb_1\f_1+\big(1-2/N\big) \fb_2\f_2\ .
\end{align}
\end{subequations}
The eigenvalues of the matrix \eqref{Beta:Cartan.mass.bg} 
are $m_1^2$ and $m_\pm^2$ with the corresponding orthonormal eigenvectors
\begin{align}
	e_{I<N-2}\,, \qquad
	v_\pm = \frac{(0,\ldots,0,a-c\pm\s,2b^*)}{\sqrt{
	2\s(\s\pm(a-c))}}	\,,
\end{align}
where $e_I$ is the standard basis vector 
(with a one in the $I^{\text{th}}$ position and zero everywhere else)
and $\s=\sqrt{(a-c)^2+4|b|^2}$.   
Note that \eqref{Beta:Cartan.mass.bg} is diagonal when $\f_2=0$ 
(including the $SU(2)$ case) and in the planar limit, when $N\to\infty$.

The one-loop K\"ahler potential is simply read from \eqref{Beta:1loop.eqn1}:
\begin{align}\label{Beta:1loop.bg1}
	K^{(1)} &= \scJ(m_+^2) + \scJ(m_-^2)-
		2(N-2)\big(\scJ(m_{02}^2) + \scJ(m_{03}^2)\big) \non\\
		&- 2\scJ(m_{04}^2) + (N^2-2N-1)\scJ(m_1^2)\\\non
		&+ (N-2)\big(\scJ(m_2^2) + \scJ(m_{\tilde2}^2)
		+ \scJ(m_3^2) + \scJ(m_{\tilde3}^2)\big)
		+ \scJ(m_4^2) + \scJ(m_{\tilde4}^2)\ .
\end{align}
The effect of enforcing the finiteness condition 
is to replace $(4\p)^2\scJ(x)$ by $x\log(M^2/x)$ 
for an arbitrary field dependent mass term $M^2$.

Similarly, the two-loop K\"ahler potential is read from \eqref{Beta:K2.all}:
\begin{align}\label{Beta:K2.2dbg} 
	K^{(2)} &= \non
	(N-2)\big(|hq|^2-g^2\big)\Big[ (N-3)(N-4)I(0,1,1)\\\non 
	 &\quad+(N-3)\big(2I(2_0,2,1)+2I(3_0,3,1)
	 	+I(0,\tilde2,\tilde2)+I(0,\tilde3,\tilde3)\big)\\\non
	 &\quad+2I(2_0,\tilde3,\tilde4)+2I(3_0,\tilde2,4) + 2I(4_0,2,3)
	 \Big] + \big[q\to\qm\big] \\\non
	&+(|h|^2g_q-g^2)\Big[(N-2)\Big((N-3)I(0,1,1)\\
	  &\quad+2I(0,2,2)+2I(0,3,3)\Big)+2I(0,4,4)\Big]+\big[q\to\qm\big]\\\non
	&+2\sum_{i\neq j}\Big[ \big(2|h|^2g_q-2g^2-{\varpi'\!}_{\qb,2ij}
		-{\varpi'\!}_{\qb,1ij}\big) I(|m^{ij}_0|^2,|m^{ij}|^2,1)\\\non
	&\quad+{\varpi'\!}_{\qb,2ij}I(|m^{ij}_0|^2,|m^{ij}|^2,+)
		+{\varpi'\!}_{\qb,1ij}I(|m^{ij}_0|^2,|m^{ij}|^2,-) \Big] 	\\\non
	&+|h(q-\qm)|^2\Big[\big((1-2/N)(N-5)+{\bm\h'\!}_{22}
		+2{\bm\h'\!}_{21}+{\bm\h'\!}_{11}\big)I(0,1,1)\\\non
	&\quad+2(1-2/N-{\bm\h'\!}_{22}-{\bm\h'\!}_{21})I(0,+,1)
		+2(1-2/N-{\bm\h'\!}_{12}-{\bm\h'\!}_{11})I(0,-,1) \\\non
	&\quad+{\bm\h'\!}_{22}I(0,+,+)+2{\bm\h'\!}_{21}I(0,-,+)
		+{\bm\h'\!}_{11}I(0,-,-)\Big],
\end{align}
where we have introduced a condensed notation for the masses
\[ m^2_\pm\sim\pm\,,\quad m^2_i\sim i \quad\mbox{and}\quad m^2_{0i}\sim i_0 \]
with $i=1,\tilde1,\ldots,4,\tilde4$ and defined
\[ {\varpi'\!}_{q,Kij}=|h|^2\varpi_{q,(N-K)ij}-g^2\varpi_{1,(N-K)ij}\,,
\quad {\bm\h'\!}_{IJ}=\bm\h_{(N-I)(N-J)}\ .
\]
We've also used \eqref{Beta:sums} to make the expression only dependent
on ${\varpi'\!}_{q,Iij}$ and ${\bm\h'\!}_{IJ}$ for $I,J=1,2$.
The coefficients, $\varpi_{q,Kij}$ and $\bm\h_{IJ}$, 
are then calculated using the results
\begin{align*} 
	\bH_I &= H_I\,,\quad I<N-2\,,\\
	\quad \bH_{N-2}&=(v_+)_{N-2}H_{N-2}+(v_-)_{N-2}H_{N-1}\,,\\
	\quad \bH_{N-1}&=(v_+)_{N-1}H_{N-2}+(v_-)_{N-1}H_{N-1}\ .
\end{align*}
We emphasise that $\bH_I$ and therefore $\varpi_{q,Kij}$ and $\bm\h_{IJ}$ 
are in general field dependent quantities.

We now examine the two limiting cases, $\f_2\to0$ and $N\to3$.  
In both of these limits, 
we find that the coefficients ${\varpi'\!}_{q,Iij}$ and ${\bm\h'\!}_{IJ}$ 
are independent of the background fields, 
which is not representative of the general case.

\subsection{\texorpdfstring{$SU(N) \to SU(N-1)\times U(1)$}
			{SU(N) -> SU(N-1) * U(1)}}
In the case where $\f_2\to0$ the entire mass matrix is diagonal, 
so that the unitary, diagonalising matrix is just the unit matrix.
Thus the coefficients $\varpi$ and $\bm\h$ are background independent, 
and can be calculated in closed form. 
Also, similarly to \eqref{Beta:K2.2dbg}, 
we can write $K^{(2)}$ such that we only need to know 
$\bm\h_{(N-1)(N-1)}$ and $\varpi_{q,(N-1)ij}$, 
which further eases the calculational load.
If we enforce finiteness, then the sole mass scale, $\fb_1\f_1$,
must cancel in all of the mass ratios, so that the full, quantum corrected, 
K\"ahler potential is just a deformation dependent
rescaling of the classical K\"ahler potential \cite{Kuzenko2007}.
Finally, if we choose a real deformation,
the limit of our two-loop result reproduces equation (6.5) of 
\cite{Kuzenko2007} exactly, which is a good check of our method.

\subsection{\texorpdfstring{$SU(3)\to U(1)^2$}{SU(3) -> U(1) * U(1)}}
When the gauge group is $SU(3)$ the terms with the mass $m_1^2$ 
no longer appear in the summations, 
$m_\pm^2$ is compactly written as $|h(q-\qm)(\f_1\mp\frac\rmi{\sqrt3}\f_2)|^2$ 
and the rest of the masses take the obvious limits. 
We will assume that we are on the conformal surface and set $g^2=f_q|h|^2$.
The one-loop K\"ahler potential does not simplify much, 
choosing $M^2=f_q|h\f|^2$ where $|\f|^2=\tr\F^\dag\F\neq0$, 
we have
\begin{align*}
	K^{(1)}_{SU(3)}
	&= \frac{|h|^2}{(4\p)^2}\Bigg[
		2f_q\left(\BetaMaz\log\frac\BetaMaz{|\f|^2}
		+\BetaMbz\log\frac\BetaMbz{|\f|^2}\right) \\
	&\hspace{-2em}-
		|q-\qm|^2\left(|\f_1-\frac\rmi{\sqrt3}\f_2|
		\log\frac{|\f_1-\rmi\f_2/\sqrt3\,|}{f_q|q-\qm|^{-2}|\f|^2} \right)\\
	&\hspace{-2em}-
		\Bigg(\BetaMa\log\frac\BetaMa{f_q|\f|^2}\\
	&\hspace{-2em}+
		\BetaMb\log\frac\BetaMb{f_q|\f|^2}+\left(q\to\frac1q\right)\Bigg)\\
	&\Bigg]+[\f_2\to-\f_2] \ .
\end{align*}
Although we can combine the logarithms and explicitly remove all reference to
$|\f|^2$, the analytic structure and the various limits are simpler
to examine in the above form.

To find the two-loop K\"ahler potential, 
we choose the diagonalising unitary matrix to be
\[ 
	U=\frac1{\sqrt2}\bem 1 & \rmi \\ -\rmi & -1\eem 
	\quad \implies \bH_2=(\rmi\bH_1)^*=\frac{-1}{\sqrt3}\diag(r_+,r_-,-1) \,,
\]
where $-1$ and $r_\pm=\half(1\pm\rmi\sqrt3)$ are the cube roots of minus one.
Then it is straightforward to compute $\bm\h_{IJ}=\frac13\d_{IJ}$ and
\begin{align*}
	\varpi_{q,Kij} &=\frac13\left[\bem 
	|q-\qm|^2 & |qr_--\qm r_+|^2 & |qr_-+\qm|^2
	\\|qr_+-\qm r_-|^2 & |q-\qm|^2 & |qr_++\qm|^2
	\\|q+\qm r_-|^2 & |q+\qm r_+|^2 & |q-\qm|^2
	\eem \!, \ q\to \qm \right] \ . 
\end{align*}
We split the two-loop K\"ahler potential into
$K_{SU(3)}^{(2)}=(K_{\rm A}+K_{\rm B})+(q\to\qm)$
where the labelling follows the decomposition \eqref{Beta:decomp.CW}.
Note that in the case being considered
$K_{C}=0$, since it only contributes terms of the
form $\dsI(0,x,x)$ which are zero from \eqref{defn:2Loop_dsI}.
This is also true for the integrals that come from the first terms in 
$K_{\rmI B}$ and $K_{\rmII B}$.
Substituting in the masses and using the fact that $\dsI(x,y,z)$ 
is a homogeneous function of order one to pull out a factor of $|h|^2$,
we find
\begin{align*} 
	K_{\text{A}}^{(2)}&= 2|h|^4(|q|^2-f_q)\Big[ \\
	&\times\dsI\big(\BetaMaZ,\BetaMbt,\BetaMct\big)\\
	&+\dsI\big(\BetaMbZ,\BetaMat,\BetaMc\big)\\
	&+\dsI\big(\BetaMcZ,\BetaMa,\BetaMb\big) \Big]
\end{align*}
and
\begin{align*} 
	K_{\text{B}}^{(2)}&=2|h|^4\Bigg[ \big(\third|qr_--\qm r_+|^2-f_q\big)\\
	&\times\Big[\dsI\Big(\BetaMaZ,\BetaMa,\BetaMp\Big)\\
		&\quad+\dsI\Big(\BetaMaZ,\BetaMat,\BetaMm\Big)\Big]\displaybreak[1]\\
	&+\big(\third|q r_-+\qm|^2-f_q\big) \\
	&\times\Big[\dsI\Big(\BetaMbZ,\BetaMb,\BetaMp\Big)\\
		&\quad+\dsI\Big(\BetaMbZ,\BetaMbt,\BetaMm\Big)\Big]\displaybreak[1]\\
	&+\big(\third|qr_+ +\qm|^2-f_q\big)\\
	&\times\Big[\dsI\Big(\BetaMcZ,\BetaMc,\BetaMp\Big)\\
		&\quad+\dsI\Big(\BetaMcZ,\BetaMct,\BetaMm\Big)\Big] \Bigg]
\end{align*}
From the expressions for $\dsI$ given in appendix \ref{A:2LoopVac}, 
we see that the above form is scale invariant. 
We note that taking the deformation to be real does not provide much
simplification, except when $\f_2=0$ and a real deformation
makes the tilded masses equal to their nontilded counterparts.

\section{Conclusion}\label{conc.sect}
The above calculations show that although it is conceptually straightforward
to calculate the loop corrections to the K\"ahler potential of $\b$-deformed
$\cN=4$ SYM on the Coulomb branch, 
the details of the calculation are quite involved for an arbitrary background.  
This is because not only do the $\half(3N-2)(N-1)$ eigenmasses enter the 
result, but also the field dependent eigenvectors.

To help reveal the general structure of the K\"ahler potential it is useful
to use the idea of matrix valued loop integrals 
(see e.g., \cite{Nibbelink2005}) discussed in section \ref{sect:Beta:2loop}. 
Then all field dependence is in the loop integrals, for example
\[ 
	\sum_{IJ}\bm\h_{IJ}I(0,m_I^2,m_J^2)
 	=\sum_{IJKL}\h_{IJKL}I(0,(\cM^\dag\cM)^{IJ},(\cM\cM^\dag)^{KL})\ .
\]
Thus we see that, assuming the finiteness condition is enforced,
the general conformally invariant structure of the 
K\"ahler potential can be written in terms of a function of 
the $\half(5N-2)(N-1)$ components of the mass matrix \eqref{Beta:mass2}
\begin{align*} 
	K(\F^\dag,\F) = |\f|^2 \, F\Bigg(
	&\frac{|g(\F^i-\F^j)|^2}{|\f|^2},
	\frac{|h(q\F^i-\qm\F^j)|^2}{|\f|^2},
	\\&
	\frac{|h(q-\qm)|^2\fb^L\f^M\h_{IJLM}}{|\f|^2}\Bigg) ,
\end{align*}
where we remember that we have chosen the background to be 
$\F=H_I\f^I=E_{ii}\F^i$.
For definiteness, we have inserted the nonvanishing 
$|\f|^2=\tr\F^\dag\F=\sum_I|\f^I|^2=\sum_i|\F^i|^2$ into all
terms in the above expression, but in general this is not necessary.
The loop corrections to the K\"ahler potential are identically zero in 
the limit of vanishing deformation, thus $F$ can always be written as one 
(for the tree level term) plus the
difference between two terms that become identical as the deformation is 
switched off.

\chapter{Goldstino actions}\label{ch:Goldstino}





Since unbroken supersymmetry is not observed in the low-energy aspects of nature
that we see around us, if four-dimensional supersymmetry is realised in nature, 
it must be spontaneously broken. That is, the fundamental laws 
(the underlying field theory) are supersymmetric, but the vacuum state 
is not invariant under supersymmetry rotations.
The spontaneous breaking of supersymmetry implies the existence of a 
massless spinor -- the \emph{Goldstino}.
Being massless, this field should be accounted for in any low-energy effective 
theories with spontaneously broken supersymmetry.\footnote%
{Since supersymmetry must ultimately be a local symmetry, the Goldstino 
is absorbed into the gravitino via the supersymmetric Higgs effect.
However, at low energies, 
the couplings of the  longitudinal mode of the gravitino 
dominate the transverse couplings, so the gravitino can be well accounted for
by just the goldstino \cite{Casalbuoni1988}.}
This chapter will primarily be concerned with the study of the
pure Goldstino action. 

In the first section we briefly examine supersymmetry breaking and the 
appearance of the Goldstino in the supersymmetric sigma model.
Although the Goldstino action and its low-energy interactions
are universal, it can appear in different forms that are related via nonlinear
field redefinitions. This will be studied in the second section
and is based on the papers \cite{KuzenkoTyler2010,KuzenkoTyler2011}. 
Some results used in this section are collected in appendix \ref{A:Golden}.
The third section describes the new Goldstino superfield embedding
that was first introduced in the paper \cite{KuzenkoTyler2011a}. 
We show how this new Goldstino embedding is related to the previously
known Goldstino actions and examine the structure of its interactions with 
other superfields.


\section{Supersymmetry breaking and the supersymmetric sigma model}
By definition, the vacuum state does not break supersymmetry 
if it is invariant under all of the supersymmetry generators,
\begin{align}\label{SusyInvarianceOfVac}
	Q_\a\ket\vac = \bar{Q}_\da\ket\vac = 0\ .
\end{align}
From the supersymmetry algebra \eqref{defn:SusyAlgebra},
this implies that the vacuum must also 
be invariant \wrt the energy-momentum generators. 
In particular, the energy of the vacuum, which is non-negative for any
supersymmetric theory
\begin{align}\label{NonNegativeSusyVacuum} 
	\VEV{H} &= \VEV{P_0} 
	= \frac14\VEV{\d^{\ada}\acom{Q_\a}{\bar{Q}_\da}} \\\non
	&= \frac14\big(\big\|Q_1\ket\vac\!\big\|^2 + \big\|Q_2\ket\vac\!\big\|^2
		+ \big\|\bar{Q}_1\ket\vac\!\big\|^2 
		+ \big\|\bar{Q}_2\ket\vac\!\big\|^2 \big) \geq 0\ ,
\end{align}
will vanish in a supersymmetry preserving vacuum \eqref{SusyInvarianceOfVac}.
This then implies that the condition 
$$\VEV{H}>0$$ 
only holds if supersymmetry is broken, so the 
vacuum energy can be used as an order parameter for supersymmetry breaking.

Invariance of the vacuum under supersymmetry \eqref{SusyInvarianceOfVac}
also implies that the vacuum expectation value (VEV) 
of any supersymmetry variation must vanish, 
\begin{align}\label{eqn:VEVDeltaX}
	\VEV{\d_\x X}
	=\VEV{\com{\x^\a Q_\a+\bar\x_\da\bar{Q}^\da}{X}}
	=0\ .
\end{align}
If Lorentz symmetry is to be preserved
then only scalar fields can have non-vanishing VEVs.
Only spinor fields can be taken to scalars by a supersymmetry variation,
and dimensional analysis constrains the variation of a spinor field
to be an auxiliary scalar field plus other terms that can not have a VEV.
Then, in particular, \eqref{eqn:VEVDeltaX} implies that the variation 
of the spinor components in a chiral and real (gauge) superfields
\begin{align*}
	\VEV{\d_{\x}\l_\a} \sim \x_\a \VEV{F} \,, \quad
	\VEV{\d_{\x}\j_\a} \sim \x_\a \VEV{D} \,,
\end{align*}
must vanish if supersymmetry is to preserved.
If the vacuum expectation of even a single auxiliary field 
does not vanish, then supersymmetry is broken as 
either $F$-term, $D$-term or mixed breaking.

Goldstone's theorem says that 
the spontaneous breaking of continuous bosonic global symmetries necessarily
implies the existence of a massless scalar field -- the Goldstone boson.
The spontaneous breaking of supersymmetry is similarly associated with 
a massless spinor Goldstino field -- the \emph{Goldstino}. 
The Goldstino can be thought of as the state generated by the (nonvanishing)
variation of the vacuum $G_\a \sim Q_\a\ket\vac$.
However, the proof of the existence of a Goldstino,
like the proof of Goldstone's theorem in the bosonic case, 
is best approached by examining the invariance of the supercurrent.
Below, we shall satisfy ourselves with simply 
examining the existence of the Goldstino in the specific case 
of nonlinear sigma models.

Supersymmetry breaking is a large and varied subject area.
For discussions of topics such as
the Witten index, mass sum rules, mediated breaking,
dynamical breaking, etc\dots, we refer the reader to the many
excellent discussions in text books, reviews and lectures 
\cite{Fayet1977, GGRS1983,WessBagger1992,West-Book,Zumino1981SBS,
Shadmi2000,Luty2005,Intriligator2007,Shirman2008,Dine2010,Bertolini2011}
and references therein.

\subsection{Supersymmetric nonlinear sigma model}
The most general low-energy\footnote%
{Low-energy here means that the equations of motion have, at most, 
two space-time derivatives.}
effective action that can be constructed from solely chiral superfields
$\F^i$  ($\Db_\da\F^i=0$) and their complex conjugates
is the supersymmetric sigma model
\begin{align}\label{defn:SusySigmaModel}
	S_\s[\Fb,\F] = \intz K(\Fb, \F) + \intc P(\F) + \cc\,,
\end{align}
where $K(\Fb,\F)$ is the K\"ahler potential and $P(\F)$ is the superpotential.

To examine the conditions for supersymmetry breaking, 
we need to extract the corresponding component action.
To this end, we define the component projections
\begin{align}\label{FlatComponentProjPhi}
	\F^i| = \f^i\,,  \qquad
	D_\a\F^i| = \sqrt2 \j^i_\a\,, \qquad
	-\frac14 D^2\F^i| = F^i\,.
\end{align}
These component fields have the following supersymmetry transformations
\begin{align} \label{eqn:FreeSusy}
\begin{aligned}
	\d_\x\f&=\sqrt2\x\j\,, \quad
	\d_\x F=-\sqrt2\rmi(\pd_a\j\s^a\bar\x)\,,\\
	\d_\x\j_\a&=\sqrt2\big(\x_\a F + \rmi(\s^a\bar\x)_\a\pd_a\f\big)\,.
\end{aligned}
\end{align} 
The projection of the superpotential is straightforward,
\begin{align}\label{FlatComponentsOfSuperpotential}
 	S_P =	\intc P(\F) &\deq \intx\frac{D^2}{-4}P(\F)\Big| \non\\
	&= \intx \left(F^iP_i(\f) - \frac12 (\j^2)^{ij} P_{ij}(\f)\right) . 
\end{align}
The component expression for the K\"ahler potential takes a little more work:
\begin{align}\label{FlatComponentsOfKahlerPotential}
	S_K &= \intz K(\Fb, \F) 
	\deq \intx\frac{D^\a\Db^2D_\a}{16}K(\Fb, \F)\Big|
	\non\\
	&= \intx\Big(-\big(
		\pd^a\f^i \pd_a\fb^\iu + \frac\rmi2 \j^{i\a}\oLRa{\pd}_\ada\jb^{\iu\da} 
		- F^i\bar{F}^\iu \big) K_{i\iu}	 	\\\non
	&+\frac12\Big(\big(\j^{j\a}(\rmi\pd_\ada\f^i)\jb^{\iu\da} 
		-(\j^2)^{ij}\bar{F}^\iu\big) K_{ij\iu}  + \cc\!\Big)
	+\frac{1}{4}(\j^2)^{ij}(\jb^2)^{\iu\ju} K_{ij\iu\ju}\Big)\ .
\end{align}

Just like (nonsupersymmetric) sigma models 
are associated with Riemannian geometry,
supersymmetric sigma models are associated with a K\"ahler geometry.
Since this does not play any significant r\^ole in our considerations,
we leave a detailed discussion to the standard textbooks
\cite{GGRS1983,WessBagger1992,West-Book,BK}. 
Instead, we'll just take advantage of the convenient language and notations.
The K\"ahler metric is 
$K_{i\iu} = \frac{\d}{\d\f^i}\frac{\d}{\d\fb^\iu}K(\fb,\f)$ 
and $K^{\iui}$ is its matrix inverse, 
where the scalar fields $\f^i$ and $\f^\iu$ are thought of as
the complex coordinates of the target space K\"ahler geometry.
The nonvanishing Christoffel symbols are 
\begin{align}
	\G^i_{jk} = K^{\iui}K_{\iu jk}\,,\qquad
	\G^\iu_{\ju\ku} = K^{\iui}K_{i\ju\ku}\,,
\end{align}
which lets us define the target space covariant derivatives
\begin{align}\label{TargetSpaceCovDeriv}
	\nabla_j V^i &= V^i_{;j} = V^i_{,j} + \G^i_{jk}V^k \,,\quad
	\nabla_j V_i = V_{i;j} = V_{i,j} - \G^k_{ij}V_k \,,
\end{align}
for some target space vector and covectors $V^i$.  
The spinors $\j_\a^i$ and $\bar\j_\da^\iu$ transform covariantly,
however their space-time derivatives need to be made covariant
\begin{align}\label{SpaceTimeTargetSpaceCovDeriv}
	\nabla_b\j^i_\a &= \pd_b\j^i_\a + \G^i_{jk}(\pd_b\f^j)\j^k_\a \,,\quad
	\nabla_b\jb^\iu_\da = \pd_b\jb^\iu_\da 
		+ \G^\iu_{\ju\ku}(\pd_b\fb^\ju)\jb^\ku_\da \ .
\end{align}
Finally, the naive auxiliary fields $F^i$ and $\bar{F}^\iu$ transform 
inhomogeneously, but a covariantly transforming auxiliary field
can be defined as
\begin{align}\label{defn:KahlerCovAuxField}
	F_\text{cov}^i = F^i - \frac12 \G^i_{jk}(\j^2)^{jk} 
	\quad\text{and }\cc
\end{align}

Using the above definitions, the component Lagrangians of the actions
\eqref{FlatComponentsOfSuperpotential} 
and \eqref{FlatComponentsOfKahlerPotential},
simplify to
\begin{align}\label{CovComponentSusySigmaModel}
	\cL_P &= \big(F_\text{cov}^iP_i 
	- \frac12 (\j^2)^{ij} P_{;ij} + \cc \big)\,, \\
	\cL_K &= -K_{i\iu}\big(\pd^a\f^i \pd_a\fb^\iu 
		+ \frac\rmi2 \j^{i\a}\oLRa{\nabla}_\ada\jb^{\iu\da} 
		- F_\text{cov}^i\bar{F}_\text{cov}^\iu  \big) 	
		+\frac{1}{4}(\j^2)^{ij}(\jb^2)^{\iu\ju} \cR_{ij\iu\ju} \ .
\end{align}
The covariant auxiliary fields' equations of motion are simply 
$P_i = K_{i\iu}\bar{F}_\text{cov}^\iu$, 
so the eliminated component Lagrangian is
\begin{align}\label{CovEliminatedSusySigmaModel}
	\cL 
	&= -\pd_a\fb^\iu K_{i\iu} \pd^a\f^i  
	- \frac\rmi2 K_{i\iu} \j^{i\a}\oLRa{\nabla}_\ada\jb^{\iu\da} 
	- \bar{P}_\iu K^{\iui} P_i 	\non\\
	&- \frac12(\j^2)^{ij} P_{;ij} - \frac12(\jb^2)^{\iu\ju} \bar{P}_{;\iu\ju}
	+\frac{1}{4}(\j^2)^{ij}(\jb^2)^{\iu\ju} \cR_{ij\iu\ju} \ .
\end{align}

The scalar potential, obtained by setting the derivatives
and spinors to zero, is simply
\begin{align}
	V = \bar{P}_\iu K^{\iui} P_i \geq 0\ .
\end{align}
It is positive semi-definite, since we required 
that the scalar fields have canonical kinetic terms 
(so $K_{i\iu} = \d_{i\iu} + \text{higher order terms}$).
Supersymmetry is preserved iff the VEV of the scalar potential is zero,
and since the K\"ahler metric is positive definite, this implies that
supersymmetry is preserved iff there is a solution to the simultaneous equations
\begin{align}
	P_i(\f) = \bar{P}_\iu(\fb) = 0 \ .
\end{align}
Note that if such a solution exists, then it is 
automatically a minimum of the potential, so a possible vacuum.
Since $P_i = K_{i\iu}\bar{F}_\text{cov}^\iu$ 
and $\expt{F^i_\text{cov}}=\expt{F^i}$, 
we see once again that supersymmetry is broken
only if there is a nonvanishing VEV for one or more of the auxiliary fields 
$F^i$. 

Vacua occur at the minima of the scalar potential, where
\begin{align}\label{CovFirstDerivOfV}
 0  = V_j 
 	= \bar{P}_\iu K^{\iui} P_{i,j} -  \bar{P}_\iu K^{\iu k}\G^i_{kj} P_i
	= \bar{P}_\iu K^{\iui} P_{;ij}\ .
\end{align}
If supersymmetry is not broken, then $\bar{P}_\iu=0$ and we are automatically
at the global minimum of the potential.
If supersymmetry is broken, it tells us that 
the nonvanishing vector $\bar{P}_\iu K^{\iui}$ is a null vector for $P_{;ij}$. 

Now note that the fermion mass term is proportional to $\j^{i\a}P_{;ij}\j^j_\a$.
Assuming that the vacuum breaks supersymmetry, equation \eqref{CovFirstDerivOfV}
implies that the fermion mass matrix has the null vector $\VEV{F^i}$
and so there must exist a massless fermion, the Goldstino.
If we think about changing the fermion basis diagonalise the mass matrix, 
then we see that the Goldstino is proportional to
\begin{align*}
	G_\a &= \l^i_\a k_{i\iu}\bar{f}^\iu |f|^{-1} \,,	\\
\intertext{where}
	f^i = \VEV{F^i}\,, \quad k_{i\iu}&=\VEV{K_{i\iu}}\,,\quad
	|f|^2 = f^i k_{i\iu} \bar{f}^\iu\ .
\end{align*}
Then, the vacuum exception value of the 
supersymmetry variation of the Goldstino is easily seen to be
\begin{align*}
	\VEV{\d_\x G_\a} = \VEV{\d_\x \l^i_\a} k_{i\iu}\bar{f}^\iu |f|^{-1}
	= \sqrt{2} |f| \x_\a\ .
\end{align*}
If new fields are introduced that have 
their vacuum expectation value subtracted, 
i.e., $\f^i \to \f^i - \VEV{\f^i}$ and $F^i \to F^i - \VEV{F^i}$,
where $\pd_a\VEV{\f^i}=\pd_a\VEV{F^i}=0$,
then the full supersymmetry transformation of the Goldstino is
the inhomogeneous 
\begin{align}\label{eqn:SusyXform_Goldstino_SigmaModel}
	\d_\x G_\a 
	&= \sqrt{2} |f| \x_\a + \sqrt2\big(\x_\a F^i + \rmi(\s^a\bar\x)_\a\pd_a\f^i\big)
		k_{i\iu}\bar{f}^\iu |f|^{-1}\ .
\end{align}

Finally, we note that all of the above considerations are straight forward 
to generalise to gauged nonlinear supersymmetric sigma models with
optional Fayet-Iliopoulos terms, e.g., \cite{WessBagger1992,Argyres2001}. 
For nonlinear sigma models with $\cN = 2$ supersymmetry the associated geometry
is even more restrictive, see, e.g., the review \cite{Kuzenko2010a}.

\subsection{Models with a single chiral superfield}
The simplest model that breaks supersymmetry is the Polonyi model
\cite{Polonyi1977}, 
\begin{align}\label{PolonyiModel}
	S[\Fb,\F] = \intz \Fb\F + f\intc\F + \bar{f}\intac\Fb\,,
\end{align}
which is normally coupled to a larger model of interest 
in order to break the larger system's supersymmetry.
In the above action, the derivative of the superpotential is $\pd_\F P(\F) = f$,
which is always non-zero, so 
supersymmetry is broken.
The component Lagrangian is
\begin{align*}
	\cL = - \pd^a\fb\pd_a\f -\frac12\l\oLRa{\pd}\bar\l 
		+ \bar{F}F + f F + \bar{f}\bar{F}
		= -\bar{f}f - \pd^a\fb\pd_a\f -\frac12\l\oLRa{\pd}\bar\l  \,,
\end{align*}
where the second equality follows from putting the auxiliary fields on-shell.
Clearly this describes a free theory with broken supersymmetry and a 
vacuum energy density of $|f|^2$. 
Both the goldstino $\l_\a$ and its superpartner, 
the sgoldstino $\f$ are massless.

The sgoldstino $\f$ does not appear in the effective potential,
so its vacuum expectation value parameterises a (non-supersymmetric) 
moduli space of vacua. 
However, since supersymmetry is broken, 
this moduli space is not protected from quantum corrections 
(which require a coupling to other sectors of the action)
and is lifted by higher order terms in the K\"ahler potential,
see, for example \cite{Zumino1981SBS,Komargodski2009,Bertolini2011}.

The simplest modification to the K\"ahler potential that could come 
from a low-energy effective action due to the model's coupling
to a larger system is
\begin{align*}
	K(\Fb,\F) = \Fb\F + \frac{c}{4} \Fb^2\F^2 \ .
\end{align*}
Then the scalar potential becomes 
$$ V = |f|^2(1+c\fb\f)\,, $$
so the squared mass of the scalar field 
is $c|f|^2$ and the vacuum occurs at $\f=0$.
In this case, supersymmetry remains broken, 
but there is no moduli space of vacua.

Finally, 
if we add a mass term to the superpotential,  $P(\F) = f\F+\frac{m}2\F^2$,
then the component equation $\pd P / \pd\f=0$ is solved by $\f = -f/m$
and the scalar potential $V = |f+m\f|^2$ is zero for the vacuum.
With such a mass term 
there always exists a vacuum expectation value for the scalar field
that preserves supersymmetry.

Although the (uncoupled) model \eqref{PolonyiModel} is trivial, 
if the chiral field is constrained,
then we can obtain non-trivial Goldstino dynamics equivalent to that of
Akulov and Volkov. 
In particular, the simple constraint, $\F^2=0$ yields
the model investigated in section \ref{ssect:KS}.

\subsection{The O'Raifeartaigh model}
The simplest model of \emph{interacting} chiral superfields 
that breaks supersymmetry was found by O'Raifeartaigh \cite{O'Raifeartaigh1975}. 
It requires three chiral superfields, with the canonical K\"ahler potential
and the superpotential
\begin{align*}
	P(\F_i) = m\F_1\F_2 + g\F_0(\F_1^2-\a^2)\,,
\end{align*}
without loss of generality, the coupling constants can be chosen to be real.
For non-zero $g\a^2$, the equations 
$P_{,0} = g(\f_1^2-\a^2) = 0$ and $P_{,2} = m\f_1 = 0$ 
are not compatible, so supersymmetry is broken.
The scalar potential is
\begin{align*}
	V = |g|^2|\f_1^2-\a^2|^2 + |m\f_1|^2 + |m\f_2+2g\f_0\f_1|^2\ .
\end{align*}
Minimising the first two terms fixes $\f_1$ and the final term can 
be made zero by the appropriate choice of $\f_2$. 
This means the minimum of potential has a flat direction along $\f_0$.
The fermion mass matrix 
\[ P_{ij} = \bem0&2g\f_1&0\\2g\f_1&g\f_0&m\\0&m&0\eem\,, \]
has a vanishing determinant, for $m\neq0$ it has a null space of dimension 1, 
establishing the existence of the massless Goldstino.

\section{Goldstino actions and their symmetries}\label{sect:Goldstino}
As discussed above, the Goldstino action is a necessary part of any low-energy
model with broken supersymmetry and is, at low energies, universal 
\cite{WessBagger1992}.
This means that any two actions that describe Goldstino dynamics
must be equivalent up to a field redefinition. 
This fact that can be proven by using the general theory of 
the nonlinear realisations of $\cN = 1$ supersymmetry
\cite{Ivanov1977,Ivanov1978,Ivanov1982,Uematsu1982},
which is an extension of the Callan-Coleman-Wess-Zumino formalism 
\cite{Coleman1969,Callan1969}.
This does not mean that all Goldstino actions are of equal utility -- 
different actions emphasise different aspects of the model 
and make different types of computations either more or less difficult.
In this section, which is based on the papers
\cite{KuzenkoTyler2010,KuzenkoTyler2011}, 
we find the explicit transformations that map
the various known Goldstino actions onto one another.

We start this section by deriving the Akulov-Volkov (AV) action.
We then compute a finite-dimensional Lie group $G$ of all field transformations 
of the form $\l \to {\l}' = \l +O(\l^3)$
which preserve the functional structure of low-energy Goldstino-like actions. 
Associated with $G$ is its twelve-parameter 
subgroup $H$ of trivial symmetries of the AV action. 
The coset space $G/H$ is naturally identified with 
the space of all Goldstino actions.
We then apply our construction to study
the properties of five different Goldstino actions available in the literature.
Making use of the most general field redefinition mentioned above,
we find explicit maps between all five cases. In each case 
there is a twelve-parameter freedom in these maps
due to trivial symmetries inherent in the Goldstino actions.
Finally, by using the pushforward of the AV supersymmetry, 
we find the off-shell nonlinear supersymmetry transformations
that leave the other four actions invariant and compare to 
the supersymmetry transformations normally associated with those actions.

Many of the results in this section were derived, or at least checked,
using some \emph{Mathematica} code made to manipulate 
and canonicalise spinor expressions.
All of the code, results, checks and extra discussions can be found
in the \emph{Mathematica} notebook distributed with both the 
preprint and published versions of \cite{KuzenkoTyler2011}.


\subsection{The Akulov-Volkov model}\label{ssect:AV}
The Akulov-Volkov (AV) model \cite{Volkov1972, Volkov1973} is the second oldest 
supersymmetric theory in four space-time dimensions.
It describes the low-energy dynamics of a massless Nambu-Goldstone spin-1/2
particle which is associated with the spontaneous breaking of rigid 
supersymmetry and  is called the Goldstino. 
A derivation of the AV model using superspace techniques was given in 1973 
\cite{Volkov1974a} by its discoverers. 
Nice textbook reviews of the AV model are also available, see e.g.,
\cite{WessBagger1992, GGRS1983}.

The simplest way to derive the AV action is to consider 
the hypersurfaces in superspace
\begin{align}\label{defn:AV_hypersurface}
	 \q^\a = \k\l^\a(x)\,,
\end{align}
that transform covariantly under supersymmetry translations
\begin{align*}
	\l'^\a(x') = \l'^\a(x + \rmi \k (\l\s\bar\x - \x\s\bar\l))
	=  \l^\a(x) + \frac1\k \x^\a\ .
\end{align*}
The constant $\k$ has mass dimension equal to $2$ 
and is is introduced so that $\l_\a$ has the canonical mass dimension of
a spinor, $\frac32$. 
Looking at an infinitesimal supersymmetry translation, we see that
\begin{align}\label{eqn:AV_SUSY}
\begin{aligned}
	\d_\x \l^\a(x) &= \frac1\k\x^\a 
		- \rmi\k(\l(x)\s^a\bar\x-\x\s^a\bar\l(x))\pd_a\l^\a(x)\\
	&\deq \frac1\k\x^\a + \D_{\x}^a(x)\pd_a\l^\a(x) \,,
\end{aligned}
\end{align}
which can be compared with, e.g., the variation of the Goldstino in a 
supersymmetric nonlinear sigma model \eqref{eqn:SusyXform_Goldstino_SigmaModel}.
That \eqref{eqn:AV_SUSY} obeys the supersymmetry algebra 
\eqref{defn:SusyAlgebra} is easy to check.
First we need 
\begin{align*}
	\d_{\x_1} \D_{\x_2}^a 
	= - \rmi\k(\d_{\x_1}\l(x)\s^a\bar\x_2-\x_2\s^a\d_{\x_1}\bar\l(x))
	= -\rmi(\x_2\s^a\bar\x_1-\x_1\s^a\bar\x_2)
		+ \D_{\x_1}^b\pd_b\D_{\x_2}^a\ ,
\end{align*}
from which we see
\begin{align*}
	\d_{\x_1}\d_{\x_2}\l^\a 
	&= -\rmi(\x_2\s^a\bar\x_1-\x_1\s^a\bar\x_2)\pd_a\l^\a \\
	 &+ (\D_{\x_1}^b\pd_b\D_{\x_2}^a+\D_{\x_2}^b\pd_b\D_{\x_1}^a)\pd_a\l^\a
	 + \D_{\x_1}^b\D_{\x_2}^a\pd_a\pd_b\l^\a .
\end{align*}
Antisymmetrising gives the result
\begin{align}\label{AV:SusyAlg}
	\com{\d_{\x_1}}{\d_{\x_2}}\l^\a 
	= -2\rmi(\x_1\s^a\bar\x_2-\x_2\s^a\bar\x_1)\pd_a\l^\a\ .
\end{align}

To get an invariant action we examine the supersymmetric Cartan 1-forms 
$\rmd\z^M=(\rmd\z^m,\rmd\z^\m,\rmd\bar\z_{\dot{\m}})$,
where $\rmd\z^M D_M = \rmd z^A \pd_A$. Comparing coefficients of 
the derivatives $\pd_A$ we can read off
\begin{align*}
	\rmd\z^\a = \rmd\q^\a\,,\quad 
	\rmd\bar\z_\da = \rmd\qb_\da\,,\quad
	\rmd\z^a = \rmd x^a+\rmi\rmd\q\s^a\qb+\rmi\rmd\qb\st^a\q\ .
\end{align*}
These 1-forms are, by construction, invariant under supersymmetry translations.
So, to construct super-Poincar\'e invariants we can use the volume elements
$\rmd^2\z$, $\rmd^2\bar\z$ and $\rmd^4\z$.
We move these 1-forms onto the hypersurface \eqref{defn:AV_hypersurface},
\begin{align}
\!\!\rmd\l^\a(x) = \rmd x^a\pd_a\l^\a(x)\,, \quad 
	\rmd\z^m(x) = \rmd x^a \underbrace{\big(\d_a^m 
				  + \rmi\k^2 \l(x)\s^m\oLRa{\pd}_a\bar\l(x) \big)}_{E_a{}^m},
\!\!
\end{align}
where we've identified the inverse vierbein $E_a{}^m$.
The bosonic volume element becomes $\rmd^4\z=\rmd^4x\det(E)$ 
and since the determinant of the inverse vierbein $E_a{}^m$
transforms as a total derivative
\begin{align*}
	\d_\x \det(E) = \pd_a\big( \D^a \det(E) \big)\ .
\end{align*}
its integral, the Akulov-Volkov action,
\begin{align}\label{eqn:AV1}
	S_\AV[\l,\bar\l] 
	= -\frac1{2\k^2}\int\rmd^4x\det(E)
\end{align}
is invariant under supersymmetric transformations. 
The normalisation has been chosen to give the standard Goldstino kinetic term.

This construction of an invariant action is slightly different from
the normal construction for phenomenological  Lagrangians 
for nonlinearly realised \emph{non-spacetime} symmetries. 
This is discussed by Volkov in \cite{Volkov1994}.
Although we won't investigate it in this thesis, the vierbein $E_a{}^m$
can be used to construct Goldstino couplings to matter fields
that are manifestly invariant under nonlinear supersymmetry transformations,
this is thoroughly investigated in the standard literature of 
nonlinearly realised supersymmetry
\cite{Ivanov1977,Ivanov1978,Ivanov1982,Uematsu1982}.

To get the explicit form of $S_\AV$ we need to expand the determinant.
First we define some notation that we will use for the rest of this chapter.
We define the Lorentz indexed matrices
\begin{align}\label{defn:v}
	v_a{}^b	 &= \rmi\l\s^b\pd_a\bar\l \,, \qquad 
	\vb_a{}^b = -\rmi\pd_a\l\s^b\bar\l\,, 
\end{align}
and denote the trace of any Lorentz 
matrix $M_a{}^b$ as $\expt{M}=\tr(M)=M_a{}^a$.
We use the following result that holds for $4\times4$ matrices
\begin{align*}
\det(1+M) &= 1 + \expt{M} + \frac12 \big(\expt{M}^2-\expt{M^2}\big) \\
		&+\frac16 \big(\expt{M}^3-3\expt{M}\expt{M^2}+2\expt{M^3}\big) 
		+ \det{M}\,,
\end{align*}
to calculate the determinant of the vierbein, $E = 1 + \k^2(v + \vb)$. 
We find
\begin{align}\label{eqn:AV2}
	S_\AV[\l,\bar\l] = -\half&\intx\Bigg(\k^{-2} + \expt{v+\bar v}
	+2\k^2\Big(\expt{v}\expt{\bar v}-\expt{v\bar v}\Big) \\\non
	&+{\k^4}\Big(\expt{v^2\vb}-\expt{v}\expt{v\vb}
		-\half\expt{v^2}\expt{\vb}+\half\expt{v}^2\expt{\vb}+\cc\Big)\Bigg)\,.
\end{align}
As first noticed in \cite{Kuzenko2005e} the 8th-order terms vanish identically.
This has now been proved many times, but probably the easiest method 
was given in appendix A of \cite{Kuzenko2005e}, where they noted that 
\begin{align*}
	\det(v + \vb) \propto \eps^{abcd}\eps_{klmn}
		v_a^{\,k}v_b^{\,l}\vb_c^{\,m}\vb_d^{\,n}
	= \l^2\bar\l^2\eps^{abcd}\eps_{klmn}
		(\pd_a\l\s^{kl}\pd_b\l)(\pd_c\bar\l\st^{mn}\pd_d\bar\l)\,,
\end{align*}
which vanishes due to the sigma-matrix identities
\begin{align*}
	\eps_{abcd}\st^{cd}=2\rmi\st_{ab}\,,
	\quad\text{and}\quad
	(\s^{ab})_\a{}^\b (\st_{ab})^\da{}_\db = 0\ .
\end{align*}
The vanishing of the 8th-order terms can also be seen by simply
writing them in a unique basis, 
such as that defined in appendix \ref{sect:bases}.

\subsection{General Goldstino action}\label{ssect:GenGoldstino}
The general structure of $S_\AV$ and any other low-energy Goldstino action is 
schematically%
\footnote{%
Here we ignore higher-derivative corrections to the Goldstino actions.
}
\begin{equation} \label{eqn:GoldStruct}
  S_{\text{Goldstino}} \sim \intx \sum_{n=0}^4  \k^{2n-2}\l^n\bar\l^n\pd^n\, .
\end{equation}
This follows from dimensional counting and the fact that
a Goldstino field must parametrise 
a coset space of the $\cN=1$ super-Poincar\'e group
and thus always occur in the combination $\k\,\l= \q$. 
The most general field redefinition that preserves such a structure is
\begin{align} \label{eqn:GeneralFieldRedef}
	\l_\a \to \l'_\a &= \l_\a+{\k^2}\l_\a\expt{\a_1 v + \a_2\vb} 
		+\rmi\a_3{\k^2}(\s^a\bar\l)_\a(\pd_a\l^2)  \\\non  					
	&+{\k^4}\l_\a\big(\b_1\expt{v\vb}+\b_2\expt{v}\expt{\vb}
		+ \b_3\expt{\vb^2} + \b_4 \expt{\vb}^2
		+ \b_5\pd^a\l^2\pd_a\bar\l^2 \\\non 
	&\quad+\b_6\bar\l^2\square\l^2\big) 
	+\rmi{\k^4}(\s^a\bar\l)_\a(\pd_a\l^2)\expt{\b_7v +\b_8\vb} \\\non 		
	&+ {\k^6}\l_\a\big(\g_1\expt{v\vb^2}+\g_2\expt{v\vb}\expt{\vb}
		+ \g_3\expt{v}\expt{\vb^2} + \g_4\expt{v}\expt{\vb}^2 \\\non
	&\quad +\g_5\expt{\vb}\pd^a\l^2\pd_a\bar\l^2 \big) \,.
\end{align}
The coefficients can be complex and we denote their real and imaginary parts as
\begin{align} \label{eqn:CoeffReIm}
	\a_i = \a_i^\rmr + \rmi\a_i^\rmi\,,\quad
	\b_j = \b_j^\rmr + \rmi\b_j^\rmi\,,\quad
	\g_k = \g_k^\rmr + \rmi\g_k^\rmi\ .
\end{align}
This field redefinition is equivalent to that given in \cite{Kuzenko2005e} 
up to some 7-fermion identities.
The proof that it is a minimal basis of all possible terms preserving 
\eqref{eqn:GoldStruct} is provided in 
the \emph{Mathematica} program distributed with \cite{KuzenkoTyler2011}.
All Goldstino actions of the form \eqref{eqn:GoldStruct} 
are invariant under rigid chiral  transformations
\begin{equation}
  \l_\a \to \rme^{\rmi \vf  } \l_\a \, ,\qquad 
  \bar\l_{\dot \a} \to \rme^{-\rmi \vf  } \bar\l_{\dot \a}\,.
\end{equation}
Without enforcing this symmetry, 
one can introduce a more general field redefinition than
the one defined by equation \eqref{eqn:GeneralFieldRedef}.

The set of all transformations \eqref{eqn:GeneralFieldRedef} 
forms a 32-dimensional Lie group $G$.
The composition rule for the elements of $G$ is spelled out in appendix 
\ref{sect:composition}.

By applying the field redefinition \eqref{eqn:GeneralFieldRedef} to 
the AV action we generate the most general Goldstino action.
This can then be compared against other Goldstino actions to find the 
maps that relate them to the AV action.  
The general result, written in the basis of appendix \ref{sect:bases}, is
{\allowdisplaybreaks
\begin{align} \label{eqn:GeneralAV}
	&S_\AV[\l'(\l,\bar\l),{\bar\l'}(\l,\bar\l)] 	
	= -\half\intx\Bigg(\k^{-2} + \expt{v+\bar v}    
\allowdisplaybreaks\\ \non
	&+\k^2\Big[ \big\{(2\a_1+2\a_3+1)\expt{v}^2 - (2\a_3+1)\expt{v^2}+\cc\big\}
		+4\a_ 2^\rmr\expt{v}\expt{\vb}-2\a_3^\rmr\pd^a\l^2\pd_a\bar\l^2\Big] 
\allowdisplaybreaks \\ \non
	&+{\k^4} \Big[\Big\{
		(2|\a_3|^2+4\a_2\a_3^*-2\a_2-3\a_3+\a_3^*+2\b_3^*+4\b_6^*-\tfrac12)
			\expt{v^2}\expt\vb   \\ \non
	&\qquad-(4|\a_3|^2+8\a_3\a_2^*+4\a_1\a_3^*+4\a_1+4\a_3^\rmr
		  -2\b_1+8\b_6+4\b_8^*+1)\expt{v}\expt{v\vb} \\ \non
	&\qquad+(4|\a_3|^2+4\a_3^\rmr+1)\expt{v^2\vb}
		+(|\a_1|^2+|\a_2|^2+2|\a_3|^2+\a_1\a_2^*+2\a_2^*\a_3\\ \non
	&\quad\qquad+2\a_3^*\a_1+2\a_1+4\a_2^\rmr+6\a_3^\rmr
		+2\b_2+2\b_4^*+2\b_7+2\b_8^* +\tfrac12)	\expt{v}^2\expt\vb  \\ \non
	&\qquad -(2\a_2^*\a_3+\a_1-\a_2^*-2\a_3^*-2\b_5+4\b_6+2\b_8^*)
			\expt{v}\pd^a\l^2\pd_a\bar\l^2  + \cc \Big\}   \\ \non
	&\quad+(|\a_1|^2-|\a_2|^2-4|\a_3|^2
			-8(\a_2^\rmr\a_3^\rmr+\a_2^\rmi\a_3^\rmi+\b_6^\rmr))
			\rmi\l\s^a\bar\l (\expt{v}\oLRa{\pd}_a\expt\vb)\Big] 
\allowdisplaybreaks\\ \non
	&+{\k^6}\Big[ \Big\{
	  2\big(\a_1(\b_1^*-2\b_5^*-2\a_3)+\a_2^*(3+8\a^\rmr_3-\b_1+2\b_5) 
	  +4\a_3(\a_3^*+\b_5^*) \\\non
	&\quad\qquad +2\a_3^*(3-\b_1+4\b_5+2\b_7+4\a_3^*) 
	  -2\rmi\b_1^\rmi+4\b_5+2\b_7+\g_1\big)\expt{v}\expt{v\vb^2} \\\non
	&\qquad+\big(-\a_2^*(3+2\a_1+\a_2^*-\b_1-2\b_3-4\b_6-2\b_7)
		-2\b_1^*+2\b_3-\a_1\b_1^* \\\non
	&\quad\qquad +2\a_3(1+2\a_1+2\a_3-\b_1^*+2\b_4^*+4\b_6^*+2\b_8^*)
	   	-2\b_4^*+8\rmi\b_6^\rmi \\ \non
	&\quad\qquad -2\a_3^*(2+\a_1+2\a_2^*+2\a_3^*-2\b_3-4\b_6)-2\b_8^*+2\g_3\big)
	  \expt{v}^2\expt{\vb^2} + \cc \Big\} \\\non
	&\quad+2\Re\big(2\a_3(-\b_1^*+2\b_3^*+2\b_5^*+4\b_6^*)
		-\b_1-2\b_3+2\b_5-4\b_6\big)\expt{v^2}\expt{\vb^2} \\\non
	&\quad+4\Re\big(2\a_3(\b_1^*-2\b_5^*-2\a_3-1)+\b_1-2\b_5\big)
		\expt{v\vb v\vb} \\\non
	&\quad-4\Re\big(\a_1(\a_2+2\a_3+2\b_5^*+2\b_7^*)
		+\a_2(3+2\a_2^*-6\a_3-\b_1^*)-\b_1 + \b_2 + 4\b_5\\\non
	&\quad\qquad
		+2\a_3(10\a_2^\rmr+8\a_3^\rmr-\b_1^*+\b_2^*+2\b_5^*+2\b_7^*+3)
		+ 4\b_7 - \g_2 + 2\g_5 \big)
			\expt{v}\expt{\vb}\expt{v\vb} \\\non
	&\quad+2\Re\big(\a_1(2\a_3+\b_2^*+2\b_7^*)
		+\a_2(3+\a_2+4\a_2^*-8\a_3+\b_2^*+2\b_4^*+4\b_6^*+2\b_7^*+2\b_8^*) 
		 \\\non
	&\quad\qquad
		+2\a_3(3+8\a_3^\rmr-\b_1^*+\b_2^*+2\b_4^*+2\b_5^*+4\b_6^*+2\b_7^*+2\b_8^*)
		-\b_1+2\b_2+2\b_4\\\non
	&\quad\qquad+2\b_5+4\b_6+4\b_7+2\b_8+2\g_4) 
		+4(\a_1^\rmr\a_2^\rmr+6\a_2^\rmr\a_3^\rmr+\a_3^\rmr\a_1^\rmr)\big)
		\expt{v}^2\expt{\vb}^2
	\Big]	\Bigg)\,.
\end{align}}
Note that only the real parts of $\g_4$, $\g_2$ and $\g_5$ occur and that
the latter two only appear in the combination $2\g_5-\g_2$. This corresponds
to the fact that the field redefinitions generated by $\g_2^\rmi$, $\g_4^\rmi$, 
$\g_5^\rmi$ and $\g_5^\rmr=2\g_2^\rmr$ are symmetries of the free action.

The general action \eqref{eqn:GeneralAV} has a nonlinear supersymmetry 
that can be derived from the
pushforward of the AV supersymmetry \eqref{eqn:AV_SUSY}
\begin{align} \label{eqn:AV_SUSY_Pushforward}
\begin{aligned}
\d_\x \l_\a &= \d_\x \l_\a(\l',\bar\l')\Big|_{\l'=\l'(\l,\bar\l)} \\
		&= \d_\x\l'^\b\cdot\frac{\d}{\d\l'^\b}\l_\a(\l',\bar\l')
		+ \d_\x\bar\l'_\db\cdot\frac{\d}{\d\bar\l'_\db}\l_\a(\l',\bar\l')
			\Big|_{\l'=\l'(\l,\bar\l)}\,,
\end{aligned}
\end{align}
where $\l_\a(\l',\bar\l')$ is the inverse of \eqref{eqn:GeneralFieldRedef}
that can be found using the results of appendix \ref{sect:composition}.
The explicit, all order expression for this supersymmetry is very long, 
but the leading order is easily calculated
\begin{align} \label{eqn:Gen_SUSY}
	\d_\x \l_\a &= \frac{\x_\a}{\k}
	 + \rmi\k\big((1+2\a_3)(\x\s^a\bar\l)  		
	 	-(1+\a_2)(\l\s^a\bar\x)\big)\pd_a\l_\a 	\non \\ 
	 &\quad  - \rmi\k\a_1(\x\s^a\pd_a\bar\l)\l_\a
	 - \k \expt{\a_1 v + (\a_2+2\a_3)\vb}\x_\a 	\non \\ 
	 &\quad- \frac{\rmi\k}2(\a_2+2\a_3)(\s^a\bar\x)_\a\pd_a\l^2 
	 + O(\k^3) \ .
\end{align}

By writing the AV action in the basis of appendix \ref{sect:bases} 
we can compare \eqref{eqn:AV2} with \eqref{eqn:GeneralAV}. 
We find that there is a twelve-dimensional family of
symmetries of the form \eqref{eqn:GeneralFieldRedef}:
\begin{align} \label{eqn:AV2AV}
	\l_\a \to \l'_\a &= \l_\a+\rmi\a_2^\rmi{\k^2}\l_\a\expt{\vb} 
	 +\rmi{\k^4}(\s^a\bar\l)_\a(\pd_a\l^2)\expt{\b_7v +\b_8\vb} \\\non
	&+{\k^4}\l_\a\Big( 2(2\b_6+\b_8^*) \expt{v\vb} 
	 +(4\b_6^\rmr-\b_4^*-\b_7-\b_8^*)\expt{v}\expt{\vb} \\\non
	&- (2\b_6+\rmi\a_2^\rmi)\expt{\vb^2} 
	 + \b_4\expt{\vb}^2
	 + (\tfrac\rmi2\a_2^\rmi+2\b_6+\b_8^*)\pd^a\l^2\pd_a\bar\l^2 
	 + \b_6\bar\l^2\square\l^2\Big) \\\non
	&+ {\k^6}\l_\a\Big( \g_5\expt{\vb}\pd^a\l^2\pd_a\bar\l^2
	 +(\rmi\a_2^\rmi-2\b_7-4\b_8^\rmr)\expt{v\vb^2} \\\non
	&-(2\a_2^\rmi (2 \b_6^\rmi-\b_8^\rmi)
		+\b_4^\rmr+8\b_6^\rmr-3\b_7^\rmr-\b_8^\rmr
		-\rmi\g_2^\rmi-2\g_5^\rmr)\expt{v\vb}\expt{\vb} \\\non
	&+ (\tfrac\rmi2 \a_2^\rmi(4\b_6+2\b_7+2\b_8^*-1)
		+\b_4^*+2\b_6^\rmr-6\rmi\b_6^\rmi+3\b_8^\rmr+\rmi\b_8^\rmi)
		\expt{v}\expt{\vb^2} \\\non
	&+ (-\tfrac12 \a_2^\rmi(3\b_4^\rmi+4\b_6^\rmi+\b_7^\rmi+3\b_8^\rmi)
		+6\b_6^\rmr-\b_7^\rmr+\rmi\g_4^\rmi)\expt{v}\expt{\vb}^2 \Big) \,,
\end{align}
where, the sake of compactness, we write $\b_6^\rmr=-\tfrac18(\a_2^\rmi)^2$. 
The free parameters in the above field redefinition are
\begin{align} \label{eqn:ReImFreeParams}
	\a_2^\rmi,\; \b_4^\rmr,\; \b_4^\rmi,\; \b_6^\rmi,\; 
	\b_7^\rmr,\; \b_7^\rmi,\; \b_8^\rmr,\; \b_8^\rmi,\; 
	\g_2^\rmi,\; \g_4^\rmi,\; \g_5^\rmr,\; \g_5^\rmi\,.
\end{align}
The set of such transformations is a 12-dimensional subgroup $H$ 
of the group $G$ introduced above. 
In section \ref{ssect:Trivialities} we will show that all of the
transformations \eqref{eqn:AV2AV} are \emph{trivial} symmetries.%
\footnote{The definition of a trivial symmetry is given at the beginning of 
section \ref{ssect:Trivialities}.}
Such trivial symmetries appear in all of the mappings from one Goldstino
action to another and we will always choose the above set of free parameters.

Although the trivial symmetries \eqref{eqn:AV2AV} 
preserve the structure of the action,
they do not preserve the off-shell form of the nonlinear supersymmetry.
We can restrict the parameters of the 
pushforward supersymmetry \eqref{eqn:AV_SUSY_Pushforward}
to the trivial symmetry parameters of \eqref{eqn:AV2AV}. 
This generates a 12 parameter family of (on-shell equivalent) 
nonlinear supersymmetries for the AV action.
In general, these supersymmetry transformations are quite unwieldy, 
e.g., \eqref{eqn:KS_SUSY}, 
but the full result is available in 
the \emph{Mathematica} code distributed with \cite{KuzenkoTyler2011}.

\subsection{\texorpdfstring{Ro\v{c}ek's}{Rocek's} Goldstino action}
\label{ssect:Rocek}
The first paper to construct a field redefinition between different
realisations of Goldstino actions was by Ro\v{c}ek in 1978, \cite{Rocek1978}.
He assumed that the Goldstino was 
contained in a chiral superfield with the free action 
\begin{align} \label{eqn:free_WZ_action}
	S[\F, \bar \F]&=\int {\rm d}^4 x {\rm d}^4\q\, 
	\bar\F\F = \intx(\f\square\bar\f-\rmi\j\pd\bar\j+F\bar F)\,,
\end{align}
where we use the component projections \eqref{FlatComponentProjPhi}.
Ro\v{c}ek then looked for a transformation 
\( \big(\f, \j, F\big) \to \big(\f(\l), \j(\l), F(\l)\big) \)
that mapped the corresponding linear supersymmetry transformation 
\eqref{eqn:FreeSusy}
onto the AV supersymmetry transformation \eqref{eqn:AV_SUSY}.
This yielded a unique solution that we reproduce below. 
This solution was then recast in terms of the supersymmetric constraints
\begin{subequations}\label{eqn:Roceks_Constraints}
\begin{align} 
	\F^2 &=0 \,,   \label{eqn:Roceks_Constraint1} \\
	-\frac14 \F\Db^2\bar\F &= f\F \,, \label{eqn:Roceks_Constraint2}
\end{align}
\end{subequations}
where $f$ is a dimensional constant inversely proportional to $\k$ 
and is chosen to be real.%
\footnote{The sign of $f$ in the above equation differs from that given 
by Komargodski and Seiberg \cite{Komargodski2009} .}

We approach the problem the other way around, i.e., we start with the 
free action \eqref{eqn:free_WZ_action} and the constraints 
\eqref{eqn:Roceks_Constraint1} and \eqref{eqn:Roceks_Constraint2}.
We then derive the consequent Goldstino action $S_{\rm R}[\j,\bar\j]$ 
which is compared to the general Goldstino action \eqref{eqn:GeneralAV}
in order to find the map $\l\to\l(\j)$ that takes the AV action to 
$S_{\rm R}$. This map is then inverted to reproduce Ro\v{c}ek's results
$\big(\f(\l), \j(\l), F(\l)\big)$.

As noticed by Ro\v{c}ek (in his discussion of the 2D analogue of the AV model), 
the constraints \eqref{eqn:Roceks_Constraint1} 
and \eqref{eqn:Roceks_Constraint2}
mean that an arbitrary low-energy action
\begin{align}\label{effective-action1}
	S_\mathrm{eff} = \int  {\rm d}^4 x {\rm d}^4\q\, K(\bar\F,\F) +
	\left( \intx {\rm d}^2\q\, P(\F) + \cc \right)\,,
\end{align}
can always be reduced to a functional proportional to the free action. 
The first constraint \eqref{eqn:Roceks_Constraint1} allows the reduction
of \eqref{effective-action1} to
\begin{align} \label{effective-action2}
	\tilde{S}_{\rm eff} = \intx\rmd^4\q\, \bar\F\F 
		+ \left( \h \intx\rmd^2\q\,\F + {\rm c.c.}
		\right)\,,
\end{align}
modulo a trivial rescaling of the superfields 
and for some constant parameter $\h$. 
Imposing the second constraint \eqref{eqn:Roceks_Constraint2} 
makes all three structures in \eqref{effective-action2} completely equivalent,
so that the action can be written as either a pure kinetic term or 
a pure $F$-term.%
\footnote{In the  approach of Komargodski and Seiberg \cite{Komargodski2009}, 
which is discussed in the next section,
only the constraint \eqref{eqn:Roceks_Constraint1} is imposed. 
As a result, they work with an action of the form \eqref{effective-action2}.} 

The constraint  \eqref{eqn:Roceks_Constraint1} can be solved explicitly 
in terms of the component fields \cite{Casalbuoni1989,Komargodski2009}. 
This amounts to the fact that the scalar component of the chiral superfield 
becomes a function of the other fields, 
\begin{align} \label{eqn:SolnTo1stConst}
	\F^2=0 \quad  \iff  \quad \f = \frac{\j^2}{2F} \,.
\end{align}
The second constraint, $\F \bar D^2 \bar \F=-4f\F$,
is used to write the auxiliary field in terms of the spinor.
The simplest approach is to use the highest component of the constraint
to get an implicit equation for $F$
\begin{align} \label{eqn:Fbar1}
	F 
	  =	f + \bar F^{-1}\expt{\ub} 
		- \frac14 \bar F^{-2}\bar\j^2\square(F^{-1}\j^2)\,,
\end{align}
Here and below we use the notation
\begin{align} \label{eqn:defn of u}
	u=(u_a{}^b)=(\rmi\j\s^b\pd_a\bar\j )\,, \qquad
	\bar u = (\ub_a{}^b)=(-\rmi\pd_a\j\s^b\bar\j)\,,
\end{align}
and the same convention for matrix trace as in section \ref{ssect:AV}.
Equation \eqref{eqn:Fbar1}
can be solved by repeated substitution. After some work, we find
\begin{align} \label{eqn:F-final}
	F &= f \Big(1 + f^{-2}\expt{\ub} 
		- f^{-4}\big(\expt{u}\expt{\ub} + \frac14\bar\j^2\square\j^2 \big)
		+ f^{-6}(\expt{u}^2\expt{\ub} + \cc) \non\\
		&\quad + \frac14 f^{-6}\big(\expt\ub\j^2\square\bar\j^2
			+ 2\expt{u}\bar\j^2\square\j^2 + \bar\j^2\square(\j^2\expt\ub)\big) \\
		&\quad -3f^{-8}\big(\expt{u}^2\expt\ub^2
			+\frac14\j^2\bar\j^2\square(\expt{u}^2-\expt{u}\expt\ub+\expt\ub^2) 
			+ \frac{1}{16}\j^2\bar\j^2\square\bar\j^2\square\j^2\big) \Big) .\non
\end{align}

To get an action that maps onto the AV action,
we set $S_{\rm R}=-f\intx F$ with $F$ given by \eqref{eqn:F-final} 
and choose $f$ such that $f^{-2}=2\k^{2}$.
This yields
\begin{align} \label{eqn:ActionR}
	S_{\rm R} &= -\frac12\intx\!\Big(\k^{-2} + \expt{u+\ub}
		+\k^2\big(\pd^a\j^2\pd_a\bar\j^2-4\expt{u}\!\expt\ub\!\big) \\ \non
		&+4\k^4\big(\!\expt{u}\!\big(\bar\j^2\square\j^2+2\expt{u}\!\expt\ub\!\big)
			+\cc\big) \\\nonumber
		&+24\k^6\big(\expt{u^2}\expt{\ub^2}-3\expt{u}^2\expt\ub^2
			-2\expt{u}\expt\ub\expt{u\ub} 
			-\frac38\j^2\bar\j^2\square\j^2\square\bar\j^2\big)
	\Big)\,,
\end{align}
where we added surface terms to make $S_{\rm R} $ manifestly real.
This action has a nonlinearly realised supersymmetry that follows from the
linear supersymmetry transformations \eqref{eqn:FreeSusy}
and the solutions to the constraints given above, 
eqs.\ \eqref{eqn:SolnTo1stConst} and  \eqref{eqn:F-final},
\begin{align} \label{eqn:Rocek_NL_Susy}
	\d_\x\j_\a 
	&= \frac1\k\x_\a \Big(1 + 2\k^2\expt{\ub} 
		- 4\k^4\big(\expt{u}\expt{\ub} + \frac14\bar\j^2\square\j^2 \big)
		+ 8\k^6(\expt{u}^2\expt{\ub} + \cc) \non\\
		&+ 2\k^6\big(\expt\ub\j^2\square\bar\j^2
			+ 2\expt{u}\bar\j^2\square\j^2 + \bar\j^2\square(\j^2\expt\ub)\big) \\
		&-48\k^8\big(\expt{u}^2\expt\ub^2
			+\frac14\j^2\bar\j^2\square(\expt{u}^2-\expt{u}\expt\ub+\expt\ub^2) 
			+ \frac{1}{16}\j^2\bar\j^2\square\bar\j^2\square\j^2\big)\Big) \non\\
	&+ \rmi\k(\s^a\bar\x)_\a\pd_a
		\Big(\j^2
		\Big(1 - 2\k^2\expt{\ub} + 4\k^4\expt{\ub}^2 
			+ \k^4\bar\j^2\square\j^2\Big)
		\Big)\,. \non
\end{align}

Comparing the action \eqref{eqn:ActionR} to \eqref{eqn:GeneralAV}, 
we find the map that takes $S_\AV$ to $S_{\rm R}$:
\begin{align} \label{eqn:AV2R} \non
	\l_\a &= \j_\a-\k^2(1-\rmi\a_2^\rmi)\j_\a\expt{\ub}
		-\rmi{\k^2}(\s^a\bar\j)_\a(\pd_a\j^2)\Big(\tfrac12					
		-{\k^2}\expt{\b_7 u + \b_8 \ub}	\Big)  \\\non						
	&+{\k^4}\j_\a \Big(2(\rmi\a_2^\rmi+2\b_6+\b_8^*-3)\expt{u\ub}
		+ (1-\tfrac\rmi2\a_2^\rmi-\b_4^*+4\b_6^\rmr-\b_7-\b_8^*)
			\expt{u}\expt{\ub} \\ \non
	&+ (\tfrac32-2\rmi\a_2^\rmi-2\b_6)\expt{\ub^2} 
		+ \b_4 \expt{\ub}^2
		+ (\rmi\a_2^\rmi+2\b_6+\b_8^*-\tfrac52)\pd^a\j^2\pd_a\bar\j^2 
		+ \b_6\bar\j^2\square\j^2\Big) \\ \non
	&+ {\k^6}\j_\a \Big(\expt{u\ub^2}
		- (16 + 2\a_2^\rmi(2\b_6^\rmi-\b_8^\rmi) 
			- 8\b_6^\rmr - 2\b_7^\rmr - 4\b_8^\rmr 
			- \rmi\g_2^\rmi - 2\g_5^\rmr)\expt{u\ub}\expt{\ub} \\ \non
	&- (18 - 2\b_4^* -4\b_6^* - \b_7 + 2\rmi \b_8^\rmi - 4\b_8^\rmr 
			-\rmi\a_2^\rmi(2\b_6+\b_7+\b_8^*-3))\expt{u}\expt{\ub^2} \\ \non
	&- (45 -\b_4^\rmr -12 \b_6^\rmr -\b_8^\rmr -\rmi \g_4^\rmi
			+ \tfrac12\a_2^\rmi(3\b_4^\rmi+4\b_6^\rmi+\b_7^\rmi+3\b_8^\rmi) 
			)\expt{u}\expt{\ub}^2 \\ 
	&+ \g_5\expt{\ub}\pd^a\j^2\pd_a\bar\j^2 \Big) \,,  
\end{align}
where $4\b_6^\rmr=5-\tfrac12(\a_2^\rmi)^2$.
The inverse of this map can be found using the inversion formula of section 
\ref{sect:composition} and it only matches the result presented in 
\cite{Rocek1978} when the twelve free parameters \eqref{eqn:ReImFreeParams} 
are set to
\begin{align} \label{eqn:Roceks_coeffs}
	\a_2^\rmi=\b_4=\b_6^\rmi=0\,,\quad
	\b_7=-\frac12\,,\quad 
	\b_8=\frac32\,,\quad
	\g_2^\rmi=\g_4^\rmi=0\,,\quad
	\g_5=-3\,.
\end{align}
By using the composition rules of appendix 
\ref{sect:composition} it can be checked that all of the extra freedom is due 
to the trivial symmetries of the AV action \eqref{eqn:AV2AV}.

Inverting the above field redefinition with the specific coefficients 
\eqref{eqn:Roceks_coeffs} we obtain the solutions to the constraints 
on the AV side: 
\begin{subequations} \label{eqn:Roceks_Results}
\begin{align} 
	f\f &= \frac12\l^2\left(1+\k^2\expt{\vb}
		+\k^4\bar\l^2(\pd^a\l\s_{ab}\pd^b\l)\right) \,, \\
	f F &= \tfrac1{2}\k^{-2}+\expt{\vb} + \tfrac14\k^2\bar\l^2\square\l^2  \\ \non
		&+\k^2\big(\bar\l^2(\pd^a\l\s_{ab}\pd^b\l)
			-(\l\s_a\bar\l)(\pd^b\l\s_b\pd^a\bar\l)
			-(\pd^b\l\s_a\bar\l)(\l\s_b\pd^a\bar\l)\big) \\ \non
		&+\rmi\k^4\bar\l^2\big((\l\s_c\pd^c\bar\l)(\pd^a\l\s_{ab}\pd^b\l)
			-2(\l\s_a\pd^b\bar\l)(\pd^a\l\s_{bc}\pd^c\l)\big) \,,  \\
	\j_\a &= \l_\a + \k^2\l_\a\expt{\vb} + \tfrac\rmi2\k^2(\s^a\bar\l)_\a
		\pd_a\l^2\big(1-\k^2\expt{v-\vb}\big) \non\\ 
		&\quad+\k^4\l_\a\big(\expt{v}\expt\vb-\expt{\vb^2}
			+\tfrac12\pd^a\l^2\pd_a\bar\l^2+\tfrac14\bar\l^2\square\l^2\big)\\\non
		&\quad+\k^6\l_\a\big(\expt{v\vb^2}-\expt{v\vb}\expt\vb
			-\tfrac12\expt{v}(\expt{\vb^2}-\expt\vb^2)\big)\,.
\end{align}
\end{subequations}
These match Ro\v{c}ek's results (upon setting his parameter  $\a$ to zero) 
up to a couple of small typographical errors in his version of eq.\
(\ref{eqn:Roceks_Results}c).\footnote%
{The calculation with $0\neq\a\in\dsR$ has also been performed 
and the conclusion is identical.
}
Note the absence of any 8-fermion terms in (\ref{eqn:Roceks_Results}b) implies
their absence in the AV action \eqref{eqn:AV2} -- a fact 
rediscovered in \cite{Kuzenko2005e}.

Now that we have the mapping between $S_\AV$  and $S_{\rm R}$, we can calculate
the pushforward \eqref{eqn:AV_SUSY_Pushforward} of the AV supersymmetry,
which yields a 12 parameter family of supersymmetry transformations. 
In general, they are quite unwieldy, e.g., \eqref{eqn:KS_SUSY}, but
the full result has been calculated and is available in the code distributed
with \cite{KuzenkoTyler2011}.
We find that the pushforward of the AV supersymmetry only reduces to the 
supersymmetry \eqref{eqn:Rocek_NL_Susy}
when the free parameters are fixed
to \eqref{eqn:Roceks_coeffs}. 
This explains the uniqueness of Ro\v{c}ek's results.

\subsection{Casalbuoni-De~Curtis-Dominici-Feruglio-Gatto\\ and 
Komargodski-Seiberg action}\label{ssect:KS} 
The action that we analyse in this section was introduced by 
Casalbuoni {\it et al.}\ 
in 1989 \cite{Casalbuoni1989}, a work that has 
unfortunately remained largely unnoticed.
The same action has recently been rediscovered and very effectively utilised
by Komargodski and Seiberg \cite{Komargodski2009}. 
The novelty of the Komargodski-Seiberg (KS) approach is, 
in particular, that they related the 
Goldstino dynamics to the superconformal anomaly multiplet 
$X$ corresponding to the Ferrara-Zumino supercurrent \cite{Ferrara1975}.
Under the renormalisation group, the multiplet of anomalies $X$, 
defined in the UV, flows in the IR to a chiral superfield $X_{NL}$
obeying the constraint $X_{NL}^2=0$.
This type of constraint was first introduced by Ro\v{c}ek \cite{Rocek1978} 
and is discussed in the previous section. 
Finally, one of the crucial results of \cite{Komargodski2009} is that Komargodski and Seiberg
showed how to generalise their Goldstino action
to include higher-derivative interactions
and couplings to supersymmetric matter.
In this work, as we are only interested in the equivalence of 
the various Goldstino models, we will not consider such interactions. 
The Goldstino model of \cite{Casalbuoni1989,Komargodski2009} 
will be called the KS action for brevity.

The model is described by a single chiral superfield constrained by 
\begin{align}\label{F^2}
	\F^2=0\,. 
\end{align}
As discussed in section \ref{ssect:Rocek}, 
the most general low-energy action that can be constructed from $\F$ is
\begin{align} \label{eqn:KS0}
	S[\F,\bar\F] = \intx {\rm d}^4\q\, \bar\F\F 
	+ \left(f\intx {\rm d}^2\q\, \F + \cc\right)\,,
\end{align}
where, without loss of generality, we can choose the coupling constant 
$f$ to be real. 
Apart from the constraint, 
this is exactly the Polonyi model \eqref{PolonyiModel}.
As in the previous section, 
we find that for $S_\KS$ to match $S_\AV$ then $f$ must be 
such that $2f^2=\k^{-2}$.
As in Ro\v{c}ek's model, the nilpotent constraint is used to solve for the 
scalar component field \eqref{eqn:SolnTo1stConst}, 
with the component fields of $\F$ given in \eqref{FlatComponentProjPhi}.
This leaves the component action
\begin{align} \label{eqn:KS0.5}
\!\!S[\j,\bar\j,F,\bar F]
	= \intx \big( -\frac12\expt{u+\ub}
	  + \frac{\bar\j^2}{2\bar F}\square\frac{\j^2}{2F}
	  + f F + f \bar F + F \bar F
	\big) , \!\!
\end{align}
where $u_a{}^b$ and ${\bar u}_a{}^b$ are defined in \eqref{eqn:defn of u}.
In Ro\v{c}ek's model the second constraint \eqref{eqn:Roceks_Constraint2} 
is used to eliminate the auxiliary complex field.  
In the KS model one does not have such a constraint, 
and both terms in the action \eqref{eqn:KS0} remain  essential.
The auxiliary scalar is removed from \eqref{eqn:KS0.5} 
using its equations of motion, 
leaving the fermionic action \cite{Komargodski2009}
\begin{equation}\label{eqn:KS1}
\!\!S_{\KS}[\j,\bar\j] = -\half\intx\Big(\k^{-2} + \expt{u+\ub}+
		\k^2\pd^a\bar{\j}^2\pd_a \j^2
		+\k^6\j^2\bar\j^2\square\j^2\square\bar\j^2\Big) ,\!\!
\end{equation}
Comparing \eqref{eqn:ActionR} with \eqref{eqn:KS1} clearly shows that 
the two actions are different.
The KS action appears to have the simplest form among all the Goldstino models.

In both \cite{Casalbuoni1989} and \cite{Komargodski2009}, 
the action \eqref{eqn:KS0} was analysed using a Lagrange multiplier field 
to enforce the constraint $\F^2=0$.
This analysis was used in \cite{Komargodski2009} to show the \emph{on-shell} 
equivalence of the KS action with the Ro\v{c}ek action \eqref{eqn:ActionR}; 
to show the \emph{off-shell} equivalence takes a little more work, 
we present the explicit mapping at the end of section \ref{ssect:Trivialities}.
The Lagrange multiplier analysis also has some interesting aspects 
in and of itself that were investigated in detail in \cite{KuzenkoTyler2011}.

By writing the KS action in the basis of appendix \ref{sect:bases} 
and comparing with the general action \eqref{eqn:GeneralAV} we find the mapping
that takes the AV action onto the KS action
\begin{align} \label{eqn:AV2KS} \non
	\l_\a &= \j_\a+\rmi\a_2^\rmi{\k^2}\j_\a\expt{\ub} 
		-\tfrac\rmi2{\k^2}(\s^a\bar\j)_\a(\pd_a\j^2)
	+ \rmi{\k^4}(\s^a\bar\j)_\a(\pd_a\j^2)\expt{\b_7u +\b_8\ub} \\\non  					
	&+{\k^4}\j_\a\Big(
	2(\rmi\a_2^\rmi+2\b_6+\b_8^*)\expt{u\ub}
	+ \tfrac12(3-2\b_4^*-2\b_8^*+8 \b_6^\rmr-2\b_7
		-\rmi\a_2^\rmi)\expt{u}\expt{\ub} \\\non
	&- (\tfrac12+2\rmi\a_2^\rmi+2\b_6)\expt{\ub^2} 
	+ \b_4 \expt{\ub}^2
	+ (\tfrac{1}{2}+\rmi\a_2^\rmi+2\b_6+\b_8^*)\pd^a\j^2\pd_a\bar\j^2 
	+ \b_6\bar\j^2\square\j^2\Big) \\\non
	&+\k^6\j_\a\Big( \expt{u\ub^2}
	+ (2-2\a_2^\rmi(2\b_6^\rmi-\b_8^\rmi)+4\b_6^\rmr+2\b_7^\rmr+2\b_8^\rmr
		+\rmi\g_2^\rmi+2 \g_5^\rmr)\expt{u\ub}\expt{\ub} \\\non
	&+ (1+\rmi\a_2^\rmi(2\b_6+\b_7+\b_8^*-2)+2\b_4^*+2\b_6^\rmr-6\rmi\b_6^\rmi
		+3\b_8^\rmr-\rmi\b_8^\rmi)\expt{u}\expt{\ub^2} \\\non
	&+ (1-\tfrac12\a_2^\rmi(3\b_4^\rmi+4\b_6^\rmi+\b_7^\rmi+3\b_8^\rmi)
		+\tfrac12(\b_4^\rmr+16\b_6^\rmr-\b_7^\rmr+\b_8^\rmr+2\rmi\g_4^\rmi))
		\expt{u}\expt{\ub}^2 \\
	&+ \g_5\expt{\ub}\pd^a\j^2\pd_a\bar\j^2 \Big) \,,
\end{align}
where $-8\b_6^\rmr = {1 + (\a_2^\rmi)^2}$.
By using the composition rules of appendix \ref{sect:composition} 
it can be checked that all of the freedom in \eqref{eqn:AV2KS}
is due to the trivial symmetries of the AV action.

Alternatively, we can attribute the freedom in \eqref{eqn:AV2KS} 
to the trivial symmetries of $S_\KS$. 
These are found in a similar manner to those of
$S_\AV$ and are given by the field redefinition 
(with  $-8\b_6^\rmr=(\a_2^\rmi)^2$)
\begin{align} \label{eqn:KS2KS}
	\j_\a \to \j'_\a &= \j_\a+\rmi\a_2^\rmi\k^2\j_\a\expt{\ub}  
	+\rmi{\k^4}(\s^a\bar\j)_\a(\pd_a\j^2)\expt{\b_7u +\b_8\ub}\\\non  					
	&+{\k^4}\j_\a\big(
		 2(2\rmi\a_2^\rmi+2\b_6+\b_8^*)\expt{u\ub}
		- (\b_4^*-4\b_6^\rmr+\b_7+\b_8^*)\expt{u}\expt{\ub}\\\non  	
		&- 2(\rmi\a_2^\rmi+\b_6)\expt{\ub^2} 
		+ \b_4 \expt{\ub}^2
		+ (2\rmi\a_2^\rmi+2\b_6+\b_8^*)\pd^a\j^2\pd_a\bar\j^2 
		+ \b_6\bar\j^2\square\j^2\big) \\\non
	&+ \k^6\j_\a\big( \g_5\expt{\ub}\pd^a\j^2\pd_a\bar\j^2
		+ (2\a_2^\rmi(\b_8^\rmi-2\b_6^\rmi)+\rmi\g_2^\rmi+2\g_5^\rmr)
			\expt{u\ub}\expt{\ub} \\\non
		&+ (2 (\b_4^*+6\b_6^\rmr-2\rmi\b_6^\rmi+\b_8^*)
			+\rmi\a_2^\rmi(2\b_6+\b_7+\b_8^*-4))\expt{u}\expt{\ub^2} \\\non
		&+ (-\tfrac12\a_2^\rmi(3\b_4^\rmi+4\b_6^\rmi+\b_7^\rmi+3\b_8^\rmi)
		+2\b_4^\rmr+12\b_6^\rmr+2\b_8^\rmr+\rmi\g_4^\rmi)\expt{u}\expt{\ub}^2
		 \big) \,.
\end{align}
These symmetries are completely equivalent to those given in \eqref{eqn:AV2AV}.
However, due to the simplicity of KS action and its equations of motion, 
it is easiest to prove the triviality of the above transformations rather
than \eqref{eqn:AV2AV}.
This is done in section \ref{ssect:Trivialities}. 

Finally, we come to the question of  how the supersymmetry algebra is
realised on the fields $\j_\a$ of the KS action.  
The original off-shell linear
supersymmetry of the chiral field $\F$ becomes nonlinear 
in the action \eqref{eqn:KS0.5} and is only realised on-shell after 
the auxiliary equations of motion are enforced to yield \eqref{eqn:KS1}.
This was discussed in \cite{Casalbuoni1989} and the structure of 
an \emph{off-shell} nonlinearly realised supersymmetry that $S_\KS$ 
should possess has been an open question since. 
Now, more than twenty year later, we are in a position to address the problem!

We can calculate the pushforward of the AV supersymmetry
to give us the supersymmetry transformations that leave the KS action invariant
\begin{align} 
\d_\x \j_\a &= \d_\x \j_\a(\l,\bar\l)\Big|_{\l=\l(\j,\bar\j)}
		= \d_\x\l^\b\cdot\frac{\d}{\d\l^\b}\j_\a(\l,\bar\l)
		+ \d_\x\bar\l_\db\cdot\frac{\d}{\d\bar\l_\db}\j_\a(\l,\bar\l)
			\Big|_{\l=\l(\j,\bar\j)}\,, \non
\end{align}
where $\j=\j(\l,\bar\l)$ and $\l=\l(\j,\bar\j)$ are exact inverse mappings.
In \cite{KuzenkoTyler2010} we used the map obtained from 
\eqref{eqn:AV2KS} by setting all twelve free parameters to zero 
and its inverse calculated using the inversion formula of 
appendix \ref{sect:composition}, 
to find the leading order terms to the KS supersymmetry.
By generating a basis for all possible supersymmetry terms, 
which is available and proved to be minimal in
the \emph{Mathematica} code distributed with \cite{KuzenkoTyler2011},
it was possible to automate the rest of the calculation. 
The full result is
{\allowdisplaybreaks
\begin{align}\label{eqn:KS_SUSY} 
	\d_\x\j_\a &= \frac1\k \x_\a 
	+ \k\Big( 
		\x_a\expt{\ub} - (\rmi\j\s^a\bar\x)\pd_a\j_\a
		+\tfrac12(\rmi\s^a\bar\x)_\a\pd_a\j^2 \Big) 	
		\\\non
	&+ \k^3\Big( 
		\x_\a\big\{
			  \tfrac12\expt{u\ub} - \expt{u}\expt{\ub}
			- \tfrac14\expt{\ub^2} 
			- \tfrac14 \pd^\a\j^2\pd_a\bar\j^2 
			+ \tfrac18 \bar\j^2\square\j^2 \big\}\\\non
	&\;	+ \j_\a\big\{\square\j^2\bar\x\bar\j 
			- \tfrac32\pd^a\j^2\pd_a\bar\x\bar\j 
			- \tfrac32(\j\square\j)\bar\x\bar\j + \tfrac34\square(\j\x)\bar\j^2
			 \\ \non 
	&\quad	- \tfrac12\pd^a(\j\x)\pd_a\bar\j^2
			+ \tfrac12(\x\s^a\pd_b\bar\j)(\pd_a\j\s^b\bar\j) 
			+ (\x\s^a\bar\j)(\pd_b\j\s^b\pd_a\bar\j)\big\} \\\non
	&\;	+ \pd_a\j_\a\big\{
			  (\rmi\j\s^a\bar\x)\expt{\tfrac12\ub-u}
			+ (\j\s^a\pd_b\bar\j)(\x\s^b\bar\j)
			+ (\x\s^a\pd_b\bar\j)(\j\s^b\bar\j)\\\non
	&\quad	- \tfrac34 (\rmi\s^a\bar\x)_\a\pd_a\j^2\expt{\ub} 
			- \tfrac34(\s^a\pd_b\bar\j)_\a\pd_a\j^2(\j\s^b\bar\x)
			- \j_\a(\pd_a\pd_b\j\s^a\bar\x)(\j\s^b\bar\j)\big\}
			\Big)\allowdisplaybreaks\\\non
	&+\k^5\Big( 
		\x_a\big\{\pd^a\j^2\pd_a\bar\j^2\expt{\tfrac34\ub-u}
			+ \tfrac18\expt{\ub}\j^2\square\bar\j^2 
			- \tfrac18\expt{u}\bar\j^2\square\j^2  
			- \expt{u}^2\expt{\ub}	\\\non
	&\quad	+ \expt{u\ub}\expt{\tfrac32\ub-u}
			- \tfrac34\expt{v^2}\expt{\vb} - \tfrac14\expt{v}\expt{\vb^2}
			+ \tfrac12\expt{v^2\vb} + \tfrac12\expt{v\vb^2}
			\big\}\\\non
	&\;+ \j_\a\big\{
			  \tfrac14\expt{u}\bar\j\bar\x\square\j^2
			+ 4\expt{\ub}\pd^b\j^2\pd_b\bar\j\bar\x
			+ \tfrac34(\rmi\j\s^a\bar\x)\square\j^2\pd_a\bar\j^2
			 \\\non
	&\quad	- \tfrac12 (\rmi\pd_a\j\s^a\bar\x)\big(
			  \expt{u}\expt{\ub} + \expt{u\ub}
			+ \tfrac34\bar\j^2\square\j^2 - \tfrac12(\pd\j)^2\bar\j^2 
			+ \tfrac32\pd^b\j^2\pd_b\bar\j^2\big) \\\non
	&\quad	- \tfrac12 (\rmi\x\s^a\pd_a\bar\j)\big( 
			  \expt{u\ub}
			+ \tfrac32(\pd\j)^2\bar\j^2 
			+ \pd^b\j^2\pd_b\bar\j^2 \big)\\ \non 
	&\quad	+ \tfrac14(\rmi\x\s^a\bar\j)\big( 
			  \pd_a\j^2\square\bar\j^2
			- \pd_a\j^2(\pd\j)^2	
			- \pd_a\pd_b\j^2\pd^b\bar\j^2	\big) \\\non
	&\quad-	(\rmi\x\s^a\pd_b\bar\j)\big( 
			  \tfrac14\pd^b\j^2\pd_a\bar\j^2
			+ (\pd_a\j\s^c\pd_c\bar\j)(\j\s^b\bar\j)
			- \tfrac12(\pd_a\pd_c\j\s^b\bar\j)(\j\s^c\bar\j) \big)\\\non
	&\quad+ \expt{\ub}\big(
			  3(\pd_b\j\s^a\bar\x)(\j\s^b\pd_a\bar\j)
			+ 2(\pd_a\pd_b\j\s^a\bar\x)(\j\s^b\bar\j) \big)
			\big\}\\\non
	&\;+\pd_a\j_\a\big\{ \tfrac12(\rmi\j\s^a\bar\x)\big(
			  \expt{u}\expt{\ub} + \expt{u\ub}
			- 3\expt{\ub}^2 - \tfrac54\bar\j^2\square\j^2 
			- (\pd\j)^2\bar\j^2 \big) \\\non
	&\quad	+ \tfrac12(\rmi\x\s^a\bar\j)\expt{u\ub} 
			+ \tfrac12(\rmi\pd_b\j\s^a\bar\x)(\j\s^b\pd_c\bar\j)(\j\s^c\bar\j)  
			- \tfrac74(\rmi\j\s_b\bar\x)\bar\j^2\pd^a\pd^b\j^2\\\non
	&\quad	+ \tfrac12(\rmi\x\s^c\pd_b\bar\j)\big(
			  (\pd_c\j\s^b\bar\j) - (\j\s^b\pd_c\bar\j)\big)(\j\s^a\bar\j)
			\big\}			 \\\non
	&\;	+ \tfrac32\pd_a\pd_b\j_\a (\rmi\pd^b\j\s^a\bar\x)\j^2\bar\j^2
		- \tfrac14(\rmi\s^a\bar\j)_\a\j^2\pd^b(\j\x)\pd_a\pd_b\bar\j^2 \\\non
	&\;	-\tfrac18(\rmi\s^c\st^b\pd_b\j)_\a\j^2\big(
			  \bar\j^2(\x\s^a\pd_a\pd_c\bar\j)
			+ 2(\pd_a\bar\j\pd_c\bar\j)(\x\s^a\bar\j)\big)
	\Big)\allowdisplaybreaks\\\non
	&+ \tfrac18\k^7\Big( 
	\x_\a\big\{2\expt{u^2\ub^2} + 13\expt{(u\ub)^2}
			- 10\expt{u}\expt{u\ub^2} - 4\expt{u^2\ub}\expt\ub\\ \non 
	&\quad	- 5\expt{u^2}\expt{\ub^2} + 5\expt{u\ub}^2
	 		+ \tfrac{11}4\j^2\bar\j^2\pd^a\pd^b\j^2\pd_a\pd_b\bar\j^2 \big\} \\\non
	&\;+ \j^2\bar\j^2\big\{ 
			  \tfrac14(\s^b\st^a\pd_b\j)_\a\pd_a(\j\x)\square\bar\j^2 \\\non	 
	&\quad	+ \pd^b\j_\a(\x\s^c\pd_c\bar\j)\big(
			  (\pd_a\j\s^a\pd_b\bar\j)
			+ 6(\pd_b\j\s^a\pd_a\bar\j) \big) \\\non
	&\quad	+ 5(\s^a\pd_b\bar\j)_\a\big(
			  (\pd_a\j\s^b\pd_c\bar\j)\pd^c(\j\x)
			- (\pd_a\j\s^c\pd_c\bar\j)\pd^b(\j\x) \big)\\\non
	&\quad	+ \tfrac54(\s^a\st^c\x)_\a\square(\pd_a\j^2\pd_c\bar\j^2)
			+ \tfrac72(\s^c\pd_c\bar\j)_\a\pd_a\pd_b\j^2(\x\s^a\pd^b\bar\j)\\\non
	&\quad	- \pd_b\j_\a\big(
			  5(\pd_a\j\s^d\st^c\s^a\pd_d\bar\j)(\x\s^b\pd_c\bar\j)
		  	- 4(\pd_a\j\s^d\st^b\s^a\pd_d\bar\j)(\x\s^c\pd_c\bar\j)
		  	\big)\\\non
	&\quad  -\tfrac12\pd_a\j_\a\big(
			  13\pd^a(\bar\j\bar\x)\square\j^2
			- 37\square(\j\s^a\bar\j)(\pd_b\j\s^b\bar\x) \big)\\\non
	&\quad	- 16(\s^b\bar\x)_\a\pd_a\pd_b\j^2  (\pd_c\j\s^c\pd^b\bar\j) 
			- 15(\s^b\st^c\pd_a\j)_\a (\pd_c\j\s^a\pd_d\bar\j)(\pd_b\s^d\bar\x)
			\\\non
	&\quad	+\tfrac{37}{2}\pd_b\pd_c\j^2\big(
			  (\s^b\st^a\pd_a\j)_\a\pd^c(\bar\j\bar\x)
			- (\s^b\st^a\pd^c\j)_\a\pd_a(\bar\j\bar\x) \big) \\\non
	&\quad	+ \pd_a\pd_b\j^2\big(
			  4(\s^c\bar\x)_\a(\pd^b\j\s^a\pd_c\bar\j)
			+ 24(\s^a\bar\x)_\a(\pd^b\j\s^c\pd_c\bar\j) \big) 	\big\}\Big)\,.
\end{align}}%
So we see that the cost of the simple action $S_\KS$ is the 
complicated supersymmetry transformation $\d_\h\j_\a$. 
When using the above basis 
(which is not necessarily optimal for describing the KS supersymmetry), 
there does not seem to be much simplicity to be gained 
in choosing different trivial symmetry parameters in \eqref{eqn:AV2KS}.
The full 12-parameter family of KS supersymmetry transformations 
is available in the \emph{Mathematica} code distributed with \cite{KuzenkoTyler2011}, 
but the structure is too unwieldy to reproduce here.
It has explicitly been checked that this mapping satisfies 
the supersymmetry algebra \eqref{AV:SusyAlg} 
and leaves the action \eqref{eqn:KS1} invariant.

The rest of the Goldstino actions considered in this paper have a natural 
nonlinear supersymmetry that is either the starting point for the model
or follows from the combination of a linear supersymmetry 
and some supersymmetric constraints. 
For such actions there is a specific choice of the 
12 trivial symmetry parameters that allows for an organising and simplifying 
of the nonlinear supersymmetry. It is not clear if such set of parameters
and consequent simplification can be found for the supersymmetry of the 
KS action.


\subsection{The chiral Alkulov-Volkov action}\label{ssect:SW}
The AV supersymmetry transformation  \eqref{eqn:AV_SUSY} 
mixes the fields $\l$ and $\bar\l$.  It was
Zumino \cite{Zumino1974LondonConf} who
introduced an alternate form of nonlinearly realised  
supersymmetry that does not have such a mixing
\begin{align} \label{eqn:chAV_SUSY}
	\d_\x\tilde\l_\a = 
		\frac1\k\x_\a - 2\rmi\k(\tilde\l\s^a\bar\x)\pd_a\tilde\l_\a \,.
\end{align}
This lack of mixing simplifies many types of calculations, 
a fact that was first noticed and exploited
by Samuel and Wess \cite{SamuelWess1983}.
This new supersymmetry is related to the AV one via the simple 
field redefinition \cite{Ivanov1978,SamuelWess1983}
\begin{align} \label{eqn:AVtoChAV_Simp}
	\tilde\l_\a(x) 
	= \l_\a(y)\,, \qquad 
		y=x-\rmi\k^2\l(y)\s\bar\l(y)\,,
\end{align}
which is essentially a nonlinear version of the relations defining 
the chiral superspace coordinates.
The above field redefinition has been explicitly expanded many times in 
the literature and the result can be written in terms of the general
field transformation \eqref{eqn:GeneralFieldRedef} with the parameters
\begin{equation}\label{eqn:AVtoChAV_Simp_Params}\begin{gathered} 
	\a_1=\b_1=\b_3=\g_5=0\,, \quad
	\a_2=\b_2=-\b_4=-\g_1=\g_2=-1\,, \\
	\a_3=-\b_5=\b_7=-\b_8=-\g_3=\g_4=1/2\,, \qquad
	\b_5=-1/4\,.
\end{gathered}\end{equation}

The action for this model is normally constructed in terms of  the superfield
\begin{gather} \label{eqn:chAV_Superfield}
	\tilde\L_\a(x,\q,\bar\q) = \exp(\q Q+\bar\q\bar Q)\tilde\l_\a(x)\,,
\end{gather}
where $(\q Q+\bar\q\bar Q)\lt_\a=\d_\q\lt_\a$ is the transformation 
\eqref{eqn:chAV_SUSY} using the parameter $\q$ instead of $\x$.
The action is then
\begin{align}\label{eqn:chAV_Action}
	S_\SW &= -\frac{\k^2}2\intx \rmd^4 \q\tilde\L^2\bar{\tilde\L}^2
		= -\frac{\k^2}2\intx\rmd^4\q\frac1{4!}\d_\q^4
			(\tilde\l^2\bar{\tilde\l}^2) \non\\\non
		&=-\frac12\intx\Big(\k^{-2}+\expt{\vt+\vbt}
			+\k^2\big(\pd^a\lt^2\pd_a\bar\lt^2+4\expt{\vt}\expt{\vbt}\big)\\\non
		&\qquad	+\k^4\big(\expt{\vt}\big(2\pd^a\lt^2\pd_a\bar\lt^2
				+4\expt{\vt\vbt}+4\expt{\vbt}^2-2\expt{\vbt^2}
				-\bar\lt^2\square\lt^2\big) + \cc\big)	\\
		&\qquad	+\k^6\big(\lt^2\bar\lt^2\square\lt^2\square\bar\lt^2
				-8\expt{\vt}^2\expt{\vbt^2}-8\expt{\vt^2}\expt{\vbt}^2\big)
		\Big)\,.
\end{align}
A similar superfield approach can also be used to reproduce the normal 
AV action \eqref{eqn:AV2} \cite{WessBagger1992}.
In section \ref{sect:mCL} 
we show how this action and superfield have 
an equivalent description in terms of a constrained complex linear superfield 
that is, in some ways, the more fundamental object.

We can also use the superfield \eqref{eqn:chAV_Superfield} 
to solve Ro\v{c}ek's constraints 
\eqref{eqn:Roceks_Constraint1} and \eqref{eqn:Roceks_Constraint2}.
Following \cite{SamuelWess1983}, it can be shown that  
\begin{align} \label{eqn:SW_Soln_of_Roceks_Constraints}
	\F  = -\frac{\k^2}{8f} \Db^2(\tilde\L^2\bar{\tilde\L}^2)
\end{align}
solves both constraints 
\eqref{eqn:Roceks_Constraint1} and \eqref{eqn:Roceks_Constraint2} 
and immediately gives the relationship between
Ro\v{c}ek's model and the chiral AV Goldstino.
A similar construction starting with the normal AV Goldstino can be used
to reproduce \eqref{eqn:Roceks_Results} with minimal effort.
This approach is related to the general approach \cite{Ivanov1978} 
based on nonlinear representation theory 
and will not be further investigated in this thesis.

By writing the action \eqref{eqn:chAV_Action} in the basis of appendix 
\ref{sect:bases} and comparing to \eqref{eqn:GeneralAV}, 
we find the map that takes $S_\AV$ to $S_\SW$ 
\begin{align} \label{eqn:AV2SW} \non
	\l_\a &= \lt_\a+{\k^2}(1+\rmi\a_2^\rmi)\lt_\a\!\expt{\vbt} 
	-\frac{\rmi\k^2}2 (\s^a\bar\lt)_\a(\pd_a\lt^2)\big(1						
		-2{\k^2}\expt{\b_7 \vt + \b_8 \vbt}\!	\big)  \\\non						
	&+{\k^4}\lt_\a\big(2(1+\rmi\a_2^\rmi+2\b_6+\b_8^*)\expt{\vt\vbt}
		+ (2-\tfrac\rmi2\s_2^\rmi-\b_4^*+4\b_6^\rmr-\b_7-\b_8^*)
			\expt{\vt}\expt{\vbt} \\ \non
		&- (\tfrac12+2\rmi\a_2^\rmi+2\b_6)\expt{\vbt^2} 
		+ \b_4 \expt{\vbt}^2
		+ (\tfrac32+\rmi\a_2^\rmi+2\b_6+\b_8^*)\pd^a\lt^2\pd_a\bar\lt^2 
		+ \b_6\bar\lt^2\square\lt^2\big) \\
	&+ {\k^6}\lt_\a\big(\expt{\vt\vbt^2}
		+ (2\b_7^\rmr+\rmi\g_2^\rmi+2\g_5^\rmr-2\a_2^\rmi(2\b_6^\rmi-\b_8^\rmi))
			\expt{\vt\vbt}\expt{\vbt} \\\non
		&- (2-2\b_4^*+8\rmi\b_6^\rmi+\b_7-2\b_8^\rmr 
			+ \rmi\a_2^\rmi(1-2\b_6-\b_7-\b_8^*))\expt{\vt}\expt{\vbt^2} \\\non
		&+ (1+4\b_6^\rmr-\b_7^\rmr - \tfrac{\a_2^\rmi}2(3\b_4^\rmi
			+4\b_6^\rmi+\b_7^\rmi+3\b_8^\rmi)+\rmi\g_4^\rmi)
			\expt{\vt}\expt{\vbt}^2
		+ \g_5\expt{\vbt}\pd^a\lt^2\pd_a\bar\lt^2 \big) \,,
\end{align}
with $-8\b_6^\rmr = 2+(\a_2^\rmi)^2$.

Using this map we can then calculate the pushforward of the AV supersymmetry.
As expected, it only matches \eqref{eqn:chAV_SUSY} for a single choice of the 
free parameters:
$\b_4^\rmr=1$, $\b_7^\rmr=\b_8^\rmr=-1/2$ 
with all other free parameters set to zero. 
This choice of parameters also 
reduces \eqref{eqn:AV2SW} back to the inverse of 
\eqref{eqn:AVtoChAV_Simp}.

\subsection{The supersymmetric Born-Infeld action}\label{ssect:BI}
The $\cN=1$ supersymmetric Born-Infeld (SBI) action
was originally introduced in \cite{Cecotti1987,Deser1980}
as a supersymmetric extension of the Born-Infeld theory \cite{Born1934},
and as such it is not unique.
Bagger and Galperin \cite{Bagger1997a}, 
and  later Ro\v{c}ek and Tseytlin \cite{Rocek1998}, 
using alternative techniques, discovered that the action 
given in \cite{Cecotti1987} describes a Goldstone-Maxwell multiplet
associated with partial  $\cN=2 \to \cN=1$ supersymmetry breaking.
The constrained superfield formulations of \cite{Bagger1997a,Rocek1998} 
were later shown to be derivable from the superembedding of 
a D3-brane in $\cN=2$ superspace \cite{bandos2001space}. 
This is an example of how the superembedding approach can be used as 
as generic covariant covariant method for the description of 
(partial) \emph{spontaneous} supersymmetry breaking 
\cite{pasti2000superembeddings,sorokin2000superbranes}.

Although this Goldstone SBI action was argued to be unique 
\cite{Bagger1997a,Rocek1998}, 
there exists, in fact, a two-parameter deformation of the theory 
\cite{Kuzenko2009} which also 
describes partial  $\cN=2 \to \cN=1$  supersymmetry breaking.
The SBI action is also known to be invariant under U(1) duality rotations 
\cite{Kuzenko2000,Kuzenko2000a,BMZ1999}.

The most elegant way to formulate the SBI theory is as the vector Goldstone 
action for partially broken $\cN=2$ supersymmetry \cite{Bagger1997a,Rocek1998}.
The approaches developed in \cite{Bagger1997a} and \cite{Rocek1998} are rather 
different from the conceptual point of view but both yield 
a manifestly $\cN=1$ supersymmetric nonlinear theory 
of an Abelian vector multiplet. Its action is given 
in terms of a constrained chiral superfield $X$ constructed in terms of the 
vector-multiplet field strength 
$W_\a$ and its conjugate ${\bar W}_{\dot \a}$
\begin{align} \label{eqn:BI-constrained}
	S[W,\bar W] = \frac14\int  {\rm d}^4 x {\rm d}^2\q\, X + \cc\,, \qquad
	X + \frac{\k^2}4 X \bar D^2 \bar X = W^2\,.
\end{align}
The constraint is solved by \cite{Rocek1998}
\begin{gather} \label{eqn:BI-constraint-soln}
	X = W^2 - \frac{\k^2}{2}\bar D^2 (W^2\bar W^2 f(A,B))\,,\;
	f(A,B)^{-1} = 1+\frac12A+\sqrt{1+A+\frac14B^2}\,, \non\\
	A = \frac{\k^2}2(D^2W^2+\Db^2\Wb^2)\,,\quad
	B = \frac{\k^2}2(D^2W^2-\Db^2\Wb^2)\,.
\end{gather}
This gives the SBI action
\begin{align} \label{eqn:BI-Susy}
	S[W,\bar W] = 
	\frac{1}{2}  \int  {\rm d}^4 x {\rm d}^2\q\, W^2 
		+ \k^2\int  {\rm d}^4 x {\rm d}^4\q\, W^2\Wb^2 f(A,B)\,. 
\end{align}
The action is also invariant under the nonlinearly realised 
(non-manifest) $\cN=2$ supersymmetry transformation
\begin{align} \label{eqn:BI-SUSY0}
\!\!\!\!\!	
	\d_\h W_\a = \frac1\k \left(\h_\a + \frac{\k^2}4\h_\a\Db^2\bar X 
		+ \rmi\k^2(\s^a\bar\h)_\a\pd_aX \right), \quad
	\d_\h X = \frac2\k \h^\a W_\a\,. \!\!\!
\end{align}

Projection to the fermionic action is consistent with both the equations
of motion and the second supersymmetry \cite{Kuzenko2005e}. 
We use the component projections 
\begin{align} \label{eqn:BI-project}
	W_\a| = \c_\a\,,\quad 
	\frac1{2\rmi}D_{(\a}W_{\b)}|=F_{\a\b}\to0 \,, \quad
	-\frac12 D^\a W_\a| = D \to 0\,,
\end{align}
to find the Goldstino action
\begin{align} \label{eqn:BI}
	S_\BI[\c,\bar\c] &= -\frac12\intx\Big(	\expt{w+\bar w}
	+ \k^2\big(\pd^a\c^2\pd_a\bar\c^2 - 4\expt{w}\expt{\wb}\big) \\\non
	&+ 8\k^4\big(\expt{w}^2\expt\wb+\frac12\expt{w}\bar\c^2\square\c^2+\cc\!\big) 
		-12\k^6\big(\expt{w}^2\bar\c^2\square\c^2 + \cc\! \big) \\\non
	&- 48\k^6\big(\expt{w}^2\expt{\wb}^2 
		-\frac12\expt{w}\expt{\wb}\pd^a\c^2\pd_a\bar\c^2
		+\frac1{16}\c^2\bar\c^2\square\c^2\square\bar\c^2\big)\Big)\,.
\end{align}
where $w_a{}^b=\rmi\c\s^b\pd_a\bar\c$. 
The fermionic sector of the general $\cN=1$ vector self-dual model 
considered in \cite{Kuzenko2005e} 
only differs by a rescaling of the last line above,
but only those with the fermionic sector given above can be mapped 
to the Akulov-Volkov action \cite{Kuzenko2005e}.
The supersymmetry \eqref{eqn:BI-SUSY0} is projected to 
\begin{align} \label{eqn:BI-SUSY}
\!	\d_\h \c_\a &= \tfrac{\h_\a}\k 
	+ 2\k\h_\a\Big(\!\expt{\wb}-2\k^2(\expt{w}\expt\wb+\tfrac14\bar\c^2\square\c^2)
	+4\k^4\big(\!\expt{w}^2\expt\wb+\expt{w}\expt\wb^2 \non\\
	&-\tfrac12\expt{\wb}\pd^a\c^2\pd_a\bar\c^2+\tfrac12\expt{w}\bar\c^2\square\c^2
	+\tfrac14\expt\wb\c^2\square\bar\c^2+\tfrac14\c^2\bar\c^2\square\expt{\wb}\!\big)
\non\\\non
	&-24\k^6\big(\expt{w}^2\expt{\wb}^2
	-\tfrac12\expt{w}\expt{\wb}\pd^a\c^2\pd_a\bar\c^2
	+\tfrac1{16}\c^2\bar\c^2\square\c^2\square\bar\c^2 \\
	&\qquad+\tfrac14(\expt{w}^2\bar\c^2\square\c^2+\cc)\big)\Big) \non\\
	&+\rmi\k(\s^a\bar\h)_\a\pd_a\Big(\c^2\big(
	1-2\k^2\expt\wb+4\k^2\expt\wb^2+\k^4\bar\c^2\square\c^2\big)\Big)\,.
\end{align}

By writing the action \eqref{eqn:BI} in the basis of appendix \ref{sect:bases} 
and comparing to \eqref{eqn:GeneralAV}, 
we find the map that takes $S_\AV$ to $S_\BI$:
\begin{align} \label{eqn:AV2BI}
\!\!\!\l_\a &= \c_\a-\k^2\big((1-\rmi\a_2^\rmi)\c_\a\expt{\wb}  
		+\tfrac\rmi2(\s^a\bar\c)_\a(\pd_a\c^2)\big)
	+\rmi{\k^4}(\s^a\bar\c)_\a(\pd_a\c^2)\expt{\b_7w +\b_8\wb} \non\\\non  					
	&+{\k^4}\c_\a\big(
	2(\rmi\a_2^\rmi+2\b_6+\b_8^*-3)\expt{w\wb}
	- (\tfrac\rmi2\a_2^\rmi+\b_4^*-4\b_6^\rmr+\b_7+\b_8^*-1)
		\expt{w}\expt{\wb} \\\non
	&+ (\tfrac32-2\rmi\a_2^\rmi-2\b_6)\expt{\wb^2} 
	+ \b_4 \expt{\wb}^2
	+ (\rmi\a_2^\rmi+2\b_6+\b_8^*-\tfrac52)\pd^a\c^2\pd_a\bar\c^2 
	+ \b_6\bar\c^2\square\c^2\big) \\\non
	&+ {\k^6}\c_\a\big(
	  \expt{w\wb^2}
	- (2\a_2^\rmi(2\b_6^\rmi-\b_8^\rmi)-8\b_6^\rmr-2\b_7^\rmr-4\b_8^\rmr
		-\rmi\g_2^\rmi-2\g_5^\rmr+16)\expt{w\wb}\expt{\wb}\\\non
	&+ (\rmi\a_2^\rmi(2\b_6+\b_7+\b_8^*-3)+2\b_4^*+4\b_6^*
		+\b_7+4\b_8^\rmr-2\rmi\b_8^\rmi-18)\expt{w}\expt{\wb^2} \\\non
	&- (\tfrac12\a_2^\rmi(3\b_4^\rmi+4\b_6^\rmi+\b_7^\rmi+3\b_8^\rmi)
		-\b_4^\rmr-12\b_6^\rmr-\b_8^\rmr-\rmi\g_4^\rmi+45)
		\!\expt{w}\!\expt{\wb}^2 \\
	&+ \!\g_5\!\expt{\wb}\!\pd^a\c^2\pd_a\bar\c^2 \big)\,,
\end{align}
where we defined $8\b_6^\rmr=10-(\a_2^\rmi)^2$.
Once again, there are twelve free parameters \eqref{eqn:CoeffReIm} 
that correspond to the trivial symmetries of either action.

The pushforward of the AV supersymmetry using the map \eqref{eqn:AV2BI}
matches the supersymmetry \eqref{eqn:BI-SUSY} provided 
$\b_8^\rmr=\frac12=-\b_7^\rmr$ and all other free coefficients are zero.
In \cite{Hatanaka2003}, the theory of nonlinear realisations of supersymmetry
\cite{Ivanov2001}
was used to construct a scheme for finding the map from $S_\AV$ to $S_\BI$.
When explicitly carried out, this should reproduce \eqref{eqn:AV2BI} 
with the above choice of parameters.

\subsection{The chiral-scalar Goldstino action}\label{ssect:CS}
In \cite{Bagger1997} the $\cN=1$ tensor multiplet \cite{Siegel1979t} 
was used to construct a Goldstone action for partial supersymmetry breaking. 
The tensor multiplet is described by a real linear scalar $L$ such that 
$D^2L=\Db^2L=0$. The authors of \cite{Bagger1997} associated with $L$
the spinor superfield $\j_\a=\rmi D_\a L$, which, up to a switch in chirality, 
satisfies constraints and has a free action identical
to the field strength $W_\a$ used in the SBI action given above.
This correspondence allowed them to construct a Goldstone action by following 
the analogy with the SBI action. 
In \cite{Rocek1998} the same action was derived via a 
nilpotency constraint on the $\cN=2$ tensor multiplet.
The analogy with the SBI action is so close that the pure fermionic part of
the actions are exactly the same \cite{McCarthy2005}, 
and thus there is no need to further examine it in this thesis.

However, the tensor multiplet action can be dualised to obtain a Goldstone
action for partial supersymmetry breaking constructed from a chiral superfield.
The action obtained from this procedure is%
\footnote{We have rescaled relative to the conventions of Bagger and Galperin
in order to have an explicit dimensional coupling constant $\k$, 
a canonical fermion kinetic term  
and a canonical leading order Goldstino supersymmetry transformation.}
\begin{align} \label{eqn:BG-sf}
\begin{aligned}
	S[\f,\bar\f] &= \int  {\rm d}^4 x {\rm d}^4\q\, \cL(\f,\bar\f)\,,\\
	\cL(\f,\bar\f) &= 2\bar\f\f
		+\k^2(D^\a\f D_\a\f)(\Db_\da\bar\f\Db^\da\bar\f)f(A,B)\,,
\end{aligned}
\end{align}
where%
\footnote{Note that we use the opposite signature to that of \cite{Bagger1997}.}
\begin{gather}\label{eqn:BG-f,A,B}
	f(A,B)^{-1} = 1 + \frac12A+\sqrt{1+A+B}\,,\;\; 
	A = 16\k^2\left(\pd_m\f\pd^m\bar\f-\frac1{16}D^2\f\Db^2\bar\f\right) ,
	 \non\\
	B = 2^6\k^4\left((\pd_m\f\pd^m\bar\f)^2
		-(\pd_m\f\pd^m\f)(\pd_n\bar\f\pd^n\bar\f)\right)\,.
\end{gather}
In \cite{Gonzalez-Rey1998a} 
it was shown how the $D^2\f\Db^2\bar\f$ term may be removed
by a field redefinition of $\f$.  By a different field redefinition 
\cite{Bagger1997}
it can also be shown that this action matches the leading order 
expression given in \cite{Bagger1994}.
The action \eqref{eqn:BG-sf} is invariant under the nonlinear supersymmetry
transformation
\begin{align} \label{BG-sf-susy}
	\k\d_\h\f = \q\h + \frac{\k^2}4\h^\a\Db^2D_\a\cL\,.
\end{align}

Once again, projection to the fermion action is consistent with both 
the equations of motion \cite{McCarthy2005} and the second supersymmetry.
We use the projection
\begin{align} \label{eqn:BG-proj}
	\f|=0\,,\quad D_\a\f|=\c_\a\,,\quad D^2\f|=0\,,
\end{align}
to obtain the fermionic action
\begin{align} \label{eqn:BG}
	S_\BG[\c,\bar\c] &= -\frac12\intx\Big(\k^{-2} + \expt{w+\wb} \non\\\non
	&-2\k^2\big((\expt{w}^2+\expt{w^2}+\cc)+2\expt{w}\!\expt{\wb}
		+\tfrac12\pd^a\c^2\pd_a\bar\c^2\big) \\\non
	&+ 2\k^4\Big(3(\expt{w^2}+3\expt{w}^2)\expt\wb+6\expt{w}\expt{w\wb}
		+2\expt{w^2\wb}\\\non
	&-2\expt{w}\bar\c^2\square\bar\c^2+\cc\Big) 
	-8\k^6\Big((\expt{w^2}\expt{\wb^2}+\cc) \\\non
	& +\expt{w\wb w\wb}+4\expt{w^2\wb^2}+10\expt w^2\expt\wb^2 \non\\
	&\qquad+14\expt{w}\expt\wb\expt{w\wb}-\expt{w\wb}^2\Big)\Big)\;,
\end{align}
which is invariant under the projection of \eqref{BG-sf-susy} to
its fermionic components.
%
Then, by comparing the above action to \eqref{eqn:GeneralAV}, 
the map that takes $S_\AV$ to $S_\BG$ is found to be
\begin{align} \label{eqn:AV2BG}
	\l_\a  &= \c_\a - \k^2\c_\a\expt{2w + (1-\rmi\a_2^\rmi)\wb} 
		+\frac\rmi2{\k^2}(\s^a\bar\c)_\a(\pd_a\c^2) \non\\\non
	&+\rmi{\k^4}(\s^a\bar\c)_\a(\pd_a\c^2)\expt{\b_7w +\b_8\wb} 				
	+{\k^4}\c_\a\big(
	 2(\left(2\b_6+\b_8^*+4\right)-\rmi\a_2^\rmi)\expt{w\wb} \\\non
	&- (\b_4^*-4\b_6^\rmr+\b_7+\b_8^*+\tfrac\rmi2\a_2^\rmi-14)
			\expt{w}\expt{\wb} \\\non
	&- (\tfrac{1}{2}+2\b_6)\expt{\wb^2} 
	+ \b_4 \expt{\wb}^2
	+ (2\b_6+\b_8^*+\tfrac52)\pd^a\c^2\pd_a\bar\c^2 
	+ \b_6\bar\c^2\square\c^2\big) \\  
	&+ {\k^6}\c_\a\big(
	(4\rmi\a_2^\rmi-4\b_7-8\b_8^\rmr-11)\expt{w\wb^2} 
	+ \g_5\expt{\wb}\pd^a\c^2\pd_a\bar\c^2\\\non
	&- (2\a_2^\rmi(2\b_6^\rmi-\b_8^\rmi)+2\b_4^\rmr+24\b_6^\rmr+2\b_8^\rmr
		-\rmi\g_2^\rmi-2\g_5^\rmr+36)\expt{w\wb}\expt{\wb} \\\non
	&+ (\rmi\a_2^\rmi(\b_8^*+2\b_6+\b_7+4)+\b_7+2\b_8^\rmr+3)
		\expt{w}\expt{\wb^2} \\\non
	&+ (-\frac{1}{2}\a_2^\rmi(3\b_4^\rmi+4\b_6^\rmi+\b_7^\rmi+3\b_8^\rmi)
		-\b_4^\rmr+8\b_6^\rmr-\b_8^\rmr+\rmi\g_4^\rmi-9)\expt{w}\expt{\wb}^2
	 \big) \,,
\end{align}
where we have used $-8\b_6^\rmr=10+(\a_2^\rmi)^2$.

The pushforward of the AV supersymmetry using the map \eqref{eqn:AV2BG}
matches the projection of the supersymmetry \eqref{BG-sf-susy} provided 
$\b_4^\rmr=5$, $\b_7^\rmr=\frac12$, $\b_8^\rmr=-\frac32$, $\g_5^\rmr=3$
and all other free coefficients are zero.

\subsection{Trivial symmetries and field redefinitions}
\label{ssect:Trivialities}
A trivial symmetry of a field theory is a symmetry transformation that reduces
to the identity transformation on-shell, i.e., 
\begin{align} \label{eqn:trivial}
	\vf^i \to \vf'^i = f^i(\vf,\dots) \xrightarrow{\text{on-shell}} \vf^i
	\quad \text{ such that} \quad S[\vf'] = S[\vf]\ .
\end{align}
It is well known (see, e.g., \cite{GitmanTyutin1990,HenneauxTeitelboim1994}) 
that an infinitesimal symmetry transformation,
\begin{align} \label{eqn:infinitesimal_trivial}
	\vf^i \to \vf^i + \d\vf^i\,, 
\qquad S_{,i} [ \vf ] \,\d \vf^i =0\,, 
\end{align}
is trivial if and only if it can be written, 
using  DeWitt's condensed notation,  as
\begin{align} \label{eqn:infinitesimal_trivial2}
	\d\vf_i = S_{, j}[\vf ] \, \L^{ j i} [\vf ]\,, 
	\qquad \L^{ji} = -(-1)^{ij} \L^{ij}\,, 
\end{align}
for some super-antisymmetric matrix  $\L^{j i}$.
More generally, a transformation $\vf^i \to \vf'^i = f^i(\vf,\dots)$ 
is said to be trivial if it reduces to the identity map on the mass shell, 
\begin{align} \label{eqn:trivial_transform}
	\vf^i \to \vf'^i = f^i(\vf,\dots) \xrightarrow{\text{on-shell}} \vf^i\,.
\end{align}

The bulk of this section is dedicated to showing that the symmetries
of the KS action found in  section \ref{ssect:KS} are all trivial.
We note that when two actions are related by a field redefinition, 
then trivial symmetries of one action are mapped onto trivial symmetries
of the other.  
Thus the triviality of the 12-parameter family of symmetries of $S_\KS$,
\eqref{eqn:KS2KS}, implies
the triviality of the same family of symmetries in any Goldstino action,%
\footnote{%
The free Majorana fermion action is not related to Goldstino action by the
field redefinitions \eqref{eqn:GeneralFieldRedef}. 
Apart from the four universal $O(\k^6)$ trivial symmetries,
it has only one other symmetry of the form \eqref{eqn:GeneralFieldRedef}. 
That symmetry is of $O(\k^4)$, trivial and, 
since there are no interaction terms, it has no higher order corrections terms.
} 
including the AV action.

The equations of motion that follow from \eqref{eqn:KS1} are
\begin{align} \label{eqn:EOM_KS}
	\rmi(\s_a\pd^a\bar\j)_\a 
	= \k^2\j_\a \square\bar\j^2 (1-2\k^4\bar\j^2\square\j^2)
	+ \k^6(\pd^a\j_\a)\j^2\pd_a(\bar\j^2\square\bar\j^2)\,,
\end{align}
and its complex conjugate. 
It's useful to contract the above with $\j^\a$ to get
\begin{align} \label{eqn:EOM_KS2}
	\expt{u} &= \k^2\j^2 \square\bar\j^2 (1-2\k^4\bar\j^2\square\j^2) 
	\text{  and c.c.}
\end{align}
We first apply these equations of motion to the general field redefinition
\eqref{eqn:GeneralFieldRedef} (with $\l\to\j$) to see when it is
trivial with respect to KS action. 
We can then specialise to the case of the symmetries of $S_\KS$.

Initially, we only use the contracted equation of motion \eqref{eqn:EOM_KS2}.
It is easy to see that this sends the terms associated with 
$\a_1$, $\b_2$, $\b_4$, $\b_7$, $\b_8$ and all $\g_{i\neq1}$ to zero.
While the $\a_2$ term becomes
\[ \j_\a\expt{\ub} \xrightarrow{\eqref{eqn:EOM_KS2}} 
	\k^2\j_\a\bar\j^2\square\j^2\,, 
\]
and is thus mapped up to the $\b_6$ term. This leaves the field redefinition
\begin{align} \label{eqn:FieldRedefOS}
	\j_\a \to \tilde\j_\a \xlongequal{\!\eqref{eqn:EOM_KS2}\!} \,
	&\j_\a+\a_3\rmi{\k^2}(\s^a\bar\j)_\a(\pd_a\j^2) 
	+ \g_1{\k^6}\j_\a\expt{u\ub^2} \\\non
	&\hspace{-4ex}+{\k^4}\j_\a\big(\b_1\expt{u\ub}	+ \b_3\expt{\ub^2} 
		+ \b_5\pd^a\j^2\pd_a\bar\j^2 + (\a_2+\b_6)\bar\j^2\square\j^2\big) .
\end{align}
Looking at the 
symmetries defined by \eqref{eqn:KS2KS},
we see that combinations with $\b_1=2\b_5$ and $\b_3=-2(\a_2+\b_6)$ often occur.
Some spinor gymnastics shows that these combinations and no others
vanish on-shell
\begin{gather*} 
	\j_\a(2\expt{u\ub}+\pd^a\j^2\pd_a\bar\j^2)
	\xrightarrow{\eqref{eqn:EOM_KS}}	0\,, \\
	\j_\a\left(\expt{\ub^2}-\tfrac12\bar\j^2\square\j^2\right)
	\xrightarrow{\eqref{eqn:EOM_KS}} - \k^2\j_\a\expt{u}\bar\j^2\square\j^2
	\xrightarrow{\eqref{eqn:EOM_KS2}} 0 \,.
\end{gather*}

In summary, the general field redefinition \eqref{eqn:GeneralFieldRedef} 
(with $\l\to\j$) is trivial with respect to $S_\KS$
if the following conditions%
\footnote{The corresponding conditions for triviality of the field redefinition 
with respect to $S_\AV$ are a little more complicated:
\(\a_3=0\,,\; \b_1=2\b_5-\a_1-\a_2\,,\;
	\b_3=-(\a_2+2\b_6)\,,\; \g_1=\a_1+2\a_2-2(\b_5-2\b_6+\b_7+\b_8)\ .
\)
This can be used to directly show that the AV symmetries \eqref{eqn:AV2AV}
are trivial.
}  
on its coefficients hold
\begin{align} \label{eqn:KS_Trivial}
	\a_3=\g_1=0\,,\quad \b_1=2\b_5 \,,\quad \b_3=-2(\a_2+\b_6) \ .
\end{align}
These conditions specify the 24-parameter group 
of trivial transformations with respect to $S_\KS$.
It is now easy to check that all of the symmetries given by
\eqref{eqn:KS2KS} 
(and thus all Goldstino symmetries of the form \eqref{eqn:GeneralFieldRedef}) 
are trivial.

The above result can also be used to prove 
the triviality of any transformation relating 
the Ro\v{c}ek and the KS actions of sections \ref{ssect:Rocek}
and \ref{ssect:KS} respectively. 
First we use the composition and inversion rules 
of appendix \ref{sect:composition},
to compose \eqref{eqn:AV2R} and \eqref{eqn:AV2KS} 
and find the set of maps that take $S_{\mathrm R}$ to $S_\KS$.
These maps can be parameterised as 
\begin{align} \label{eqn:RtoKS}
\!\!
	\j_\a&\to 
	\j_\a + \k^2\overbrace{\rmi \a_2^\rmi}^{\a_2}\j_\a\expt{\ub} 
		+\rmi{\k^2}(\s^a\bar\j)_\a(\pd_a\j^2)\big(\overbrace{0}^{\a_3}+				
		{\k^2}\expt{\b_7 u + \b_8 \ub}\big) \  \non\\\non						
	&+\k^4\j_\a\big(
	      \underbrace{2(2\rmi\a_2^\rmi+2\b_6+\b_8^*)}_{\b_1}\expt{u\ub}
		- (\b_4^*-4\b_6^\rmr+\b_7+\b_8^*)\expt{u}\expt{\ub} \\\non
		&\;\underbrace{- 2(\rmi\a_2^\rmi+\b_6)}_{\b_3}\expt{\ub^2} 
		+ \b_4 \expt{\ub}^2
		+ \underbrace{(2\rmi\a_2^\rmi + 2\b_6 + \b_8^*)}_{\b_5}
			\pd^a\j^2\pd_a\bar\j^2 
		+ \b_6\bar\j^2\square\j^2\big) \\\non
	&+ {\k^6}\j_\a\big(\underbrace{0}_{\g_1}\expt{u\ub^2}
		+ (\rmi\g_2^\rmi - 2\a_2^\rmi(2\b_6^\rmi-\b_8^\rmi)
			+2\g_5^\rmr)\expt{u\ub}\expt{\ub} 
		+ \g_5\expt{\ub}\pd^a\j^2\pd_a\bar\j^2\\
		&\;+ (\rmi\a_2^\rmi(2\b_6+\b_7+\b_8^*-4) 
			+ 2(6\b_6^\rmr-2\rmi\b_6^\rmi+\b_4^*+\b_8^*))\expt{u}\expt{\ub^2} \\
		&\;+ (\rmi\g_4^\rmi 
			- \tfrac12\a_2^\rmi(3\b_4^\rmi+4\b_6^\rmi+\b_7^\rmi+3\b_8^\rmi)
			+2\b_4^\rmr+12\b_6^\rmr+2\b_8^\rmr)\expt{u}\expt{\ub}^2
		 \big) \,, \non
\end{align}
where $-8\b_6^\rmr=(\a_2^\rmi)^2$.
This directly demonstrates the off-shell equivalence of the two models.
It is then easy to check that the conditions of \eqref{eqn:KS_Trivial} are
satisfied (see the bracketing above), 
and so \eqref{eqn:RtoKS} reduces to the identity map when on-shell
with respect to $S_\KS$. The fact that the models have identical dynamics
can also be seen from their Lagrange multiplier construction that was
discussed in \cite{Casalbuoni1989,Komargodski2009} and further clarified in 
\cite{KuzenkoTyler2011}.


\pagebreak
\section{Goldstino dynamics from a constrained complex linear superfield}
\label{sect:mCL}

In the previous section we saw two classes of Goldstino actions:
\tightlists
\begin{enumerate}
\item[1)] Actions that come from nonlinear $\cN=1$ supersymmetry considerations, 
both the nonlinear realisation and constrained superfield approaches.
These form subsections \ref{ssect:AV}, \ref{ssect:Rocek}, \ref{ssect:KS} and
\ref{ssect:SW};
\item[2)] Actions that are the fermionic sector of partially broken
$\cN=2 \to \cN=1$ supersymmetry, subsections \ref{ssect:BI} and \ref{ssect:CS}.
\end{enumerate}
Of the first class, 
there are three main Goldstino actions that are constructed from 
constrained superfields constructions in the literature:
\begin{enumerate}
\item[(i)] Ro\v{c}ek's model \cite{Rocek1978} realised in terms of
	a constrained chiral superfield;
\item[(ii)] the Lindstr\"om-Ro\v{c}ek model	\cite{Lindstrom1979} 
	realised in terms of a constrained real scalar superfield;
\item[(iii)] the Samuel-Wess model \cite{SamuelWess1983} which is formulated using
	a constrained spinor superfield.
\end{enumerate}
The only one of these that we did not mention in the previous section is
the Lindstr\"om-Ro\v{c}ek model, since it has an identical component action 
to the Ro\v{c}ek action.
What is missing in this list of constrained superfield Goldstino models 
is a realisation involving a complex linear superfield.
This section, based on \cite{KuzenkoTyler2011a}, fills this gap.

\subsection{Constrained complex linear superfield}\label{ssect:Constraints}
A complex linear superfield $\G$ obeys  the only constraint ${\bar D}^2 \G =0$,  
and can be used to provide an off-shell description for the scalar multiplet 
({\it non-minimal scalar multiplet}) \cite{BK,GGRS1983}.
A modified 
complex linear superfield, $\S$,
is defined \cite{KuzenkoTyler2011a} to satisfy the constraint
\begin{align} \label{mCL:constraint}
	-\frac14 \Db^2 \S = f\,, \qquad f = {\rm const}\,.
\end{align}
Here $f$ is a parameter of mass dimension 2 which, 
without loss of generality, can be chosen to be real.
The above constraint naturally occurs if one introduces a dual
formulation for the chiral scalar model
\begin{align} \label{mCL:effective-action2}
	S[\F,\bar\F]  = \intx\!\rmd^2\q\rmd^2\bar\q\,\bar\F\F
		+ \Big(  f \intx\rmd^2\q\,\F + {\rm c.c.} \Big)			\,,
\end{align}
with  $\F$ being  chiral. 
The general solution to the  constraint \eqref{mCL:constraint} is
\begin{align} \label{mCL:components}
	\S (\q,\bar\q)= \rme^{\rmi\q\s^a\bar\q\pd_a}
		\left(\f + \q\j + \sqrt2\bar\q\bar\r
		+ \q^2F + \bar\q^2f + \q^\a\bar\q^\da U_\ada + \q^2\bar\q\bar\c
		\right)\ .
\end{align}

The free action for the complex linear superfield is 
\begin{align}
	S[\S,\bar\S] 
		&= - \intx\!\rmd^2\q\rmd^2\bar\q\, \S\bar\S  \\\non 
		&= -\intx\left(f^2 + \bar F F - \f\square\bar\f + \rmi\r\pd\bar\r 
		- \frac12\x {\bm \c} - \frac12\bar\x\bar{\bm \c}
		- \frac12 \bar{\bm U}^a\bm{U}_a
		 \right) \,,
\label{mCL-free-action}
\end{align}
where we have introduced
\begin{align}
	{\bm U}^a = U^a + 2\rmi\pd^a\f \,, \qquad 
	{\bm \c} = \c-\frac\rmi2\pd\bar\j\,.
\end{align}
It is seen from the component expression for  $S[\S,\bar\S]$ that $\f$ and $\r$ 
are physical fields while the rest of the fields are auxiliary.

It turns out that the above action is suitable for describing 
the Goldstino dynamics provided $\S$ is subject to 
the following nonlinear constraints:
\begin{align}
	\S^2 &= 0~, \label{1st constraint} \\
	-\frac{1}{4} \S\Db^2D_\a\S &= f D_\a\S~. \label{2nd constraint}
\end{align}
The constraints 
(\ref{mCL:constraint}, \ref{1st constraint}, \ref{2nd constraint}) 
are easily seen to be compatible.  
Using \eqref{mCL:constraint}, 
the final constraint can be rewritten in the form:
\begin{align}
	\rmi\S\pd_\ada\Db^\da\S = -fD_\a\S\,.
\end{align}
Any low-energy action of the form
\begin{align}
	S_\mathrm{eff} = 
	\intx\!\rmd^2\q\rmd^2\bar\q\, K(\bar\S,\S) 
\end{align}
reduces to \eqref{mCL-free-action} if $\S$ is subject to the nilpotent 
condition  \eqref{1st constraint}.

The general solution to the constraint  \eqref{1st constraint}
fixes $\f$ and two of the auxiliary fields as the following 
functions of the remaining components
\begin{align} \label{soln-to-1st-constraint}
	f \f   = \frac12\bar\r^2\,, \quad
	f\j_\a = \frac1{\sqrt2} U_\ada\bar\r^\da\,, \quad
	f F    = \frac1{\sqrt2}\bar\c\bar\r + \frac14 U^a U_a \ .
\end{align}
Taking into account the second constraint, equation \eqref{2nd constraint}, 
fixes all of the components of $\S$ as functions of the Goldstino $\bar\r$
\begin{gather}\label{solns-to-both-constraints}
\!\!	f \f		= \frac12\bar\r^2\,, \quad
	\sqrt2f^2\j_\a	= -\rmi\bar\r^2(\pd\bar\r)_\a \,, \quad
	f^{3}F 			= \bar\r^2(\pd_a\bar\r\st^{ab}\pd_b\bar\r)\,, \!\\\non
	f U_\ada		= 2\rmi(\s^a\bar\r)_\a\pd_a\bar\r_\db\,,  \quad
	f^2\bar\c_\da 	= \sqrt2\big((\bar\r\st^a\s^b\pd_b\bar\r)\pd_a\bar\r_\db
						- \frac12 (\square\bar\r^2)\bar\r_\db \big)\ .
\end{gather}
The simplicity of these solutions follows from the fact that 
the two supersymmetric constraints depend only on $\S$ and not $\bar\S$, 
i.e., the constraints are holomorphic. 
This is in contradistinction to Ro\v{c}ek's Goldstino action
discussed in section \ref{ssect:Rocek}. 
The constraints for the chiral field in that model 
(\ref{eqn:Roceks_Constraint1}, \ref{eqn:Roceks_Constraint2})
are not holomorphic and lead to the complicated solutions 
\eqref{eqn:Roceks_Results}.

The Goldstino action that follows from
\eqref{mCL-free-action} and \eqref{solns-to-both-constraints} is
\begin{align}\label{mCL:chAV_Action}
	S[\r,\bar\r] 
		&=-\frac12\intx\Big(\k^{-2}+\expt{\w+\wb}
			+\k^2\big(\pd^a\r^2\pd_a\bar\r^2+4\expt{\w}\expt{\wb}\big) \\\non
		&\qquad	+\k^4\big(\expt{\w}\big(2\pd^a\r^2\pd_a\bar\r^2
				+4\expt{\w\wb}+4\expt{\wb}^2-2\expt{\wb^2}
				-\bar\r^2\square\r^2\big) + \cc\!\big)	\\\non
		&\qquad	+\k^6\big(\r^2\bar\r^2\square\r^2\square\bar\r^2
				-8\expt{\w}^2\expt{\wb^2}-8\expt{\w^2}\expt{\wb}^2\big)
		\Big)\,,
\end{align}
where, to ease the comparison with the standard literature on
nonlinearly realised supersymmetry, 
we have used the coupling constant $\k$ defined by $2\k^2=f^{-2}$.
We also used the same notation as in the previous section for
matrix trace $\expt{M}$ of Lorentz indexed matrices $M=(M_a{}^b)$ 
and have defined the matrices
\begin{align}
	\w_a{}^b=\rmi\r\s^b\pd_a\bar\r\,,\qquad
	\wb_a{}^b=\rmi\bar\r\st^b\pd_a\r\ .
\end{align}
The above action is identical to 
the component action described by Samuel and Wess \cite{SamuelWess1983}, 
see equation \eqref{eqn:chAV_Action}.
The proof of why this is so is given in the next subsection.


Naturally associated with $\S$ and $\bar \S$ are 
the spinor superfields ${\bar \X}_\da$ and $\X_\a$ defined by 
\begin{align} \label{SW_as_Derivative}
\X_\a = \frac{1}{\sqrt2}D_\a \bar \S\,, \qquad
	\bar\X_\da = \frac{1}{\sqrt2}\Db_\da\S\ .
\end{align}
Making use of the constraints \eqref{mCL:constraint}, \eqref{1st constraint} 
and \eqref{2nd constraint}, we can readily uncover 
the constraints that the above spinor superfields obey. 
They are
\begin{align}
\label{SW_Constraint1}
	{\bar D}_\da {\bar \X}_\db 
		&= \k^{-1} \eps_{\da \db}\,, \\
\label{SW_Constraint2}
	D_\a {\bar \X}_\da
		&= 2 \rmi \k {\bar \X}^\db \pd_{\a \db} {\bar \X}_\da  \ ,
\end{align}
where, as above,  $2\k^2=f^{-2}$.
These are exactly the constraints given in \cite{SamuelWess1983},
so we recognise $\X_\a$ as the Samuel-Wess superfield.
This connection is discussed in more detail in the following subsections.
It appears that the Goldstino realisation in terms of $\S$ and $\bar \S$ 
is somewhat more fundamental than the one described by equations 
(\ref{SW_Constraint1}) and (\ref{SW_Constraint2}).

\subsection{Comparison to other Goldstino models}\label{ssect:comparisons}
The two most basic Goldstino models start with the nonlinear 
Akulov-Volkov (AV) supersymmetry \eqref{eqn:AV_SUSY}
\begin{align} \label{AV:SUSY}
	\d_\h\l_\a &= \frac1\k\h_\a 
			-\rmi\k \big(\l\s^b\bar\h-\h\s^b\bar\l\big)\pd_b\l_\a \,,
\end{align}
and the chiral nonlinear AV supersymmetry \eqref{eqn:chAV_SUSY}
\begin{align} \label{chAV_SUSY}
	\d_\h\x_\a = 
		\frac1\k\h_\a - 2\rmi\k(\x\s^a\bar\h)\pd_a\x_\a \ .
\end{align}
The Goldstino actions associated with these nonlinear realisations
were discussed in the previous section.

As discussed in \cite{Luo2010c}, the AV supersymmetry is naturally associated
with a real scalar superfield 
(also known as ``vector superfield'' in the early supersymmetry literature), 
while the chiral AV supersymmetry
is associated with a chiral scalar.
Constraints that eliminate all fields but the Goldstino have previously been 
given for both of these types of superfields.
The first was for the chiral scalar, $\F$, where Ro\v{c}ek \cite{Rocek1978}
introduced the constraints 
\begin{align} \tag{\ref{eqn:Roceks_Constraints}}
	\F^2 = 0\,, \quad \F\Db^2\Fb = -4 f \F\ .
\end{align}
The appropriate constraints for the real scalar, 
\begin{align} \label{RL_Constraints}
	V^2=0\,,  \quad \quad VD^\a\Db^2D_\a V=16fV \,,
\end{align}
were given by Lindstr\"om and Ro\v{c}ek \cite{Lindstrom1979}.
The first constraint in both of these sets is a nilpotency constraint, 
while the second is such that the standard kinetic term is equivalent to 
a pure $F$- or $D$-term respectively.
This latter property is not one possessed by the second constraint 
\eqref{2nd constraint} for the complex linear superfield.

The constraints for both the chiral and real scalar superfields were solved in
\cite{SamuelWess1983} in terms of the spinor Goldstino superfield 
\begin{align} \label{chiral-Goldstino-SF}
	\X_\a(x,\q,\bar\q) = \rme^{\d_\q}\x_\a \  .
\end{align}
The actions of the supercovariant derivatives $D_\a$ and $\Db_\da$ on $\X_\a$
follow from the supersymmetry transformation \eqref{chAV_SUSY}
and are exactly the constraints presented in
\eqref{SW_Constraint1} and (\ref{SW_Constraint2}).
The solutions for the constrained superfields that were given in 
\cite{SamuelWess1983} are
\begin{align} \label{SW-solns}
	2f\F = -\frac{\k^2}{4}\Db^2\big(\X^2\bar{\X}^2\big)\,,\qquad
	2fV = {\k^2}\X^2\bar{\X}^2 \ .
\end{align}
From these solutions, it is straightforward to check that $fV=\F\bar\F$.

It is interesting to note that exactly the same solutions work, 
\begin{align} \label{SW-solns2}
	2f\F = -\frac{\k^2}{4}\Db^2\big(\L^2\bar{\L}^2\big)\,,\qquad
	2fV = {\k^2}\L^2\bar{\L}^2 \ ,
\end{align}
if we replace $\X$ with the spinor
Goldstino superfield that follows from the normal AV supersymmetry
(see, e.g., \cite{WessBagger1992})
\begin{align} \label{AV_Superfield}
	{\L}_\a(x,\q,\bar\q) = \rme^{\d_\q}{\l}_\a \ .
\end{align}
Using (\ref{AV:SUSY}), 
the actions of the supercovariant derivatives on this superfield are 
\cite{WessBagger1992}
\begin{align} \label{AV_Constraints}
	D_\a\L_\b = \frac1\k\eps_{\b\a} + \rmi\k \bar\L_\da\pd_\a^\da\L_\b\,,
	\qquad 
	\Db_\da\L_\b = -\rmi\k\L^\a\pd_\ada\L_\b \ .
\end{align}
The projection to the components of \eqref{SW-solns2} immediately reproduces 
the results of \cite{Rocek1978} and gives the relation between 
the constrained superfield Goldstino models and the (chiral) AV Goldstino.

${}$For the complex linear superfield $\S$,
the solution to the constraints \eqref{mCL:constraint}, \eqref{1st constraint} 
and \eqref{2nd constraint} in terms of $\bar \X_\da$ 
is very simple:
\begin{align} \label{mCL-SW-soln}
	2f\S = \bar{\X}^2 \ .
\end{align}
Projection to components yields $\r_\a=\x_\a$ and the component solutions
\eqref{solns-to-both-constraints}. So we see that the model discussed in this
section is the natural constrained superfield to associate with the chiral AV 
Goldstino and the Samuel-Wess superfield \eqref{chiral-Goldstino-SF} 
can be considered derivative \eqref{SW_as_Derivative}.
The Ro\v{c}ek and Lindstr\"om-Ro\v{c}ek superfields can both be constructed
from the complex linear scalar as
\begin{align}
	\F = -\frac12 f\k^2 \Db^2(\bar\S \S) \quad\text{and}\quad
	V  = 2 f \k^2 \bar\S \S \ .
\end{align}

Unlike the chiral and real superfield cases, the solution of the
complex linear constraints in terms of the superfield $\L_\a$ is
different from that using $\X_\a$. Some work gives
\begin{align} \label{mCL-AV-soln}
\begin{aligned}
	2f\S &= 4\big(\bar\L^2+\frac\k2 D^\a(\L_\a\bar\L^2) 
			- \frac{\k^2}{16} D^2(\L^2\bar\L^2)\big)  \\
		 &= \bar\L^2\big(1-\rmi\k^2(\L\s^a\pd_a\bar\L)
		 	+ \k^4\L^2(\pd_a\bar\L\st^{ab}\pd_b\bar\L)\big) \,.
\end{aligned}
\end{align}	

See the recent paper \cite{McArthur2010} for an different and insightful 
investigation of the relationship between the complex linear
and Samuel-Wess constrained superfields, starting from the 
chiral superspace coset construction 
and the general theory of nonlinear realisations.

\subsection{Couplings to matter and supergravity}\label{ssect:couplings}

The constraints \eqref{mCL:constraint} 
and \eqref{2nd constraint} admit nontrivial generalisations such as 
\begin{align}
\label{coupled mCL 1}
 	-\frac14 \Db^2 \S &= X\,, \qquad {\bar D}_\da X =0\,,  \\
\label{coupled mCL 2}
	-\frac14 \S\Db^2D_\a\S &= X D_\a\S\,, 
\end{align}
for some (composite) chiral scalar $X$
possessing a non-vanishing expectation value. 
The solution of these constraints and the resultant action are analysed in the 
next subsection.
Such constraints%
\footnote{Modified linear constraints of the form 
(\ref{coupled mCL 1}) were first introduced in \cite{Deo1985} 
and naturally appear, e.g., when one considers ``massive'' 
off-shell $\cN=2$ sigma-models \cite{Kuzenko2006} 
in projective superspace \cite{LindstromRocek1988,LindstromRocek1990}.
}
are compatible with the nilpotency condition \eqref{1st constraint}.  
This makes it possible to construct couplings of the Goldstino to matter fields. 
For example, 
we can choose $X = f + G_1(\vf) + G_2(\vf) \tr(W^\a W_\a)$, 
where $G_1$ and $G_2$ are arbitrary holomorphic functions of 
some matter chiral superfields $\vf$,
$W_\a$ is the field strength of a vector multiplet 
and the trace is over the gauge indices. 
The resulting Goldstino-matter couplings can be compared 
with those advocated recently by Komargodski and Seiberg \cite{Komargodski2009}. 
In the approach of \cite{Komargodski2009}, the Goldstino is described by 
a chiral superfield $\F$ subject to the nilpotent constraint 
\eqref{eqn:Roceks_Constraint1}.
Matter couplings for the Goldstino in \cite{Komargodski2009}  are generated simply 
by adding suitable interactions to the Lagrangian.\footnote%
{The complex auxiliary field $F$ contained in $\F$ 
is to be eliminated using its resulting equation of motion, 
which renders the supersymmetry {\it on-shell}.}
In our case, the Goldstino superfield $\S$ also obeys 
the nilpotency condition $\S^2=0$, along with the differential constraints 
(\ref{mCL:constraint}, \ref{2nd constraint}).
Matter couplings can be generated by deforming the latter constraints to 
the form given by eqs.\ \eqref{coupled mCL 1} and \eqref{coupled mCL 2}.
Similarly to the analysis in section \ref{ssect:Constraints}, 
the constraints \eqref{1st constraint} and \eqref{coupled mCL 1}
can be solved in terms of the Goldstino ${\bar \rho}_\da$ 
and two more independent fields $U_{\a \da} $ and  ${\bar \chi}_\da$. 
The latter fields become functions of the Goldstino and matter fields 
upon imposing the constraint (\ref{coupled mCL 2}). 
During this process, the supersymmetry remains off-shell! 

We also note that the constraints 
\eqref{coupled mCL 1} and \eqref{coupled mCL 2}
can be further generalised to allow for 
a coupling to an Abelian vector multiplet;
this requires replacing the covariant derivatives in 
\eqref{coupled mCL 1} and \eqref{coupled mCL 2}
by gauge-covariant ones and turning $X$ into a covariantly chiral superfield,
with $X$ and $\S$ having the same U(1) charge.

The constraints \eqref{mCL:constraint} and \eqref{2nd constraint} 
can naturally be generalised to supergravity%
\footnote{Our conventions for $\cN=1$ supergravity correspond to \cite{BK}.
}
as
\begin{align}
\label{mCL:constraint SuGra}
 	-\frac14 (\cDb^2 - 4R) \S &= X\,, \qquad 	\cDb_\da X = 0\,,  \\
\label{2nd constraint SuGra}
	-\frac14 \S  (\cDb^2 - 4R) \cD_\a\S &= X \cD_\a\S\ , 
\end{align}
for some covariantly chiral scalar $X$. 
Here $\cD_A =(\cD_a , \cD_\a , \cDb^\da)$ 
denote the superspace covariant derivative corresponding to 
the old minimal formulation \cite{WessZumino1978,StelleWest1978,FerraraVanN1978} 
for $\cN=1$ supergravity, 
and $R$ the covariantly chiral scalar component of the superspace torsion  
described in terms of $R$, $G_{\a \da}$ and $W_{\a \b \g}$
(see \cite{BK,GGRS1983,WessBagger1992} for reviews). 
The constraints \eqref{mCL:constraint SuGra} and \eqref{2nd constraint SuGra} 
have to be accompanied by the nilpotency condition \eqref{1st constraint}. 
As an example, consider the simplest case when $X$ is constant. 
We represent $X = (\sqrt2 \k)^{-1} = \const$, 
where $\k$ can be chosen to be real.
As a minimal generalisation of (\ref{SW_as_Derivative}), 
we now introduce spinor superfields
$\X_\a = \frac{1}{\sqrt2}\cD_\a \bar \S$ and 
$	\bar\X_\da = \frac{1}{\sqrt2} \bar\cD_\da\S $.
Using the constraints (\ref{1st constraint}), (\ref{mCL:constraint SuGra}) 
and (\ref{2nd constraint SuGra}), 
we can derive closed-form constraints obeyed, e.g., by ${\bar\X}_\da$. 
They are
\begin{align}
\label{SW SuGra 1}
	\cDb_\da {\bar \X}_\db 	&=  \eps_{\da\db} 
		\Big( \frac1\k -\k R \,{\bar \X}^2 \Big) \, , \\
\label{SW SuGra 2}
	\cD_\a {\bar \X}_\da &=  \k \Big( 2\rmi  \,{\bar \X}^\db \cD_{\a \db} 
		{\bar \X}_\da  -G_{\a \da } {\bar \X}^2 \Big)\ ,
\end{align}
where $G_{\a\da}$ is the supergravity extension of the traceless Ricci tensor 
(see \cite{BK,GGRS1983,WessBagger1992} for more details). 
The constraints (\ref{SW SuGra 1}) and (\ref{SW SuGra 2}) were introduced by 
Samuel and Wess \cite{SamuelWess1983} as a result of nontrivial guess work
and are a non-minimal extension of 
\eqref{SW_Constraint1} and \eqref{SW_Constraint2}. 
In our approach, these constraints are trivial consequences 
of the formulation in terms of the complex linear Goldstino superfield.

Since the publication of \cite{KuzenkoTyler2011a}, 
the \emph{modified} complex linear superfield has been renamed 
the \emph{improved} complex linear superfield 
\cite{KuzenkoGTM2011Three, ButterKuzenko2012}.
In \cite{ButterKuzenko2012}, it was
used to construct a novel non-minimal four dimensional supergravity 
that has $\cN=1$ anti de Sitter superspace as its maximally symmetric solution.
This overturned a thirty year old belief that such a construction
was impossible.    

\subsection{Coupling to a matter sector}%
\label{CoupledMCLConstraints}

In the previous subsection, we discussed how the constrained
complex linear superfield realisation of the Goldstino could be coupled
to various other sectors.
It was proposed that coupling to a matter sector could be obtained  
by using the modified constraints (\ref{coupled mCL 1}, \ref{coupled mCL 2}).
In this subsection we solve those constraints and investigate the resultant
component supersymmetry transformations and component action.

\subsubsection{The constraints}
We start with a constrained complex linear superfield $\S$
interacting with some chiral superfield $X$:
\begin{align}
\label{eqn:mCLConstr1}
 	-\frac14 \Db^2 \S &= X\,, & {\bar D}_\da X &= 0\,,  \\
	\S^2 &= 0\,,& -\frac14 \S\Db^2D_\a\S &= X D_\a\S\, .
\label{eqn:mCLConstr2}
\end{align}
The first line contains the standard constraints for interacting complex linear 
and chiral superfields \cite{Deo1985}.
The second line is what makes $\S$ the Goldstino superfield.
The superfield $X$ should have a nonvanishing vacuum expectation value
and is in general of the form
$X = f + G_1(\F) + G_2(\F) \tr(W^\a W_\a)$,
where $\F$ are some chiral matter superfields and $W_\a$ is the field
strength of a vector multiplet.

We use the component projections 
\begin{gather}
	\S| = \f\,, 			D_\a\S| = \j_\a\,, 
	\Db_\da\S| = \sqrt{2}\bar\r_\da\,, 					\\\non
	-\frac14D^2\S| = F\,, 	D_\a\Db_\da\S| = U_\ada\,,
	-\frac14D^2\Db_\da\S| = \bar\c_\da\,,				\\\non
	-\frac14\Db^2\S| = X| = x\,,
	-\frac14D_\a\Db^2\S| = D_\a X = \vf_\a\,, 			\\\non
	\frac1{16}D^2\Db^2\S| = -\frac14 D^2 X| = \dsF\ .
\end{gather}
When $X = f$ (so that $\vf_\a=\dsF=0$), the analysis has to reduce to
that of subsection \ref{ssect:Constraints}.
After some work, the final solutions to the constraints 
(\ref{eqn:mCLConstr1}, \ref{eqn:mCLConstr2}) are
\begin{align}\label{eqn:mCLConstrSolns}
	x \phi &= \frac{1}{2}\bar\r^2\, ,\qquad 
	x \tilde{U}_{\alpha \db}=-2\rmi \left(\pd \bar\r\right)_\a
		\bar\r_\db+\sqrt{2}\vf_\a\bar\r_{\db}\,, \non\\\non
	x^2 \psi_\a &= \frac{1}{2}\bar\r^2
		\left(-\rmi \sqrt{2}\left(\pd \bar\r\right)_\a+\vf_\a\right) \,,\\ 
	x^3 F& = \bar\r^2 \left(\pd^a\bar\r\st_{ab}\pd^b\bar\r
		+\frac{\rmi}{2\sqrt{2}}\vf \pd \bar\r
		+\frac{1}{2}x \dsF\right)\,, \\\non
	x^2\overline{\tilde{\chi }}_{\db} 
		&=-\frac1{\sqrt{2}}\left(\left(\pd_b\bar\r\st^a\s^b\bar\r\right)\pd_a\bar\r_\db
		+\left(\pd_a\bar\r\st^a\s^b\pd_b\bar\r\right)\bar\r_{\db}
		+\frac{1}{2}\left(\square \bar\r^2\right)\bar\r_{\db}\right) \\\non
		&-\rmi \left(\vf \s^a\bar\r\right)\pd_a\bar\r_{\db} 
		+\frac\rmi4\pd_{\db}^\a\left(\bar\r^2\vf_\a\right)
		+\sqrt{2}x \dsF\bar\r_{\db}
		+\sqrt{2} \left(x^{-1}\pd_ax\right)
		\left(\pd^a\bar\r^2\right)\bar\r_{\db}\ .
\end{align}
It is easily checked that these solutions reduce to 
\eqref{solns-to-both-constraints} when $X\to f$.


\subsubsection{Supersymmetry transformations}
The Goldstino $\bar\r_\da$  inherits its supersymmetry transformation from 
the complex linear superfield
\begin{align}
	\sqrt2\d_\x\bar\r_\da 
	&= \Db_\da\d_\x\S|  			\\ \non
	&= 2\bar\x_\da x + x^{-1}\big(2\rmi(\x\s^a\bar\r)\pd_a\bar\r_\da
	+ \rmi\bar\r^2\x^\a\pd_\ada\log(x) + \sqrt2(\x\j)\bar\r_\da\big)\ .
\end{align}
The components of the chiral superfield transform as normal
\begin{align*}
	\d_\x x = \x\j\,, \quad
	\d_\x\j_\a = 2\x_\a \dsF - 2\rmi\bar\x_\db\pd_a^\db x\,,\quad
	\d_\x \dsF = \rmi(\bar\x\pd\j)\ .
\end{align*}
It is a straightforwad, but tedious exercise to check that these transformations
give the supersymmetry algebra.

What's interesting about these supersymmetry transformations 
compared to the canonical approach to nonlinear supersymmetry, 
is that the matter fields transform linearly like there was no Goldstino, 
while the Goldstino transforms into a mix of itself and the matter fields.  

Using these supersymmetry transformations we can construct 
a Samuel-Wess type superfield. Define
\begin{align}
	\bar\X_\da(x,\q,\bar\q) = \rme^{\d_\q}\bar\r_\da(x)\ .
\end{align}
This is a constrained superfield satisfying constraints
that are a generalisation of (\ref{SW_Constraint1}, \ref{SW_Constraint2}):
\begin{align}
	D_\a\bar\X_\db &= 2^{-1/2}X^{-1} \big(2\rmi\bar\X^\da \pd_\ada\bar\X_\db
		+ \rmi\bar\X^2\pd_{\a\db}\log(X) 
		+ 2^{1/2}(D_\a X)\bar\X_\db\big)\,, \non\\
	\Db_\da\bar\X_\db &= 2^{1/2} \eps_\dadb X\ .
\end{align}
The superfield $\bar\X_\da$ is a description of the interacting Goldstino 
that is equivalent to the original constrained complex linear superfield $\S$.
In fact, we can reconstruct the latter 
as
\begin{align*}
	\S = \frac12 X^{-1} \bar\X^2\ .
\end{align*}
It is not hard to 
check that this construction satisfies the 
original constraints (\ref{eqn:mCLConstr1}, \ref{eqn:mCLConstr2}).
It also yields the same component projections \eqref{eqn:mCLConstrSolns}
and the first term, at least, is easy
\begin{align*}
	\f = \S| = \frac12X^{-1}\bar\X^2| = \frac12 x^{-1}\bar\r^2\ .
\end{align*}

\subsubsection{The action}
The superfield action consists of 
a pure matter part and a Goldstino/matter part,
$S[\S,\bar\S,X,\bar{X}] = S[X,\bar{X}] + S[\S,\bar\S]$.
When the nonlinear constraints are enforced, 
the latter 
decomposes into a pure Goldstino part, 
equal to the Samuel-Wess action 
and a part that contains the Goldstino/matter coupling terms.

Projecting out the components of the action
\begin{align*}
	S[\S,\bar\S] = -\intz \bar\S \S\,,
\end{align*}
and using the contraint solutions \eqref{eqn:mCLConstrSolns} yields
\begin{align}
\!	S[\S,\bar\S] 
	&= -\intx \Bigg[|x|^2
	+ \frac{\rmi}{2} \r \oLRa\pd \bar{\r} 
	+\frac{1}{4}|x|^{-2}\big((\pd_a\r^2)(\pd^a\bar\r ^2)
		+4 (\r \pd \bar\r)(\bar\r \pd \r)\big) \non\\\non
	&-\frac{1}{4}|x|^{-4}\Big(\rmi\bar\r^2
		\big((\pd_b\r\s^a\st^b\r)(\pd \bar\r)^\a\pd_a\r_\a
		+ (\pd_a\r\s^a\st^b\pd_b\r)(\r \pd \bar\r) \\\non
	&+\frac12 \r^2 (\r \pd \bar\r)\big)+\cc\Big)
	+|x|^{-6} \r^2\bar\r^2 \left((\pd^a\r \s_{ab}\pd^b\r)
		(\pd^a\bar\r \tilde{\s }_{ab}\pd^b\bar\r)\right) 
									\allowdisplaybreaks\\ 
	&-\left(\frac{1}{\sqrt{2}}\bar{\vf }\bar\r 
		+\frac{1}{2}x^{-1}\bar\r^2\bar{\dsF}+\cc\right)\\\non
	&+\frac{1}{4}|x|^{-2}\Big(2(\r\vf )(\bar\r\bar\vf)
		+\r^2\bar\r^2\pd^a\log (x)\pd_a\log (\bar{x})\\\non
	&+\left(2\sqrt{2}\rmi(\r\vf )(\bar\r \pd \r)
		-\r^2(\pd^a\bar\r^2)\pd_a\log(\bar{x})+\cc\right)\Big) 
						\allowdisplaybreaks\\\non
	&+\frac{1}{16}|x|^{-4}\Bigg(\bar\r^2\Big(4\sqrt{2}
		(\r\s^a\bar{\vf})(\pd \bar\r)^\a\pd_a\r_\a 
	+ \sqrt{2}(\pd \bar\r)^\a\pd_{\a\dot\a}(\r^2\bar\vf^\da)\\\non
	&+8\rmi\bar{x}\bar{\dsF}(\r\pd \bar\r)
		+8\rmi(\bar{x}^{-1}\pd_a\bar{x})(\pd^a\r^2)(\r \pd \bar\r)
		+2\sqrt{2}(\pd_b\r\s^a\st^b\r)(\vf \pd_a\r) \\\non
	&+2\sqrt{2}(\pd_a\r\s^a\st^b\pd_b\r)(\vf\r)
		+\sqrt{2}(\square\r^2)(\vf\r)
		+4\rmi(\r\s^a\bar{\vf})(\vf \pd_a\r) \\\non
	&+\rmi\vf^\a\pd_{\a\dot\a}(\r^2\bar{\vf}^{\dot\a}) 
		-4\sqrt{2}\bar{x} \bar{\dsF}(\vf\r )
		-4\sqrt{2}\pd_a\log(\bar{x})(\pd^a\r^2)(\vf\r)\Big)\ +\cc\!\Bigg) 
					\allowdisplaybreaks\\\non
	&+\frac{1}{4}|x|^{-6} \r^2\bar\r^2 \Big(
		-\frac{1}{2}(\bar\vf\pd\r)(\vf\pd\bar\r) + |x \dsF|^2
		+\big\{\rmi\sqrt{2}(\pd^a\r\s_{ab}\pd^b\r)(\vf \pd \bar\r) \\\non
	&+2(\pd^a\r\s_{ab}\pd^b\r)(x \dsF)
		+\frac{\rmi}{\sqrt{2}}(\bar{\vf }\pd\r)(x\dsF) +\cc\! \big\}
		\Big)\Bigg]\,.
\end{align}
In the limit $X\to f$, only the first three lines survive 
and can be checked to match the Samuel-Wess action 
\eqref{mCL:chAV_Action} (or equivalently \eqref{eqn:chAV_Action}).

This action is quite complicated, but it does represent the most general
coupling of matter to the complex linear Goldstino.
Once again, it is in contradistintion to the other approaches to 
matter-Goldstino coupling. Normally, the Goldstino action is untouched and 
the matter action is modified to introduce the couplings.
In the above action, the matter sector remains untouched 
and the Goldstino action contains the couplings 
as a consequence of the modified contraints.


\chapter{Conclusion}
In this thesis we have examined the low-energy effective actions 
of various supersymmetric theories.
 
In chapter \ref{Ch:Neq1Quant} we examined the background field quantisation of 
a general $\cN=1$ super-Yang-Mills theory. 
We concentrated specifically on backgrounds that allow for calculations of 
low-energy effective actions 
and discussed the limitations and alternatives to the choices made.
In section \ref{sect:Neq1-1Loop} we used the results of the preceding discussion
to examine the one-loop effective action, in particular the K\"ahler potential
and leading terms to the Euler-Heisenberg effective potential.
These results were then used to derive the well-known 
one-loop finiteness conditions for $\cN=1$ super-Yang-Mills theory.
This chapter formed the base for the calculations 
of loop corrections to supersymmetric effective actions
in the subsequent three chapters.

Chapter \ref{Ch:WZ} examined the Wess-Zumino model. 
We rederived known results about the K\"ahler potential up to two-loops
and the leading corrections to the auxiliary potential at one-loop.
The two original contributions in this chapter were 
the completion of the programme given in \cite{Buchbinder1994a,Buchbinder1993} 
with the presentation of the full one-loop auxiliary potential
and the examination of the conditional convergence in its leading term.
To the best of our knowledge, 
these results have not appeared in the literature before this thesis.

Based on the paper \cite{Kuzenko2007a}, 
chapter \ref{Ch:SQED} examined the K\"ahler and Euler-Heisenberg
sectors of the $\cN=1$ SQED effective action.
We extended the K\"ahler potential analysis of \cite{Kuzenko2007a}
to an arbitrary 2-parameter $R_\xi$ gauge. 
This exposed the gauge dependence of the effective action, 
emphasising that it is not a physical quantity 
and showed the strengths and weaknesses of the various common gauge choices,
including their infrared behaviours.
This chapter also included a calculation of the two-point function using 
propertime and functional techniques.

In chapter \ref{Ch:BetaDef}, the two-loop correction to the K\"ahler potential
of $\b$-deformed super-Yang-Mills was examined. This work
was based on \cite{Tyler2008} and was a continuation of the work 
in \cite{Kuzenko2005b,Kuzenko2007}. Since the K\"ahler potential receives
no corrections in $\cN=4$ SYM and only at one-loop in $\cN=2$ SYM, its
existence at two loops in the $\b$-deformed theory is purely 
a product of the deformation and is thus of interest. 
This chapter showed the utility of the background field approach to calculations
and made use of the new, clean form of the two-loop Feynman integral described
in appendix \ref{A:2LoopVac}.

In chapter \ref{ch:Goldstino}, we switched from studying the quantum corrections
to low-energy effective actions and turned to the universal sector of all 
models with broken (global) supersymmetry, the Goldstino action.
After reviewing the general properties of supersymmetry breaking and 
its realisation in supersymmetric sigma models,
the original results of 
\cite{KuzenkoTyler2010,KuzenkoTyler2011} were presented.
First, using the most general field redefinition that preserves 
the structure of Goldstino actions,
the most general (pure) Goldstino action was constructed and its 
nonlinear supersymmetry transformation was examined. 
This calculation was aided by \emph{Mathematica} code that the author
wrote and published in \cite{KuzenkoTyler2011}.
The general Goldstino action was then compared to 
all of the other Goldstino actions found in the literature
allowing for the construction of explicit field redefinitions that map
between the various Goldstino realisations. 
It was observed that all of these maps have 12 free parameters that are due
to trivial symmetries inherent in any Goldstino action. 
Finally, we presented a new embedding of the Goldstino into 
an improved complex linear superfield \cite{KuzenkoTyler2011a}.
Its relationship to the other Goldstino superfield embeddings was examined
and its possible interactions with matter and gravity were elucidated.
In the final section \ref{CoupledMCLConstraints}, the unpublished
results for the component reduction and action of the new Goldstino 
coupled to arbitrary matter sector were presented.

\section*{Outlook} 
Since this thesis covered a wide range of topics and calculations, there
are many possible directions for further work.


In chapter \ref{Ch:WZ} the conditional convergence issue has not been
satisfactorily resolved. Although the final one-loop effective potential result
agrees with the component calculations, both were performed using dimensional 
regularisation, which is known to be problematic in supersymmetric theories.
The lack of a simple and consistent regularisation for superfield calculations
is an ongoing issue. Since the Wess-Zumino model is the simplest 
four dimensional supersymmetric field theory, this conditional convergence 
problem could be a good place to examine the regularisation issues more closely.

The conditional convergence of terms that should be finite by power counting
and the corresponding ambiguities 
that can't be fixed by renormalisation conditions
is not restricted to the leading term in the Wess-Zumino 
one-loop auxiliary potential.
A similar problem occurs in the two-loop $F^4$ terms in
superconformal $\cN=2$ SYM theories.%
\footnote{This was observed in some unpublished work of the author
that was a continuation of the investigations started in 
\cite{Kuzenko2004b, Kuzenko2004c}.}
The careful calculation of the coefficients of these terms 
is important in testing non-renormalisation theorems such as the Dine-Seiberg
conjecture \cite{Dine1997} and its proposed refinement \cite{Kuzenko2004b}.

A natural extension of the work in chapter \ref{Ch:SQED} is to extend
the calculation to three-loops. There is currently some work at performing
the three-loop calculation in non-supersymmetric QED \cite{Huet2011,Huet2009}
using worldline formalism. Supersymmetric theories have the advantage
of fewer diagrams to evaluate, and $\cN=2$ SQED is an especially attractive
target for a three-loop calculation due to the comparatively simple structure 
of its two-loop Euler-Heisenberg effective action.
A three-loop, four-dimensional Euler-Heisenberg calculation will 
be of use in testing various all-order conjectures about the structure
of QED effective actions \cite{Huet2011,Schubert2011}. For example,
the Lebedev-Ritus exponentiation conjecture and related mass renormalisation
\cite{Lebedev1984,Ritus1998},
the AAM conjecture \cite{Affleck1982Pair}, 
and the functional structure for effective actions in a self-dual background 
\cite{Dunne2001}. 

The analysis of Goldstino actions in section \ref{sect:Goldstino}
left open some questions about the Komargodski-Seiberg action. 
In particular, if its off-shell nonlinear supersymmetry transformation has 
a simple form that can be related to some 
nice property of the supersymmetric action.
Another interesting avenue for exploration, 
which might also provide some insight to the above problem, 
would be to examine how the field redefinitions
described in section \ref{sect:Goldstino} translate to the superembedding
approach of \cite{pasti2000superembeddings} 
(and reviewed in the broader context of \cite{sorokin2000superbranes}).
Finally, the new embedding of the Goldstino in a complex linear superfield
may have some advantages over other Goldstino embeddings in the coupling
to supersymmetric matter and supergravity. 
It would be worthwhile examining 
its behaviour in some phenomonolgically interesting models 
as well as 
the superhiggs mechanism when the theory is coupled to supergravity.

Deriving the two-loop results used in chapters 
\ref{Ch:WZ} and \ref{Ch:BetaDef}
lead to some new insights into the structure of the differential equations
satisfied by Feynman diagrams as outlined in appendix \ref{A:2LoopVac}.
These insights could lead to improved methods of performing the reduction to 
and the evaluation of master integrals for multiloop Feynman diagrams,
as well as some new insight to their analytic structure.
Other questions of interest are: 
which Lie algebras (and representations) are associated to which diagrams; 
and the relationship between the contraction/deletion structure 
that generates the inhomogeneous terms in the differential equations
and the contraction/deletion structure inherent in 
the forest formula (and Hopf algebra) of the renormalisation procedure.
The standard methods of generating difference equations for reducing
Feynman integrals to a minimal set of master integrals 
produces an overcomplete set of relations that have to be reduced 
using techniques such as the construction of Gr\"obner bases \cite{Smirnov2007}. 
The method described in appendix \ref{A:2LoopVac} emphasises 
the Lie algebra structure of the differential equations from the start, 
so does not produce extraneous relations.

\appendix  
\chapter[Background field propagators]%
		{The derivation of some background field propagators}
		\label{A:Propagators}

In this appendix we derive some explicit results for the propagators 
in the presence of specific background fields.
These propagators will be used throughout the first half of this thesis.
We start in section \ref{sect:RegByDimlRed} 
with a quick description of regularisation by dimensional reduction,
as it will be our primary regularisation method. 
Section \ref{sect:PropsCovConst} then describes the derivation of
the heat kernel $K(z,z'|s)$ for the operator $\BoxV$ \eqref{defn:BoxV}.
For on-shell backgrounds, 
the operators $\BoxV$ and $\Box_\pm$ are related, so we find the chiral 
heat kernels $K_\pm(z,z'|s)$ by taking derivatives of $K(z,z'|s)$.
We also present a simple derivation of $K_{\pm\mp}(z,z'|s)$.
We round off the section by looking at the self-dual limits and the
expressions suitable for calculating $F^2$ and $F^4$ type terms.
Many parts of the discussion in this section closely parallel that of 
\cite{Kuzenko2003a}.
In section \ref{sect:WZProp} we derive the heat kernel for the Wess-Zumino
model in a background that is constant in spacetime $\pd_a\f=0$
but allowed to vary in the Grassmann directions $D_\a\f\neq0$.
This derivation, which is central to chapter \ref{Ch:WZ}, 
first appeared with some typographical errors in 
\cite{Buchbinder1994a}.

\section{Regularisation by dimensional reduction}
\label{sect:RegByDimlRed}
To regularise by dimensional reduction 
\cite{Siegel1979,Siegel1980,Capper1980}, 
we stay in the $(4|4)$-dimensional 
superspace, spanned by $z=(x,\q,\qb)$, 
but we let the superfields only depend on $(d|4)$-dimensions, where $d<4$.  
That is, we divide the bosonic coordinates into 
$x_a=(x_\au,x^{\abar})$ or $x=(\un{x},\Bar{x})$ where
$\au=0,\ldots,d-1$ and $\abar=d-1,\ldots,3$.  
Then all superfields satisfy
\begin{align}
	\J(z)=\J(\un x,\q,\q)
	\quad\implies\quad
	\pd_\abar\J=0~, 
\end{align}
and thus have normal expansions in the Grassmann coordinates, 
with ordinary, but dimensionally reduced, component fields.
In particular, the vector component
$V_a(x) \sim (D\s_a\Db)\J(z)|$ 
has a four dimensional index and thus has extra $\eps$-scalars 
compared to the normal dimensionally regularised vector fields 
which are truly $d$-dimensional.
This means that the supersymmetry algebra stays in four dimensions,
but the spacetime derivatives have split
\begin{align}
	\{D_\a,\Db_\da\}&=-2\rmi\s^a_\ada\pd_a,,&
	\pd_a&=\d_a^\au \pd_\au + \d_a^\abar \pd_\abar \ . &
\end{align}

For gauge theories (discussed in chapter \ref{Ch:Neq1Quant}),
requiring that $\cD_A\J$ is also constant over the final $2\eps$ dimensions 
implies that the gauge prepotential, potentials and field strength must 
also be constant over the final $2\eps$ dimensions, 
but the field strengths have the same number of components as the normal
4-dimensional theory.

As we don't want to integrate over the $4-d=2\eps$ 
dimensions where the fields are constant, 
we define our integration measure as 
\begin{equation}
	\int\!\rmd z = \m^{-2\eps}\int\!\rmd^dx\rmd^4\q \,,\quad 
	\int\!\rmd^4\q\,(\cdots)=\frac1{16}\pd^2\bar\pd^2(\cdots)\big|_{\q=\qb=0}\,,
\end{equation}
where the renormalisation mass $\m$ must be introduced for dimensional reasons.
We also define the Dirac delta functions on superspace:
\begin{align}\label{defn:DRedDelta}
	\un\d(z,z')=\m^{2\eps}\d^d(x-x')\d^4(\q-\q')I(z,z')
\end{align}
where $I(z,z')$ is the supersymmetric parallel displacement propagator, 
described below. 
As in the main text, we define the functional inner product
\begin{equation}
  \J_1\cdot \J_2 = \int\!\rmd z \J_1(z)\J_2(z)\,,\quad \text{s.t.}
  \quad \J\cdot\un\d = \un\d\cdot \J = \J \ . 
\end{equation}
The same considerations carry over to momentum space, 
starting with the identity
\begin{align*}
	\m^{2\eps}\d^d(x-x') = \m^{2\eps}\intk\rme^{\rmi k (x-x')}\ .
\end{align*}

Of course, regularisation by dimensional reduction is known to be inconsistent,
see, e.g., 
\cite{Siegel1980, Avdeev1981, Avdeev1983a, Avdeev1983, 
	Delbourgo1981, Bonneau1980, Jack1997, Stockinger2005}. 
This inconsistency follows from the mixing of $4$- and $d$-dimensional objects,
and thus, in component field calculations, can be avoided by 
not using $4$-dimensional identities such as the Fierz identity. 
However, this discards much of the calculational advantage 
of keeping the $4$-dimensional objects and, since the Fierz identity
is needed in proving the invariance of a supersymmetric Lagrangian,
it means that scattering amplitudes calculated from such a regularisation 
do not necessarily respect supersymmetry.
Since superfields and $D$-algebra are inherently $4$-dimensional,
the consistent dimensional reduction schemes mentioned in the references
above can not be used.
Of course, the ambiguities 
can be fixed using appropriate
finite counterterms found be enforcing Ward-Takahashi-Slavnov-Taylor identities. 
See the textbook \cite{GGRS1983} for more discussion.

The ambiguity can only arise in diagrams containing enough interactions 
to generate the right algebraic objects. 
In \cite{Avdeev1983a} a table is given showing the number of loops where 
ambiguities could arise or supersymmetry could be broken 
in the naive and consistent dimensional reduction schemes 
for 2, 3 and 4 point functions in the Wess-Zumino model.
However, when calculating background field dependent propagators
an infinite number of interactions must be summed over, thus the problems
could arise even at the level of the propagators and one-loop effective action.
This is related to the difficulties around equation \eqref{IBP_Kernel1} 
and why we choose to use regularisation 
by a propertime cut-off in chapter \ref{Ch:SQED}.


\section{Parallel displacement propagator}
The parallel displacement propagator is vital 
in performing fully covariant background field calculations.
In this section, we content ourselves with simply stating its main properties.
For more detailed treatments, see e.g., 
\cite{Barvinsky1985, Kuzenko2003a, Kuzenko2005Heidelberg}
and references within.
It is an operator that 
parallel transports a superfield along a straight line in superspace
$z^M(t)=(z-z')^M t+z'^M$.  
To define it, we need the supersymmetric intervals (Cartan 1-forms)
$\z^A=(\r^a,\z^\a,\z_\da)$ 
which follow from 
$ (z^A-z'^A)\pd_A = \z^A D_A$, 	
which implies that
\begin{equation}\label{defn:SusyIntervals}
	\r^a=(x-x')^a-\rmi\q\s^a\qb'+\rmi\q'\s^a\qb \,,\quad
	\z^\a=(\q-\q')^\a~,\quad \zb_\da=(\qb-\qb')_\da \ .
\end{equation}
Then we can write the defining properties of the 
parallel displacement propagator as
\begin{subequations}\label{defn:ParallelDispOp}
\begin{align}
	I(z,z')&\to\rme^{\rmi \cK(z)}I(z,z')\rme^{-\rmi \cK(z')}\,,\\
	\z^A\cD_AI(z,z')&=0 \ ,\\
	I(z,z)&=\ds1 \,,
\end{align}
\end{subequations}
where the first line describes its properties under the gauge transformation
\eqref{eqn:GaugeXformPhi_Quant}
and the second uses a gauge covariant derivative defined in 
\eqref{defn:Neq1CovD}.
Equations \eqref{defn:ParallelDispOp} imply the important result
\begin{align}\label{eqn:IAnnihilate} 
	I(z,z')I(z',z) = \ds1 \ .
\end{align}

From a naive point of view, the parallel displacement propagator
is needed so that the right hand side of \eqref{defn:BoxVGreen}
has the correct gauge transformation properties.\footnote%
{
The parallel displacement propagator in the $\d$-function \eqref{defn:DRedDelta}
can be relaxed to any function that satisfies the first and last conditions
of \eqref{defn:ParallelDispOp}. 
This said, the specific choice of the parallel displacement propagator is 
not only geometrically natural, but its nice properties 
help to arrange the covariant derivatives into field strengths
in the heat kernels and Schwinger-DeWitt expansions below.
It is also essential for higher-loop calculations where it is needed for
the covariant Taylor series used in moving all background dependence to a single
spacetime point.
See \cite{Kuzenko2003a} for a more detailed discussion.
}
The equations in \eqref{defn:ParallelDispOp} 
have a solution in terms of a path ordered exponential
\[ 
	I(z,z')=\mathrm{Pexp}\left(-\rmi\int_{z'}^z\G(z'')\cdot\rmd z''\right)\,,
\]
showing that the parallel displacement propagator is also 
a dimensionally reduced quantity. 
This also could have been argued from the requirement that $\J\cdot\un\d=\J$.

Further properties of the supersymmetric parallel displacement propagator
and its use in covariant Taylor series are given in \cite{Kuzenko2003a}.
The two results needed in the subsections below are 
the general covariant Taylor series result 
and its specific application to the the covariant derivative of $I(z,z')$.
The covariant Taylor series for a superfield,  $\J$,  
in some representation of the gauge group,
$\J(z) \to \rme^{\rmi\cK(z)} \J(z)$,
is
\begin{align}\label{eqn:covTaylor}
	\J(z) = I(z,z')\sum_{n=0}^\infty \frac{1}{n!}\z^{A_n}\dots\z^{A_1}
		\cD'_{A_1}\dots\cD'_{A_n} \J(z') \ .
\end{align}
Applying the above series to the superfield $\cD_\b I(z,z')$ yields
\begin{align}\label{eqn:covTaylorDaI}
	\cD_\b I(z,z') &= \Big(
	\frac{1}{12}\zb^\db\z^\ada F_{\ada\bdb}(z)
	- \frac\rmi2\z_\bdb\big\{ \Wb^\db(z)+\frac13\zb^\da\cDb_\da\Wb^\db(z)\big\}
	\non\\
	&+ \frac13\zb^2\big\{W_\b(z)-\frac12\z^\a\cD_\a W_\b(z) 
		+ \frac14\z_\b\cD^\a W_\a(z)\big\} \non\\
	&+ \frac13\z_\b\zb_\db\Wb^\db(z)
	\Big)I(z,z')\ .
\end{align}


\section{Free bosonic heat kernels}\label{sect:BosonicProp}
For the sake of later comparison, 
it is worth quickly reviewing the derivation
of the heat kernel for a free massive Klein-Gordon field.
Since this theory only contains scalar fields, 
dimensional reduction and dimensional regularisation 
are equivalent provided $d=4-2\eps<4$.

The action for a free real scalar field is 
\[ S[\f] = \frac12\intdx \f (\Box-m^2) \f  \ .\]
The corresponding Feynman propagator $G(x,x')=\rmi\expt{\f(x)\f(x')}$ 
is defined by $(\Box-m^2)G(x,x')=-\d^d(x,x')$ and the 
causal (Feynman) boundary conditions. Formally, we can write
\begin{align} \label{FreeKGProp}
\begin{aligned}
	G(x,x') &= \frac{-1}{\Box-m^2+\rmi\e} \d^d(x,x') 
	= \rmi\int_0^\infty\!\!\rmd s\,\rme^{\rmi s (\Box-m^2+\rmi\e)}\d^d(x,x') \\
	&= \rmi\int_0^\infty\!\!\rmd s\,\rme^{-\rmi s (m^2-\rmi\e)}K(x,x'|s)  \ .
\end{aligned}
\end{align}
The heat kernel $K(x,x'|s)$ can then be evaluated by going to momentum space,
completing the square and performing the Gaussian integral
\begin{align} \label{FreeKGHeatKern}
	K(x,x'|s) &= \rme^{\rmi\Box s}\d^d(x,x')
	=  \m^{2\eps}\intk \rme^{-\rmi k^2 s + \rmi k(x-x')} \non \\
	&= \frac{\rmi\m^{2\eps}}{(4\p\rmi s)^{d/2}}
		\rme^{\rmi\frac{(\un x - \un x')^2}{4s}}\ .
\end{align}


\section[Covariantly constant background]%
		{Propagators in a covariantly constant background gauge field}
\label{sect:PropsCovConst}


For the Laplacian \eqref{defn:BoxV}
\( \BoxV=\cD^a\cD_a-\cW^\a\cD_\a+\cWb_\da\cDb^\da \)
and some constant mass-squared matrix, $m^2$, we introduce the auxiliary action 
\begin{equation} 
	S[\J]=\int\rmd z\J^\dag(\BoxV-m^2)\J \,,
\end{equation}
where the superfield $\J$ is in some representation of the gauge group,
\begin{align} \label{eqn:AuxFieldXform_A1}
	\J(z) \to \rme^{\rmi \cK(z)}\J(z) \ .
\end{align}
The corresponding Green's function 
$G(z,z')=\rmi\expt{\J(z)\J^\dag(z')}$ 
is defined by
\begin{equation}\label{defn:BoxVGreen}
	(\BoxV-m^2)G(z,z')=-\un\d(z,z') \ .
\end{equation}
and transforms under the background gauge transformations
\eqref{KLgaugeXform2a} as
\begin{align}\label{GaugeXformG} 
	G(z,z') \to \rme^{\rmi \cK(z)}G(z,z')\rme^{-\rmi \cK(z')} \ .
\end{align}

Assuming a constant mass matrix, we use the ansatz
\begin{equation}
	G(z,z')=\rmi\int_0^\infty\!\!\rmd s\,\rme^{-\rmi m^2s} K(z,z'|s) \,,
\end{equation}
where the heat kernel $K(z,z'|s)$ satisfies
\begin{equation}
	(\rmi\frac\rmd{\rmd s}+\BoxV)K(z,z'|s)=0
	\quad\text{and}\quad
	K(z,z'|0)=\un\d(z,z') \ ,
\end{equation}
and has the same gauge transformation as $G(z,z')$. 
This means that the heat equation has the formal solution
\begin{equation}\label{KeqExpBox}
	K(z,z'|s)=\rme^{\rmi(\BoxV+\rmi\e)s}\un\d(z,z')\,,
	\quad \e\to0_+ \ .
\end{equation}

Assuming that the gauge field is covariantly constant
\begin{equation}
	\cD_a\cW_\b=0  \quad \implies \quad 
	\left[\cD_a, \cW^\b\cD_\b-\cWb_\db\cDb^\db\right]=0\,,
\end{equation}
allows the exponential in \eqref{KeqExpBox} to be factorised as 
\cite{Kuzenko2003a}
\begin{gather}\label{KeqExpFact}
	K(z,z'|s)=U(s)\rme^{\rmi(\cD^a\cD_a+\rmi\e)s}\un\d(z,z')
	= U(s)\tilde K(z,z'|s)\,, \\ 
\intertext{where} 
	U(s) = \rme^{-\rmi s(\cW\cdot\cD-\cWb\cdot\cDb)} \ . 
\end{gather}
The heat kernel, $\tilde K(z,z'|s)$, obeys the equations
\begin{align}
	\label{KtildeEqn}
	\left(\rmi\frac\rmd{\rmd s}+\cD^a\cD_a\right)\tilde K(z,z'|s)=0 
	\quad \text{and} \quad
	\tilde K(z,z'|0)&=\un\d(z,z')\ .
\end{align}
To find its explicit representation, we first
Fourier transform the bosonic part of the delta function
\begin{equation}
	\un\d(z,z')=\m^{2\eps}\intk\rme^{\rmi k^\au\r_\au}\z^2\zb^2I(z,z')\,,
\end{equation}
where $\r_\au = \d^a_\au \r_a$, to get
\begin{align}
	\label{KtildeDef}
	\tilde K(z,z'|s)&=\m^{2\eps}\intk\rme^{\rmi k^\au\r_\au}
		\rme^{\rmi s(\cD+\rmi k)^2}\z^2\zb^2 I(z,z')  \ .
\end{align}
Then, the exponential operator above can be shown to collapse to a function
of field strengths by integrating by parts the identity
\begin{align*} 
	0 =  \m^{2\eps}\intk\frac\pd{\pd k^{\un a}}
		\Big(\rme^{\rmi k^\au\r_\au}
		\rme^{\rmi s(\cD+\rmi k)^2}\z^2\zb^2 I(z,z')\Big) \,,
\end{align*}
and using
\( 
	[(\cD+\rmi k)^2,\diamond]^n(\cD_a+\rmi k_a)
	=(-2\rmi)^n(\cF^n)_a^{~b}(\cD_b+\rmi k_b) 
\)
to get 
\begin{align} \label{IBP_Kernel1}
	0 = \d^a_\au \Big(\rmi\r_a\tilde K(z,z'|s)	
	-2s\Big(\frac{1-\rme^{-2s\cF}}{2s\cF}\Big)_a{}^b\,\cD_b
	\tilde K(z,z'|s)\Big) .
\end{align}
The term inside the brackets only vanishes up to some $\e$-scalar term
$X_a = \d_a^\abar X_\abar$
that obeys $\d _\au^a X_a=0$. 
We will avoid the complication that this extra freedom introduces
by performing the rest of this calculation in four dimensions, 
where \eqref{IBP_Kernel1} implies 
\begin{equation}\label{IBP_Kernel}
	\cD_a\tilde K(z,z'|s)
	=\rmi\big(\frac{\cF}{1-\rme^{-2s\cF}}\big)_{a}^{~b}\r_b
		\tilde K(z,z'|s)
\end{equation}
Taking the derivative of \eqref{IBP_Kernel}
and using the heat equation \eqref{KtildeEqn} gives
\begin{align*}
	\frac\rmd{\rmd s}\tilde K(z,z'|s)
	&=\frac\rmd{\rmd s}\left(\frac{-1}{2}\tr\log\big(1-\rme^{-2s\cF}\big)
		+\frac\rmi4 \r\cF\coth s\cF\r\right)\tilde K(z,z'|s)\ .
\end{align*}
This is then integrated to give
\begin{equation}\label{KtildeC}
	\tilde K(z,z'|s)=
		\frac{-\rmi}{(4\p)^2}
		\det\!\Big(\frac{2\cF}{1-\rme^{-2s\cF}}\Big)^{\half}
		\,\rme^{\frac\rmi4\r\cF\coth s\cF\r}\z^2\zb^2 C(z,z') \,,
\end{equation}
where the $\det$ is over the 
4-dimensional Lorentz matrix 
and the integration constant
\begin{align*} 
	\frac{-\rmi}{(4\p)^2}
		\det(2\cF)^{\half}\z^2\zb^2 C(z,z')\,,
		\text{ where } C(z,z') \xrightarrow{\cF\to0} \ds1\,,
\end{align*}
has been chosen so that we recover the free heat kernel 
when the background gauge field is switched off. 
There is still some freedom in the choice of $C(z,z')$, 
which is $s$-independent and 
must have the same gauge transform as $\tilde K$ and $G$.
This freedom is fixed by comparing the the boundary condition \eqref{KtildeEqn}
$s\to0$ limit of \eqref{KtildeC}
\begin{align*} 
	\tilde K(z,z'|s\to0) 
		&\approx \frac{-\rmi}{(4\p s)^2}
			\rme^{\frac\rmi{4s}\r^2}\z^2\zb^2 C(z,z')
		\to \d^4(x-x')\z^2\zb^2 C(z,z') \,,
\end{align*}
which requires that $C(z,z')=I(z,z')$.
This result can also be obtained by comparing to the Schwinger-DeWitt expansion 
(see e.g., \cite{Kuzenko2003a})
\begin{equation*}
	\tilde K(z,z'|s)
	=
	 \frac{-\rmi}{(4\p s)^{2}}
		\rme^{\frac\rmi{4s}\r^2} \sum_{n=0}^\infty a_n(z,z')(\rmi\,s)^n \,,
	\quad a_0(z,z')=\z^2\zb^2I(z,z') \ .
\end{equation*}

So, the full heat kernel \eqref{KeqExpBox} becomes
\begin{equation}\label{full_heat_kernel4D} 
	K(z,z'|s) = \frac{-\rmi}{(4\p s)^{2}}
		\det\!\Big(\frac{s\cF}{\sinh s\cF}\Big)^{\half}\,U(s)\,
		\rme^{\frac\rmi4\r\cF\coth s\cF\r}\z^2\zb^2 I(z,z')\,,
\end{equation}
which, by comparison with \eqref{FreeKGHeatKern}, we can modify in order to 
approximate regularisation by dimensional reduction
by making the replacement 
\begin{align}
	\frac{-\rmi}{(4\p s)^{2}} \to \frac{\rmi\m^{2\e}}{(4\p\rmi s)^{d/2}}\ .
\end{align}
Finally, the action of $U(S)$ in \eqref{full_heat_kernel4D} can be evaluated
by introducing the $U$-shifted fields $\J(s)=U(s)\J U(-s)$
and calculating
\begin{align}\label{shifted-objects}
	\cW^\a(s)&=(\cW \rme^{-\rmi s\cN})^\a\,,
		&\cWb_\da(s)&=(\cWb \rme^{\rmi s\bar\cN})_\da~\,,		\non \\ \non
	\z^\a(s)&=\z^\a+(\cW\frac{\rme^{-\rmi s\cN}-1}\cN)^\a\,,
		&\zb_\da(s)&=\zb_\da+(\cWb\frac{\rme^{\rmi s\cNb}-1}{\cNb})_\da\,, \\
	\r_\daa(s)&=\r_\daa-2\int_0^s\!\rmd t
		(\cW_\a(t)\zb_\da(t)+\z_\a(t)\cWb_\da(t))\,,\hspace{-10em}	\\\non
	I(z,z'|s)&=\exp\left\{\int_0^s\rmd t~\Xi(\z,\cW,\cWb | t) \right\} 
		I(z,z') \,, \hspace{-10em}
\end{align}
where we use the abbreviations for derivatives of the field strength
\begin{align} \label{defn:cN} 
\begin{gathered}
	\cN_\a^{~\b}=\cD_\a\cW^\b\,, \quad \cNb^\da_{~\db}=\cDb^\da\cWb_\db\,,\\
	\cF_{\ada,\bdb}=\rmi\eps_\ab\cNb_{(\dadb)}+\rmi\eps_\dadb\cNb_{(\ab)}\,,
\end{gathered}
\end{align}
which are invariant under $U$-shifts: $[U(s),\cN]=[U(s),\cF]=0$,
and we have introduced the symbol
\begin{align} 
\begin{split}
\Xi(\z,\cW,\cWb)=\frac1{12}\r^\daa(\cW^\b\zb^\db-\z^\b\cWb^\db)
	(\eps_\ba\cDb_\db\cWb_\da-\eps_\dbda\cD_\b\cW_\a)\\
	-\frac{2\rmi}{3}\z\cW\zb\cWb-
	\frac\rmi3\big(\z^2(\cWb^2-\quart\zb\cDb\cWb^2)	+\cc\!\big) \ .
\end{split}
\end{align}
The heat kernel then take the form
\begin{equation}\label{full_heat_kernel} 
	K(z,z'|s)
	=\frac{\rmi\m^{2\e}}{(4\p\rmi s)^{d/2}} 
		\det\!\Big(\frac{s\cF}{\sinh s\cF}\Big)^{\half}\,
		\rme^{\frac\rmi4\r(s)\cF\coth s\cF\r(s)}\z(s)^2\zb(s)^2 I(z,z'|s)\,,
\end{equation}

Having a covariantly constant background 
forces the field strength to be in the Cartan subalgebra, 
which we can think of as $\rnk(G)$ 
linearly independent diagonal matrices in the Lie algebra.
The 
Lorentz determinant in \eqref{full_heat_kernel} 
can be evaluated by using the eigenvalues (\wrt the Lorentz structure) 
of $\cF=\Big[\cF_a{}^b\Big]$, which are the four diagonal matrices
\begin{align} \label{eqn:covConstF-eigenvalues}
	\l_\pm = \pm\frac\rmi2(B\pm\Bb)\,,
\end{align}
where 
\begin{align} \label{defn:B^2}
	B^2   = \frac12\tr \cN^2\,, \quad 
	\Bb^2 = \frac12\tr\cNb^2 \,,
\end{align}
with the traces only over the Lorentz spinor indices.
This is completely analogous to the standard eigenvalue analysis 
of a constant electromagnetic background,
but now there are two invariants for each of the $\rnk(G)$ 
directions in the Cartan subalgebra.
The Lorentz determinant in \eqref{full_heat_kernel} becomes
\begin{align} \label{eqn:LorentzDet-CovConstBG}
	\det\!\Big(\frac{s\cF}{\sinh s\cF}\Big)^{\half}
	= 
	\frac{s \l_+}{\sin(s \l_+)}\frac{s \l_-}{\sin(s \l_-)} \ .
\end{align}

If the background is also on-shell, 
$\cD^\a\cW_\a=\tr\cN=0 \iff \cD_\a\cW_\b = \cD_\b\cW_\a$,
then we have the trace identities
\begin{align} \label{eqn:covConstOnShell-TraceN}
	\tr\cN^{2n}=2B^{2n}\,,
	\quad
	\tr\cN^{2n+1}=0\ .
\end{align}
We also have the simple covariant Taylor expansions \eqref{eqn:covTaylor}
\begin{align} \label{eqn:CovTaylor-OnShell}
\begin{gathered}
	I(z,z')\cN(z')I(z',z)=\cN(z)\,,	\quad	
	I(z,z')\cF(z')I(z',z)=\cF(z)\,, \\
	I(z,z')\cW(z')I(z',z)=(1-\z\cD)\cW(z) \ ,
\end{gathered}
\end{align}
that are needed in two- and higher-loop calculations.

Note that the supergraph $D$-algebra, the ability to move
covariant derivatives from one side of a propagator to another,
is deformed in the presence of a background gauge field.
For a covariantly constant background, 
the heat kernel satisfies the identity 
\begin{align} \label{HK-cD-algebra}
	\cD'_\a K(z,z'|s) = -\big(\rme^{\rmi s\cN}\big)_\a{}^\b\,\cD_\b K(z,z'|s)\,,
\end{align}
and similarly for $\cDb_\da$.
This leads to a non-trivial result for the propagator.
For a self-dual background (see subsection \ref{ssect:Self-dual})
we recover ``half'' of the supergraph $D$-algebra,
while in the $\cD_A\cW_\b=\cD_A\cWb_\db=0$ (see subsection \ref{ssect:cDcW=0}) 
limit we recover the full supergraph $D$-algebra.

\subsection{Chiral/antichiral propagators}
Assuming that the background gauge field is on-shell implies the 
following relations between the d'Alembertians 
$\BoxV$ and $\Box_\pm$ 
(\ref{defn:BoxV} and \ref{defn:Box+-})
\begin{align}
	\cD^2\Box_+ = \cD^2\BoxV = \BoxV\cD^2 \qquad
	\cDb^2\Box_- = \cDb^2\BoxV = \BoxV\cDb^2 \ .
\end{align}
This implies that the (anti)chiral propagators 
can be derived from the full propagators
and similarly for the heat kernels 
\begin{align}\label{defn:ChiralHKs}
\begin{aligned}
K_+(z,z')&=-\frac14\cDb^2K(z,z')=-\frac14\cDb'^2K(z,z') \ , \\
K_-(z,z')&=-\frac14\cD^2K(z,z')=-\frac14\cD'^2K(z,z') \ .
\end{aligned}
\end{align}
The Feynman rules of sections \ref{sect:PropsFeynRules} and \ref{sect:SQEDQuant}
also require the antichiral-chiral heat kernel
\begin{align} 
	K_{-+}(z,z') = \frac1{16}\cD^2\cDb'^2 K(z,z') 
	= \frac1{16}\cD^2\cDb^2 K(z,z') \,,
\end{align}
and its conjugate $K_{+-}$. 
These were explicitly given in \cite{Kuzenko2003},\footnote%
{In the $U(1)$ case, 
	the chiral heat kernel was first derived in a special gauge in
	\cite{Buchbinder1985}.
}
but without derivation. 
We fill that gap by rederiving $K_-$ and $K_{-+}$ below.
The other two heat kernels, $K_+$ and $K_{+-}$, 
can then be obtained via complex conjugation.

First we examine $K_-$:
\begin{align*} 
	K_-(z,z'|s) &= -\frac14\cD^2 K(z,z'|s) = -\frac14\cD'^2 K(z,z'|s) \\
	&=\frac{\rmi\m^{2\e}}{(4\p\rmi s)^{d/2}}
		\det\!\Big(\frac{s\cF}{\sinh s\cF}\Big)^{\half}\,U(s)
		\zb^2\rme^{\frac\rmi4\r\cF\coth s\cF\r}(-\frac14\cD'^2)\z^2 I(z,z')\,,
\end{align*}
where we can take the primed derivatives past the exponential because
$$\zb^2{\cD'_\a}{\rme^{\frac\rmi4\r\cF\coth s\cF\r}} = 0 \ .$$
Note that equation \eqref{eqn:covTaylorDaI} reduces to 
\begin{gather*} 
	\zb^2\cD'_\a I(z,z') = \zb^2 A_\a I(z,z')\,,
	\quad A_\a = -\frac\rmi2 \r_\ada \bar{W}^\da\,,
\end{gather*}
so that it's straightforward to calculate
\begin{align*} 
	\zb^2(-\frac14\cD'^2)\z^2 I(z,z')
	&= \zb^2\left(1+\z A-\frac14\z^2A^2-\frac14\z^2\cD'^\a A_\a\right)I(z,z')\\
	&= \zb^2 \rme^{-\frac\rmi2\z\r\bar{W}}I(z,z') \ .
\end{align*}
Thus we have the antichiral and chiral heat kernels
\begin{subequations}\label{chiral_heat_kernel}
\begin{align}
\label{eqn:K-}
	K_-(z,z'|s) 
	&=\frac{\rmi\m^{2\e}}{(4\p\rmi s)^{\frac d2}}
		\det\!\Big(\frac{s\cF}{\sinh s\cF}\Big)^{\!\half}\,U(s)\zb^2
		\rme^{\frac\rmi4\r\cF\coth s\cF\r-\frac\rmi2\z\r\bar{W}}I(z,z')\,, \\
\label{eqn:K+}
	K_+(z,z'|s) 
	&=\frac{\rmi\m^{2\e}}{(4\p\rmi s)^{\frac d2}}
		\det\!\Big(\frac{s\cF}{\sinh s\cF}\Big)^{\!\half}\,U(s)\z^2
		\rme^{\frac\rmi4\r\cF\coth s\cF\r-\frac\rmi2W\r\zb}I(z,z') \ .
\end{align}
\end{subequations}

The derivation of the antichiral-chiral kernel is similar, 
just a little more messy. First we write
\begin{align*} 
	&K_{-+}(z,z'|s)
	=\frac1{16}\cDb'^2 \cD^2 K(z,z'|s)
	= -\frac14\cDb'^2 K_-(z,z'|s) \\
	=&\frac{\rmi\m^{2\e}}{(4\p\rmi s)^{\frac d2}}
		\det\!\Big(\frac{s\cF}{\sinh s\cF}\Big)^{\!\half}\,
		U(s)(-\frac14\cDb'^2)\zb^2	\underbrace{
		\rme^{\frac\rmi4\r\cF\coth s\cF\r-\frac\rmi2\z\r\bar{W}}I(z,z')
		}_{\dsE}\,,
\end{align*}
where we've named the last term $\dsE$ since 
$\cDb'^\da \dsE = \bar{A}^\da \dsE$, 
where the conjugate of \eqref{eqn:covTaylorDaI} (see \cite{Kuzenko2003a}),
implies that
\begin{align*} 
\begin{split}
	\bar{A}^\da = \frac12(\st\z)^\da\cF\coth s\cF\r'+\z^2\cWb^\da
	-\frac1{12}\z_\a\r^\dbb\cF_\bdb{}^\daa \\
	+\frac\rmi2 \zb_\da\r^\daa(1-\frac23\z\cD)\cW_\a
	+\frac13\z\cW\zb^2+\frac13\z^2\zb\cWb\ .
\end{split}
\end{align*}
We observe that the action of the derivatives in $K_{-+}$ exponentiates as
\begin{align*} 
	-\frac14\cDb'^2\zb^2\dsE 
	= (1+\zb\bar{A}-\frac14\zb^2\bar{A}^2-\frac14\zb^2\cDb'\bar{A})\dsE
	= \rme^{\zb\bar{A}-\frac14\zb^2\cDb'\bar{A}}\dsE \ .
\end{align*}
So the exponentials can be combined and we write the final result as
\begin{align} \label{K-+}
	K_{-+}(z,z'|s)
	=\frac{\rmi\m^{2\e}}{(4\p\rmi s)^{\frac d2}}
		\det\!\Big(\frac{s\cF}{\sinh s\cF}\Big)^{\!\half}\,
		U(s)\rme^{\frac\rmi4\rt\cF\coth s\cF\rt+R(z,z')}I(z,z')\,,
\end{align}
where, following \cite{Kuzenko2003}, 
we've introduced the antichiral-chiral coordinates
\begin{align} \label{tilde_rho}
	\rt^a = \r^a-\rmi\z\s^a\zb\,,\quad
	\cD_\b\rt^a=\cDb'_\db\rt^a=0 \,,
	\cDb_\db\rt_\ada = -4\rmi\eps_\dbda \z_\a \,,
\end{align}
and the notation
\begin{align} \label{R(z,z')}
\begin{split}
	R(z,z') = \frac13(\z^2\zb\cWb-\zb^2\z\cW) 
	- \frac\rmi2(\cW\rt\zb+\z\rt\cWb) \\
	- \frac\rmi{12}\rt_\ada(\z^\a\zb\cDb\cWb^\da + 5\zb^\da\z\cD\cW^\a)\ .
\end{split}
\end{align}

\subsection{Self-dual limit}
\label{ssect:Self-dual}
In this section we describe the simplifications that occur in the above
heat kernels when the background gauge field satisfies a
relaxed self-dual condition \cite{Kuzenko2004a,Kuzenko2004c}
\begin{align*} 
	\cW_\a\neq0\,, \quad
	\cD_\a\cW_\b=0\,, \quad
	\cDb_{(\da}\cWb_{\db)}\neq0\ .
\end{align*}
As described in \cite{Kuzenko2004a,Kuzenko2004c},
these conditions are not compatible with the structure of a single, real
vector multiplet in Minkowski space, rather they should be considered
as a formal restriction on the field strength in order to calculate
a particular sector of the low-energy effective action.

Note that the results of this section are not used in this thesis,
as our analysis of the effective action for $\cN=1$ SQED in a self-dual 
background in section \ref{sect:SelfDualSQED} 
is performed by taking the self-dual limit of the full result.


A self-dual background simplifies calculations by both simplifying 
the background dependent Green's functions and by allowing
``half'' of the standard supergraph D-algebra to take place 
\cite{Kuzenko2004c}.
From \eqref{HK-cD-algebra} we see that in the self-dual limit,
the heat kernel obeys
\begin{align*} 
	\cD'_\a K(z,z'|s) = -\cD_\a K(z,z'|s)\,,
\end{align*}
which now integrates up to the full Green's function as
\begin{align*} 
	\cD'_\a G(z,z'|s) = -\cD_\a G(z,z'|s)\,,
\end{align*}
which is the standard supergraph $D$-algebra result.
The $\cDb_\da$ derivatives do not share this property.

With the notation (introduced above)
\begin{align*} 
	\cNb_\da{}^\db = \cDb_\da\cWb^\db\,, \quad 
	\Bb^2=\frac12\tr\cNb^2 = \frac14\cDb^2\cWb^2\,,
\end{align*}
where the trace is over the Lorentz indices and not the group indices. %
The Lorentz determinant in the heat kernels 
(\ref{full_heat_kernel}, \ref{chiral_heat_kernel}, \ref{K-+})
reduces to
\begin{align*} 
	\det\!\Big(\frac{s\cF}{\sinh s\cF}\Big)^\half
	=\left(\frac{s\Bb/2}{\sin(s\Bb/2)}\right)^2 . 
\end{align*}
The shifted variables simplify to
\begin{align}
	\cW^\a(s)&=\cW^\a\,,
		&\cWb_\da(s)&=(\cWb \rme^{\rmi s\bar\cN})_\da~\,,		\non \\ \non
	\z^\a(s)&=\z^\a-\rmi s \cW^\a\,,
		&\zb_\da(s)&=\zb_\da+(\cWb\frac{\rme^{\rmi s\cNb}-1}{\cNb})_\da\,, \\
	\r_\daa(s)&=\r_\daa-2\int_0^s\!\rmd t
		(\cW_\a\zb_\da(t)+\z_\a(t)\cWb_\da(t))\,,\hspace{-10em}	\\\non
	I(z,z'|s)&=\exp\left\{\int_0^s\rmd t~\Xi(\z,\cW,\cWb | t) \right\} 
		I(z,z') \ . \hspace{-20em}
\end{align}

\subsection{The simplest non-trivial background} 
\label{ssect:cDcW=0}
If, as in section \ref{ssect:Neq1F2F4}, 
we are only interesting in calculating $F^2$ and $F^4$ terms 
in the effective action then we can use the following conditions 
on the background gauge field
\begin{align} \label{eqn:cDcW=0}
	\cD_A\cW_\b=\cD_A\cWb_\db=0\ .
\end{align}
The determinant term in the heat kernels now reduces to the identity, 
and the common term in the exponentials reduces as $\cF\coth s\cF \to \ds1/s$.
The field strength now becomes invariant under the action of $U(s)$,
while the shifted Cartan 1-forms 
\begin{align}
	\z_\a(s)&=\z_\a-\rmi s\cW_\a\,,& \zb_\da(s)&=\zb_\da+\rmi s\cWb_\da\,,
\end{align}
form Grassmann delta-functions in the kernels $K$ and $K_\pm$.
\begin{align}\label{HK_DWeq0}
	K(z,z'|s)&=\frac{\rmi\m^{2\eps}}{(4\p\rmi s)^{d/2}}
		(\z-\rmi s \cW)^2(\zb+\rmi s\cWb)^2\rme^{\rmi\r^2/4s} I(z,z')\,, \\
	K_+(z,z'|s)&=\frac{\rmi\m^{2\eps}}{(4\p\rmi s)^{d/2}}(\z-\rmi s \cW)^2
		\rme^{\rmi\r^2/4s+\rmi s\cW^2(\zb+\rmi s\cWb)^2/6} I(z,z')\,,\\
	K_-(z,z'|s)&=\frac{\rmi\m^{2\eps}}{(4\p\rmi s)^{d/2}}(\zb+\rmi s \cWb)^2
		\rme^{\rmi\r^2/4s+\rmi s\cWb^2(\z-\rmi s\cW)^2/6} I(z,z')\ .
\end{align}
The antichiral-chiral kernel does not have any Grassmann delta-functions
to help simplify things. It becomes
\begin{align} 
	K_{-+}(z,z'|s)
	=\frac{\rmi\m^{2\e}}{(4\p\rmi s)^{\frac d2}}
		U(s)\rme^{\frac\rmi4\rt^2
		+ \frac13(\z^2\zb\cWb-\zb^2\z\cW) 
		- \frac\rmi2(\cW\rt\zb+\z\rt\cWb)}I(z,z')\, .
\end{align}

\section[Wess-Zumino propagator]%
		{Wess-Zumino propagator in space-time constant background}
\label{sect:WZProp}

As we saw in chapter \ref{Ch:WZ}, all propagators that occur in the 
Wess-Zumino model with arbitrary background chiral fields $\J$ and $\Jb$ 
can be obtained as different chiral projections 
of the Green's function of the operator 
\begin{align*} \D 
	=\Box-\quart(\J\Db^2+\Jb D^2)\ .
\end{align*}
The dimensionally regularised (see section \ref{sect:RegByDimlRed})
heat kernel of this operator obeys the differential equation (DE)
\begin{align} \label{eqn:WZHeatEqn}
	\left(\rmi\frac\rmd{\rmd s}+\D
		\right)U_V^{(\J)}(z,z'|s)&=0\,,\\
	U_V^{(\J)}(z,z'|0)
	&=\m^{2\eps}\d^{4}(\q-\q')\d^d(x-x') \ .
\end{align}
If we assume that the background is constant over space-time, 
$\pd_a\J=\pd_a\Jb=0$, then the heat kernel factorises as \cite{Buchbinder1994a}
\begin{align} \label{eqn:WZ:HeatKernFactorize} 
	U_V^{(\J)}(z,z'|s)
	= \rme^{-\frac{\rmi s}4\left(\Jb D^2+\J\Db^2\right)}U_V^{(0)}(z,z'|s)
	\deq \O(s)U_V^{(0)}(z,z'|s) \,,
\end{align}
where 
\( U_V^{(0)}(z,z'|s)
	=\d^4(\q-\q')U(x,x'|s)
\)
and 
\begin{equation}
	U(x,x'|s) 
	= \exp(\rmi s\Box)\d^d(x-x')
	= \frac{\rmi\m^{2\eps}}{(4\p\rmi s)^{d/2}}\rme^{\frac{\rmi}{4}(x-x')^2/s}\,,
\end{equation} 
is the $d$-dimensional free bosonic heat kernel.
To find the full heat kernel \eqref{eqn:WZ:HeatKernFactorize}
we need only obtain an explicit form of the operator $\O$.

The heat equation \eqref{eqn:WZHeatEqn} implies 
that the operator $\O(s)$ satisfies
\begin{align} \label{eqn:WZHeatEqn2}
	\rmi\frac\rmd{\rmd s}\O(s) = \frac14\O(s)\big(\J\Db^2+\Jb D^2)\,,
	\quad\O(0)=1 \ .
\end{align}
To solve this, following \cite{Buchbinder1994a}, 
we expand the operator $\O$ as
\begin{align}\label{eqn:decompOasABC}
	\O(s) &= \frac1{16}A(s)D^2\Db^2+\frac1{16}\tilde{A}(s)\Db^2D^2
	+\frac18B^\a(s)D_a\Db^2+\frac18\tilde{B}_\da(s)\Db^\da D^2 \non\\
	&\qquad +\frac14C(s)D^2+\frac14\tilde{C}(s)\Db^2 + 1\ .
\end{align}
Note that only $A$ and $\tilde A$ can contribute to the 1-loop potential.

\subsection{Calculating the heat kernel}
At this point, it is convenient to introduce some notation
\begin{align} \label{def:WZPropShortHands}
\begin{split}
	a=(D^\a\J)(D_\a\J)\,,\quad  b=(D^2\J)\,,\quad  
	\m=(D^\a\J)\pd_\ada(\Db^\da\Jb)\,, \\
	u^2=\Jb\J\Box\,,\quad
	\scF^2=\bbar b/64\,,\quad \scG^2=u^2+\scF^2 \,, \quad
	\b = \frac18\bem 0&\bbar\\ b& 0\eem .
\end{split}
\end{align}
Note that to move between the tilded and non-tilded symbols 
in \eqref{eqn:decompOasABC}, we make the replacements 
$D^\a\leftrightarrow\Db_\da$, $\J\leftrightarrow\Jb$ which implies
$a\leftrightarrow\abar$, $b\leftrightarrow\bbar$ and $\m\leftrightarrow-\m$.
We are also using the convention that derivatives act on all terms to their
right unless bracketed. This means that $\m$ is actually a
differential operator that obeys $\m^2=-\frac12\abar a \Box$.
The other important square to note is $\b^2=\scF^2\ds1$.

Now the heat equation \eqref{eqn:WZHeatEqn2} decomposes as 
\begin{subequations}
\begin{align}
 	\frac\rmd{\rmd s} \bem A\\\tilde A\eem
 	&= -\rmi	\bem\J&0\\0&\Jb\eem  \bem C\\\tilde C\eem ,\label{Aeqn}\\
 \left(\frac\rmd{\rmd s} + \bem 0&\J\pd^\daa\\\Jb\pd_\ada&0\eem \right)
 	\bem B^\a\\\tilde B_\da\eem
 	&= -\rmi\bem(D^\a\J)C\\(\Db_\da\Jb)\tilde{C}\eem , \label{Beqn} \\
 \left(\frac\rmd{\rmd s} + 2\rmi \b \right) \bem C\\\tilde{C} \eem
	 	+\rmi\, \bem\Jb (\Box A+1)\\\J(\Box\tilde A+1)\eem
  	&=-\half\bem B^\a\pd_\ada(\Db^\da\Jb)\\
  		\tilde{B}_\da\pd^\daa(D_\a\J)\eem  , \label{Ceqn}
\end{align}
with $A(0)=\tilde{A}(0)=B^\a(0)=\tilde{B}_\da(0)=C(0)=\tilde{C}(0)=0$.
We can eliminate $A$ and $\tilde A$ from the equation for $C$ and $\tilde C$
by moving to the second order DE
\begin{align} \label{Ceqn2} 
 \left(\frac{\rmd^2}{\rmd s^2}+2\rmi\b 
 \frac\rmd{\rmd s}+\Jb\J\Box\right) \bem C\\\tilde{C}\eem
 =-\half\frac\rmd{\rmd s}
 \bem B^\a\pd_\ada(\Db^\da\Jb)\\ \tilde{B}_\da\pd^\ada(D_\a\J)\eem  ,
\end{align}
\end{subequations}
where we need the initial ``velocity''
$\pd_s\big(C,\tilde{C}\big)\big|_{s=0}=-\rmi\big(\Jb,\J\big)$.

We can solve the coupled equations (\ref{Beqn}, \ref{Ceqn2}) for $B$ and $C$
by expanding \wrt the Grassmann parameters $D_\a\J$ and $\Db_\da\Jb$, 
\begin{subequations}\begin{align}\label{cexpn}
\begin{split}
	C&=C_0+a C_{20}+\abar C_{02}+\m C_{11}+\abar a C_{22}\,,\\
	\Ct&=\Ct_0+\abar\Ct_{20}+a\Ct_{02}-\m\Ct_{11}+\abar a \Ct_{22}\,,
\end{split}\\
\label{bexpn}
\begin{split}
	B^\a&=(D^\a\J)(\hat{B}_0 + \abar\hat{B}_2)
	+(\Db_\da\Jb)\pd^\daa(\check{B}_0 + a \check{B}_2)\,,\\
  \Bt_\da&=(\Db_\da\Jb)(\hat\Bt_0 + a\hat\Bt_2)
	+(D^\a\J)\pd_\ada(\check\Bt_0 + \abar\check\Bt_2) \ .	
\end{split}
\end{align}
\end{subequations}
This allows us to step through the DEs order by order 
in $D_\a\J$ and $\Db_\da\Jb$.  

In the following, we first solve the zeroth order DE 
(which is the only one to have non-zero initial conditions).
We then list all of the higher order DEs before noting their common structure
and providing the solutions.

\subsubsection{Zeroth order differential equations}
Keeping all terms independent of $D_\a\J$ and $\Db_\da\Jb$ 
in \eqref{Ceqn2} gives
\begin{equation}\begin{gathered}
	\left(\frac{\rmd^2}{\rmd s^2}+2i\b\frac{\rmd}{\rmd s}+u^2\right)
	\bem C_0\\\Ct_0\eem =0 \,,\\
	\bem C_0\\\Ct_0\eem\Big|_{s=0}=0\,,	\quad
	\frac{\rmd}{\rmd s}\bem C_0\\\Ct_0\eem\Big|_{s=0}=-\rmi\JbJm \,,
\end{gathered}\end{equation}
which has the solution
\begin{align} \label{soln:WZPropC0}
	\hspace{-4mm}
	\bem C_0\\\Ct_0\eem
	=-\rmi\frac{\sin{s\scG}}{\scG}\rme^{-\rmi s\b}\JbJm
	=-\rmi\frac{\sin s\scG}\scG
	\left(\cos{s\scG}-\rmi\b\frac{\sin s\scF}\scF\right)\JbJm.
\end{align}

\subsubsection{First order differential equations}
Keeping only the first order terms in \eqref{Beqn} gives
\begin{align*}\non
\left(\frac\rmd{\rmd s}+\bem 0&\J\pd^\ada\\\Jb\pd_\ada&0\eem \right)
 \bem (D^\a\J)\hat{B}_0+(\Db_\da\Jb)\pd^\ada\check{B}_0
  \\(\Db_\da\Jb)\hat\Bt_0+(D^\a\J)\pd_\ada\check\Bt_0\eem
 =-\rmi\bem(D^\a\J)C_0\\(\Db_\da\Jb)\tilde{C}_0\eem  .
\end{align*}
Extracting the coefficients of $D_\a\J$ and $\Db_\da\Jb$ leads to two equations
that can be recombined to give the second order DE
\begin{align} 
	\left(\frac{\rmd^2}{\rmd s^2}+u^2\right)
	\bem\check{B}_0 \\ \check\Bt_0\eem
	&= \rmi \bem 0&\J\\ \Jb&0\eem\bem C_0\\\Ct_0\eem \,,
\intertext{and the relation}
	\bem\hat{B}_0\\\hat\Bt_0\eem
	&=\frac{-1}{\Jb\J}\bem 0&\J\\\Jb&0\eem
	\frac\rmd{\rmd s}\bem\check{B}_0\\\check\Bt_0\eem .
\end{align}

\subsubsection{Second order differential equations}
At this order, we simply read off the equation
\[ 
 \left(\frac{\rmd^2}{\rmd s^2}+2i\b\frac{\rmd}{\rmd s}+u^2\right)
 \bem a C_{20}+\abar C_{02}+\m C_{11}\\
 	  \abar\Ct_{20}+a\Ct_{02}-\m\Ct_{11}\eem
 	  =\frac12\frac\rmd{\rmd s}
 	  \bem -\m \hat B_0 + \abar\Box\check B_0 \\
 	  	\phantom{-}\m \hat{\tilde B}_0 + a\Box\check{\tilde B}_0\eem\,,
\]
from \eqref{Ceqn2}.  This can be split into the two 2nd order DEs
\begin{align}
	\left(\frac{\rmd^2}{\rmd s^2}+2i\b\frac{\rmd}{\rmd s}+u^2\right)
	 \bem a C_{20}+\abar C_{02}\\\abar\Ct_{20}+a\Ct_{02}\eem 
	 &=\frac\Box2\frac\rmd{\rmd s} 
	 	\bem \abar \check B_0 \\ a \check\Bt_0\eem , \\
	\left(\frac{\rmd^2}{\rmd s^2}+2i\b\frac{\rmd}{\rmd s}+u^2\right)
		 \bem \phantom{-}C_{11} \\ -\Ct_{11}\eem 
	 &=-\frac12\frac{\rmd}{\rmd s} 
		 \bem\phantom{-}\hat B_0 \\ -\hat\Bt_0 \eem .
\end{align}
Although we could separate the DEs for $C_{20}$ and $C_{02}$, it is simpler
(more symmetric) to leave them in their entangled form.

\subsubsection{Third order differential equations}
Keeping only the third order terms in \eqref{Beqn} gives
\begin{align}\begin{split}
	\left(\frac\rmd{\rmd s}+\bem 0&\J\pd^\daa\\\Jb\pd_\ada&0\eem \right)
 	\bem \abar (D^\a\J)\hat{B}_2+a(\Db_\db\Jb)\pd^\dba\check{B}_2\\
  		a(\Db_\da\Jb)\hat\Bt_2+\abar (D^\b\J)\pd_\bda\check\Bt_2\eem \\
    =-\rmi\bem(D^\a\J)(\abar C_{02}+\m C_{11})\\
  			(\Db_\da\Jb)(a\Ct_{02}-\m\Ct_{11})\eem .
\end{split}\end{align}
Using 	\( (D^\a\J)\m=\half a(\Db_\da\Jb)\pd^\daa \)
and		\( (\Db_\da\Jb)\m=-\half\abar(D^\a\J)\pd_\ada \)
we can split the above to get the second order DE
\begin{align}
\left(\frac{\rmd^2}{\rmd s^2}+u^2\right)\bem\check{B}_2\\\check\Bt_2\eem
	&=\rmi\bem 0&\J\\\Jb&0\eem\bem C_{02}\\\Ct_{02}\eem
	 -\frac\rmi2\frac{\rmd}{\rmd s}\bem C_{11}\\\Ct_{11}\eem \,,
\intertext{and the relation}
	\bem\hat{B}_2\\\hat\Bt_2\eem&=\frac{-1}{\Jb\J}\bem 0&\J\\\Jb&0\eem\left(
	\frac\rmd{\rmd s}\bem\check{B}_2\\\check\Bt_2\eem 
	+\frac\rmi2\bem C_{11}\\\Ct_{11}\eem\right)  . 
\end{align}

\subsubsection{Fourth order differential equations}
The final DE is easily read from \eqref{Ceqn2},
\begin{align}
 	\left(\frac{\rmd^2}{\rmd s^2}+2i\b\frac{\rmd}{\rmd s}+u^2\right)
	\bem C_{22}\\\Ct_{22}\eem
	=\frac\Box2\frac\rmd{\rmd s}\bem\check{B}_2\\\check\Bt_2\eem~.
\end{align}

\subsubsection{Results for the heat kernel}
The DEs that need to be solved to find the terms of order one and greater 
are all second order, inhomogeneous DEs with `zero' initial conditions, 
i.e.,
\begin{subequations}\begin{align}
	\left(\frac{\rmd^2}{\rmd s^2}+2i\b\frac{\rmd}{\rmd s}
		+u^2\right)\c_C(s) &= v_C(s)\,,
	\qquad \c_C(0)=\dot\c_C(0)=0 \,, \\
	\left(\frac{\rmd^2}{\rmd s^2}+u^2\right)\c_B(s) &= v_B(s) \,,
	\qquad \c_B(0)=\dot\c_B(0)=0 \,,
\end{align}\end{subequations}
where the $\c_{B,C}$ are component 2-vectors of $B$ or $C$ respectively 
and the inhomogeneous terms $v_{B,C}$ depend on 
the solutions to lower order components.
Using variation of parameters on the general solutions to the 
associated homogeneous differential equations 
yields
\begin{subequations}\begin{align}
	\c_C(s)&=\rme^{\rmi s\b(\scG/\scF-1)}
	\int_0^s\!\!\rmd{t}\,\rme^{-2\rmi t\b\scG/\scF}
	\int_0^t\!\!\rmd{\t}\,\rme^{\rmi\t\b(\scG/\scF+1)}v_C(\t) \,, \\
	\c_B(s)&=\rme^{\rmi su}\int_0^s\!\!\rmd{t}\,\rme^{-2\rmi tu}
	\int_0^t\!\!\rmd{\t}\,\rme^{\rmi\t u}v_B(\t) \ .
\end{align}\end{subequations}
The following solutions have all been found by hand 
and checked that they satisfy the original DEs 
and boundary conditions using Mathematica \cite{WZNotebook}.
The solutions for the components of $\big(C(s),\Ct(s)\big)$ are
\addcontentsline{toc}{subsection}{Results for the heat kernel}
\begin{subequations} \label{WZ-HKC-Solutions}
\begin{align}
\bem C_0\\\Ct_0\eem 
	&= -\rmi\frac{\sin(s\scG)}{\scG}\rme^{-\rmi s\b}\JbJm ,\\
\bem C_{11}\\\Ct_{11}\eem 
	&= \frac{s}{8\scF^2}\left(\frac{\sin(su)}{su}
		-\frac{\sin(s\scG)}{s\scG}\cos(s\scF)\right)\JbJm ,
 \allowdisplaybreaks\\
\bem C_{20}\\\Ct_{20}\eem 
 	&= \frac{\Box\b}{8\scF^2}\bem 0&\J\\\Jb&0\eem \!\! \Bigg[\
		\frac{\rmi s}{2u^2}\left(\frac{\sin(s\scF)}{s\scF} \cos(s\scG)
		-\cos(s\scF)\frac{\sin(s\scG)}{s\scG}\right) \non \\  \Bigg.
		& +\frac{\b}{u^2}\left(
		\frac{\cos(su)}{\scF^2}-\frac{\sin(s\scF)\sin(s\scG)}{\scF\scG}-
		\frac{\cos(s\scF)\cos(s\scG)}{\scF^2}\right) \\ \non
		& -\frac{\rmi s}{2\scG^2}\left(\cos(s\scG)
		-\frac{\sin(s\scG)}{s\scG}\right)
		\rme^{-\rmi s\b} \Bigg]\JbJm ,  
 \allowdisplaybreaks\\
\bem C_{02}\\\Ct_{02}\eem 
	&= \frac{\Box}{16}\bem 0&\J\\\Jb&0\eem\frac{-\rmi\b}{\scF^2}\Bigg[
		\frac{\sin(s\scG)\cos(s\scF)}{u^2\scG}
		-\frac{\sin(s\scF)\cos(s\scG)}{u^2\scF} \\\non
		&+\frac{s\rme^{-\rmi s\b}}{\scG^2}
		\left(\cos(s\scG)-\frac{\sin(s\scG)}{s\scG}\right)\!\Bigg]\!\JbJm ,
 \allowdisplaybreaks\\
\bem C_{22}\\\Ct_{22}\eem
	&= \frac{-\rmi\Box}{128\scF^2}\Bigg[
		\frac{\rmi s^2\b}{\scF^2}
		  \left(\frac{\sin(s\scF)}{s\scF}\frac{\sin(s\scG)}{s\scG}
		  -\frac{\sin(su)}{su}\right)
		+ s\,\rme^{-\rmi s\b} \times \\ \non
		&\times\!\left(\!\frac{\sin(s\scG)}{s\scG}\left(
			\frac{1+\rmi s\b}{\scF^2}
			-\frac{3\scF^2+(1+s^2u^2)\scG^2}{2\scG^4} \right) 
			+ \cos(s\scG)\frac{3\scF^2+\scG^2}{2\scG^4}\right)\\ \non
		&-\frac{s}{u^2}\left(\frac{u^2}{\scF^2}\frac{\sin(su)}{s u}
			+\frac{\sin(s\scF)}{s\scF}\cos(s\scG)
			-\cos(s\scF)\frac{\sin(s\scG)}{s\scG}\right) \Bigg]\JbJm ,
\end{align}\end{subequations}
and the solutions for the components of $\big(B^\a(s),\Bt_\da(s)\big)$ are
\begin{subequations} \label{WZ-HKB-Solutions}
\begin{align}
\bem \check{B}_0\\\check{\Bt}_0\eem
	&= \frac{\rmi s}{2u^2}\bem 0&\J\\\Jb&0\eem\left(
		\frac{\b}{\scF}\frac{\cos(su)}{s\scF}
 		-\rme^{-\rmi s\b}\left(\rmi\frac{\sin(s\scG)}{s\scG}
		+\frac{\b}{\scF}\frac{\cos(s\scG)}{s\scF}\right)\!\right)\!\JbJm ,\\ 
\bem \hat{B}_0\\\hat{\Bt}_0\eem
	&= \frac{\rmi s\b}{2\scF^2}\left(\frac{\sin(su)}{su}
		-\frac{\sin{s\scG}}{s\scG} \rme^{-\rmi s\b}\right)\JbJm ,
 \allowdisplaybreaks\\
\bem \check{B}_2\\\check{\Bt}_2\eem
	&= \frac{-\rmi s^2}{32\scF^2}\Bigg[
		\frac{2\rmi\b}{su^2}\Big(
			\frac{\scG^2}{\scF^2}\frac{\sin(s\scG)}{s\scG}\cos(s\scF)
			-\cos(s\scG)\frac{\sin(s\scF)}{s\scF}\Big)		\\\non
	&+ \frac{\sin(su)}{su}\Big(1-\frac{2\rmi\b}{s\scF^2}\Big)
		-\rme^{-\rmi s\b}\Bigg(\!\!
			\Big(1+\frac{\rmi\b}{s\scG^2}\Big)\frac{\sin(s\scG)}{s\scG}
			-\frac{\rmi\b}{\scG}\frac{\cos(s\scG)}{s\scG}\Bigg)\!
		\Bigg]\!\JbJm \!, 
 \allowdisplaybreaks\\
\bem \hat{B}_2\\\hat{\Bt}_2\eem
	&= \frac{-\rmi}{\Jb\J}\frac{1}{16\scF^2}\bem 0&\J\\\Jb&0\eem\Bigg[
		\frac{\sin(su)}{2u}-\cos(su)\frac{s\scF^2-2\rmi\b}{2\scF^2} \\ \non
		&\quad -\left(	\frac{\sin(s\scG)}\scG
		+\frac{\rmi\b}{\scF^2}\cos(s\scG)\right)\cos(s\scF) \\ \non
		&\quad +\frac{1}{2\scG^2}\left(u^2s\cos(s\scG)+(\scF^2+\scG^2)
		\frac{\sin(s\scG)}\scG \right) \rme^{-\rmi s\b}\Bigg]\JbJm \ .
\end{align}
\end{subequations}
The solution for $\big(A(s),\At(s)\big)$ is just 
a term-by-term integration of the solution for $\big(C(s),\Ct(s)\big)$ 
given in \eqref{WZ-HKC-Solutions} above. 

From the above results, the initial condition $\O(0)=1$ 
is easily checked.
It is also worth checking that the initial velocity
$\O'(0)=-\frac\rmi4(\Jb D^2 + \J\Db^2)$ is satisfied,
which implies that only $\big(C_0,\Ct_0\big)$ 
has a non-vanishing first derivative at $s=0$. 
The above results have the correct initial velocity.%
\footnote{We note that while the results presented in \cite{Buchbinder1994a} 
do satisfy the initial condition $\O(0)=1$, 
the typograpical errors cause the higher components to not satisfy
the initial velocity condition $\O'(0)=-\frac\rmi4(\Jb D^2 + \J\Db^2)$.} 

\subsection{K\"ahler approximation}\label{ssect:WZProp:Kahler}
The condition $\J=\const$ can be used, e.g., 
to calculate the K\"ahler potential of the
Wess-Zumino model (see chapter \ref{Ch:WZ}).
In this limit we have 
\(a=b=\m=\cF=0\)
and \(u^2=\cG^2=\Jb\J\Box\), 
which implies that
\(B=\Bt=0$, $C=-\rmi\Jb\sin(su)/u\) 
and $A=\tilde{A}=\Jb\J(\cos(su)-1)/{u^2}$.
This means that $\O(s)$ has the expansion
\begin{align} \label{WZ:O(s)-in-Kahler-approx}
	\O(s) 
		  = 1 - \frac{\rmi}{4}\frac{\sin su}{u}\left(\Jb D^2 + \J\Db^2\right)
		  	  + \frac{\Jb\J}{16}\frac{\cos su - 1}{u^2}\acom{D^2}{\Db^2} \ .
\end{align}

This result can also be derived directly from \eqref{eqn:WZHeatEqn2}.
In the limit of $\J=\const$, we can take a second propertime derivative
to find the inhomogeneous harmonic oscillator equation%
\footnote{
In deriving this differential equation 
we used the standard superspace projectors
\begin{gather*} 
	P_+ = \frac{\Db^2D^2}{16\Box}\,,\quad
	P_- = \frac{D^2\Db^2}{16\Box}\,,\quad
	P_0 = \frac{D^\a\Db^2D_\a}{-8\Box}
		= \frac{\Db_\da D^2 \Db^\da}{-8\Box}\,,\\
	P_iP_j=\d_{ij}\,,\quad P_0+P_++P_-=1\ .
\end{gather*}
}
\begin{align*} 
	\O''(s) = -\frac1{16}\O(s)(\Jb D^2 + \J\Db^2)^2
		= -u^2\O(s)(P_++P_-)
		= -u^2\O(s) + u^2 P_0\,,
\end{align*}
with the initial conditions $\O(0)=1$, $\O'(0)=-\frac\rmi4(\Jb D^2 + \J\Db^2)$.
This is easily solved to give \eqref{WZ:O(s)-in-Kahler-approx}.

\subsection{Expansion up to four derivatives}\label{ssect:WZProp:4DExp}
The leading order term in the auxiliary potential contains 
four Grassmann derivatives. In this subsection we expand the solution
for $C(s)$, given above, up to four derivatives and then integrate to 
find $A(s)$ that is used in the one-loop effective action calculation.

The basic series expansions needed are
\begin{align*} 
	\scG &= 
		u \big(1+\frac{\scF^2}{2u^2}+\dots\big)\,, &
	\rme^{-\rmi s \b} &= 
		1-\rmi s \b -\frac12 s^2\scF^2 + \dots \\
	\cos(s\scF) &= 1-\frac12s^2\scF^2+\dots \,,&
	\cos(s\scG) &= \cos(s u)\Big(1-\frac{su\tan(su)}{2u^2}\scF^2+\dots\Big) \\
	\frac{\sin(s\scF)}{s\scF} &= 1 - \frac16 s^2\scF^2 + \dots \,, &
	\frac{\sin(s\scG)}{s\scG} &= \frac{\sin(su)}{su}
		\Big(1 - \frac{1-su\cot(su)}{2u^2}\scF^2 + \dots\Big) .
\end{align*}
We can then quickly read off the components of $C(s)$ to the relevant orders
\begin{align*}
 \bem C_0\\\Ct_0\eem 
 &=	-\rmi\frac{\sin(su)}{u} \left(1-\rmi s \b 
 	- \frac{1+s^2u^2-su\cot(su)}{2u^2}\scF^2\right)\JbJm \,,\\
 \bem C_{11}\\\Ct_{11}\eem 
 &=\frac{\sin(su)}{u}\left(\frac{1+s^2u^2-su\cot(su)}{2^4u^2}\right)
 	\JbJm \,,\\
 \bem C_{20}\\\Ct_{20}\eem 
 &=	\frac{\rmi s}{128u^2}\left(\big(1+\frac13s^2u^2\big)\cos(su)
 	-\frac{\sin(su)}{su}\right)\bem\bbar\Jb/\J\\ b\J/\Jb\eem \,,\\
 \bem C_{02}\\\Ct_{02}\eem 
 &= \frac{s^2}{16}\Big(\frac{\sin(su)}{su}-\cos(su)\Big)\bem\J/\Jb\\\Jb/\J\eem
	+\bem0&\Jb/\J\\\J/\Jb&0\eem								
	\bem C_{20}\\\Ct_{20}\eem \,,\\
 \bem C_{22}\\\Ct_{22}\eem
 &= \frac{\rmi s}{2^{10}u^2}\Big(\big(\frac23s^2u^2-7\big)\cos(su)
 	- \frac{\sin(su)}{su}\big(s^4u^4+3s^2u^2-7\big)\Big)
 	\bem1/\J\\1/\Jb\eem \ .
\end{align*}

Using the relation \eqref{Aeqn}, we integrate the above to find
\begin{align*}
 \bem A_0\\\At_0\eem 
 &=	-\frac{\Jb\J}{u^2}\Big(1-\cos(su)\Big)\bem1\\1\eem
 	-\frac{\rmi\b}{u^3}\Big(su\cos(su)-\sin(su)\Big)\bem\Jb^2\\\J^2\eem \\
 	&-\frac{\Jb\J s\scF}{2u^3}\Big(su\cos(su)-\sin(su)\Big)\bem1\\1\eem\,,\\
 \bem A_{11}\\\At_{11}\eem 
 &= \frac{-\rmi s\Jb\J}{16u^3}\big(\sin(su)-su\cos(su)\big)\bem1\\1\eem \,,\\
 \bem A_{20}\\\At_{20}\eem 
 &= \frac{s}{128u^3}\Big(su\cos(su)-\big(1-\frac13s^2u^2\big)\sin(su)\Big)
 	\bem\bbar\Jb\\ b\J\eem \,,\\
 \bem A_{02}\\\At_{02}\eem 
 &= \frac{-\rmi}{16u^3}\Big(\big(3-s^2u^2\big)\sin(su)-3su\cos(su)\Big)
 	\bem\J^2/\Jb\\\Jb^2/\J\eem + \bem \At_{20}\\ A_{20}\eem \,,\\
 \bem A_{22}\\\At_{22}\eem 
 &= \frac{s}{1024u^3}\Big(\big(7-\frac{10}{3}s^2u^2\big)\sin(su)
 	+su\big(s^2u^2-7\big)\cos(su)\Big)\bem1\\1\eem \ .
\end{align*}



\chapter[Effective potential for WZ]
		{Effective potential for the Wess-Zumino model in components}
\label{A:WZCompEffPot}

In this appendix we derive the well-known 
\cite{Fujikawa1975,O'Raifeartaigh1976a,Huq1977,Amati1982,Grisaru1983a,
	Miller1983a,Miller1983,Fogleman1983}
form of the one-loop effective potential \eqref{WZCompEffPot-MtmInt} 
using the component formalism of the Wess-Zumino model. 
We then evaluate the integrals to get the resultant 
component effective potential in a form that aids the comparison 
with the component projections of the superfield effective potential 
in section \ref{ssect:WZCompProj}.

\section{Background field quantisation}
Projecting the classical Wess-Zumino action \eqref{Action:WZ} 
to components using 
\begin{align}
	\F| = \f\,,  \qquad
	D_\a\F| = \sqrt2 \j_\a\,, \qquad
	-\frac14 D^2\F| = F\,,
\end{align}
yields the component action
\begin{align}\label{Action:WZ-Components}
\begin{aligned}
	S[\f,\j,F] = \intx&\Big(-(\pd_a\fb)(\pd^a\fb) 
	+ \frac\rmi2\big(\j\pd\bar\j-\bar\j\pd\j\big) + \bar{F}F \\
	&+ \Big\{F\big(m\f+\frac\l2\f^2\big)-\frac12\j^2(m+\l\f)\Big\} + \cc
	\Big) \ .
\end{aligned}
\end{align}
We want to examine the purely bosonic effective potential, so make the 
quantum-background split
\begin{align*}
	\f \to \f + a\,, \quad 
	F \to F + f\,, \quad
	\pd_b a = \pd_b f = 0\ .
\end{align*}
With this background, the action becomes
\begin{align*}
	S[\f+a,\j,F+f] 
	= S[a, f] + S_\text{lin} + S_\text{quad} + S_\text{int}\,,
\end{align*}
where the tree level potential is
\begin{align*}
	S[a,f] = \intx\Big( \bar{f}f + f(ma+\frac\l2f^2) + \cc \Big)
		  \xrightarrow{\text{on-shell}} -\intx \big|ma+\frac\l2a^2\big|^2\ .
\end{align*}
The linear term, $S_\text{lin}$, does not contribute to the effective action,
while the interaction term, 
\begin{align*}
	S_\text{int} = \intx\Big(\frac\l2 \f^2F + \cc\Big)\,,
\end{align*}
only contributes at higher loops.
Finally, the terms quadratic in the quantum fields are
\begin{align*}
	S_\text{quad} = \intx\Big(
	\bar\f\Box\f + \rmi\j\pd\bar\j + \bar{F}F
	+ \big(m' \f F + \frac{\l f}2\f^2 -\frac{m'}2\j^2 + \cc\big)
	\Big)\,,
\end{align*}
where \(m' \deq m + \l a\).
This can be written in the functional form
\begin{align}\label{WZCompQuad}
	S_\text{quad} &= 
	\frac12 \bem\rmi\j & -\rmi\bar\j\eem
		\bem m'\ds1 & \rmi\pd \\ \rmi\tilde\pd & \bar{m}'\ds1\eem
		\bem\rmi\j \\ -\rmi\bar\j\eem \\
	&+\frac12 \bem\f&\fb&F&\bar{F}\eem
		\bem \l f & \Box & m' & 0 \\ \Box & \bar\l\bar{f} & 0 & \bar{m}'\\
			m' & 0 & 0 & 1 \\ 0 & \bar{m}' & 1 & 0 \eem
		\bem\f\\\fb\\F\\\bar{F}\eem ,
\end{align}
from which we read of the fermionic and bosonic Hessians $H_F$ and $H_B$.

\section{One-loop effective potential}
From \eqref{WZCompQuad}, we can immediately write down 
the one-loop effective action
\begin{align*}
	\G^{(1)} = \G_B^{(1)} + \G_F^{(1)}
			= \frac\rmi2\log\Det H_B - \frac\rmi2\log\Det H_F\ .
\end{align*}
Both of the contributions are easily evaluated 
through repeated use of the identity
\begin{align*}
	\Det\bem A & B \\ C & D \eem 
	= \Det(D)\Det(A-BD^{-1}C) = \Det(AD-BD^{-1}CD) \ .
\end{align*}
We find
\begin{align*}
	\G_F^{(1)} 
	= -\frac\rmi2\log\Det\big(|m'|^2\ds1 - (\rmi\pd)(\rmi\tilde\pd)\big)
	= -\rmi\log\Det(|m'|^2 - \Box)\,,
\end{align*}
and
\begin{align*}
	\G_B^{(1)} = \frac\rmi2\log\Det\Big(
			\bem \Box & \bar\l\bar{f} \\ \l f & \Box\eem - |m'|^2\ds1\Big)
		= \frac\rmi2\log\Det\big((\Box-|m'|^2)^2 - |\l f|^2\big)\ .
\end{align*}
These are then recombined to get the result quoted in 
\eqref{WZCompEffPot-MtmInt}
\begin{align}
	\G^{(1)} = \frac{\rmi}{2}\log\Det
		\Big(1-\frac{|\l f|^2}{(\Box-|m+\l a|^2)^2}\Big)\ .
\end{align}

This can be evaluated by going to momentum space
\begin{align*}
	\G^{(1)} &= \frac{\rmi}{2}\intx\intk
	\log\Big(1-\frac{|\l f|^2}{(k^2+|m+\l a|^2)^2}\Big) \,,
\end{align*}
and factorising the argument of the logarithm
\begin{align}
	V^{(1)} = -\frac{\rmi}{2}\intk&\Big[\!
	\log(k^2+|m'|^2-|\l f|)+\log(k^2+|m'|^2+|\l f|)	\nonumber\\
	&-2\log(k^2+|m'|^2) \Big]\ .
\end{align}
Each of the terms in the above integral is UV divergent 
(converging for $1<\eps<2$)
\begin{align*}
	(4\p)^2 V^{(1)} 
	&= \frac{-\G(\eps)(4\p)^{\eps}}{2(2-3\eps+\eps^2)}\Big[
		(|m'|^2-|\l f|)^{2-\eps}+(|m'|^2+|\l f|)^{2-\eps}-2(|m'|^2)^{2-\eps}
		\Big]\\
	&= \frac{|\l f|^2}{4}\Big(-\frac2\eps - 3 + 2\log\frac{|m'|^2}{\mub^2}\Big)
		  + \frac{(|m'|^2 + |\l f|)^2}{4}\log\frac{|m'|^2 + |\l f|}{|m'|^2}\\
	&\quad+ \frac{(|m'|^2 - |\l f|)^2}{4}\log\frac{|m'|^2 - |\l f|}{|m'|^2} 
	+ O(\eps)\ .
\end{align*}
This expression separates into a term quadratic in $|f|$, 
which lifts to the K\"ahler potential,
and terms of order $|f|^4$ and higher that produce the auxiliary potential:
\begin{align*}
	(4\p)^2 V^{(1)} 
	&= \frac{|\l f|^2}{2}\Big(-\frac1\eps+\log\frac{|m'|^2}{\mub^2}\Big)\\
	&-\Big(\frac34|\l f|^2 - |\l f||m'|^2\tan^{-1}\frac{|\l f|}{|m'|^2}
	 - \frac{|\l f|^2+|m'|^4}{4}\log\big(1-\frac{|\l f|^2}{|m'|^4}\big)\Big) .
\end{align*}
Using the results of the discussion in section \ref{ssect:WZCompProj}, 
we see that the above expression lifts to reproduce 
the full superfield effective potential 
(\ref{WZ-1loop-Kahler}, \ref{WZ-fullAuxPot1}).



\chapter{Integer relation algorithms}\label{A:IntegerRelations}
In this appendix we use integer relation algorithms to examine the integrals 
that occur at two-loops in (S)QED with a self-dual background.
In this thesis, 
such integrals are found in section \ref{sect:SelfDualSQED}.

\section{Introduction}
The advent of quick and powerful integer relations algorithms
 is one of the major advances in modern computer based mathematics.
In fact, in special issue \cite{Top10Algorithms2000} 
of the ``Computing in Science and Engineering'' magazine, 
named the PSLQ  algorithm \cite{PSLQ1992, PSLQ1999} as one of the 
``top 10 algorithms'' of the 20th century.

An integer relation algorithm is a procedure that, 
given a list of real (complex) numbers 
$x = (x_1, x_2, \dots, x_n)$, 
will say whether there exists a set of integers (Gaussian integers) 
$a = (a_1, a_2, \dots, a_n)$
such that
\begin{align} \label{eqn:IntegerRelation}
	a_1 x_1 + a_2 x_2 + \dots + a_n x_n = 0 \ .
\end{align}
For such an algorithm to be useful it needs to be
fast, numerically stable and 
able to track precision or give bounds on the final output.

The first known integer algorithm is Euclid's algorithm 
which has been known since at least the time of the ancient Greeks. 
Given a pair of numbers $x$ and $y$, 
chosen such that (without loss of generality) $x>y>0$,
you can write $x = q_0 y + r_0$
for some integer quotient $q_0>0$ and some remainder $0 \leq r_0 < y$. 
This is rewritten as
$$ \frac{x}{y} = q_0 + \frac{r_0}{y} = q_0 +\cfrac1{{y}/{r_0}} \ .$$
Now since $y>r_0$, we must have $y/r_0=q_1+r_1/r_0$ 
for integer $q_1>0$ and some $0 \leq r_1 < r_0$. 
This process continues and yields the continued fraction
\[
	\frac{x}{y} = q_0 + \cfrac1{q_1+\cfrac1{q_2+\dots\cfrac1{q_n+r_n/r_{n-1}}}}
	\ .
\]
As argued in Euclid, Book VII,
for integer (or equivalently rational) $x$ and $y$, the remainders must also 
be integers.  Then, since it is a strictly decreasing sequence bounded by zero,
the algorithm terminates in a finite number steps, the final nonzero remainder 
being the greatest common divider of $x$ and $y$. 
In Euclid, Book X, the algorithm was given for line segments, or equivalently 
their real number lengths. Now the remainders are themselves line segments 
and so algorithm only terminates if the original lines are commensurable,
i.e., if their lengths $x$ and $y$ satisfy the relation
\[ 
	a_1 x + a_2 y = 0\,,\quad a_1,a_2\in\dsZ \ .
\]
In many cases, the real numbers $x$ and $y$ can only be calculated to
a certain precision (number of significant decimal digits).
Then the algorithm should terminate if the remainder is zero to the precision
at that step or when the remainder has zero precision.
In the first case, the numbers have an integer relation at that precision,
in the second case they do not. 

Euclidean-type algorithms are used in many areas of mathematics, 
however, for this appendix, the relevant generalisation is to
an algorithm on integer lattices.
Any two non-parallel vectors $\vect x, \vect y \in \dsZ^d$ 
generate a two dimensional lattice 
\( 	\L = \left\{ a_1 \vect x + a_2 \vect y | a_i\in\dsZ \right\} \).
The problem of \emph{basis reduction} is to find a \emph{short basis} 
for for $\L$, i.e., a pair $(\vect x', \vect y')$ that generates
the same lattice and is as orthogonal as possible.
For a 2D lattice, this means that the angle $\a$ between 
$\vect x'$ and $\vect y'$ must be such that $\sin(\a) \geq 1/2$.
Such a basis may be created by repeated application of the map
$(\vect x, \vect y) \to (\vect y, \vect x - q \vect y)$
where we assume that $|\vect x| \geq |\vect y|$ and define
$q = \left\lfloor\frac{\vect x\cdot \vect y}{\vect y\cdot \vect y}\right\rceil$,
where $\lfloor z \rceil = \sgn(z)\left\lfloor|z|+\frac12\right\rfloor$.
This is known as Gauss' algorithm, despite it 
first being described by Lagrange and only later by Gauss
(see \cite{Nguyen2009,Nguyen2009a} and references within).

The problem of basis reduction for higher dimensional lattices was 
solved in 1850 by Hermite, 
but since it contains the shortest vector problem as a subproblem, 
the algorithm is necessarily NP.
The first polynomial time algorithm was given in 1982 by
Lenstra, Lenstra and Lov\'asz (LLL) \cite{LLL}.
It does not yield a completely reduced basis, 
but only guarantees a ``LLL reduced'' basis --
which is good enough for most applications.

Any lattice reduction algorithm can be used as an integer relation algorithm
by taking the vector of real numbers, $\vect x\in\dsR^n$,
multiplying them by the $m\approx10^p$, where $p$ is the lowest precision, 
and truncating to get a vector of (large) integers, $\vect y\in\dsZ^n$.
The lattice is then spanned by the rows of the augmented identity matrix
$B=(\cI_n|\vect y)$. Elements in the reduce basis will be of the form
$(\vect a \,|\, \vect a \cdot \vect y)$, and since $|\vect y|$ is large, 
the smallest reduced basis element will be the smallest integer vector $\vect a$
such that $\vect a \cdot \vect y \approx 0$.
If this relation is stable under both varying the multiplier $m$
and/or increasing the precision then the result 
can be considered to be more than an numerical artifact.
If the algorithm does not terminate before the precision is exhausted, 
then bounds can be placed on the existence of an integer relation.

Finding multidimensional generalisations of the Euclidean 
integer relation algorithm has a long history, 
with attempts made by Jacobi, Hermite, Poincar\'e and others 
\cite{Ferguson1979,PSLQ1999}.
All of these attempts were without proofs and counterexamples were found.
The first proven multidimensional algorithm was given in 1977 
by Ferguson and Forcade \cite{Ferguson1979}.
The PSLQ algorithm \cite{PSLQ1992, PSLQ1999}, 
like its forbear the Ferguson-Forcade algorithm,
is an integer relation algorithm that does not rely on lattice reduction, 
although is based on lattice manipulations.
Since it can use machine arithmetic in many intermediate steps, it is faster
than other integer relation algorithms. 
It was made famous by its use in the discovery of the 
BBP formula \cite{BBP1997}
\[
	\pi = \sum_{k=0}^\infty \frac1{16^k} \left(
	\frac4{8k+1} - \frac2{8k+4} - \frac1{8k+5} - \frac1{8k+6} \right) \,,
\] 
which can be combined with a spigot algorithm to extract 
the binary/hexadecimal digits of $\pi$.
A large (and growing) number of such formul\ae{} are now known 
\cite{Bailey2000BBP} which can be used to extract the digits of 
a variety of constants in a number of bases.

PSLQ is now extensively used in the study of \emph{Euler sums} 
(which are also known as multiple zeta values) 
\cite{BaileyBorweinGirgensohn1994,Borwein1997,Borwein1998}
\[
 \zeta\!\begin{pmatrix} 	s_1,  & s_2  &\dots& s_r \\ 
 						\s_1, & \s_2 &\dots& \s_r \end{pmatrix}
 \deq \sum_{k_1>k_2>\dots>k_r>0} \frac{\s_1^{k_1}}{k_1^{s_1}}
 		\frac{\s_2^{k_2}}{k_2^{s_2}} \dots \frac{\s_r^{k_r}}{k_r^{s_r}} \,,
\]
where the $\s_i = \pm1$.
These sums occur, e.g., in the study of higher loop Feynman diagrams. 
PSLQ has also been used to directly solve the differential equations 
associated with Feynman diagrams \cite{Broadhurst1998a},
see appendix \ref{A:2LoopVac} for a discussion of the differential equations
associated with the two-loop vacuum integrals.



\section{Application to (S)QED with a self-dual background}
The self-dual low-energy effective action of $\cN=0,1,2$ SQED
can be written in terms of products and derivatives of the function 
\begin{align} \label{defn:xi-SD-SQED}
	\x(x) = -x\left(\frac1{2x}-\log(x)+\j(x)\right)
		  = \frac12\int_0^\infty\!\rmd{s}
		  		\left(\frac1{s^2}-\frac1{\sinh^s}\right)\rme^{-2sx}\,,
\end{align}
where $\j(x)=\G'(x)/\G(x)$ is the digamma function, $x \sim m^2/F$ 
and $F$ is the single degree of freedom left in the field strength.
See \cite{Dunne2001,Dunne2002,Dunne2002a,Dunne2004,
			Kuzenko2004a,Kuzenko2004c,Kuzenko2007a}
for these results.


The two-loop scalar and spinor QED effective actions
are both quadratic in $\x(x)$:
\begin{align*}
	\cL_\text{spinor}^{(2)} \propto \frac32 \x^2(x) - \x'(x) \,,\qquad
	\cL_\text{scalar}^{(2)} \propto -3 \x^2(x) + \x'(x) \ .
\end{align*}
However, the known two-loop results for supersymmetric theories 
prior to \cite{Kuzenko2007a} 
were all \emph{linear} in $\x(x)$ and its derivatives.
The $\cN=2$ SQED 2-loop effective action is proportional to 
\[ 1 + x^2\x'''(x)\,, \]
and the $\cN=4$ $SU(N)$ SYM effective action in a 
$U(1)$ gauge multiplet background is proportional to
\[ \frac{1}{x^2} - \frac{4}{x}\xi(x) + 4\xi'(x) - 2x\xi''(x) \ . \]

Like the non-supersymmetric calculations, 
the $\cN=1$ SQED two-loop effective action contains \emph{products} of $\x(x)$ 
and its derivatives. As seen in section \ref{sect:SelfDualSQED},
the calculation can be broken up into parts coming from the functions
$T(s,t)$, $f(s)$ and $\scF_\pm(s)$ defined in equations 
\eqref{SQED-T(s,t)-defn}, \eqref{defn:SQED f(s)} and \eqref{defn:scF+} 
respectively. The first two parts are straightforward to calculate, 
see section \ref{sect:SelfDualSQED}, 
but the contribution coming from $\scF_\pm$ is hard to integrate
and is the primary reason for this appendix.

\subsection{LLL in \emph{Mathematica}}
\lstset{language=Mathematica,tabsize=2,basicstyle=\footnotesize}
Since version 1, \emph{Mathematica} \cite{Mathematica}
has had an implementation of the LLL algorithm\footnote%
{For details, see 
\href{http://reference.wolfram.com/mathematica/ref/LatticeReduce.html}
	 {http://reference.wolfram.com/mathematica/ref/LatticeReduce.html}.}
which can be used to construct an integer relation algorithm.\footnote%
{Since version 8 (2010), 
\emph{Mathematica} also has an PSLQ based integer relation algorithm called 
\href{http://reference.wolfram.com/mathematica/ref/FindIntegerNullVector.html}
	{FindIntegerNullVector}.}
Based on some code by Prof.\ Paul Abbott \cite{TranscendentalRecognize1996},
we define
\begin{lstlisting}
Recognize[num_Real, basis_List, ord_?Positive] := Module[
  {vect, mat, lr, ans},
  vect = Round[10^Floor[ord - 1] Join[{num}, N[basis, ord]]];
  mat = Append[IdentityMatrix[Length[vect]], vect];
  lr = LatticeReduce[Transpose[mat]]; 
  While[lr[[1, 1]] === 0, lr = RotateLeft[lr]];
  ans = First[lr[[1]]]^(-1) Most[Rest[lr[[1]]]].basis;
  Sign[N@ans] Sign[num] ans]
Recognize[num_, b_List] := Recognize[num, b, Precision[num]]
\end{lstlisting}
We then define numerical values for the symbolic derivatives 
$\text{\texttt{Xi[n,x]}} = \x^{(n)}(x)$ and the basis lists 
\texttt{XiList1} and \texttt{XiList2}:
\begin{lstlisting}
Xi[x_] := -x (1/(2 x) - Log[x] + PolyGamma[x])
N[Xi[n_Integer, x_], prec_:MachinePrecision] := 
		N[Derivative[n][Xi][x], prec];
XiList1[x_, m_] := Prepend[Table[Xi[n, x], {n, 0, m}], 1]
XiList2[x_, n_] := Union@Flatten@Outer[Times,#,#]&@XiList1[x,n]
\end{lstlisting}

In the following subsections we outline how the above code is used to find the 
closed form for some of the integrals examined in 
section \ref{sect:SelfDualSQED}.
A more complete examination can be found in the attached \emph{Mathematica} 
notebook \cite{LLLNotebook}.
This notebook also contains code for the series expansion check mentioned 
at then end of section \ref{sect:SelfDualSQED}.

\subsubsection{One-loop}
As a quick check, we see if \texttt{Recognize} can identify the correct 
one-loop result. First, define the numerical integral
\begin{lstlisting}
oneLoop[x_ , prec_:MachinePrecision] := 
	NIntegrate[s(1/s^2 - 1/Sinh[s]^2)E^(-2 s x), 
	{s, 0, Infinity}, WorkingPrecision -> prec]
\end{lstlisting}
We can then check that it works for the first few integer values of $x$.
Using 20 digits of precision, 
\begin{lstlisting}
Table[Recognize[oneLoop[x, 20], XiList1[x, 4]], {x, 5}]
\end{lstlisting}
returns
\begin{lstlisting}
{-Xi[1, 1], -Xi[1, 2], -Xi[1, 3], -Xi[1, 4], -Xi[1,5]}
\end{lstlisting}
which is consistent with the analytic result
\[
\int_0^{\infty}\!\!\rmd{s}\, 
	s \left(\frac{1}{s^2}-\frac{1}{\sinh ^2(s)}\right) \rme^{-2 s x}
	= -\xi'(x) \ .
\]

\subsubsection{Two-loop: \texorpdfstring{$I_\scF$}{I_scF}}
Here we are interested in finding a closed form expression for the integral
\begin{align*}
	2x^2\int_0^\infty\!\!\!\rmd{s}\rmd{t}\,
		{\sinh^2(s+t)}\left(\frac{s^2}{\sinh^2s}
			-\frac{st\cosh(s+t)}{\sinh s \sinh t}\right)\rme^{-2x(s+t)}\ .
\end{align*}
Although we could do a two dimensional numerical integral, 
it is slow to get the required precision for the \texttt{Recognize} function.
It is best to use the change of variables \eqref{defn:2DPropTimeCOV}
and then perform the $\a$-integral to get the one-dimensional integrand
\begin{lstlisting}
intgnd[x_, t_] := 2 x^2 Csch[t]^2 E^(-2 t x) (
	t Coth[t] PolyLog[2, E^(2 t)] - PolyLog[2, E^(-2 t)] 
	- Coth[t] PolyLog[3, E^(2 t)] 
	+ (Zeta[3] + Zeta[2] t - t^2 + 1/3 t^3) Coth[t] 
	- 3 t^2 + 2 t Log[E^(2 t) - 1] + Zeta[2])
\end{lstlisting}
that is integrated over all positive $t$. 
We define the memoized, arbitrary precision numerical integral
\begin{lstlisting}
int[x_, prec_:MachinePrecision, mr_:Automatic] := 
 int[x, prec, mr] = Total[ParallelTable[Block[{NIntegrate}, 
   NIntegrate[intgnd[x, t], {t, Sequence@@pts}, 
   WorkingPrecision -> prec, MaxRecursion -> mr]], 
   {pts, Partition[{0, 1, 10, Infinity}, 2, 1]}]]
\end{lstlisting}
where, from examining the shape of \texttt{intgnd[x,t]}, 
we've broken the integral into the ranges $(0, 1), (1, 10), (10, \infty)$
to aid the convergence.

A bit of experimentation shows that the working precision of 32 decimal places
is sufficient to identify the integral. 
Running 
\begin{lstlisting}
intTable = Table[Recognize[int[x, 32], 
	Union[XiList1[x, 3], XiList2[x, 1]]], {x, 1, 7}]
\end{lstlisting}
returns a list containing the identities for the first 7 integer values of $x$. 
We can then use the \emph{Mathematica} function \texttt{Fit} to fit the
coefficients of the basis to polynomials in $x$,
\begin{lstlisting}
Transpose[List @@@ Expand[intTable]] /. Xi[_, _] -> 1;
Fit[#, {1, x, x^2, x^3}, x] & /@ % // Chop // Rationalize
\end{lstlisting}
This outputs
\begin{lstlisting}
{1/4,-2 x,-1,2 x^2,2 x,-x^2,-x^3}
\end{lstlisting}
from which we obtain the result \eqref{eqn:SQED-intscF}.

\subsubsection{Integer relations in \texttt{XiList2}}
It turns out that for specific integer $x$, the basis functions
\texttt{XiList2[x, 3]} are actually overcomplete. 
This can cause problems in using integer relations algorithms 
to identify $I_\scF$. In the case of the LLL-based 
\texttt{Recognize} algorithm above, this is problem is mostly avoided 
by the inclusion of the \texttt{RotateLeft} line to choose 
the first row of the reduced basis that has a nonvanishing dependence 
on the number of interest. 
Algorithms not based on a full lattice reduction often do not give this option 
and the use of an overcomplete basis becomes problematic. 
For each $x$, the basis would have to be reduced to a minimal one,
and then the representation of the quantity of interest would have to be
interpreted modulo the basis relation for that specific $x$. 

Although we won't examine this issue further, we will use the built-in
PSLQ-based \texttt{FindIntegerNullVector} algorithm to identify the 
integer relations obeyed by the basis used in the $I_\scF$ calculation,
\begin{lstlisting}
Union[XiList1[x, 3], XiList2[x, 1]] == 
{1, Xi[0, x], Xi[0, x]^2, Xi[1, x], Xi[0, x] Xi[1, x], 
	Xi[1, x]^2, Xi[2, x], Xi[3, x]}
\end{lstlisting}
for the first few values of $x$.
The command
\begin{lstlisting}
Xi2Rel = Table[FindIntegerNullVector[
   N[Union[XiList1[x, 3], XiList2[x, 1]], 150]], {x, 1, 10}]
\end{lstlisting}
returns the table of identities
\begin{equation}\left(
{
\begin{smallmatrix}
 -2 & -6 & -12 & 6 & 24 & -12 & 15 & -5 \\
 4 & -63 & -6 & 126 & 24 & -24 & 30 & -20 \\
 -43 & 864 & 32 & -2592 & -192 & 288 & -360 & 360 \\
 -106 & 2703 & 54 & -10812 & -432 & 864 & -1080 & 1440 \\
 -35329 & 1101504 & 13824 & -5507520 & -138240 & 345600 & -432000 & 720000 \\
 \vdots & \vdots & \vdots & \vdots & \vdots & \vdots & \vdots & \vdots 
\end{smallmatrix} }
\right)\end{equation}
this table generates \emph{exact} integer relations, as can be checked using
\begin{lstlisting}
Table[Xi2Rel[[x]].Union[XiList1[x, 3], XiList2[x, 1]], 
 {x, 1, 10}] /. Xi[n_, x_] :> Derivative[n][Xi][x] // Expand
\end{lstlisting}
which returns a list of zeros.

\subsection{PSLQ in python with mpmath}
The multi precision mathematics library for Python%
\footnote{\url{htpp://www.python.org}}, 
mpmath \cite{mpmath}, is used in both of the major
Python based computer algebra systems,
Sage \cite{sage} and SymPy \cite{sympy}.
It contains an implementation of PSLQ
that is also tied into various built-in number recognition algorithms.%
\footnote{See 
\href{http://mpmath.googlecode.com/svn/trunk/doc/build/identification.html}%
	 {http://mpmath.googlecode.com/svn/trunk/doc/build/identification.html} 
for more details.}

The following code tests the identity
\[
	\int_0^\infty s^3 \text{csch}^2(s) e^{-2 s x}\rmd{s}
	= \frac{1}{4 x^2}\left(1+x^2\xi'''(x)\right)\,,
\]
that occurs in the integral $I_\rmII$ of equation \eqref{eqn:SQED-intII}.
The code is set to use 50 digits of precision 
and will run on any computer with a working install of python and mpmath.
\begin{lstlisting}[language=Python, breaklines=true,
	basicstyle=\footnotesize, tabsize=2]
from mpmath import mp, quad, pslq, inf, diff, \
	exp, csch, log, psi
mp.dps = 50; mp.pretty = True

xi = lambda x: -x * (1/(2*x) - log(x) + psi(0,x))

def xiList1(x, i):
	return [1] + [diff(xi, x, i) for i in range(i+1)]
	
###########################
# Check the relation 
# 4 * x**2 * I2a(x) == 1 + x**2 * xi'''(x)

def I2a(x): 
	return quad((lambda s: s**3 * csch(s)**2 * exp(-2*x*s)), [0, inf])

print [pslq([I2a(x)] + xiList1(x, 4)) for x in range(1, 6)]
# [[-4,  1, 0, 0, 0, 1, 0], 
#  [-16, 1, 0, 0, 0, 4, 0], 
#  [-36, 1, 0, 0, 0, 9, 0], ...]
\end{lstlisting}

The final commented lines above are the output of the program 
and clearly demonstrate that the result \eqref{eqn:SQED-intII}
found for $I_\rmII$ is correct.

\chapter{Two-loop vacuum integrals}\label{A:2LoopVac}


\section{Introduction}
This chapter is devoted to the study of the two-loop vacuum integrals. 
For convenience we restrict our attention to 
$d=4-2\eps$ dimensional Euclidean integrals, 
the Wick rotation to Minkowski integrals being straightforward.

There are only two possible two-loop topologies: \\
1) The figure-eight graph
\begin{align} \label{defn:Fig8}
	J(x_1,x_2)=
	\raisebox{-0.4\height}{\scalebox{.8}{\begin{picture}(0,0)%
\includegraphics{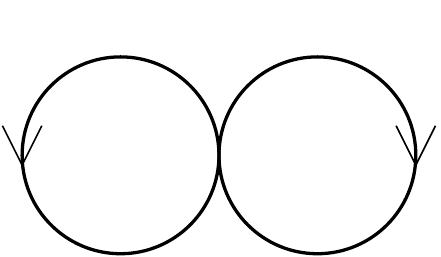}%
\end{picture}%
\setlength{\unitlength}{4144sp}%
\begingroup\makeatletter\ifx\SetFigFont\undefined%
\gdef\SetFigFont#1#2#3#4#5{%
  \reset@font\fontsize{#1}{#2pt}%
  \fontfamily{#3}\fontseries{#4}\fontshape{#5}%
  \selectfont}%
\fi\endgroup%
\begin{picture}(2004,1163)(2599,-1875)
\put(2791,-1456){\makebox(0,0)[lb]{\smash{{\SetFigFont{12}{14.4}{\rmdefault}{\mddefault}{\updefault}{\color[rgb]{0,0,0}$\vec p$}%
}}}}
\put(4411,-1456){\makebox(0,0)[rb]{\smash{{\SetFigFont{12}{14.4}{\rmdefault}{\mddefault}{\updefault}{\color[rgb]{0,0,0}$\vec k$}%
}}}}
\put(3826,-871){\makebox(0,0)[lb]{\smash{{\SetFigFont{12}{14.4}{\rmdefault}{\mddefault}{\updefault}{\color[rgb]{0,0,0}$x_2=m_2^2$}%
}}}}
\put(3376,-871){\makebox(0,0)[rb]{\smash{{\SetFigFont{12}{14.4}{\rmdefault}{\mddefault}{\updefault}{\color[rgb]{0,0,0}$x_1=m_1^2$}%
}}}}
\end{picture}%
}}
	= \int\!\!\rmd{k}\,\rmd{p}\,
		\frac1{p^2+m_1^2}\frac1{k^2+m_2^2} \,; 
\end{align}
2) The fish/sunset diagram
\begin{align}
 \label{defn:Fish}
	I(x_1,x_2,x_3)=
	\raisebox{-0.5\height}{\scalebox{.8}{\begin{picture}(0,0)%
\includegraphics{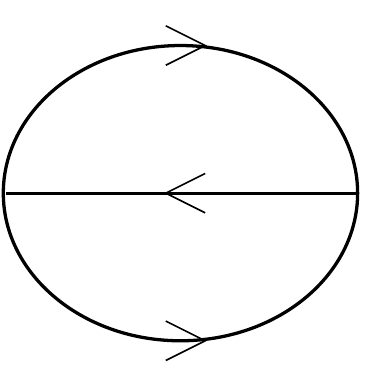}%
\end{picture}%
\setlength{\unitlength}{4144sp}%
\begingroup\makeatletter\ifx\SetFigFont\undefined%
\gdef\SetFigFont#1#2#3#4#5{%
  \reset@font\fontsize{#1}{#2pt}%
  \fontfamily{#3}\fontseries{#4}\fontshape{#5}%
  \selectfont}%
\fi\endgroup%
\begin{picture}(1653,1758)(2438,-2290)
\put(3376,-1636){\makebox(0,0)[lb]{\smash{{\SetFigFont{12}{14.4}{\rmdefault}{\mddefault}{\updefault}{\color[rgb]{0,0,0}$\vec p + \vec k$}%
}}}}
\put(3376,-961){\makebox(0,0)[lb]{\smash{{\SetFigFont{12}{14.4}{\rmdefault}{\mddefault}{\updefault}{\color[rgb]{0,0,0}$\vec p$}%
}}}}
\put(3376,-1996){\makebox(0,0)[lb]{\smash{{\SetFigFont{12}{14.4}{\rmdefault}{\mddefault}{\updefault}{\color[rgb]{0,0,0}$\vec k$}%
}}}}
\put(2521,-1321){\makebox(0,0)[lb]{\smash{{\SetFigFont{12}{14.4}{\rmdefault}{\mddefault}{\updefault}{\color[rgb]{0,0,0}$x_2=m_2^2$}%
}}}}
\put(2476,-2221){\makebox(0,0)[lb]{\smash{{\SetFigFont{12}{14.4}{\rmdefault}{\mddefault}{\updefault}{\color[rgb]{0,0,0}$x_3=m_3^2$}%
}}}}
\put(2476,-691){\makebox(0,0)[lb]{\smash{{\SetFigFont{12}{14.4}{\rmdefault}{\mddefault}{\updefault}{\color[rgb]{0,0,0}$x_1=m_1^2$}%
}}}}
\end{picture}%
}}
	= \int\!\!
		\frac{\rmd{k}\,\rmd{p}}{(p^2+m_1^2)((p+k)^2+m_2^2)(k^2+m_3^2)} \,,
\end{align}
where we've used the dimensionally regularised momentum measure 
$\rmd{k}=\m^{2\eps}{(2\p)^{-d}}{\rmd^dk}$
and similarly for $\rmd{p}$.

The figure-8 graph (like any 1-vertex reducible graph) decomposes into 
a product of separate Feynman integrals,
\begin{align} \label{defn:Fig8Split}
	J(x_1,x_2) = J(x_1) J(x_2)\ .
\end{align}
Where $J(m^2)$ is the one-loop integral defined by 
\begin{align} \label{defn:J_Integral}
	J(m^2) 
	&= \m^{2\eps}\intk\frac1{k^2+m^2} 
	= \frac{\m^{2\eps}}{(4\p)^{d/2}}\int_0^\infty
		\frac{\rmd s}{s^{d/2}}\rme^{-m^2s} \\ \non
	&= \frac{\m^{2\eps}}{(4\p)^{d/2}}\G\big(1-\frac{d}{2}\big)m^{d-2}
	\approx -\frac{m^2}{(4\p)^2}\left(\frac1\eps+1-\log\frac{m^2}{\mub^2}
		+\ord\eps\right) ,
\end{align}
and $\mub^2=4\p\rme^{-\g}\m^2$ is the 
modified minimal subtraction (\MSb) renormalisation point.
Note that this integral is related to \eqref{defn:scJ_Integral} via
\begin{align*} 
	J(m^2) = m^2 \frac{\pd}{\pd{m^2}} \scJ(m^2) \ .
\end{align*}

\section{Fish diagram}
In this section we present a derivation of a closed form for \eqref{defn:Fish}.
In the literature there are four main approaches to calculating this integral.  
It can be directly calculated, as in \cite{Davydychev1993} 
where the Mellin-Barnes representation for the propagators is used, 
or it can be calculated indirectly by exploiting the 
different 
differential equations 
\cite{Kotikov1991, Kotikov1991a, Tkachov1981, Chetyrkin1981}
that $I(x_1,x_2,x_3)$ has to satisfy.  
The first differential equation is the homogeneity equation,
\begin{equation}\label{eqn:FishHomogEqn}
  (1-2\eps - \sum_i x_i\pd_{x_i})I(x_1,x_2,x_3)=0~,
\end{equation}
and was used in \cite{VanDerBij1984,Hoogeveen1985,McDonald2003} to express
$I(x_1,x_2,x_3)$ in terms of its first derivatives, 
which have a nicer $\eps$-expansion.
The second type of differential equation is the ordinary differential 
equation of \cite{Remiddi1998}.%
\footnote{Also used in \cite{Ford1992a} 
for the special case of two equal masses.}
The final approach uses first-order partial differential equations 
and was first given in \cite{Ford1992}. 
It is the approach that we'll re-examine in this section.
All of these approaches must yield equivalent results and
we leave such comparisons to \cite{Tyler2008} and references within.

In \cite{Ford1992} and later publications, e.g., 
\cite{Nibbelink2005, Kuzenko2007, Tyler2008}, a sequence 
of first order partial differential equations was derived by using 
integration by parts identities for the momentum integrals \eqref{defn:Fish}.
These equations were then integrated using the method of characteristics.
In this appendix we present a new way of looking at these differential equations
and show that they belong to a group of continuous flows of the mass parameters.
This new point of view is enlightening in that it shows how such groups can
be derived for any Feynman diagram and how they're related to the 
contraction/deletion properties of the Symanzick polynomials%
\footnote{Mathematica code for constructing Feynman diagrams and calculating
their properties (including the Symanzick polynomials) has been made available
at \url{%
http://demonstrations.wolfram.com/ScalarFeynmanDiagramsAndSymanzikPolynomials/}.
} 
of that diagram. 
Research in this direction is ongoing 
and we leave the details to future publications.

We start by writing the fish diagram \eqref{defn:Fish} using the 
Schwinger propertime parameters 
(the Schwinger-Nambu representation \cite{Cvitanovic1974})
\begin{align}
 \label{eqn:Fish:PT}
	I(x_1,x_2,x_3) 
	= \frac{\m^{4\eps}}{(4\p)^d}\int_0^\infty \frac{\rme^{-s_1x_1-s_2x_2-s_3x_3}
		\rmd{s_1}\rmd{s_2}\rmd{s_3}}{(s_1s_2+s_2s_3+s_3s_1)^{d/2}}\,,
\end{align}
where $s_1s_2+s_2s_3+s_3s_1$ is the (first) Symanzik polynomial 
\cite{Nakanishi1971, ItzyksonZuber1980, Bogner2010} 
for the graph \eqref{defn:Fish}.
We can recast this into a vector form
\begin{align}
 \label{eqn:Fish:PT:Vec}
	I(\bfx) 
	= \frac{\m^{4\eps}}{(4\p)^d}\int_0^\infty \frac{\rme^{-\bfs^\rmT\bfx}
		\rmd^3\bfs}{(\bfs^\rmT Q \bfs)^{d/2}} 
	\quad\text{where}\quad
	Q = \frac12\bem0&1&1\\1&0&1\\1&1&0\eem \ .
\end{align}
Now let $\bfx=\bfx(t)=\rme^{A t}\bfx(0)$ for some constant matrix $A$
and consider the variation of $I(\bfx(t))$ with respect to $t$:
\begin{align} 
	\ddt{} I(\bfx) 
	= \bfx^\rmT A^\rmT \pd_\bfx I(\bfx)
	= \frac{\m^{4\eps}}{(4\p)^d}\int_0^\infty \frac{\bfs^\rmT A \pd_\bfs 
		\rme^{-\bfs^\rmT\bfx}\rmd^3\bfs}{(\bfs^\rmT Q \bfs)^{d/2}}\,,
\end{align}
integrate by parts to get
\begin{align*} 
	\ddt{} I(\bfx) 
	= \frac{-\m^{4\eps}}{(4\p)^d}\!\!\int_0^\infty\!
		\frac{s_j A_{ji}\rme^{-\bfs^\rmT\bfx}}%
		{(\bfs^\rmT Q \bfs)^{d/2}}\rmd^2\bfs\mathop{\Big|}_{\mathrlap{s_i=0}}
	- \tr(A)I(\bfx) 
	+ \frac{d\,\m^{4\eps}}{(4\p)^d}\!\!\int_0^\infty\!
		\frac{\bfs^\rmT A Q \bfs\,
		\rme^{-\bfs^\rmT\bfx}}{(\bfs^\rmT Q \bfs)^{d/2+1}}\rmd^3\bfs\,.
\end{align*}
The first term above is the sum of diagrams with the $i^{\text{th}}$ line
contracted while the final term is proportional to $I(\bfx)$ if
\begin{align} \label{eqn:flowGroupPlusHomog}
	A Q + Q A^\rmT = k_A Q \ .
\end{align}
Taking the trace of the above implies that $k_A = \frac23\tr(A)$.
So, assuming that \eqref{eqn:flowGroupPlusHomog} holds, 
we have the flow equation
\begin{align} \label{eqn:flowEqnsPlusHomog}
	\ddt{} I(\bfx) 
	= \sum_{i,j} A_{ji}\pd_{x_j}J(x_{i+1})J(x_{i+2})
	 + \frac{d-3}3 \tr(A)I(x)\ ,
\end{align}
where we're using cyclic indices on the $x_i$ in the first term.
When $A$ is proportional to the identity matrix
\eqref{eqn:flowEqnsPlusHomog} reduces to the homogeneity equation 
\eqref{eqn:FishHomogEqn}. 
To remove the homogeneity equation we can restrict $A$ to be traceless 
and we obtain the final set of flow equations
\begin{align} \label{eqn:flowEqns}
\begin{gathered}
	\ddt{} I(\bfx) 
	= \bfx^\rmT A^\rmT \pd_\bfx I(\bfx)
	= \sum_{i,j} A_{ji}\pd_{x_j}J(x_{i+1})J(x_{i+2}) \\
	\text{where}\quad
	A Q + Q A^\rmT = 0 \ .
\end{gathered}
\end{align}
This gives the variation of the integral as we change the masses
purely as a sum of contractions of the original Feynman diagram.

Equation \eqref{eqn:flowEqns} means that $A$  
must be in the Lie algebra that leaves the symmetric form $Q$ invariant. 
The eigenvectors of $Q$ are
$(1,1,1)$, $(1,-1,0)$ and $(0,1,-1)$ with the 
eigenvalues $1$, $-\half$ and $-\half$ respectively, meaning that
the Lie algebra is isomorphic to $\so(1,2)$. 
We choose our generators to be
\begin{align} \label{defn:FishGenerators}
	A_1&=\bem0&1&-1\\-1&0&1\\1&-1&0\eem  \,,&
	A_2&=\bem1&1&-1\\1&1&-1\\0&0&-2\eem  \,,&
	A_3&=\bem2&0&0\\1&-1&-1\\1&-1&-1\eem \,.
\end{align}
Expanding out \eqref{eqn:flowEqns} shows that linear combinations of 
these generators (combined with the homogeneity equation) 
yield all of the differential equations that can be obtained in the usual way
by hitting the integrand of \eqref{defn:Fish}
by the operator $\pd_{k_i}k_j\cdot$ (where $k_{i,j}\in\{k,p\}$) 
and integrating by parts.
Exponentiating these generators is easy by observing that they are 
(proportional to) periodic matrices:
$A_1^3=-3A_1$ and $A_{2,3}^3=4A_{2,3}$.

Since $A^\rmT$ leaves $\bfs^\rmT Q \bfs$ invariant, 
$A$ must leave invariant the polynomial
\begin{align} \label{defn:FishDelta}
	\D(\bfx) = \bfx^\rmT Q^{-1} \bfx 
	= 2(x_1x_2+x_2x_3+x_3x_1)-x_1^2-x_2^2-x_3^2 \ .
\end{align}
Thus our differential equations flow along surfaces of constant $\D$.
These surfaces are hyperbola centered around the vector $(1,1,1)$ 
and since $x_i\geq0$, the flows are restricted to 
either the upper sheet of the two sheeted hyperbola for $\D>0$
or three sections of the single sheeted hyperbola for $\D<0$.
We can move between the three slices for $\D<0$ using the invariance
of the integral under permutations of the masses.
The symmetric form $\D$ is exactly that which arises naturally 
(but less directly) in all of the other approaches 
to evaluating $I(\bfx)$ discussed above. 
$\D(\bfx)$ can be written as the Cayley determinant 
\begin{align*} 
 \D(x,y,z) = -\det\bem 0&1&1&1\\1&0&x&y \\ 1&x&0&z \\ 1&y&z&0 \eem ,
\end{align*}
and, since 
\begin{align*} 
	\frac{\D(m_1^2,m_2^2,m_3^2)}{(m_1+m_2+m_3)}
	= (-m_1+m_2+m_3)(m_1-m_2+m_3)(m_1+m_2-m_3) \,,
\end{align*}
$\D$ only vanishes if one of the masses is equal to the sum of the other two.
This condition naturally arises follows from the combination of 
the BPS condition $m_i=Z|e_i|$ and charge conservation
in $\cN=2$ calculations in a covariantly constant background \cite{Kuzenko2007}.

Our approach to integrating \eqref{eqn:Fish:PT} is to follow 
\cite{Ford1992} (and \cite{Nibbelink2005, Kuzenko2007, Tyler2008})
and choose a sequence of two simple flows to take an 
arbitrary mass configuration to the simple endpoints 
$I(\sqrt{-\D},0,0)$ and $I(\sqrt{\D/3},\sqrt{\D/3},\sqrt{\D/3})$ 
for $\D<0$ and $\D\geq0$ respectively.
We then relate the two using analytic continuation to get a single, 
symmetric form for the result which is then expanded around $d=4$.

We choose the first flow to be the rotation (about $(1,1,1)$) generated by $A_1$
that takes some $\bfx(0)=(X,Y,Y)$ to the general point $\bfx(t)=(x_1,x_2,x_3)$.
This rotation preserves the quantity $c =x_1+x_2+x_3$ as well as $\D$, 
so we can solve $\D = X (4 Y-X)$ and $c  = X+2Y$ to find
\begin{align*} 
	3X = c   +  \sqrt{c ^2-3\D}\,,\quad
	6Y = 2c  -  \sqrt{c ^2-3\D} \ .
\end{align*}
Note that we don't need the explicit forms of $X$ and $Y$, 
nor do we need to know the explicit value of $t$, we merely need
to know that we can reach $X$ and $Y$ without traveling through negative masses.
This means that for $\D<0$ we must use the permutation symmetry 
to choose $x_1 \geq x_2+x_3$.
The differential equation \eqref{eqn:flowEqns} generated by $A_1$ is
\begin{align} 
	\ddt{}I(\bfx) &= J'(x_1)(J(x_3)-J(x_2)) + \text{cyclic} \\ \non
		 &= - \G\Big(1-\frac d2\Big)\G\Big(2-\frac d2\Big)
		 \Big(\dot{x}_1\Big(\big(x_1-\frac{c }{2}\big)^2
		 	+\frac\D4\Big)^{d/2-2} + \text{cyclic} \Big)\,,
\end{align}
where in the last line we used the one-loop result \eqref{defn:J_Integral}
and the kinematic flow (using cyclic indices)
\begin{align} \label{eqn:A1_flow}
	\dot x_i = {\sum}_j A_{ij}x_j = x_{i+1}-x_{i+2}\ .
\end{align}
This can be integrated as
\begin{align} \label{FishFlow1}
\begin{aligned}
	I(x,y,z) &= I(X,Y,Y) - \G\Big(1-\frac d2\Big)\G\Big(2-\frac d2\Big)
	\times\\ &\times
	\left(\int_{X-\frac{c }{2}}^{x_1-\frac{c }{2}}
		+\int_{Y-\frac{c }{2}}^{x_2-\frac{c }{2}}
		+\int_{Y-\frac{c }{2}}^{x_3-\frac{c }{2}}\right)
	\frac{\rmd s}{\left(s^2+\D/4\right)^{2-d/2}}\ .
\end{aligned}
\end{align} 

For the second flow, we use the hyperbolic rotation generated by $A_3$
to move from $\mathbf{X}(0)=(X_0,Y_0,Y_0)$ to $\mathbf{X}(t)=(X,Y,Y)$.
As above, we can integrate the differential equation
\begin{align} 
	\ddt{} I(X,Y,Y) = 2J'(Y)(J(Y)-J(X))\,,
\end{align}
to get
\begin{align} \label{FishFlow2}
\begin{aligned}
	I(X,Y,Y) &= I(X_0,Y_0,Y_0) - \G\Big(1-\frac d2\Big)\G\Big(2-\frac d2\Big)
	\times\\ &\times
	\left(\int_{\frac{X_0}{2}-Y_0}^{\frac{X}{2}-Y}
		+\int_{-\frac{X_0}{2}}^{-\frac{X}{2}}\right)
	\frac{\rmd s}{\left(s^2+\D/4\right)^{2-d/2}} \ .
\end{aligned}
\end{align} 

We combine the two flows \eqref{FishFlow1} and \eqref{FishFlow2} 
and use the appropriate values for $X_0$ and $Y_0$ to get
\begin{subequations}\label{FishFlowTotal}
\begin{align} 
	I(\bfx) 
	\xlongequal{\D<0}\, &I(\sqrt{-\D},0,0) 
		+ \G\Big(1-\frac d2\Big)\G\Big(2-\frac d2\Big) \times  \\\non
		  &\times\Big(-F(x_1-\frac{c }{2}) + F(\frac{c }{2}-x_2) 
		  	+ F(\frac{c }{2}-x_3)\Big) \,, \\
	\xlongequal{\D\geq0} \,
		&I\Big(\sqrt{\frac{\D}{3}},\sqrt{\frac{\D}{3}},\sqrt{\frac{\D}{3}}\Big) 
		+ \G\Big(1-\frac d2\Big)\G\Big(2-\frac d2\Big) \times \\\non
		  &\times\Big(G(\frac{c }{2}-x_1) + G(\frac{c }{2}-x_2) 
		  	+ G(\frac{c }{2}-x_3)- 3G(\sqrt{\frac{\D}{12}})\Big) \,, 
\end{align}
\end{subequations}
where we've defined the integrals
\begin{align} \label{defn:GandF}
	G(w) = \int_0^w \!\!
		\frac{\rmd s}{\left(s^2+\Delta /4\right)^{2-\frac d2}} \,,\quad
	F(w) = \int_{\!\sqrt{-\frac{\D}{4}}}^w 
		\frac{\rmd s}{\left(s^2+\Delta /4\right)^{2-\frac d2}} \ ,
\end{align}
which can be written as ${}_2F_1$ hypergeometric functions.
Note that becomes simple to integrate $G(w)$ whenever $\D=0$.
Also, the integral $I(0,0,0)$ is scaleless and thus vanishes in dimensional
regularisation. This gives us the special case
\begin{align} 
	I(\bfx) \xlongequal{\D=0} \,
	\frac{\G\Big(1-\frac d2\Big)\G\Big(2-\frac d2\Big)}{2^{d-3}(d-3)}
	\left(\left({-x_1+x_2+x_3}\right)^{d-3}+\text{cyclic}\right)\ .
\end{align}
The $\D\geq0$ and $\D<0$ cases may be combined by either using the
explicit hypergeometric forms of $F$, $G$, $I(1,0,0)$ and $I(1,1,1)$
\cite{Kuzenko2007}
or by a careful analytic continuation of the $\D<0$ case.
Either way, we obtain the closed form that holds for all real $\D$
(n.b., in the following we always choose the 
principle branch for the square root and logarithm)
\begin{align} 
	I(\bfx) 
	&=	-\sin\Big(\frac{d\p}{2}\Big) \, I(\sqrt{\D},0,0) 
		+ \G\Big(1-\frac d2\Big)\G\Big(2-\frac d2\Big) \times \\\non
		  &\qquad\times\Big(G(\frac{-x_1+x_2+x_3}{2}) + \text{cyclic}\Big) \,.
\end{align}

In \cite{Tyler2008},
inspired by the work of \cite{Ford1992} and \cite{Davydychev2000}, 
it was shown that the expansion of $I(\bfx)$ can be written as
\begin{align} \label{eqn:2LoopExpansion}
	I(x_1,x_2,x_3) = \i(x_1) + \i(x_2) + \i(x_3) 
					+ \dsI(x_1,x_2,x_3) + \ord\eps\,,
\end{align}	
where all regularisation and renormalisation dependent details 
are in the the $\i$ terms
\begin{align} \label{defn:2Loop_Iota}
	\!\i(x) = -\frac{x}{2}\left(\frac{1}{\eps^2}
			-\frac{2}{\eps}\log\Big(\frac{x}{\hat\mu^2}\Big)
			+\z(2)+\frac{5}{2}+2\log^2\Big(\frac{x}{\hat\mu^2}\Big)
			+\ord\eps \!\right) ,
\end{align}
and we introduce the convenient renormalisation point
\begin{align} \label{defn:muh}
	 \muh^2 = \rme^{3/2}\mub^2 = 4\pi\rme^{3/2-\gamma}\mu^2 \ .
\end{align}
The finite, homogeneous term in \eqref{eqn:2LoopExpansion} is 
\begin{align} \label{defn:2Loop_dsI}
	\dsI(x_1,x_2,x_3)
	= \left(
	  \frac{x_1}{2} \log\Big(\frac{x_2}{x_1}\Big) \log\Big(\frac{x_3}{x_1}\Big)
		-\sqrt{\Delta} N\big(2\q_1\big) \right)
		+\text{cyclic}  \,,
\end{align}
where we defined the function 
\begin{align} \label{defn:2Loop_N}
	N(\q) &= -\int_0^\q \log\left(2\cos\left(\frac\f2\right)\right)\rmd\f  \,,
\end{align}
which is related to the Lobachevsky function used in \cite{Ford1992},
it's also related to the the log-cosine function \cite{Davydychev2000},
dilogarithm and Clausen functions 
\cite{VanDerBij1984,Hoogeveen1985,McDonald2003,Lewin1981}.
The angles appearing in \eqref{defn:2Loop_dsI} are defined by
\begin{align} \label{defn:2Loop_theta}
	\q_1 &= \arctan\left(\frac{-x_1+x_2+x_3}{\sqrt{\Delta}}\right)
	\quad\text{and cyclic} \ .
\end{align}
Comparisons of the above results with the others found in the literature
can be found in \cite{Tyler2008}.

Finally, we note that the separation of all regularisation and renormalisation
dependence into a sum of terms that depend on only one mass parameter 
is indispensable in cleanly demonstrating the 2-loop finiteness of the
$\b$-deformed super-Yang-Mills K\"ahler potential in chapter \ref{Ch:BetaDef}.

\chapter{Goldstino appendices}\label{A:Golden}
The following sections are needed for the study of the 
Goldstino actions in chapter \ref{ch:Goldstino}.
Since they are both quite short and do not stand by themselves,
they have been combined into a single appendix.

\section{Minimal basis for Goldstino actions}\label{sect:bases}
In order to easily compare different Goldstino actions, we need to be able
to write all terms in a common basis of Lorentz invariant terms.
We restrict our attention to terms that occur in Goldstino actions without
interactions or higher derivative terms and
thus have the structure given in \eqref{eqn:GoldStruct}.
Obviously, there is a lot of freedom in the choice of such a basis. 
We have chosen a basis where as many elements as possible can be 
written as traces of the matrices $v=(v_a{}^b)$ and $ \vb = (\vb_a{}^b)$
defined in  \eqref{defn:v}.

We choose the minimal basis for 4-fermion terms to be
\begin{align} \label{eqn:4fermion_basis}
	\expt{v^2},	\quad	\expt{\vb^2},	\quad 	\expt{v}\expt{\vb}, \quad
	\expt{v}^2,	\quad	\expt{\vb}^2,	\quad	\pd^a\l^2\pd_a\bar\l^2\,.
\end{align}
In this basis, the structure that occurs in the 
AV action \eqref{eqn:AV2} becomes
\begin{align}\label{<v><vb>-<vvb>} 
	2\left(\expt{v}\expt{\vb}-\expt{v\vb}\right) 
	&= \expt{v}^2-\expt{v^2}+\expt{\vb}^2-\expt{\vb^2}\ .
\end{align}

The 6-fermion basis is chosen to be
\begin{gather} \label{eqn:6fermion_basis}
	\expt{v^2\vb}, \;
	\expt{v}\expt{v\vb},\;				\expt{\vb}\expt{v\vb},\;
	\expt{v^2}\expt{\vb},\; 			\expt{v}\expt{\vb^2},\;
	\expt{v}^2\expt{\vb},\; 			\expt{v}\expt{\vb}^2,  \non\\
	\expt{v}\pd^a\l^2\pd_a\bar\l^2,\;	\expt{\vb}\pd^a\l^2\pd_a\bar\l^2,\;
	\rmi\l\s^a\bar\l(\expt{v}\oLRa\pd_a\expt{\vb})\ ,
\end{gather}
where the first term is the only one that is neither real 
nor in a complex conjugate pair 
and in the last term we use the symbol 
\( x\oLRa\pd_ay = x(\pd_ay)-(\pd_ax)y \ .\)
When writing 6-fermion expressions, we will often use the overcomplete
basis that includes the complex conjugate of the first term
\begin{align} 
	\expt{v\vb^2} = \expt{v^2\vb} + \half\Big(\expt{v}^2\expt{\vb}
		-2\expt{v}\expt{v\vb}-\expt{v^2}\expt{\vb}	-\cc \Big)\ .
\end{align}
Most actions in this paper have also been simplified
by rewriting the last basis element in \eqref{eqn:6fermion_basis} 
using the extra terms 
$\expt{v}\bar\l^2\Box\l^2$ and $\expt{\vb}\l^2\Box\bar\l^2$:
\begin{align} 
	\rmi\l\s^a\bar\l(\expt{v}\oLRa\pd_a\expt{\vb})
	&= \Big(\frac12\expt{v^2}\expt{\vb}-\expt{v}\expt{v\vb}
		-\frac14\expt{v}\bar\l^2\Box\l^2 \non\\
	&\quad - \frac12\expt{v}\pd^a\l^2\pd_a\bar\l^2\Big) + \cc
\end{align}

All 8-fermion terms can be written as traces. 
We choose the basis
\begin{equation}
\begin{gathered}
	\expt{v}\expt{v\vb^2},\quad 			\expt{\vb}\expt{v^2\vb},\quad
	\expt{v}^2\expt{\vb^2},\quad\;			\expt{v^2}\expt{\vb}^2, \\\;
	\expt{v^2}\expt{\vb^2},\quad			\expt{(v\vb)^2},\quad
	\expt{v}\expt{\vb}\expt{v\vb},\quad		\expt{v}^2\expt{\vb}^2 \ .
\end{gathered}
\end{equation}
The identities needed to show the vanishing of the $O(\k^6)$ 
terms in $S_\AV$ are 
\begin{align*}  
	\expt{v^2\vb^2} &=
		\expt{v}\expt{v\vb^2}+\expt{\vb}\expt{v^2\vb} 
		-\expt{(v\vb)^2} - \expt{v}\expt{\vb}\expt{v \vb}\\
		&+\half\expt{v}^2\expt{\vb}^2+\half\expt{v^2}\expt{\vb^2}\,, \\
	2\expt{v\vb}^2 &=
		\expt{v}^2\expt{\vb^2}+\expt{v^2}\expt{\vb}^2
		+\expt{v^2}\expt{\vb^2}
		-2\expt{(v\vb)^2}+\expt{v}^2\expt{\vb}^2\ .
\end{align*}
The 8-fermion term that occurs in the KS action is
\begin{align} 
	\frac14\l^2\bar\l^2\Box\l^2\Box\bar\l^2 = 
	\expt{v}^2\expt{\vb^2} + \expt{v^2}\expt{\vb}^2 +
	\expt{v^2}\expt{\vb^2} + \expt{v}^2\expt{\vb}^2 \ .
\end{align}

The proof that the above basis is both complete and minimal 
is based on a computer calculation that can be found within 
the \emph{Mathematica} notebook distributed with \cite{KuzenkoTyler2011}.
\section{Composition rule for field redefinitions}\label{sect:composition}
By direct calculation, 
the composition of a transformation \eqref{eqn:GeneralFieldRedef}
with coefficients $x_i$, $y_j$ \& $z_k$ followed by one with
coefficients $a_i$, $b_j$ \& $c_k$ 
is shown to be the same as the transformation with coefficients
$\a_i$, $\b_j$ and $\g_k$ where
\begin{align*}
  \a_1 &= a_ 1+x_ 1\,, \quad
  \a_2 = a_ 2+x_ 2\,, \quad
  \a_3 = a_ 3+x_ 3 
  \allowdisplaybreaks\\ 
  \b_1 &= b_ 1+y_ 1+a_ 2 x_ 2+4 a_ 2 x_ 3+4 a_ 3 x_ 3-x_ 1 a_ 1^*
  	-2 x_ 1 a_ 3^*-2 x_ 2 a_ 3^*-4 x_ 3 a_ 3^* \\  
  \b_2 &= b_ 2+y_2 +2 a_ 2 x_ 1+2 a_ 3 x_ 1+2 a_ 1 x_ 2+a_ 2 x_ 2 \non\\
  	&\quad+2 a_ 3 x_ 2+4 a_ 2 x_ 3+4 a_ 3 x_ 3+x_ 2 a_ 2^* \allowdisplaybreaks\\
  \b_3 &= b_ 3+y_ 3-\tfrac{a_ 2 x_ 2}{2}-2 a_ 2 x_ 3-2 a_ 3 x_ 3
  	+\tfrac{x_ 1 a_ 1^*}{2} \\  
  \b_4 &= b_ 4+y_ 4+\tfrac{3 a_ 2 x_ 2}{2}-2 a_ 2 x_ 3-2 a_ 3 x_ 3
  	+\tfrac{x_ 1 a_ 1^*}{2}+x_ 2 a_ 1^*  \allowdisplaybreaks \\ 
  \b_5 &= b_ 5+y_ 5+\tfrac{a_ 2 x_ 2}{2}+a_ 3 x_ 2+2 a_ 2 x_ 3+2 a_ 3 x_ 3
  	-\tfrac{x_ 1 a_ 1^*}{2}-x_ 2 a_ 3^* \\ 
  \b_6 &= b_ 6+y_ 6+\tfrac{a_ 2 x_ 2}{4}+a_ 3 x_ 2+a_ 2 x_ 3+a_ 3 x_ 3
  	-\tfrac{x_ 1 a_ 1^*}{4} \allowdisplaybreaks \\ 
  \b_7 &= b_ 7+y_ 7-\tfrac{a_ 1 x_ 2}{2}-\tfrac{a_ 2 x_ 2}{2}-a_ 3 x_ 2
  	-2 a_ 2 x_ 3-2 a_ 3 x_ 3+\tfrac{x_ 1 a_ 1^*}{2}+\tfrac{x_ 1 a_ 2^*}{2} \\\non
  &\quad+x_ 3 a_ 2^*+x_1 a_ 3^*+x_2 a_ 3^*+2 x_ 3 a_ 3^* \\ 
  \b_8 &= b_ 8+y_ 8+a_ 3 x_ 2+2 a_ 2 x_ 3+2 a_ 3 x_ 3+x_ 3 a_ 1^* \allowdisplaybreaks \\ 
  \g_1 &= c_ 1+z_ 1-4 x_ 2 a_ 3^2+2 y_ 1 a_ 3-4 y_ 3 a_ 3
  -4 y_ 5 a_ 3-4 y_ 8 a_ 3-b_ 1 x_ 2+2 b_ 5 x_ 2 \\ \non
  &\quad-2 y_ 1 a_ 3^*-4 y_ 7 a_ 3^*+x_1 b_ 1^*+2 x_ 3 b_ 1^*
  -2 x_ 1 b_ 5^*-4 x_ 3 b_ 5^* \allowdisplaybreaks \\ 
  \g_2 &= c_ 2+z_ 2-4 x_ 2 a_ 3^2-2 a_ 2 x_ 2 a_ 3+4 y_ 1 a_ 3
  	-4 y_ 4 a_ 3-4 y_ 5 a_ 3+4 y_ 7 a_ 3\\ \non
  	&\quad-4 y_ 8 a_ 3-2 x_ 2 a_ 1^* a_ 3+2 x_ 1 a_ 2^* a_ 3+2 x_ 2 a_ 2^* a_ 3
  	+4 x_ 3 a_ 2^* a_ 3+4 x_ 2 a_ 3^* a_ 3-a_ 1 a_ 2 x_ 2\\ \non
  	&\quad+b_ 1 x_ 2-b_ 2 x_ 2+2 b_ 5 x_ 2+2 b_ 7 x_ 2-2 b_ 8 x_ 2+3 a_ 2 y_ 1
  	+2 a_ 1 y_ 3\\ \non
  	&\quad-2 a_ 1 y_ 4+6 a_ 2 y_ 7-a_ 1 x_ 2 a_ 1^*+2 y_ 1 a_ 1^*+4 y_ 7 a_ 1^*
  	+2 a_ 2 x_ 1 a_ 2^*	+4 a_ 2 x_ 3 a_ 2^*-y_1 a_ 2^*\\ \non
  	&\quad+y_2 a_ 2^*+2 y_ 7 a_ 2^*+4 y_ 8 a_ 2^*-2 y_ 1 a_ 3^*+4 y_ 8 a_ 3^* 
  	+x_1 b_ 1^*+x_2 b_ 1^*+2 x_ 3 b_ 1^*+x_1 b_ 2^*\\\non
  	&\quad+2 x_ 3 b_ 2^*-2 x_ 1 b_ 5^*-4 x_ 3 b_ 5^*  \allowdisplaybreaks\\ 
  \g_3 &=c_ 3+z_ 3 + 2 x_ 1 a_ 3^2+2 x_ 2 a_ 3^2+4 x_ 3 a_ 3^2+3 a_ 1 x_ 2 a_ 3
  	+2 y_ 2 a_ 3+2 y_ 7 a_ 3+2 y_ 8 a_ 3\\ \non
	  &\quad+2 x_ 2 a_ 2^* a_ 3+2 x_ 2 a_ 3^* a_ 3+2 b_ 3 x_ 1+4 b_ 6 x_ 1
	  +\tfrac{b_ 1 x_ 2}{2}+b_ 3 x_ 2+2 b_ 6 x_ 2+b_ 7 x_ 2\\ \non
  	&\quad+4 b_ 3 x_ 3+8 b_ 6 x_ 3+2 a_ 1 y_ 3+2 a_ 1 y_ 6+a_ 1 y_ 8
  	+\tfrac{y_ 1 a_ 2^*}{2}+2 y_ 3 a_ 2^*-2 y_ 5 a_ 2^*+4 y_ 6 a_ 2^*
  	 \allowdisplaybreaks \\ \non
	&\quad-y_7 a_ 2^*+y_1 a_ 3^*+2 y_ 3 a_ 3^*-2 y_ 5 a_ 3^*
  	+4 y_ 6 a_ 3^*-\tfrac{x_ 1 b_ 1^*}{2}-x_ 3 b_ 1^* \\ 
  \g_4 &=c_ 4+z_ 4+ x_ 1 a_2^2+\tfrac{1}{2} x_ 2 a_2^2+2 x_ 3 a_2^2
  	+2 a_ 3 x_ 1 a_ 2+\tfrac{3}{2} a_ 1 x_ 2 a_ 2+2 a_ 3 x_ 2 a_ 2+2 x_ 3 b_ 7^* 
  	\\ \non
  	&\quad+4 a_ 3 x_ 3 a_ 2+3 y_ 2 a_ 2-y_ 3 a_ 2+y_ 4 a_ 2+3 y_ 7 a_ 2
  	+3 y_ 8 a_ 2+x_ 1 a_ 1^* a_ 2+\tfrac{1}{2} x_ 2 a_ 1^* a_ 2\\ \non
  	&\quad+2 x_ 3 a_ 1^* a_ 2+x_ 1 a_ 2^* a_ 2+2 x_ 2 a_ 2^* a_ 2
  	+2 x_ 3 a_ 2^* a_ 2+2 x_ 1 a_ 3^* a_ 2+2 x_ 2 a_ 3^* a_ 2
  	+4 x_ 3 a_ 3^* a_ 2\\ \non
  	&\quad+2 a_ 3^2 x_ 1+2 b_ 4 x_ 1+4 b_ 6 x_ 1
  	+2 b_ 8 x_ 1+2 a_ 3^2 x_ 2+3 a_ 1 a_ 3 x_ 2
  	+\tfrac{3 b_ 2 x_ 2}{2}+b_ 4 x_ 2\\ \non
  	&\quad+2 b_ 6 x_ 2+2 b_ 7 x_ 2+b_ 8 x_ 2	+4 a_ 3^2 x_ 3+4 b_ 4 x_ 3
  	+8 b_ 6 x_ 3+4 b_ 8 x_ 3+4 a_ 3 y_ 2-2 a_ 3 y_ 3\\ \non
  	&\quad+2 a_ 1 y_ 4+2 a_ 3 y_ 4+2 a_ 1 y_ 6+4 a_ 3 y_ 7
  	+a_ 1 y_ 8+4 a_ 3 y_ 8+a_ 3 x_ 1 a_ 1^*+\tfrac{3}{2} a_ 1 x_ 2 a_ 1^*\\ \non
  	&\quad+a_3 x_ 2 a_ 1^*
  	+2 a_ 3 x_ 3 a_ 1^*+\tfrac{y_ 1 a_ 1^*}{2}+\tfrac{3 y_ 2 a_ 1^*}{2}
  	+y_ 7 a_ 1^*+a_3 x_ 1 a_ 2^*+3 a_ 3 x_ 2 a_ 2^*+2 a_ 3 x_ 3 a_ 2^*\\ \non
  	&\quad+\tfrac{y_ 2 a_ 2^*}{2}+2 y_ 4 a_ 2^*-2 y_ 5 a_ 2^*+4 y_ 6 a_ 2^*
 	+2 y_8 a_2^*+2 a_ 3 x_ 1 a_ 3^*+2 a_ 3 x_ 2 a_ 3^*+4 a_3 x_3 a_3^* \\ \non
  	&\quad+y_2 a_ 3^*+2 y_ 4 a_ 3^*
  	-2 y_ 5 a_ 3^*+4 y_ 6 a_ 3^*+2 y_ 8 a_ 3^*+\tfrac{x_ 1 b_ 2^*}{2}
  	+x_ 2 b_ 2^*+x_3 b_ 2^*+x_1 b_ 7^*+x_2 b_ 7^*
  	\allowdisplaybreaks \\ 
  \g_5 &= c_ 5+z_ 5-\tfrac{a_ 1 a_ 2 x_ 2}{2} +a_ 2 a_ 3 x_ 2
  	-\tfrac{b_ 1 x_ 2}{2}-\tfrac{b_ 2 x_ 2}{2}+2 b_ 5 x_ 2+b_ 7 x_ 2
  	-b_ 8 x_ 2+a_ 1 y_ 3\\ \non
  	&\quad-a_ 2 y_ 3-2 a_ 3 y_ 3-a_ 1 y_ 4+3 a_ 2 y_ 5+2 a_ 3 y_ 5-4 a_ 2 y_ 6
  	-4 a_ 3 y_ 6+3 a_ 2 y_ 7\\ \non
  	&\quad+a_ 2 y_ 8-\tfrac{1}{2} a_ 1 x_ 2 a_ 1^*+\tfrac{y_ 1 a_ 1^*}{2}
  	+y_ 7 a_ 1^*+a_2 x_ 1 a_ 2^*+a_3 x_ 1 a_ 2^*+a_3 x_ 2 a_ 2^*\\ \non
  	&\quad+2 a_ 3 x_ 3 a_ 2^*-\tfrac{y_ 1 a_ 2^*}{2}+\tfrac{y_ 2 a_ 2^*}{2}
  	+y_7 a_2^*+2 y_8 a_2^*+2 a_2 x_1 a_3^*+2 a_3 x_1 a_3^*\\\non
  	&\quad+4 a_ 3 x_ 3 a_ 3^*-y_1 a_ 3^*+y_2 a_ 3^*+2 y_ 8 a_ 3^*
  	+\tfrac{x_ 1 b_ 1^*}{2}+x_ 3 b_ 1^*+\tfrac{x_ 1 b_ 2^*}{2}+x_ 3 b_ 2^*
  	+x_2 b_ 5^*\\\non
  	&\quad+x_1 b_7^*+2 x_3 b_7^*+2 a_3 y_7+2 a_ 2 x_3 a_2^*+4 a_2 x_3 a_3^* \ . 
\end{align*}
The composition of field redefinitions is the group law for the noncommutative 
group $G$ of all transformations that relate all of the Goldstino models.
When the field redefinitions are restricted to the trivial symmetries 
of any particular Goldstino action, then we obtain the group law 
for the subgroup $H$ of such trivial symmetry transformations. 

The composition rule above 
with $\a_i=\b_j=\g_k=0$ 
can be solved to give the inversion rule for field redefinitions.
The inverse of a field redefinition with coefficients $a_i$, $b_j$ \& $c_k$
is one with coefficients $x_i$, $y_j$ \& $z_k$, where
{\allowdisplaybreaks
\begin{align*} 
  a_1+x_1 &= 0 \,,\quad
  a_2+x_2 = 0 \,,\quad
  a_3+x_3 = 0 \\ 
  b_1+y_1 &= -|a_1|^2-4 |a_3|^2+a_2^2+4a_3^2+4 a_2 a_3-2 a_1 a_3^*-2 a_2 a_3^*\\
  b_2+y_2 &= |a_2|^2+a_2^2+4 a_3^2+4 a_1 a_2+2 a_1 a_3+6 a_2 a_3\,,\\
  b_3+y_3 &= \tfrac{|a_1|^2}{2}-\tfrac{a_2^2}{2}-2 a_3^2-2 a_2 a_3 \,,\\
  b_4+y_4 &= \tfrac{|a_1|^2}{2}+\tfrac{3 a_2^2}{2}-2a_3^2-2 a_2 a_3+a_2 a_1^*\\
  b_5+y_5 &= -\tfrac{|a_1|^2}{2}+\tfrac{a_2^2}{2}+2 a_3^2+3 a_2 a_3
  		-a_2 a_3^* \,,\\
  b_6+y_6 &= -\tfrac{|a_1|^2}{4}+\tfrac{a_2^2}{4}+a_3^2+2 a_2 a_3 \\
  b_7+y_7 &= \tfrac{|a_1|^2}{2}+2 |a_3|^2-\tfrac{a_2^2}{2}-2 a_3^2
  		-\tfrac{a_1 a_2}{2}-3 a_2 a_3+\tfrac{a_1 a_2^*}{2}
  		+a_3 a_2^*+a_1 a_3^*+a_2 a_3^*  \\ 
  b_8+y_8 &= 2 a_3^2+3 a_2 a_3+a_1^* a_3  \\ 
  c_1+z_1 &= 2 a_3 |a_1|^2+4 a_2 a_3^2+4 |a_3|^2 a_1-4 |a_3|^2 a_2+8 |a_3|^2 a_3
  		-2 a_2^2 a_3-a_2 b_1 \\ \non
  		&+2 a_3 b_1-4 a_3 b_3+2 a_2 b_5-4 a_3 b_5-4 a_3 b_8+4 a_3^2 a_1^*
  		-2 a_1 a_2 a_3^* \\ \non
  		&+4 |a_3|^2 a_2^*-2 b_1 a_3^*-4 b_7 a_3^*+2 a_1 a_2^* a_3^*+a_1 b_1^*
  		+2 a_3 b_1^*-2 a_1 b_5^*-4 a_3 b_5^* \\
  c_2+z_2 &= 3 a_2 |a_1|^2+2 a_3 |a_1|^2-4 a_2^* |a_1|^2-2 a_3^* |a_1|^2
  		+6 a_1 a_2^2+8 a_2 a_3^2-a_1 a_2^*{}^2 \\ \non
		&-a_2 a_2^*{}^2-2 a_3 a_2^*{}^2-4 a_1 a_3^*{}^2-4 a_2 a_3^*{}^2
		-8 a_3 a_3^*{}^2-4 |a_2|^2 a_1+4 |a_3|^2 a_1 \\ \non
  		&+|a_2|^2 a_2-8 |a_2|^2 a_3+8 |a_3|^2 a_3+10 a_2^2 a_3+2 a_1 a_2 a_3
  		+4 a_2 b_1+4 a_3 b_1\\ \non
  		&-a_2 b_2+2 a_1 b_3-2 a_1 b_4-4 a_3 b_4+2 a_2 b_5-4 a_3 b_5
  		+8 a_2 b_7+4 a_3 b_7\\ \non
  		&-2 a_2 b_8-4 a_3 b_8-4 |a_3|^2 a_1^*+4 a_3^2 a_1^*
  		+6 a_2 a_3 a_1^*+2 b_1 a_1^*+4 b_7 a_1^*\\ \non
  		&-8 |a_3|^2 a_2^*-4 a_3^2 a_2^*-2 a_1 a_3 a_2^*-b_1 a_2^*
  		+b_2 a_2^*+2 b_7 a_2^*+4 b_8 a_2^*\\ \non
  		&-4 |a_2|^2 a_3^*+2 a_2^2 a_3^*	-2 b_1 a_3^*+4 b_8 a_3^*
  		-4 a_1 a_2^* a_3^*+a_1 b_1^*+a_2 b_1^* \\ \non
  		&+2 a_3 b_1^*+a_1 b_2^*+2 a_3 b_2^*-2 a_1 b_5^*
  		-4 a_3 b_5^*-8 a_3 a_1^* a_2^* \\
  c_3+z_3 &= -4 (a_3)^3-2 a_1 a_3^2-10 a_2 a_3^2-2 a_1^* a_3^2
  		-2 a_2^* a_3^2-2 |a_1|^2 a_3-3 |a_2|^2 a_3 \\\non
  		&-4 |a_3|^2 a_3-a_2^2 a_3+a_2^*{}^2 a_3+4 a_3^*{}^2 a_3-7 a_1 a_2 a_3
  		+2 b_2 a_3+4 b_3 a_3+8 b_6 a_3\\\non
  		&+2 b_7 a_3+2 b_8 a_3-a_1 a_2^* a_3-b_1^* a_3
  		+\tfrac{1}{2} a_1 a_2^2+\tfrac{1}{2} a_1 a_2^*{}^2
  		+2 a_1 a_3^*{}^2-\tfrac{|a_1|^2 a_1}{2} \\\non
  		&-\tfrac{|a_2|^2 a_1}{2}-2 |a_3|^2 a_1x+2 a_2^2 a_3^*+\tfrac{a_2 b_1}{2}+4 a_1 b_3+a_2 b_3+6 a_1 b_6
  		+2 a_2 b_6\\\non
  		&+a_2 b_7+a_1 b_8+4 |a_3|^2 a_2^*+\tfrac{b_1 a_2^*}{2}+2 b_3 a_2^*
  		-2 b_5 a_2^*+4 b_6 a_2^*-b_7 a_2^*+b_1 a_3^*\\\non
  		&+2 b_3 a_3^*-2 b_5 a_3^*+4 b_6 a_3^*+2 a_1 a_2^* a_3^*
  		-\tfrac{a_1 b_1^*}{2} \\
  c_4+z_4 &= -3 (a_2)^3-\tfrac{23}{2} a_1 a_2^2-20 a_3 a_2^2-2 a_1^* a_2^2
  		-5 a_3^* a_2^2-\tfrac{13 |a_1|^2 a_2}{2}-\tfrac{9 |a_2|^2 a_2}{2}\\\non
  		&-14 |a_3|^2 a_2-30 a_3^2 a_2-\tfrac{1}{2} a_2^*{}^2 a_2-2 a_3^*{}^2 a_2
  		-18 a_1 a_3 a_2+\tfrac{9 b_2 a_2}{2}-b_3 a_2\\\non
  		&+2 b_4 a_2+2 b_6 a_2+5 b_7 a_2+4 b_8 a_2-10 a_3 a_1^* a_2-5 a_1 a_3^* a_2
  		-2 a_1^* a_3^* a_2\\\non
  		&+b_2^* a_2+b_7^* a_2-12 (a_3)^3-6 a_1 a_3^2-\tfrac{|a_1|^2 a_1}{2}
  		-\tfrac{5}{2}|a_2|^2 a_1-4 |a_3|^2 a_1\\\non
  		&-5 |a_1|^2 a_3-9 |a_2|^2 a_3-8 |a_3|^2 a_3+4 a_3 b_2-2 a_3 b_3
  		+4 a_1 b_4+6 a_3 b_4\\\non
  		&+6 a_1 b_6+8 a_3 b_6+4 a_3 b_7+3 a_1 b_8+8 a_3 b_8
  		-\tfrac{7 |a_2|^2 a_1^*}{2}	-2 |a_3|^2 a_1^*-8 a_3^2 a_1^*\\\non
  		&+\tfrac{b_1 a_1^*}{2}
  		+\tfrac{3 b_2 a_1^*}{2}+b_7 a_1^*-\tfrac{3 |a_1|^2 a_2^*}{2}
  		-4 a_3^2 a_2^*-2 a_1 a_3 a_2^*+\tfrac{b_2 a_2^*}{2}+2 b_4 a_2^*\\\non
  		&-2 b_5 a_2^*+4 b_6 a_2^*+2 b_8 a_2^*-3 a_3 a_1^* a_2^*
  		-|a_1|^2 a_3^*-3 |a_2|^2 a_3^*+b_2 a_3^*\\\non
  		&+2 b_4 a_3^*-2 b_5 a_3^*+4 b_6 a_3^*+2 b_8 a_3^*
  		+\tfrac{a_1 b_2^*}{2}+a_3 b_2^*+a_1 b_7^*+2 a_3 b_7^* \\
  c_5+z_5 &= \tfrac{(a_2)^3}{2}+3 a_1 a_2^2+4 a_3 a_2^2+\tfrac{|a_1|^2 a_2}{2}
  		+\tfrac{|a_2|^2 a_2}{2}-10 |a_3|^2 a_2+4 a_3^2 a_2
  		-\tfrac{1}{2} a_2^*{}^2 a_2\\\non
  		&-2 a_3^*{}^2 a_2+a_1 a_2 a_3 -\tfrac{b_1 a_2}{2}-\tfrac{b_2 a_2}{2}
  		-b_3 a_2+5 b_5 a_2-4 b_6 a_2+4 b_7 a_2\\\non
  		&-5 a_1 a_3^* a_2+b_5^* a_2-\tfrac{1}{2} a_1 a_2^*{}^2-a_3 a_2^*{}^2
  		-2 a_1 a_3^*{}^2-4 a_3 a_3^*{}^2-2 |a_2|^2 a_1\\\non
  		&-2 |a_3|^2 a_1-4 |a_2|^2 a_3-4 |a_3|^2 a_3+a_1 b_3-2 a_3 b_3
  		-a_1 b_4+2 a_3 b_5-4 a_3 b_6\\\non
  		&+2 a_3 b_7-2 |a_3|^2 a_1^*+\tfrac{b_1 a_1^*}{2}+b_7 a_1^*
  		-\tfrac{3 |a_1|^2 a_2^*}{2}-4 |a_3|^2 a_2^*-2 a_3^2 a_2^*
  		-a_1 a_3 a_2^*\\\non
  		&-\tfrac{b_1 a_2^*}{2}
  		+\tfrac{b_2 a_2^*}{2}+b_7 a_2^*+2 b_8 a_2^*-3 a_3 a_1^* a_2^*
  		-|a_1|^2 a_3^*-3 |a_2|^2 a_3^*-b_1 a_3^*\\\non
  		&+b_2 a_3^*+2 b_8 a_3^*-2 a_1 a_2^* a_3^*
  		+\tfrac{a_1 b_1^*}{2}+a_3 b_1^*+\tfrac{a_1 b_2^*}{2}+a_3 b_2^*
  		+a_1 b_7^*+2 a_3 b_7^* \ .
\end{align*}
}%





\backmatter

\providecommand{\href}[2]{#2}\begingroup\raggedright\endgroup

\end{document}